\documentclass[pdflatex,sn-mathphys-num]{sn-jnl}

\usepackage{graphicx}%
\usepackage{multirow}%
\usepackage{amsmath,amssymb,amsfonts}%
\usepackage{amsthm}%
\usepackage{mathrsfs}%
\usepackage[title]{appendix}%
\usepackage{xcolor}%
\usepackage{textcomp}%
\usepackage{manyfoot}%
\usepackage{booktabs}%
\usepackage{algorithm}%
\usepackage{algorithmicx}%
\usepackage{algpseudocode}%
\usepackage{listings}%
\usepackage[T1]{fontenc}
\usepackage[utf8]{inputenc}
\usepackage[normalem]{ulem}
\usepackage{url}
\usepackage{comment}
\newcommand\ul{\bgroup\markoverwith{\textcolor{red}{\rule[-0.5ex]{2pt}{1pt}}}\ULon}

\theoremstyle{thmstyleone}%
\newtheorem{theorem}{Theorem}
\theoremstyle{thmstyletwo}%

\theoremstyle{thmstylethree}%
\newtheorem{definition}{Definition}%

\usepackage{pdfpages}

\begin{document}

\title[Classification of Realisations of Random Sets]{Classification of Realisations of Random Sets}


\author*[1]{\fnm{Bogdan} \sur{Radovi\'c}}\email{
radovbog@fel.cvut.cz}

\author[2]{\fnm{Vesna} \sur{Gotovac \DJ oga\v{s}}}\email{vgotovac@pmfst.hr}
\equalcont{These authors contributed equally to this work.}

\author[1]{\fnm{Kate\v{r}ina} \sur{Helisov\'a}}\email{heliskat@fel.cvut.cz}
\equalcont{These authors contributed equally to this work.}

\affil*[1]{\orgdiv{Department of Mathematics}, \orgname{Faculty of Electrical Engineering, Czech Technical University in Prague}, \orgaddress{\street{Technick\'{a} 2}, \city{Prague}, \postcode{166 27}, \country{Czech~Republic}}}

\affil[2]{\orgdiv{Department of Mathematics}, \orgname{Faculty of Science, University of Split}, \orgaddress{\street{Ru\dj{}era Bo\v{s}kovi\'{c}a 33}, \city{Split}, \postcode{21000}, \country{Croatia}}}


\abstract{In this paper, the classification task for a family of sets representing the realisation of some random set models is solved. Both unsupervised and supervised classification methods are utilised using the similarity measure between two 
realisations derived as empirical estimates of $\mathcal{N}$-distances quantified based on geometric characteristics of the realisations, namely the boundary curvature and
the perimeter over area ratios of obtained samples of connected components from the realisations. To justify the proposed methodology,  a simulation study is performed using random set models. The methods are used  further for classifying histological images of mastopathy and mammary cancer tissue.}



\maketitle

\section{Introduction}\label{sec1}

Random sets have gained increasing attention in a variety of scientific disciplines.
They serve as a stochastic model for sets that occur in practice and are usually observed in a limited observation window.
Random sets can be used to model the shape of different tissue types in medicine \cite{hermann:2015}, to understand the arrangement of plants \cite{moeller:2010}, to study the microstructure of materials \cite{neumann:2016}.
The theoretical background of random sets can be found in
\cite{matheron:1975, molchanov:2005, serra:1982}.

In applications, we are usually given a collection of sets observed in the same observation window with the task of classifying these sets based on some selected~features.

Since, best to our knowledge, the classification problem for random set has not been studied in the literature, in this paper we want to pave the way for the classification methods for random sets.

The core ingredient of many classification algorithms are similarity measures between elements we wish to label.

If we consider the sets as realisations of a particular random set model, we can proceed with statistical inference and classify the sets based on the estimated parameters of the model. However, this approach is sometimes not feasible due to the high complexity of the realisations. In many cases, knowledge of the concrete model is not necessary since we want to focus on the similarity based on some specific features, e.g., we only need to distinguish between two types of cells in tissues from microscopic images based on their shapes, identify different growth tendencies of some plants based on the structure of their formations, recognise defects in materials based on the geometry of their microstructure, etc.

Recently, many similarity measures between two random sets have been proposed, focusing on different properties of the random sets \cite{gotovac:2016, gotovac:2019, debayle:2021, gotovac:2021b}.

It is reasonable to leverage given similarity measures between realisations of random sets for classification purposes.
We will use both unsupervised and supervised classification methods using the similarity measure derived from the two-step method for assessing similarity of random sets from \cite{gotovac:2021b} since this method has shown greatest power in distinguishing between random sets realisations based on the simulation~study.

For simplicity, we focus on the planar case of random sets, but the results can easily be extended to the multidimensional case.

In more detail, the plan is to divide each realisation of the random set into a sample of its connected components. In this way we obtain a family of connected sets. We represent each obtained connected set by values of a two selected functional features, namely, $C$-function representing its boundary curvature and the ratio of its perimeter over area ($P/A$-ratio).

By sampling the subsamples of the connected sets from two realisations we obtain two samples of connected sets. The similarity between the realisations is calculated as the convex combination of the two empirical estimates of $\mathcal N$-distances, the first being the distance between the distributions of $C$-functions of obtained samples of connected sets and the second being the distance between distributions of the $P/A$-ratios of obtained samples of the connected set. 

Furthermore, we use both supervised and unsupervised classification methods from \cite{ferraty:2006} with the aim to divide realisations of random sets into a fixed number of classes based on their similarity.

To justify the proposed methodology, we performed a simulation study using the same random set models as in \cite{gotovac:2016, gotovac:2019, debayle:2021, gotovac:2021b}.
We also apply the methodology for classifying mastopathy and mammary cancer tissue histological images from \cite{mrkvicka:2011}.

The remainder of the paper is organised as follows.
Section 2 provides the necessary theoretical background and a summary of existing results related to the shape characteristics of random planar sets and the concept of $\mathcal{N}$-distances. Section 3 introduces the proposed methodology, describing both the supervised and unsupervised classification approaches based on the similarity of random set realisations. Section 4 presents a comprehensive simulation study conducted on several random set models to assess the performance of the proposed methods. Section 5 demonstrates the application of the methodology to the classification of histological images of mastopathy and mammary cancer tissue. Finally, Section 6 discusses the obtained results and outlines potential directions for future research.

\section{Theoretical background and existing results}

\subsection{Characteristics of shape of a random planar set}
\label{sec:curv}

The first step for classification of realisations of random sets is to construct distances between individual realisations. 
Since we decided to use the distance from the paper~\cite{gotovac:2021b}, we need to express each realisation with a set of numerical and/or functional values that describe the main aspects of the shapes of components in the realisations. 
Following~\cite{gotovac:2021b}, we focused on two characteristics described below.

\begin{definition}
Consider a smooth 2D curve $\mathcal C$ parametrised by a parameter $\varphi \in [0,\phi] \subset \mathbb R$, i.e., $\mathcal C(\varphi)=(x(\varphi),y(\varphi))$.
Then the curvature $\kappa$ of $\mathcal C$ is defined as 
$$\kappa(\mathcal C(\varphi)) = \frac{x'(\varphi)y''(\varphi)-x''(\varphi)y'(\varphi)}{(x'^2(\varphi)+y'^2(\varphi))^{3/2}}.$$
\end{definition}

Let us assume that the curve $\mathcal C$ is continuous, closed (i.e. $\mathcal C(0)=\mathcal C(\phi)$) and it does not intersect itself (i.e. $\mathcal C(\varphi_1)=\mathcal C(\varphi_2) \Rightarrow \varphi_1=\varphi_2$). 
Consider a connected planar set $X$ whose boundary is given by the curve $\mathcal C$. 
It can be shown \cite{bullard:1995} that for the curvature $\kappa(z)$ evaluated in a given point $z \in \mathcal C$ and for a disc $b(z,r)$ with the center in $z$ and a radius $r$ small enough, it holds that
\begin{equation}   
\kappa(z) \approx \frac{3A^*_{b(z,r)}}{r^3}-\frac{3\pi}{2r} = \frac{3\pi}{r}\left(\frac{A^*_{b(z,r)}}{A_{b(z,r)}}-\frac{1}{2}\right),
\label{eq:curv}
\end{equation}
where $A_{b(z,r)}$ is the area of the disc $b(z,r)$ and $A^*_{b(z,r)}$ is the area of $b(z,r) \cap X$.

In~\cite{gotovac:2021b}, the authors consider a connected random set $\mathbf X$, i.e. the random set whose realisations are connected.
Denote $B_{\mathbf X}$ the boundary of $\mathbf X$ and $\kappa_{\mathbf X}(z)$ the (random) curvature at the point $z \in B_{\mathbf X}$.
From \eqref{eq:curv}, we can see that for a disc $b(z,r)$ with suitably chosen radius $r$, it holds that (up to a constant that can be neglected)
$$\kappa_{\mathbf X}(z) \propto \frac{A^*_{b(z,r),\mathbf X}}{A_{b(z,r)}},$$ 
where $A_{b(z,r)}$ is the area of the disc $b(z,r)$ and $A^*_{b(z,r),\mathbf X}$ is the area of $b(z,r) \cap \mathbf X$.
Therefore, we focus only on the ratio of these two areas. Denote
$$O_{\mathbf X,b(z,r)}=\frac{A^*_{b(z,r),\mathbf X}}{A_{b(z,r)}}$$
and define the function $$\tilde\kappa_{\mathbf X, r}(u) = |B_{\mathbf X}|^{-1}\int_{B_{\mathbf X}} \mathbf 1\{O_{\mathbf X,b(z,r)} \leq u \}dz, \quad u\in [\,0,1],$$
which is basically an analogy of the distribution function of the curvature at points on the boundary, but it is evaluated for all boundary points, so it describes the distribution for strongly dependent values.
The object of our interest is the function, analogous to density function, describing the distribution of the curvature along the boundary, i.e.
\begin{equation}
\label{eq:C-ftion}
t_{\mathbf X, r}(u) = \tilde\kappa_{\mathbf X, r}'(u).
\end{equation}
In the sequel, the function \eqref{eq:C-ftion}, which describes the curvature of the boundary of the set $\mathbf X$, is called the $C$-function, and it will be one of the characteristics used for the inference below.

The second characteristic of the random set $\mathbf X$ is the random variable describing the ratio of the perimeter and the area of $\mathbf X$.
It is denoted as $R_{\mathbf X}$ and called the $P/A$-ratios in the sequel.

In practice, we observe realisations $\mathbf x$ of the random set $\mathbf X$ in the form of binary images, so we need to adjust the definitions of the characteristics defined above to the realisations consisting of black and white pixels.
The pixels play the role of units in the sequel.
The $P/A$-ratio is simply given by the number of boundary pixels divided by the number of all pixels of the component.
For evaluating the $C$-function, fix a radius $r\in \mathbb N$, denote $Pix$ the set of all pixels of the binary image $X$, $z_1,\ldots,z_{n}$ all boundary pixels, and for each boundary pixel $z_i$, define
$$T(z_i)=\frac{\sharp \{p \in Pix: p \in b(z_i,r) \cap X\}}{\sharp \{p \in Pix: p \in b(z_i,r)\}}.$$
Then, the approximation of the function $t_{\mathbf X, r}(u)$ from~\eqref{eq:C-ftion} is
\begin{align}
\label{eq:C-ftion-est}
& t(u)= \frac{\sharp\{i \in \{1,\ldots,n\}: T(z_i)\in[u-1/l,u)\}}{n}
\quad \text{for } u=\frac{1}{l},\frac{2}{l},\ldots, 1,
\end{align}
where $l$ is the number of pixels that form the disc $b(.,r)$.

\subsection{$\mathcal{N}$-distance of probability measures}
\label{sec:Ndist}

The second step is to find an appropriate metric for describing the distance between the distribution  of characteristics of the realisations from the previous section and to propose it's approximation based on the samples.
Such a metric can be the $\mathcal{N}$-distance defined in~\cite{klebanov:2006} and modified for random functions in~\cite{gotovac:2021a}.
The basics of the theory of $\mathcal{N}$-distances are briefly recalled in the following paragraphs.

Let $\mathcal X$ be a non-empty set.
Consider a negative definite kernel $\mathcal L: \mathcal X \times \mathcal X \rightarrow \mathbb C,$ i.e. satisfying the property that for any $n \in \mathbb N$, arbitrary $w_1,..., w_n \in \mathbb C$ 
such that $\sum_{i=1}^n w_i = 0$ and arbitrary $x_1,..., x_n \in \mathcal X$ it holds that
$\sum_{i=1}^n \sum_{j=1}^n  \mathcal L(x_i, x_j)w_i \bar w_j \leq 0.$

\begin{definition}  
The negative definite kernel $\mathcal L$ is called strongly negative definite kernel if for an arbitrary probability measure $\mu$ and an arbitrary $f:\mathcal X \to \mathbb R$ such that $\int_{\mathcal X} f(x) d\mu( x)=0$ holds and $\int_{\mathcal X} \int_{\mathcal X} \mathcal L(x,y) f(x) f(y) d \mu (x) d \mu (y)$ exists  and is finite, the relation 
$$
\int_{\mathcal X} \int_{\mathcal X}   \mathcal L(x,y) 
f(x) f(y) d\mu( x) d\mu( y) = 0
$$
implies that $f(x)=0$ $\mu$-a.e.
\end{definition}

For a map $\mathcal L: \mathcal X \times \mathcal X \rightarrow \mathbb C$, denote by $\mathcal B_{\mathcal L}$ the set of all measures $\mu$ such that $\int_{\mathcal X} \int_{\mathcal X} \mathcal L(x,y) d\mu (x) d\mu (y)$ exists.

\begin{theorem}[Klebanov, 2006]
Let $\mathcal L(x,y) = \mathcal L(y,x)$. Then
\begin{align}
\label{eq:Ndist}
\mathcal N(\mu,\nu) = & 2\int_{\mathcal X} \int_{\mathcal X} \mathcal L(x,y) d\mu (x) d\nu (y) 
- \int_{\mathcal X} \int_{\mathcal X} \mathcal L(x,y) d\mu (x) d\mu (y) \nonumber \\
& 
-\int_{\mathcal X} \int_{\mathcal X} \mathcal L(x,y) d\nu (x) d\nu (y)
\geq 0 
\end{align}
holds for all measures $\mu, \nu \in \mathcal B_{\mathcal L}$ with 
equality in the case $\mu=\nu$ only, if and only if $\mathcal L$ is a strongly negative definite kernel.
\end{theorem}

Theorem \ref{eq:Ndist} ensures that $\mathcal N^{1/2}$ is a quasi-metric on $\mathcal{B}_{\mathcal L}$ for $\mathcal L$ being negative definite kernel, while if $\mathcal L$ is strongly negative definite, $\mathcal{N}^{1/2}$ is a metric between $\mu$ and $\nu.$ In the following text, the term $\mathcal N(\mu,\nu)$ from \eqref{eq:Ndist} is called the $\mathcal N$-distance of the measures $\mu$ and $\nu$.  

To estimate the $\mathcal N$-distance in practice, suppose that we have observations $x_1,\ldots,x_{m_1}$ from the distribution $\mu$ and  $y_1,\ldots,y_{m_2}$ from the distribution $\nu$.
The $\mathcal N$-distance of the measures $\mu$ and $\nu$ is then estimated as
\begin{align}\hat{\mathcal{N}}(\mu,\nu)= 
\frac{2}{m_1 m_2}\sum\limits_{i=1}^{m_1}\sum\limits_{j=1}^{m_2}\mathcal{L}(x_i,y_j)
-\frac{1}{m_1^2}\sum\limits_{i=1}^{m_1}\sum\limits_{j=1}^{m_1}\mathcal{L}(x_i,x_j)-\frac{1}{m_2^2}\sum\limits_{i=1}^{m_2}\sum\limits_{j=1}^{m_2}\mathcal{L}(y_i,y_j),
\label{eq:Nest1}
\end{align}

Many examples of strongly negative definite kernels $\mathcal{L}$ are introduced in~\cite{klebanov:2006} for the case that the observations are real numbers, i.e., realisations of real random variables.
One of the examples, used in this paper, is the Euclidean distance \begin{align}
\mathcal{L}(x,y)=|x-y|.
\label{eq:L-values}
\end{align}

When the measures $\mu$ and $\nu$ correspond to distributions of random functions, then we use the kernel introduced in \cite{gotovac:2021a},
constructed especially for such functions as follows.
Consider two functions $f$ and $g$ evaluated in discrete arguments $u_1,\ldots,u_n,$ $n \in \mathbb{N}$.
Then the strongly negative definite kernel is
\begin{align}
&\mathcal{L}(f,g)
=\sum\limits_{m=1}^{D}\sum\limits_{\left\{k_1,\ldots,k_m\right\} 
\subseteq \left\{1,\ldots,n\right\} }\left(\sum\limits_{l=1}^m \left(f(u_{k_l})-g(u_{k_l})\right)^2 \right)^{1/2},
\label{eq:L-ftions}
\end{align}
where $D$ is a chosen constant specifying the depth of dependence, see~\cite{gotovac:2021a} for more~details.

\subsection{Similarity of random sets and their realisations}

Consider connected random sets $\mathbf X$ and $\mathbf Y$ with the $C$-functions $t_{\mathbf X, r}$ and $t_{\mathbf Y, r}$ and the $P/A$-ratios $R_{\mathbf X}$ and $R_{\mathbf Y}$, respectively. 
Then in~\cite{gotovac:2021b}, the similarity of random sets is defined so that two connected random sets $\mathbf X$ and $\mathbf Y$ are considered to be similar if the distributions of $\lim_{r\rightarrow 0} t_{\mathbf X, r}$ and $\lim_{r\rightarrow 0} t_{\mathbf Y, r}$ as well as the distributions of $R_{\mathbf X}$ and $R_{\mathbf Y}$ are equal.
Since realisations usually consist of more than one component, the definition needs to be extended.
If we can suppose that the components in each realisation are independent and come from the same distribution, then we can define similarity of two random sets so that that they are considered to be similar, if the distribution of their components are similar in the above mentioned meaning. The similarity of the two realisations $\mathbf x$ and $\mathbf y$ was tested in~\cite{gotovac:2021b} via testing the hypothesis that the corresponding $\mathcal N$-distances between the $C$-functions $t_{\mathbf x, r}$ and $t_{\mathbf x, r}$ and the $P/A$-ratios $R_{\mathbf x}$ and $R_{\mathbf y}$, respectively, are equal to zero. 

In the method presented below, we are not so strict and simply consider two realisations to be more similar, the smaller their $\mathcal N$-distance is. 
So when we say here that realisations are similar, we mean that the empirical $\mathcal N$-distance between them is small, but not necessarily equal to zero.

\section{Methodology}
\label{sec:method}

We apply both supervised and unsupervised classification method with the same aim - to divide given realisations $\mathbf x_1, \ldots, \mathbf x_n$ of random closed sets into $k$ classes based on their similarity. As a consequence, we get the possibility to assign the respective class to a new observed realisation.

For this purposes, we calculate the  $\mathcal N$-distance between two realisations $\mathbf x_i$ and $\mathbf x_j$ as the estimate using the formulae \eqref{eq:Nest1} with the negative definite kernels \eqref{eq:L-values} and \eqref{eq:L-ftions} when considering only $P/A$-ratios
and only $C$-functions, respectively. 
When we consider both $P/A$-ratios
and $C$-functions together, we simply include the value of $P/A$-ratio as one of the points of the $C$-function and use \eqref{eq:L-ftions}.

Further, we consider a set $K=\{1,2,\ldots,k\}$, which represents the set of classes which are to be assigned to the realisations.
Let $(\mathbf X_i, Y_i), i=1,...,n$, be a sample of $n$ independent pairs, where the random variable $Y$ is valued in $K$. 
In practical situations, we use the notation $(\mathbf x_i,y_i)$ for the observation of the pair $(\mathbf X_i, Y_i), i=1,\ldots,n$. 

\subsection{Supervised classification of random sets}

The idea of supervised classification is based on the Bayes rule. 
Given a realisation $\mathbf x$ of the random set $\mathbf X$, we estimate the posterior probabilities
$$p_c(\mathbf x)=P(Y=c|\mathbf X=\mathbf x), \ c\in K.$$
The realisation $\mathbf x$ is then assigned to the class with the highest estimated posterior~probability. 

We can use the kernel-type estimator 
\begin{equation}
\hat p_c(\mathbf x)=\frac{\sum_{i=1}^{n}\mathbf 1_{[y_i=c]} \mathcal K(h^{-1} \mathcal N(\mathbf x,\mathbf x_i))}{\sum_{i=1}^{n}\mathcal K(h^{-1} \mathcal N(\mathbf 
 x,\mathbf x_i))},
 \label{eq:kernel-type_est_p_c}
\end{equation}
where $\mathcal K$ is a kernel with the support $[0,1]$ (i.e. $\mathcal K$ is positive and non-increasing in $[0,1]$ and $\int_0^1\mathcal K=1$), and $h$ is a bandwidth (a strictly positive smoothing parameter). 
It means that the closer $\mathbf x_i$ is to $\mathbf x$, the larger is the value $\mathcal K(h^{-1} \mathcal{N}(\mathbf x, \mathbf x_i))$, while only $\mathbf x_i$'s with the distance less than $h$ from $\mathbf x$ are taken into account. 
Thus, among the realisations $\mathbf x_i$’s belonging to the $c$-th class, the closer $\mathbf x_i$ is to $\mathbf x$, the larger is its effect on the $c$-th estimated posterior probability, and $\mathbf x_i$'s that are farther than $h$ have no effect at all.

As stated in~\cite{ferraty:2006}, it is efficient to set the bandwidth $h$ so that only $m$ nearest neighbours of the realisation $\mathbf x$ are taken into account to calculate the kernel estimator \eqref{eq:kernel-type_est_p_c}. 
In order to choose the optimal $m$ for each realisation $\mathbf x_{i_o}$, denote by $h_{m(\mathbf x_{i_o})}$ the bandwidth such that $\sharp\{i: \mathcal{N}(\mathbf x_{i_o},\mathbf x_i)<h_{m(\mathbf x_{i_o})}\}=m$.
Further, denote 
\begin{equation*}
    Loss(m, i_0) = \sum_{c=1}^{k}{\left( \mathbf{1}_{[y_{i_0}=c]}-p_{c,m}^{(-i_0)}(\mathbf x_{i_0})\right)^2},
\end{equation*}
where
\begin{equation*}
         p_{c,m}^{(-i_0)}(\mathbf x_{i_0})=\frac
         {\sum_{i: i\neq i_0}{\mathbf 1_{[y_i=c]} \mathcal K(h_{m(\mathbf x_{i_0})}^{-1}\mathcal{N}(\mathbf x_{i_0}, \mathbf x_i))}}
         {\sum_{i: i\neq i_0}{\mathcal K(h_{m(\mathbf x_{i_0})}^{-1}\mathcal{N}(\mathbf x_{i_0}, \mathbf x_i))}}
         .
\end{equation*}
Then the optimal number of nearest neighbours $m_{Loss}$ for $\mathbf x_{i_0}$ is
\begin{equation*}
    m_{Loss}(\mathbf x_{i_0}) = \arg \min_{m} Loss(m, i_0)
\end{equation*}
and the corresponding bandwidth is the value $h_{m_{Loss}(\mathbf x_{i_o})}$.

\subsection{Unsupervised classification of random sets}

\subsubsection{Non-hierarchical clustering}

When studying unsupervised classification, we start with the well-known $k$-medoid algorithm described e.g. in \cite{gordon:1999}.
The aim is to divide the realisations $\mathbf x_1, \ldots, \mathbf x_n$ to $k$~classes.
The algorithm works as follows.

\begin{enumerate}
\item Choose arbitrary $k$ realisations $\mathbf x_{i_1}, \ldots, \mathbf x_{i_{k}}$, which play the role of medoids.
\item For each realisation $\mathbf x_i$, $i=1,\ldots,n$, calculate $\mathcal N(\mathbf x_i,\mathbf x_{i_c})$ for all $c=1,\ldots,k$ and assign $\mathbf x_i$ into the $c$-th class with the smallest $\mathcal N(\mathbf x_i,\mathbf x_{i_c})$.
\item In each class, determine the new medoid as the realisation inside the class with the smallest sum of $\mathcal N$-distances to all the other realisations in the class,
\item Apply the step 2. with new medoids from step 3.
\item Repeat the procedure until all realisations have settled in the classes so that no realisation jumps to another class.
\end{enumerate}

The literature also addresses the issue of choosing the optimal number of clusters~$k$. 
However, we do not solve this problem in this paper, as in practice, which is what we are aiming at here, this number is usually determined by the nature of the given~situation.

\subsubsection{Hierarchical clustering}

Except for non-hierarchical clustering described in the previous section, we also apply an agglomerative hierarchical clustering, namely the Ward's method \cite{ward:1963}. It joins clusters sequentially using the Lance–Williams algorithm with suitably chosen parameters \cite{murtagh:2011}.
At the initial step, all clusters are singletons (i.e. each cluster is formed by a single realisation). Then at each step, we join two clusters that are closer to each other than any other two clusters, and after update mutual cluster distances. 
For the set of realisations $\mathbf x_1, \ldots, \mathbf x_n$, it works as follows.

\begin{enumerate}
\item Find $\mathbf x_i, \mathbf x_j$ with the smallest $\mathcal{N}(\mathbf x_i, \mathbf x_j)$ in the whole set.
\item Replace the realisations $\mathbf x_i, \mathbf x_j$ in the set by the cluster $\tilde{\mathbf x}=\mathbf x_i \cup \mathbf x_j$.
\item Calculate the distance $\mathcal{N}(\tilde{\mathbf x}, \mathbf x_l)$ for all $l\in\{1,\ldots,n\}\setminus\{i,j\}$ as
$\mathcal{N}(\tilde{\mathbf x}, \mathbf x_l) = \frac{1}{2}(\mathcal{N}(\mathbf x_i, \mathbf x_l)+\mathcal{N}(\mathbf x_j, \mathbf x_l)).$
\item Continue clustering in this way, whereby if two clusters are joined, namely $\tilde{\mathbf x}_1$ including $m_1$ realisations and $\tilde{\mathbf x}_2$ including $m_2$ realisations, then the distance of the cluster $\tilde{\mathbf x}_1 \cup \tilde{\mathbf x}_2$ to an other cluster $\tilde{\mathbf x}_3$ including $m_3$ realisations is 
$$
    \mathcal{N}(\tilde{\mathbf x}_1 \cup \tilde{\mathbf x}_2, \tilde{\mathbf x}_3) = \sqrt{\frac{m_1+m_3}{m}\mathcal{N}^2(\tilde{\mathbf x}_1,\tilde{\mathbf x}_3)+\frac{m_2+m_3}{m}\mathcal{N}^2(\tilde{\mathbf x}_2,\tilde{\mathbf x}_3)-\frac{m_3}{m}\mathcal{N}^2(\tilde{\mathbf x_1},\tilde{\mathbf x}_2)},
$$
where $m = m_1+m_2+m_3.$
\item Stop clustering when all realisations $\mathbf x_1, \ldots, \mathbf x_n$ from the original set form one cluster.
\end{enumerate}

The advantage of this method is that we do not have to determine in advance how many clusters we want to form, but for any required number of clusters $k$, we find the state when we had the original set of realisations divided into $k$ clusters in the hierarchical tree (it is $k$ steps before the end of the procedure).

\section{Simulation study}

\begin{figure}
    \centering
    \includegraphics[width=3.5cm,height=3.4cm]{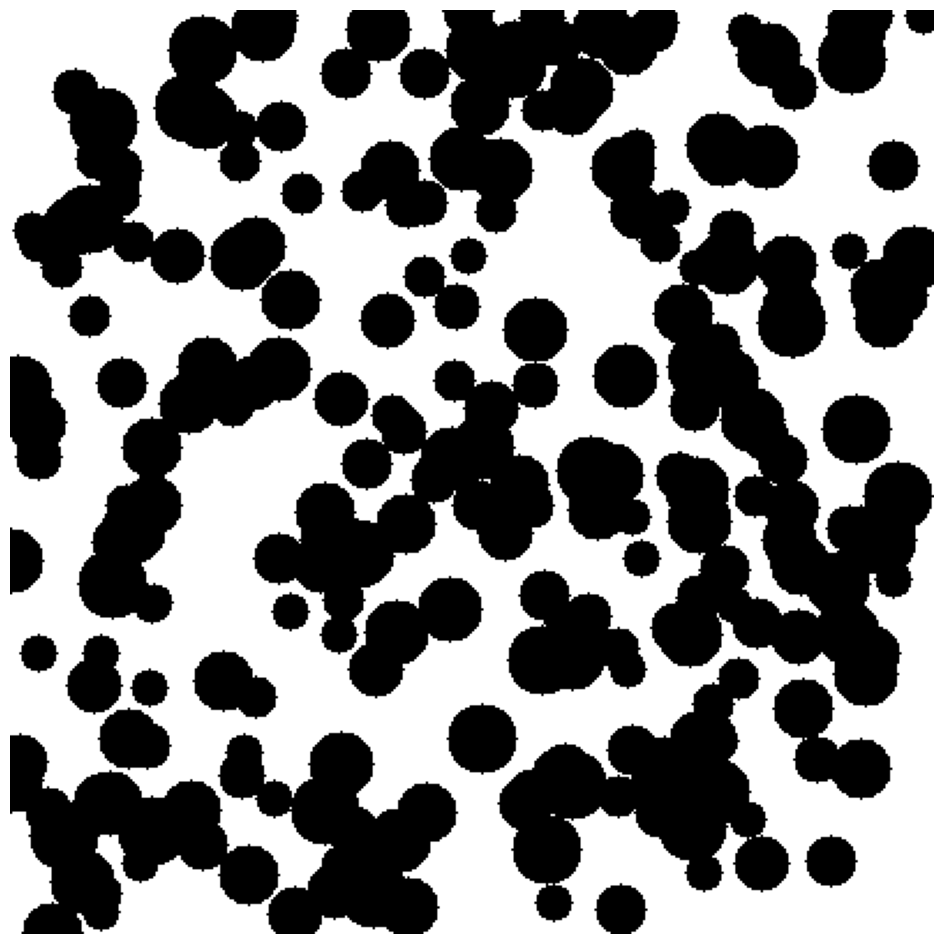}
    \hspace{0.1cm}
    \includegraphics[width=3.5cm,height=3.4cm]{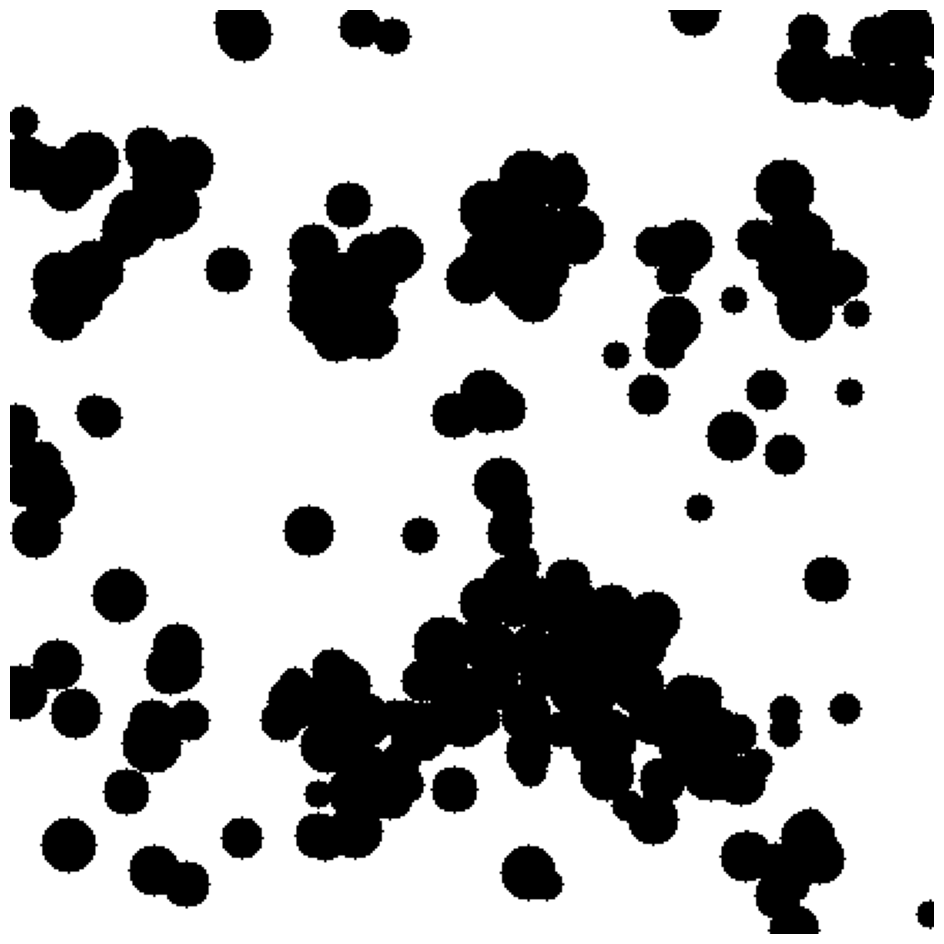}
    \hspace{0.1cm}
    \includegraphics[width=3.5cm,height=3.4cm]{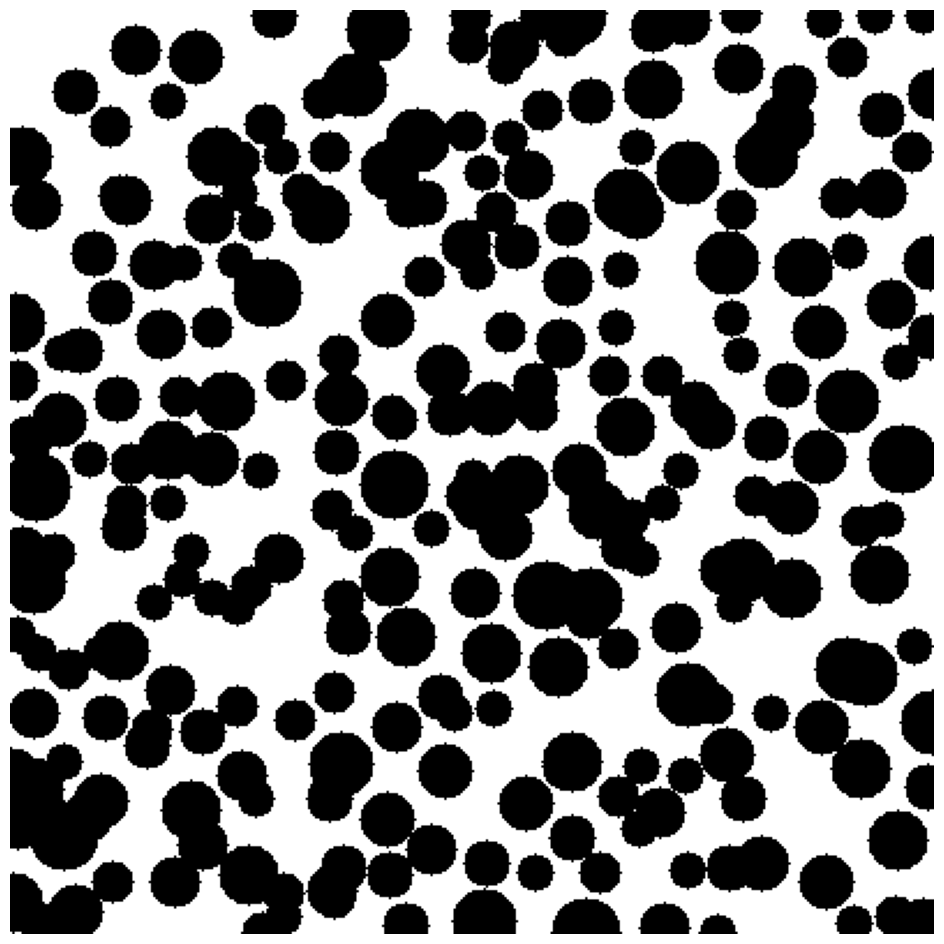}
    \hspace{0.1cm}

    \caption{Example of realisation of the Boolean, the repulsive, the cluster model, respectively.}
    \label{fig:BCR}
\end{figure}

First, we illustrate the procedure on simulated data.
We focus on models that have already been studied earlier in~\cite{gotovac:2016}, \cite{gotovac:2019}, \cite{gotovac:2021a} or in~\cite{gotovac:2021b}, namely on a Boolean model, a~cluster model and a repulsive model, where the second and the third mentioned models are simulated as Quermass-interaction processes~\cite{moeller:2008} with suitably chosen parameters. 
 Figure \ref{fig:BCR} presents examples of realisations of the models.

Note that we focus on the classification based on the distribution of the shape of the typical connected component of the realisation, without paying attention to how the connected components are arranged within the realisation.

The connected components of the repulsive model are mostly isolated discs with a~few cases of small clumps of overlapping discs. The majority of the connected components of the cluster model are also isolated discs, but the cluster model has some large clumps of closely overlapping discs. The Boolean model produces connected components that consist of clumps of overlapping discs that are elongated, and the discs are not as densely overlapping as in the case of the cluster model.

We work with 200 independently simulated realisations of each model.

In order to evaluate the $C$-functions~\eqref{eq:C-ftion}, we consider two different radii $r$, namely $r=3$ and $r=5$, just like in~\cite{gotovac:2021b}, in order to see a possible influence of this choice to obtained results. 

Since realisations of the models significantly differ in the number of components, we also study a possible influence of the number of components for the calculation of the $\mathcal N$-distance between two realisations, similarly as done in~\cite{gotovac:2021b}.
Namely, we calculate the distances using samples of 10, 20 and ’All’ components, where ’All’ means the number of components in the realisation with a smaller number of components.

\subsection{Supervised classification}

Data are split into train set and test set with a 3:1 ratio (which means that 75\% of the realisations is used for training, while 25\% is used for testing the performance of the classifier). We decided to use three settings in order to study the influence of the number of realisations on the classification: 
\begin{itemize}
\itemsep0em
    \item in the first setting we used a sample of 20 randomly chosen realisations from each model (further 'class'), Boolean (class 'B'), cluster (class 'C') and repulsive (class 'R'), meaning that in the training set we have 45 realisations (15 of each class, 'B', 'C' and 'R') in the training set and 15 realisations for testing purpose (5 of each~class)
    \item in the second setting we used a sample of 50 randomly chosen realisations from each class, meaning that in the training set we have 111 realisations (37 of each class) in~the training set and 39 realisations for testing purpose (13 of each class)
    \item in the third setting we used a sample of 100 randomly chosen realisations from each class, meaning that we have 225 realisations (75 of each class) in the training set and 75 realisations for testing purpose (25 of each class).
\end{itemize}
\vspace{-5pt}
Each of the settings mentioned above is then split into three subsettings according to~the characteristic which is used for discrimination, namely 'Ratio' (using only $P/A$-ratio), 'Curvature' (using only $C$-function) and 'Both' (using both the $P/A$-ratio and the  $C$-function). After that, the classifier is learnt three times for different numbers of components, which we use for calculating the $\mathcal N$-distance (i.e. 10, 20 and 'All', as mentioned above). After the learning stage, we use the test set and predict the labels using the posterior probabilities calculated for each class. 

To study how the choice of the radius of the osculating circle affects the performance of the classifier, we perform the same procedures for the data obtained using two different values, $r=3$ and $r=5$. The results for $r=3$ showed slightly lower classification performance (i.e., a higher misclassification rate) across all settings compared to $r=5$. It is due to the fact that the curvature is evaluated at fewer positions, leading to a smaller versatility between~classes. Therefore, we report only the results for $r=5$ here, while the results for $r=3$ are presented in the Supplementary Material.

\begin{figure}[]
    \centering
    \begin{minipage}{0.03\linewidth}\centering
        \rotatebox[origin=center]{90}{Ratio}
    \end{minipage}
    \begin{minipage}{0.93\linewidth}\centering
        \includegraphics[height=5.5cm, width=12.5cm]{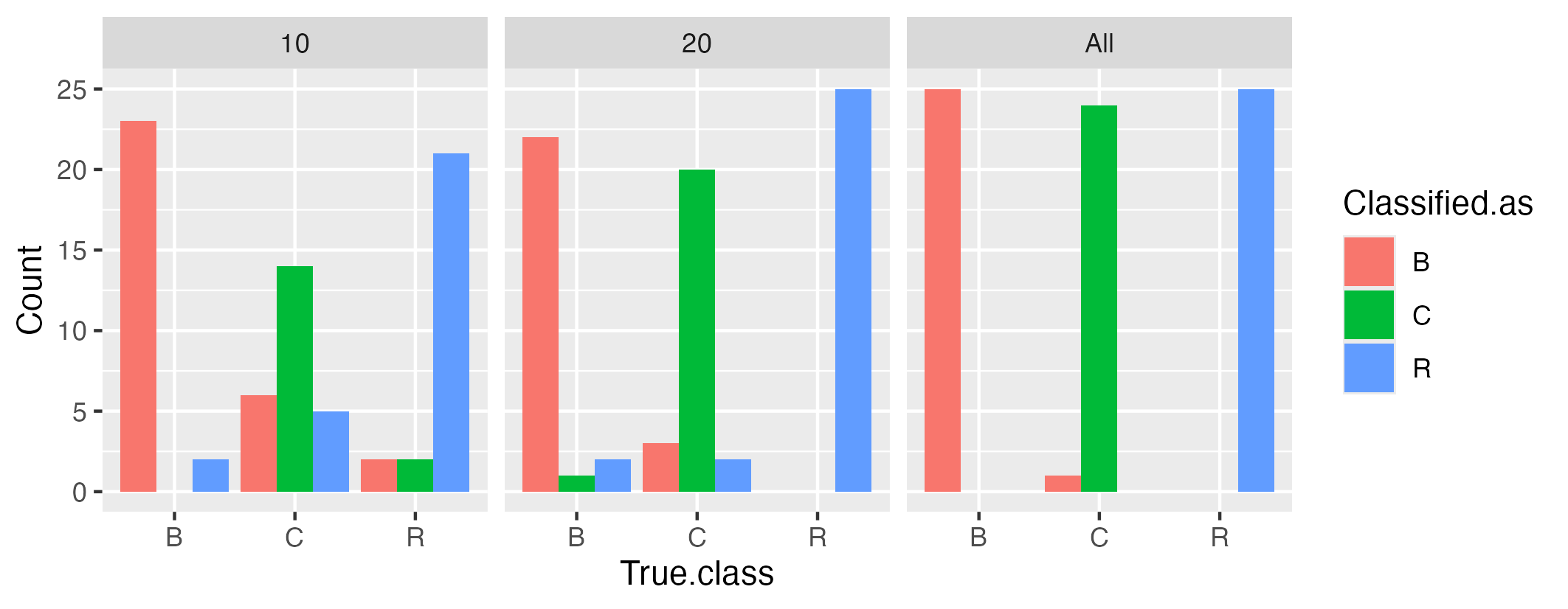}
    \end{minipage}
    \begin{minipage}{0.03\linewidth}\centering
        \rotatebox[origin=center]{90}{Curvature}
    \end{minipage}
    \begin{minipage}{0.93\linewidth}\centering
        \includegraphics[height=5.5cm, width=12.5cm]{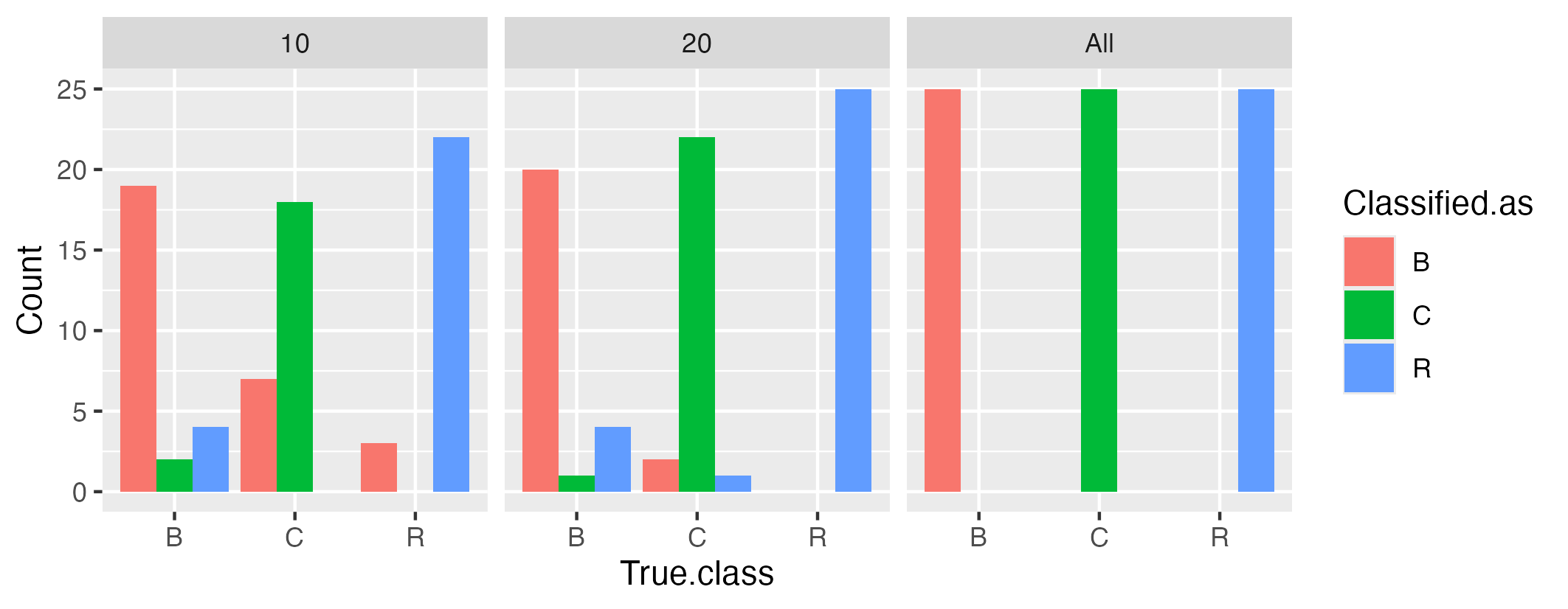}
    \end{minipage}
    \begin{minipage}{0.03\linewidth}\centering
        \rotatebox[origin=center]{90}{Both}
    \end{minipage}
    \begin{minipage}{0.93\linewidth}\centering
        \includegraphics[height=5.5cm, width=12.5cm]{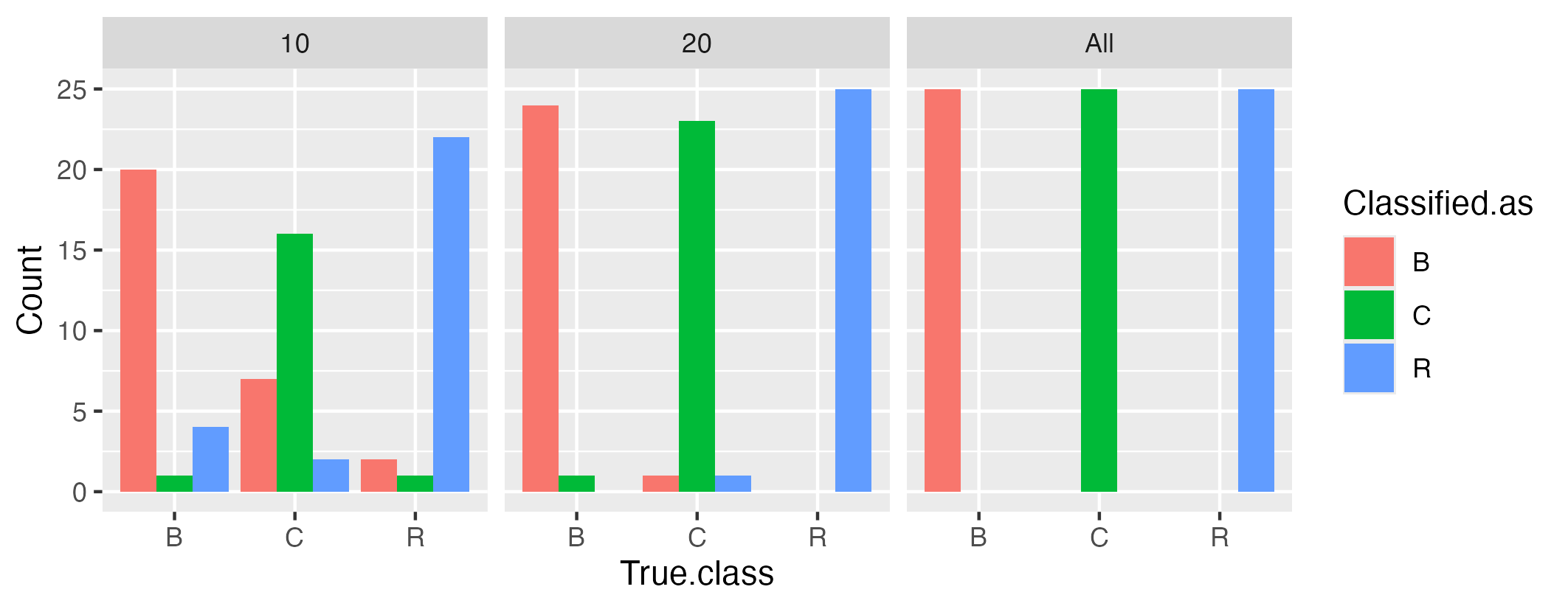}
    \end{minipage}
    \caption{Histograms of $k$-nearest neighbours classification accuracy using only the ratio, only the curvature and both ratio and curvature for discrimination when using a sample of 10, 20, and 'All' components, respectively. Misclassification rates are 7.6\%, 3.6\% and 0.4\% for 10, 20 and 'All' components, respectively, when using only the ratio, 7.1\%, 3.6\% and 0\% when using only the curvature, and 7.6\%, 1.3\% and 0\% when using both characteristics for a sample of 100 realisations that were osculated by a disc of radius $r=5$.}
    \label{fig:knn_100_best_5}
\end{figure}

The best classification results among the three settings were obtained for the highest number of realisations considered (i.e., 100). The histograms of classification accuracy are shown in Figure \ref{fig:knn_100_best_5}. We~observe that the highest overall misclassification rate was for the smallest sample size (of 10 components) as expected, it drops for a larger sample size (of 20 components), while the best performance was when considering 'All' components. Furthermore, we observe that the most problematic part was the classification of the cluster model. Similar problems occurred in~the simulation study in \cite{gotovac:2021b}. This is probably due to the fact that the cluster model contains a few larger components, a number of (Boolean-like) 2-to-10-disc components, and a greater number of (repulsive-like) single-disc components. Among the three subsettings, the lowest misclassification rate was when considering both characteristics. This reflects the results obtained in \cite{gotovac:2021b} where it was concluded that both characteristics were necessary to~correctly discriminate between different processes.

\begin{figure}[]
    \centering
    \begin{minipage}{0.03\linewidth}\centering
        \rotatebox[origin=center]{90}{20 Realisations}
    \end{minipage}
    \begin{minipage}{0.93\linewidth}\centering
        \includegraphics[height=5.5cm, width=12.3cm]{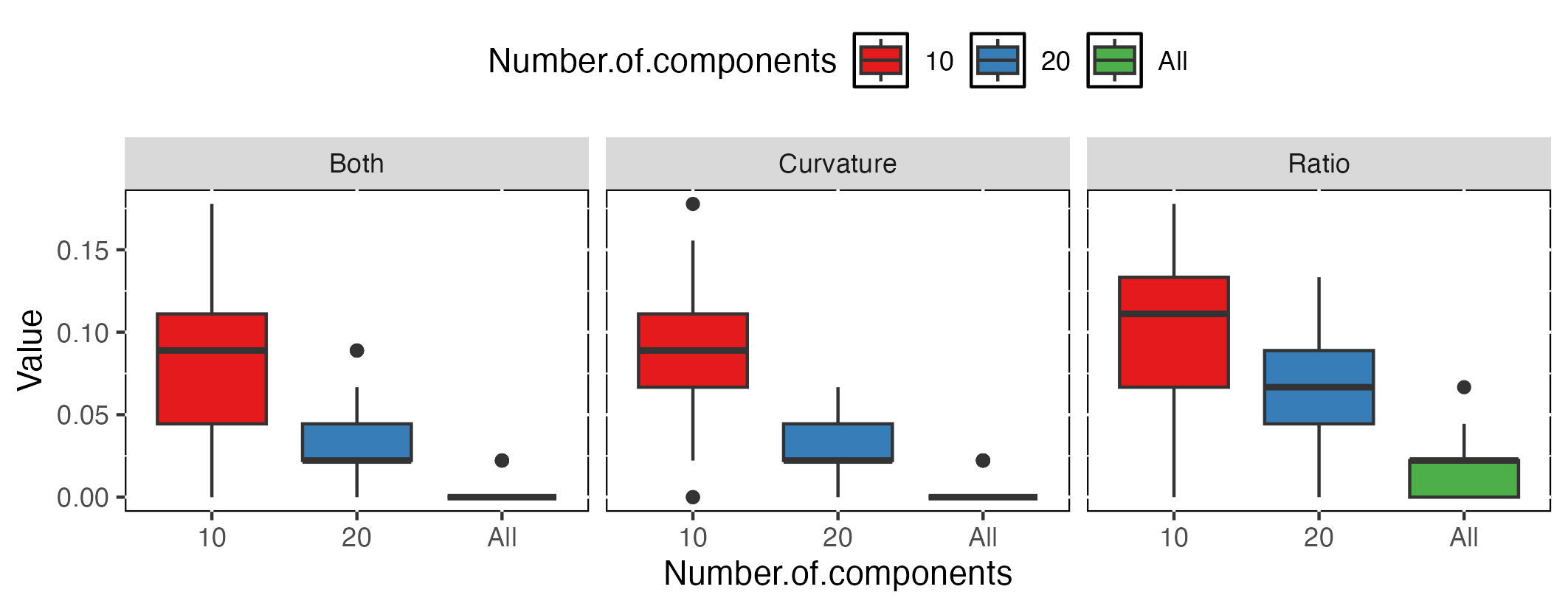}
    \end{minipage}
    \begin{minipage}{0.03\linewidth}\centering
        \rotatebox[origin=center]{90}{50 Realisations}
    \end{minipage}
    \begin{minipage}{0.93\linewidth}\centering
        \includegraphics[height=5.5cm, width=12.3cm]{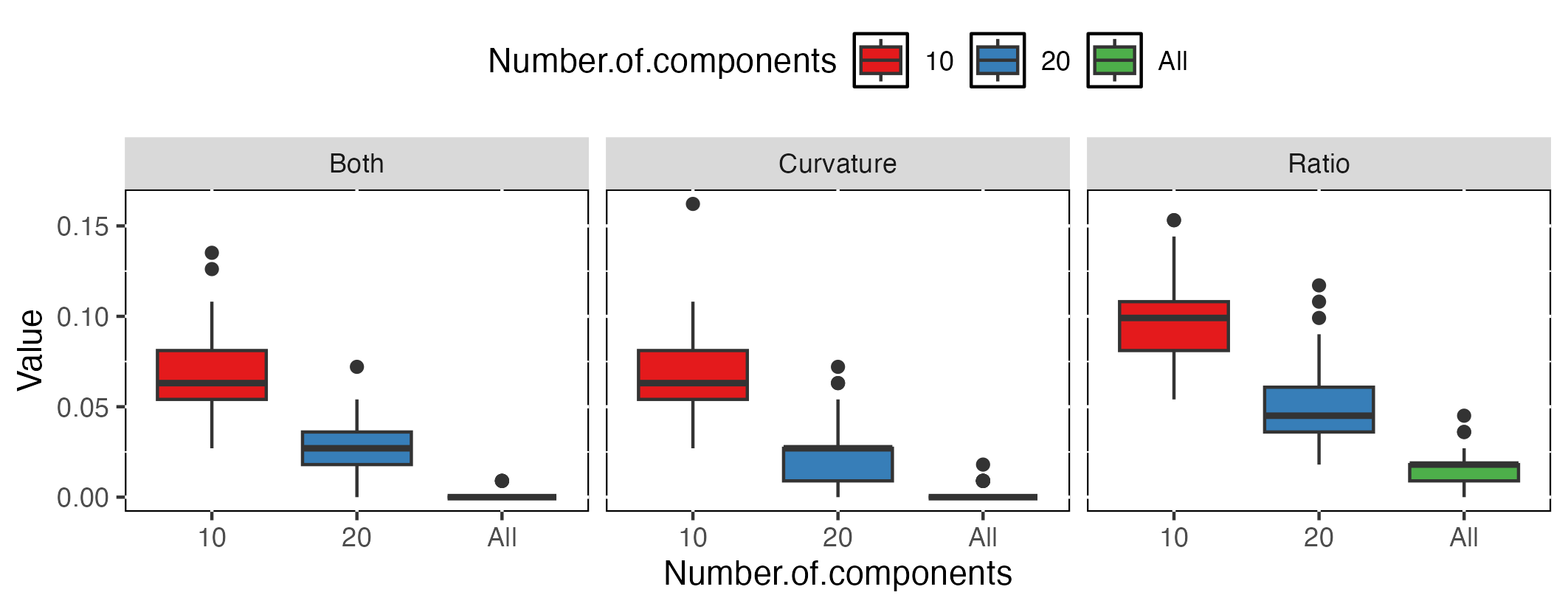}
    \end{minipage}
    \begin{minipage}{0.03\linewidth}\centering
        \rotatebox[origin=center]{90}{100 Realisations}
    \end{minipage}
    \begin{minipage}{0.93\linewidth}\centering
        \includegraphics[height=5.5cm, width=12.3cm]{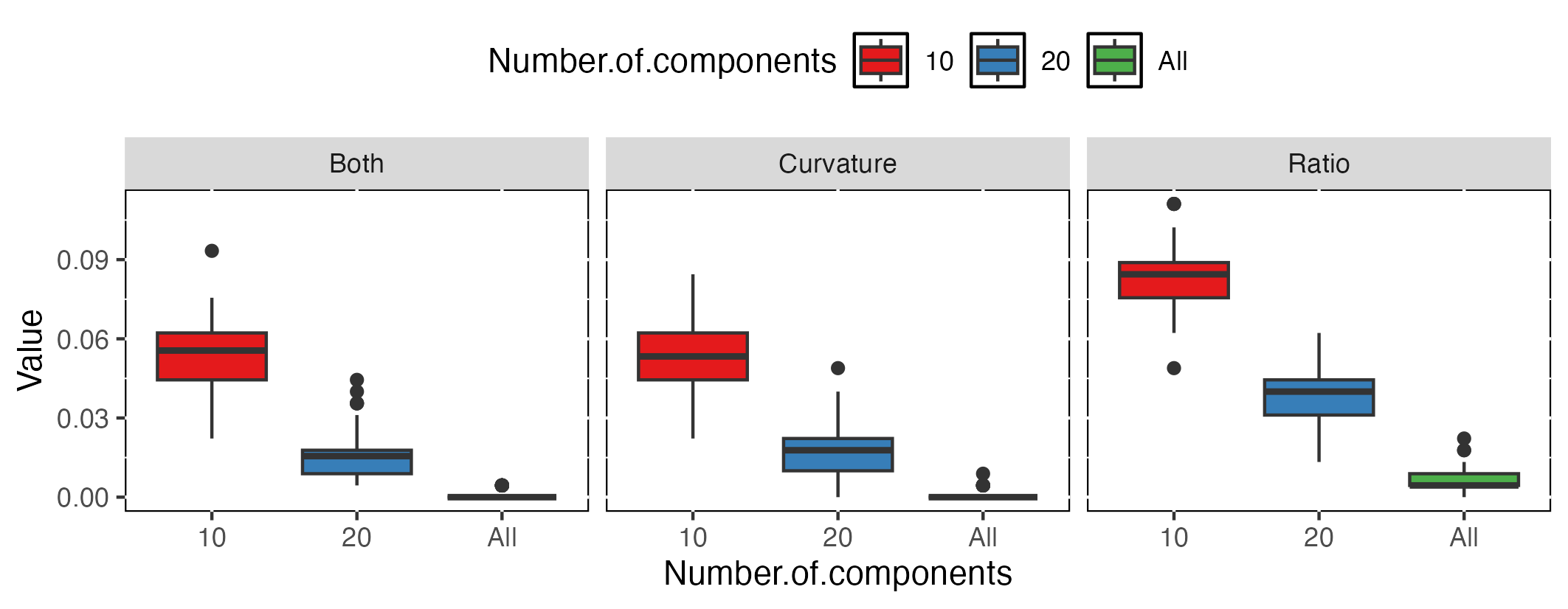}
    \end{minipage}
    \caption{Boxplots of misclassification rate for 50 runs of $k$-nearest neighbours algorithm when considering samples of 20 (top), 50 (middle) and 100 (bottom) realisations using both ratio and curvature, only the curvature and only the ratio for discrimination, respectively. For each setting, misclassification rates for different number of components considered (namely 10, 20 and 'All') are shown. Note that the characteristics were obtained using an osculating disc of radius $r=5$ on the simulated data.}
    \label{fig:knn_box_5_red}
\end{figure}

Each setting is run 50 times in order to obtain boxplots of the misclassification rate shown in Figure \ref{fig:knn_box_5_red}. The results when 20 and 50 realisations are considered show that the classifier behaves in the expected way: the misclassification rate is the highest when taking into account the smallest sample size of 10 components, it drops for a higher sample size of 20 components, and it is the lowest when considering 'All' components. Looking at results for 50 and 100 realisations, we can see that the amount of data higher than some threshold does not make the classifier significantly more precise, as the highest misclassification rate for 100 realisations is comparable to the setting working with 50 realisations, but it also indicates the dependence of the classification precision on the number of components considered. For histograms of classification accuracy, the reader is referred to the Supplementary Material.

\subsection{Unsupervised classification}

In the second part of the simulation study, we consider two clustering algorithms, non-hierarchical $k$-medoids, and agglomerative hierarchical algorithm based on Ward's method. Contrary to the supervised classification algorithms, here the data are not split into training and test sets, but are directly fed to the algorithm, which processes them until its convergence.

We decided to use the same three settings in order to study the influence of the number of realisations on the classification: 
\begin{itemize}
\itemsep0em
    \item in the first setting we used a sample of 20 randomly chosen realisations from each model Boolean (class 'B'), cluster (class 'C') and repulsive (class 'R')
    \item in the second setting we used a sample of 50 randomly chosen realisations from each~class
    \item in the third setting we used a sample of 100 randomly chosen realisations from each~class.
\end{itemize}
\vspace{-5pt}
Each of the settings is then split into three subsettings according to~the characteristic which is used for discrimination, namely 'Ratio' (using only $P/A-$ratio), 'Curvature' (using only $C-$function) and 'Both' (using both the $P/A-$ratio and the  $C-$function). 

The impact of the choice of the radius of the osculating circle on the classifier performance was again studied. The results for $r=3$, available in Supplementary Material, exhibited slightly lower performance (i.e., a higher misclassification rate) across all settings compared to $r=5$, similarly as in supervised case.

\subsubsection{Non-hierarchical clustering}

\begin{figure}[!ht]
    \centering
    \begin{minipage}{0.03\linewidth}\centering
        \rotatebox[origin=center]{90}{Ratio}
    \end{minipage}
    \begin{minipage}{0.93\linewidth}\centering
        \includegraphics[height=5.5cm, width=12.5cm]{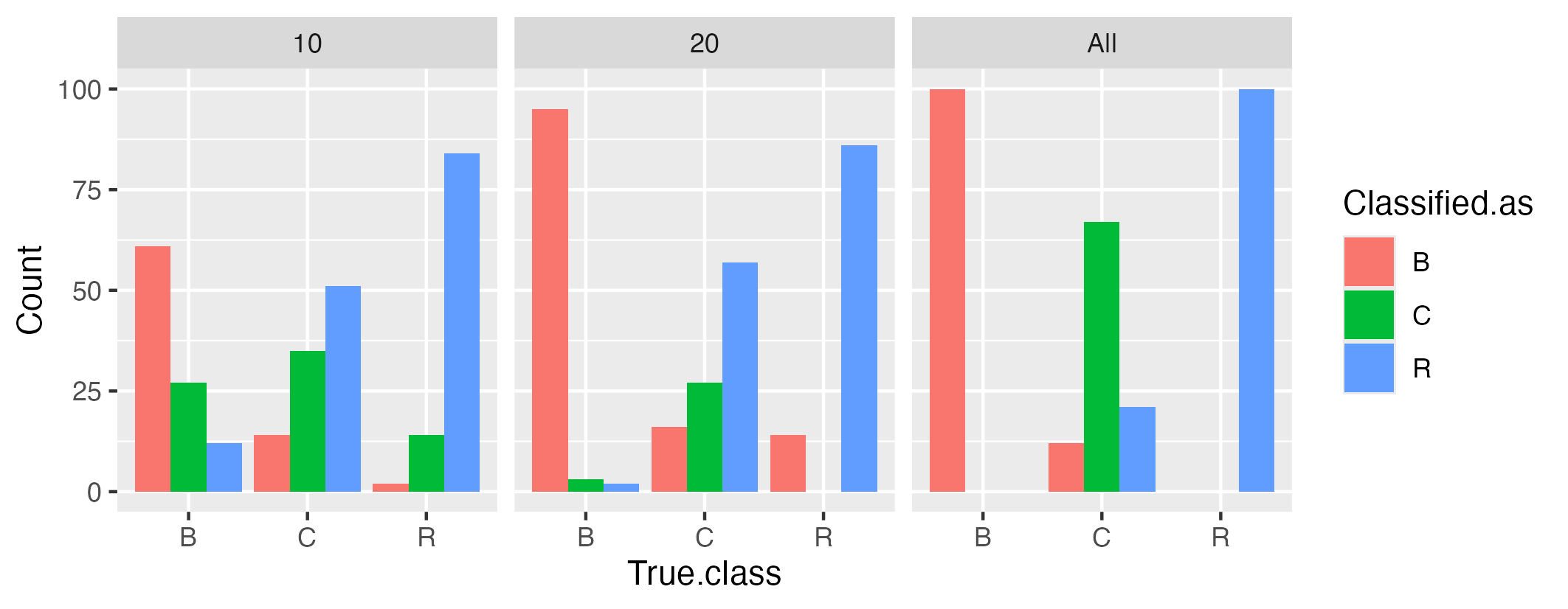}
    \end{minipage}
    \begin{minipage}{0.03\linewidth}\centering
        \rotatebox[origin=center]{90}{Curvature}
    \end{minipage}
    \begin{minipage}{0.93\linewidth}\centering
        \includegraphics[height=5.5cm, width=12.5cm]{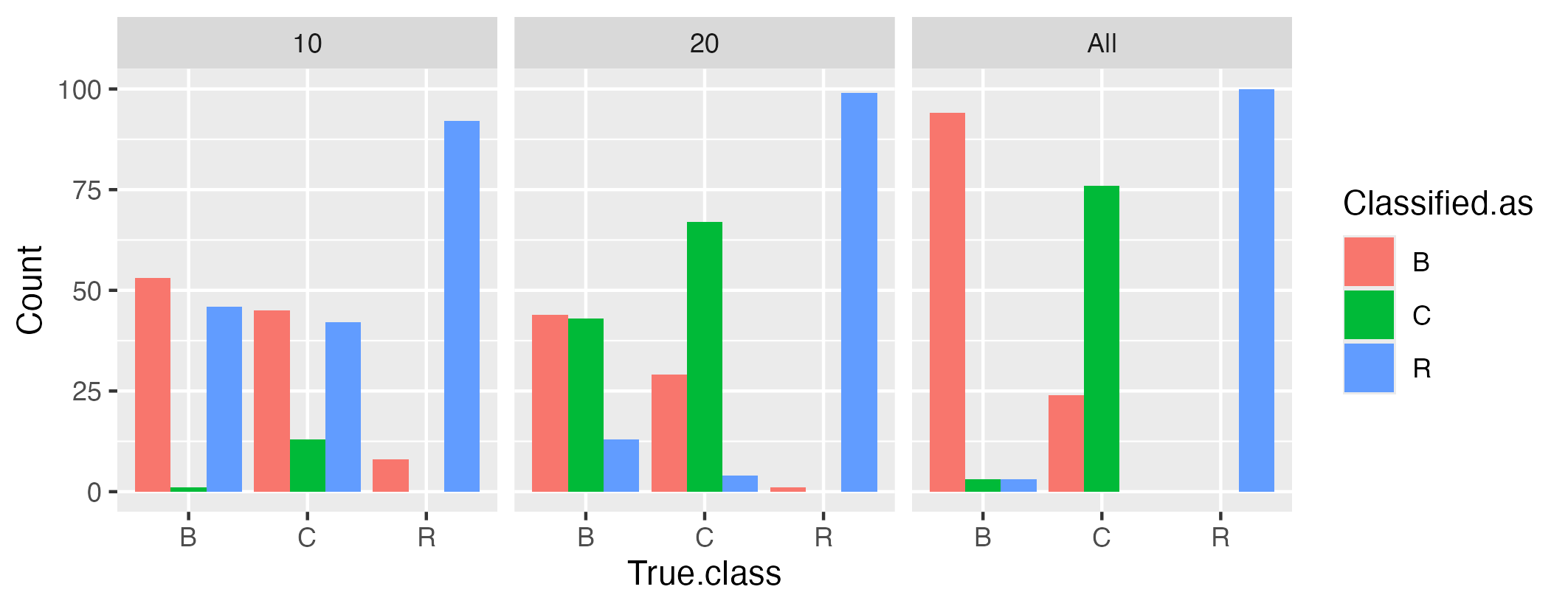}
    \end{minipage}
    \begin{minipage}{0.03\linewidth}\centering
        \rotatebox[origin=center]{90}{Both}
    \end{minipage}
    \begin{minipage}{0.93\linewidth}\centering
        \includegraphics[height=5.5cm, width=12.5cm]{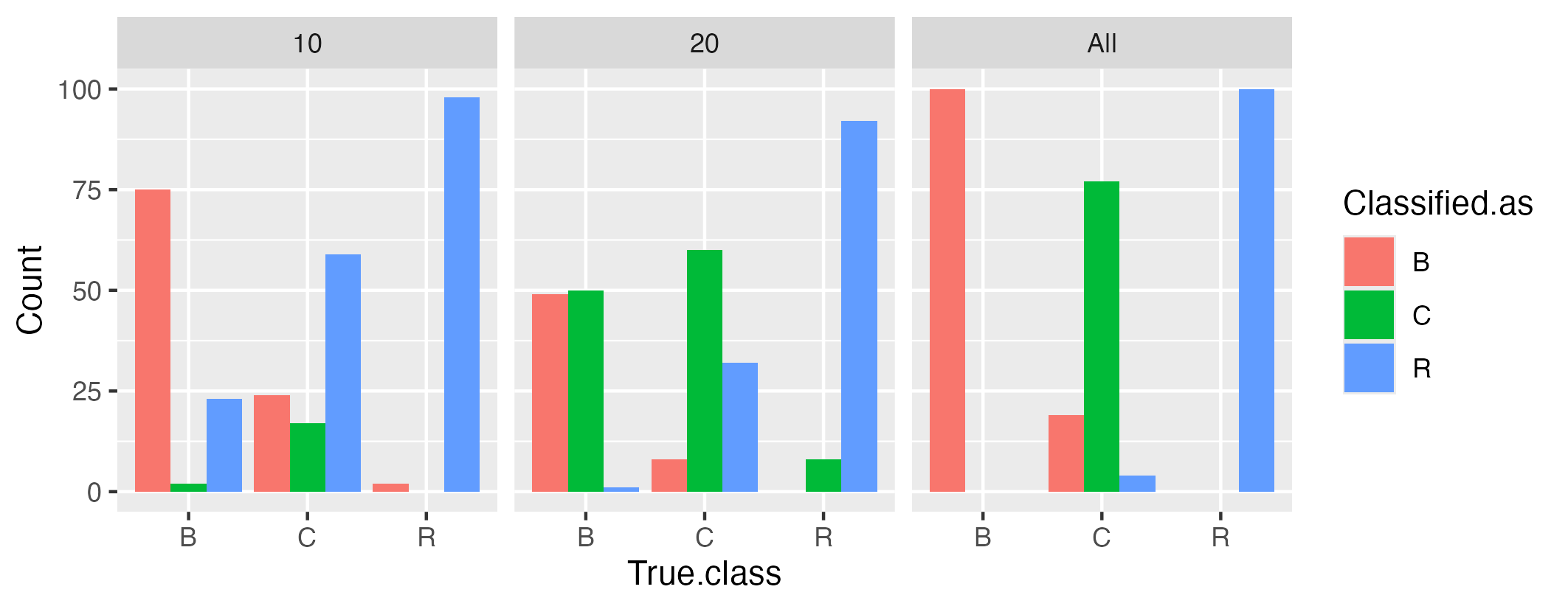}
    \end{minipage}
    \caption{Histograms of $k$-medoids classification accuracy using only the ratio, only the curvature and both ratio and curvature for discrimination when using a sample of 10, 20, and 'All' components, respectively. Misclassification rates are 40\%, 30.7\% and 11\% for 10, 20 and 'All' components, respectively, when using only the ratio, 47.3\%, 30\% and 10\% when using only the curvature, and 36.7\%, 33\% and 7.7\% when using both characteristics for a sample of 100 realisations that were osculated by a disc of radius $r=5$.}
    \label{fig:kmed_100_best_5}
\end{figure}

Focussing on the results when considering 100 realisations shown in Figure \ref{fig:kmed_100_best_5}, we~observe that the highest overall misclassification rate was again for the smallest sample size (of 10 components) as expected. The rate drops for a larger sample size (of 20 components), while the best performance was again when considering 'All' components. Furthermore, the most problematic part was the classification of the cluster model, as observed in the case of supervised classification. The results when 20 and 50 realisations are considered are presented in Supplementary Material.

\begin{figure}[!ht]
    \centering
    \begin{minipage}{0.03\linewidth}\centering
        \rotatebox[origin=center]{90}{20 Realisations}
    \end{minipage}
    \begin{minipage}{0.93\linewidth}\centering
        \includegraphics[height=5.5cm, width=12.3cm]{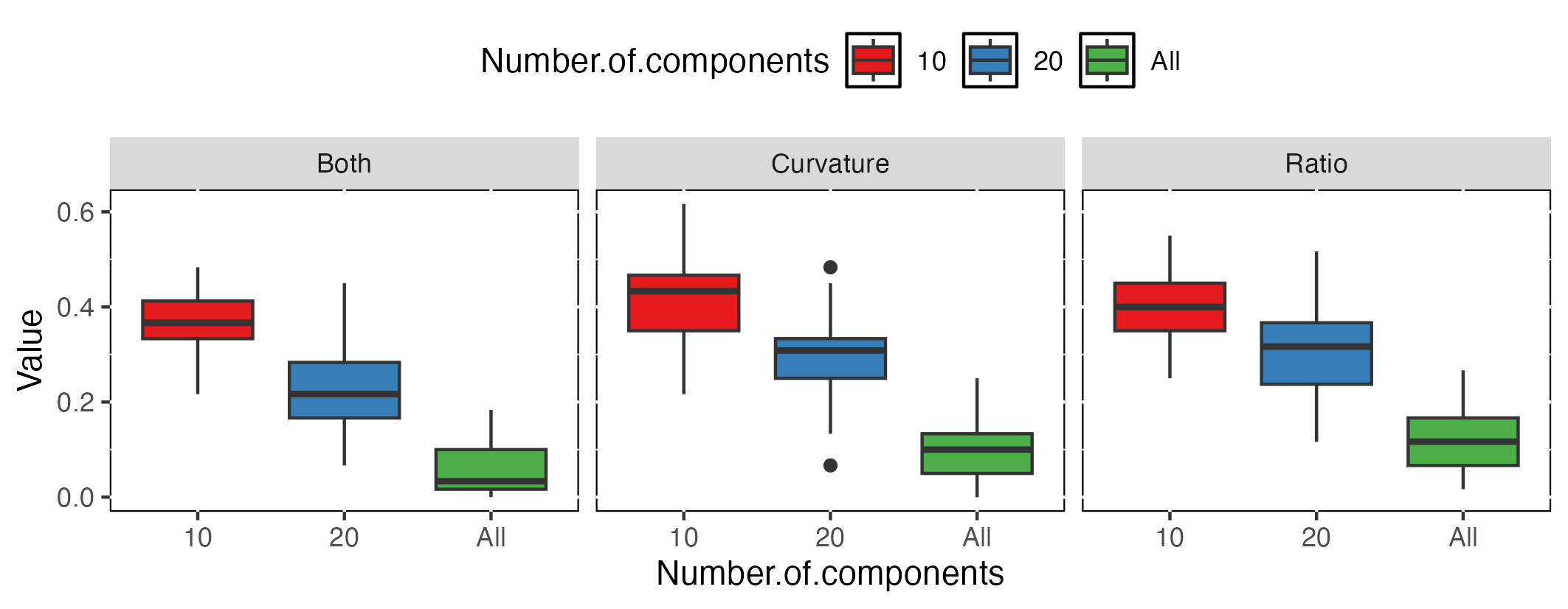}
    \end{minipage}
    \begin{minipage}{0.03\linewidth}\centering
        \rotatebox[origin=center]{90}{50 Realisations}
    \end{minipage}
    \begin{minipage}{0.93\linewidth}\centering
        \includegraphics[height=5.5cm, width=12.3cm]{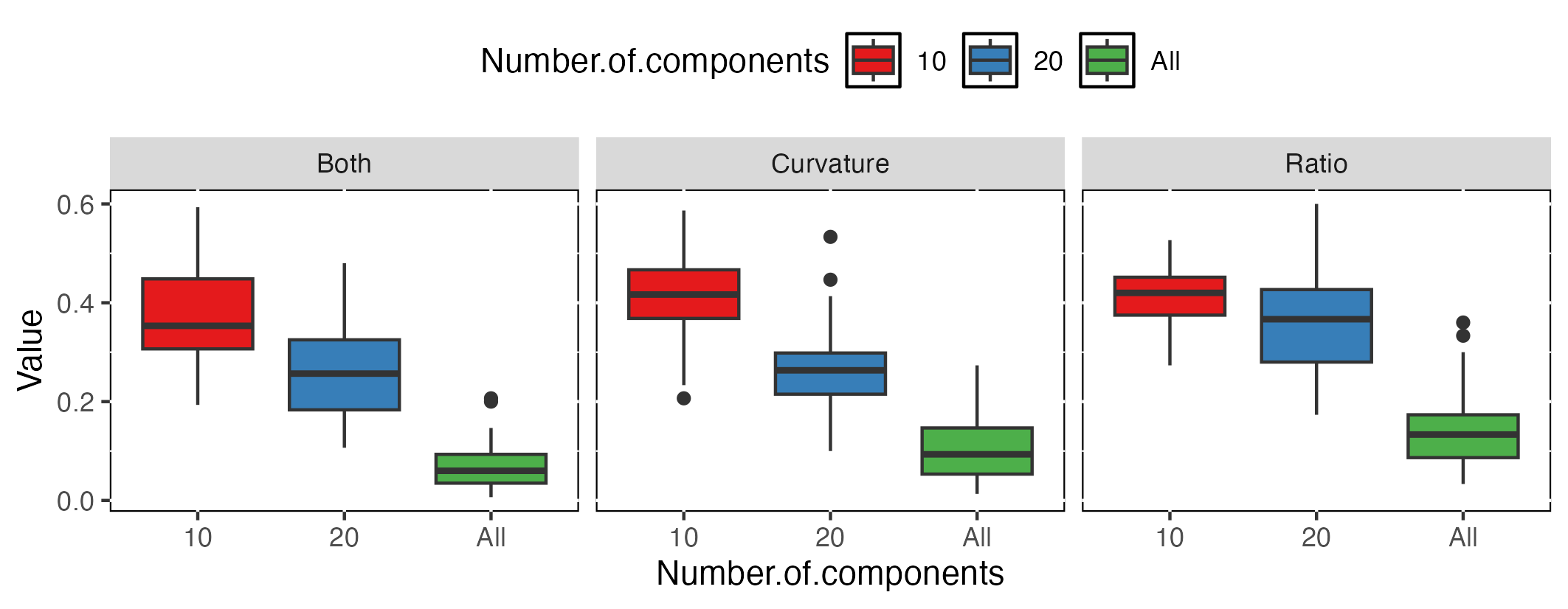}
    \end{minipage}
    \begin{minipage}{0.03\linewidth}\centering
        \rotatebox[origin=center]{90}{100 Realisations}
    \end{minipage}
    \begin{minipage}{0.93\linewidth}\centering
        \includegraphics[height=5.5cm, width=12.3cm]{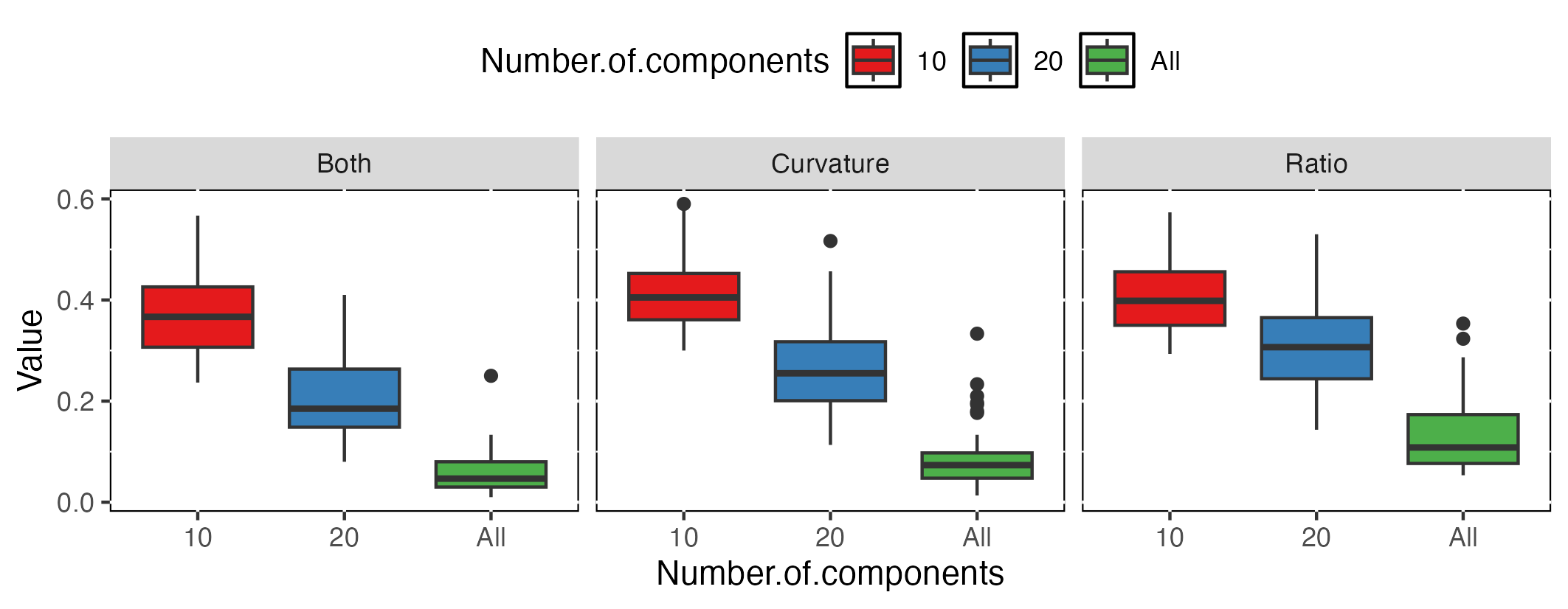}
    \end{minipage}
    \caption{Boxplots of misclassification rate for 50 runs of $k$-medoids algorithm when considering samples of 20 (top), 50 (middle) and 100 (bottom) realisations using both ratio and curvature, only the curvature and only the ratio for discrimination, respectively. For each setting, misclassification rates for different number of components considered (namely 10, 20 and 'All') are shown. Note that the characteristics were obtained using an osculating disc of radius $r=5$ on the simulated data.}
    \label{fig:kmed_box_5_red}
\end{figure}

Each setting is run 50 times in order to obtain boxplots of the misclassification rate. The results are shown in Figure \ref{fig:kmed_box_5_red}. Taking a look at the results when considering 20 and 50 realisations, we can see that the classifier does not become significantly more accurate when we feed it with more data. 

\subsubsection{Hierarchical clustering}

\begin{figure}[!ht]
    \centering
    \begin{minipage}{0.03\linewidth}\centering
        \rotatebox[origin=center]{90}{Ratio}
    \end{minipage}
    \begin{minipage}{0.93\linewidth}\centering
        \includegraphics[height=5.5cm, width=12.5cm]{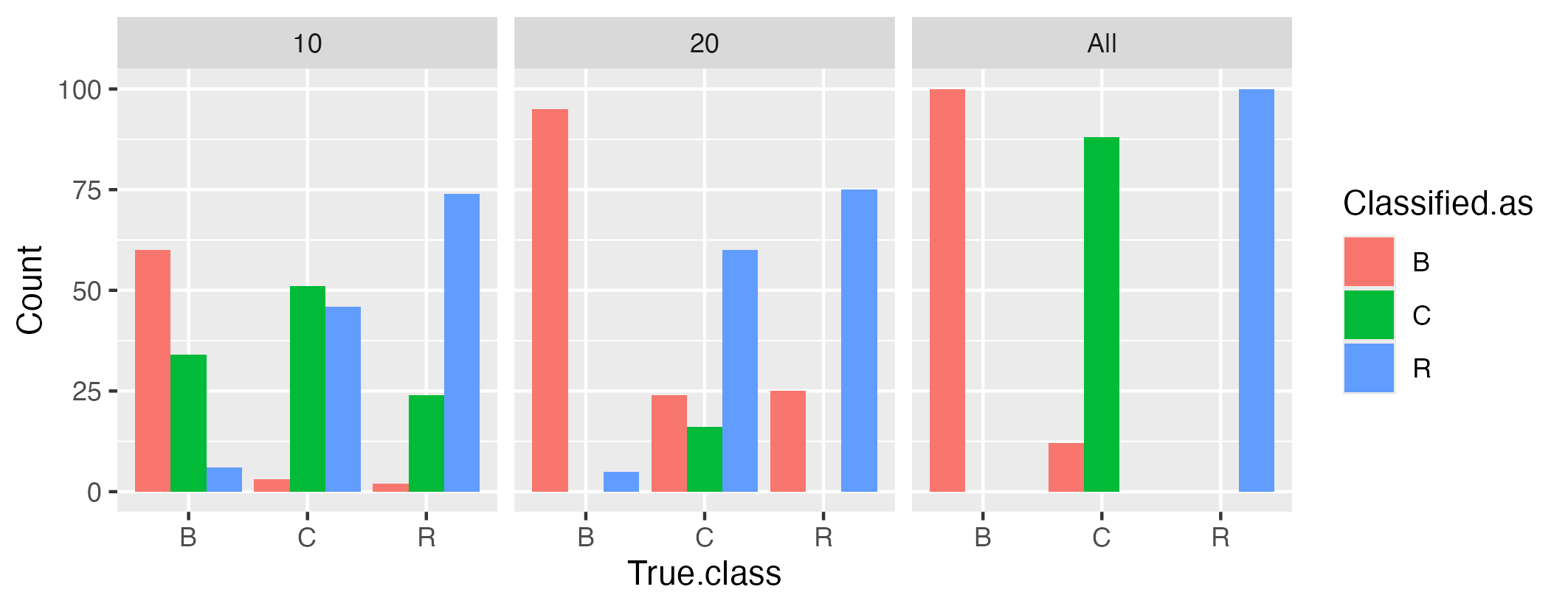}
    \end{minipage}
    \begin{minipage}{0.03\linewidth}\centering
        \rotatebox[origin=center]{90}{Curvature}
    \end{minipage}
    \begin{minipage}{0.93\linewidth}\centering
        \includegraphics[height=5.5cm, width=12.5cm]{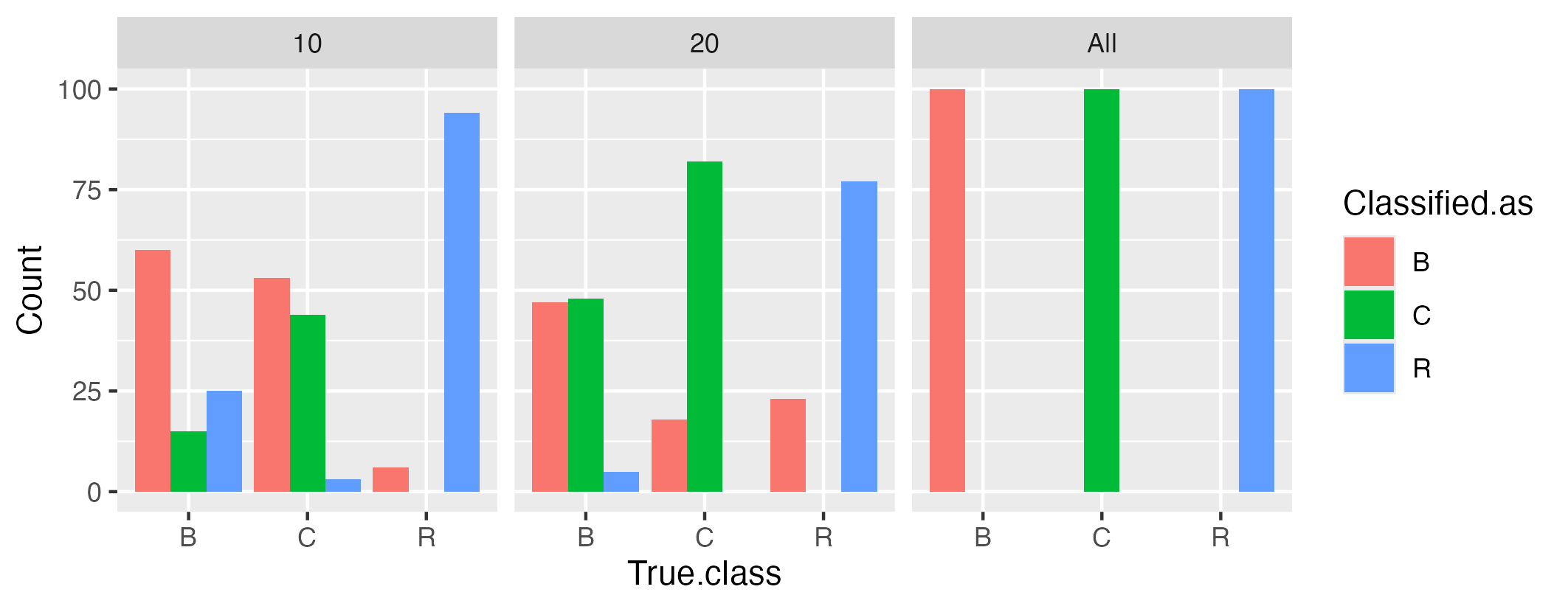}
    \end{minipage}
    \begin{minipage}{0.03\linewidth}\centering
        \rotatebox[origin=center]{90}{Both}
    \end{minipage}
    \begin{minipage}{0.93\linewidth}\centering
        \includegraphics[height=5.5cm, width=12.5cm]{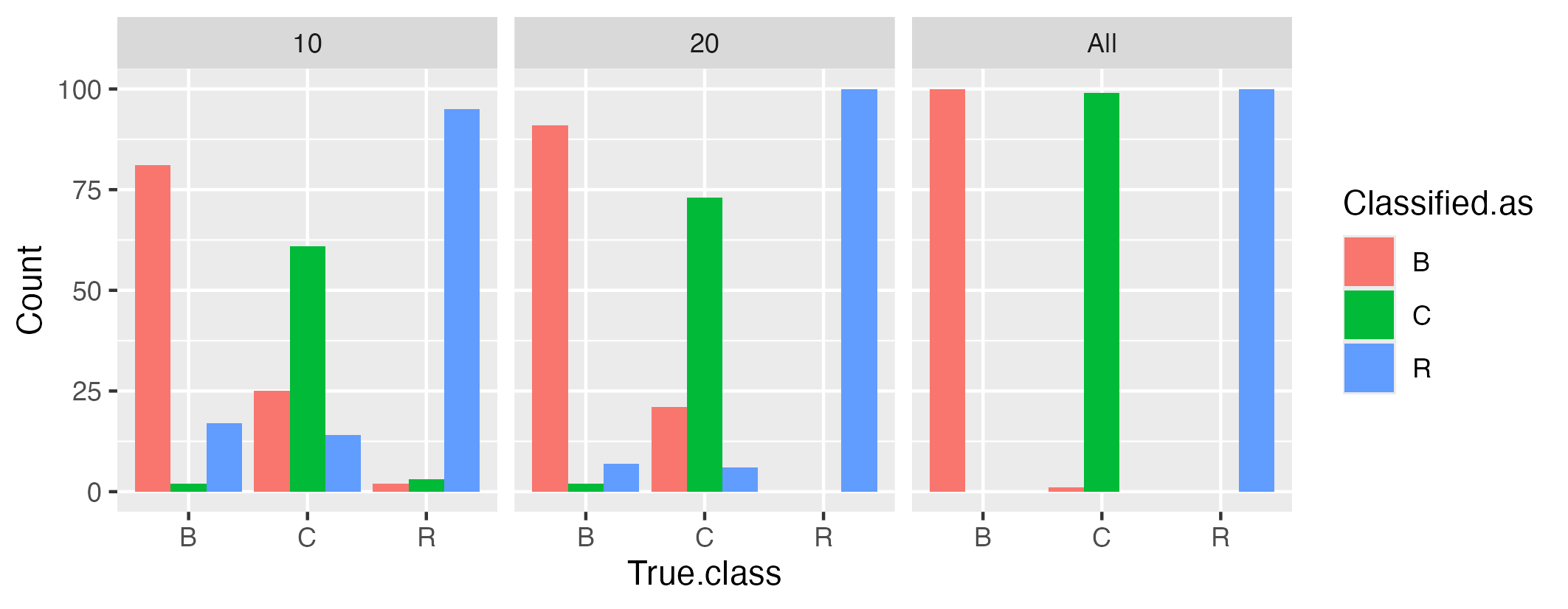}
    \end{minipage}
    \caption{Histograms of hierarchical clustering classification accuracy using only the ratio, only the curvature and both ratio and curvature for discrimination when using a sample of 10, 20, and 'All' components, respectively. Misclassification rates are 38.3\%, 38\% and 4\% for 10, 20 and 'All' components, respectively, when using only the ratio, 34\%, 31.3\% and 0\% when using only the curvature, and 21\%, 12\% and 0.3\% when using both characteristics for a sample of 100 realisations that were osculated by a disc of radius $r=5$.}
    \label{fig:hc_100_best_5}
\end{figure}

Focussing on the results when considering 100 realisations shown in Figure \ref{fig:hc_100_best_5}, we~observe that, as expected, the highest overall misclassification rate occurs for the smallest sample size (of 10 components) as expected, it decreases for a larger sample size (of 20 components), while the best performance was achieved when considering 'All' components. The most problematic aspect remains the classification of the cluster model, similar to the findings from the supervised classification case. The histograms of classification accuracy when 20 and 50 realisations are considered are presented in the Supplementary Material.

\begin{figure}[!ht]
    \centering
    \begin{minipage}{0.03\linewidth}\centering
        \rotatebox[origin=center]{90}{20 Realisations}
    \end{minipage}
    \begin{minipage}{0.93\linewidth}\centering
        \includegraphics[height=5.5cm, width=12.3cm]{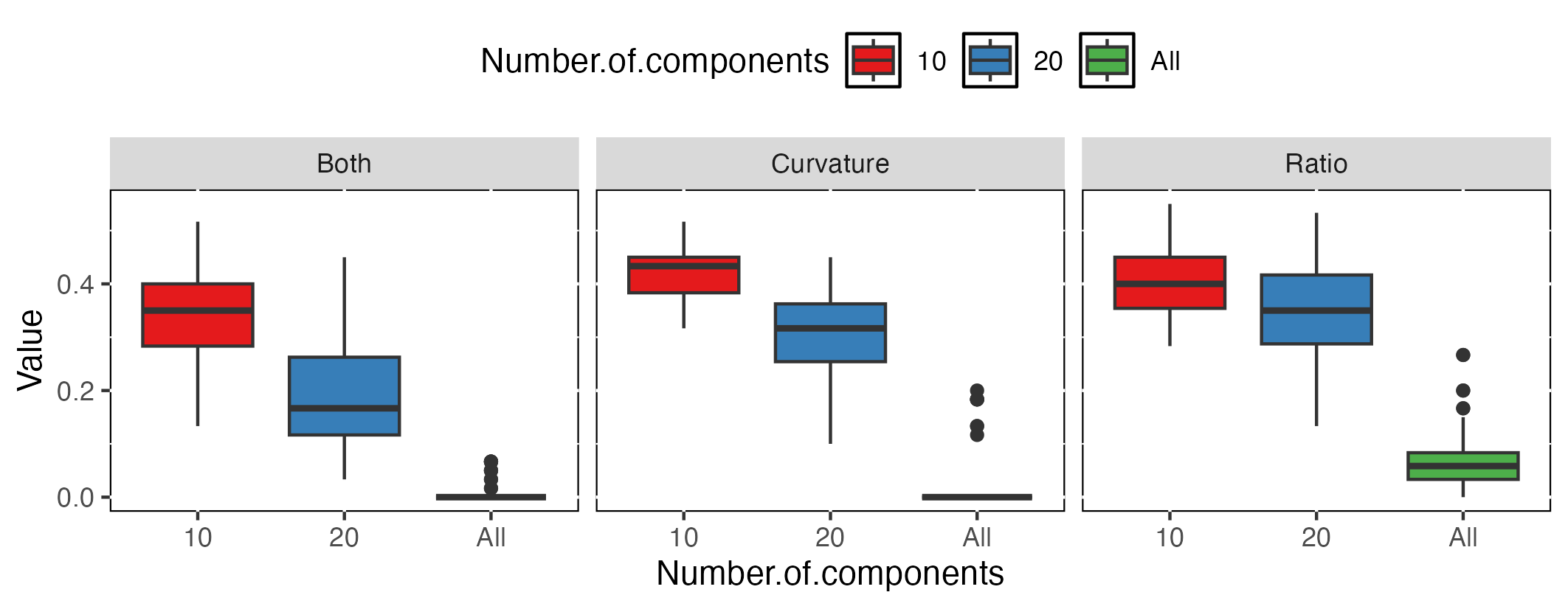}
    \end{minipage}
    \begin{minipage}{0.03\linewidth}\centering
        \rotatebox[origin=center]{90}{50 Realisations}
    \end{minipage}
    \begin{minipage}{0.93\linewidth}\centering
        \includegraphics[height=5.5cm, width=12.3cm]{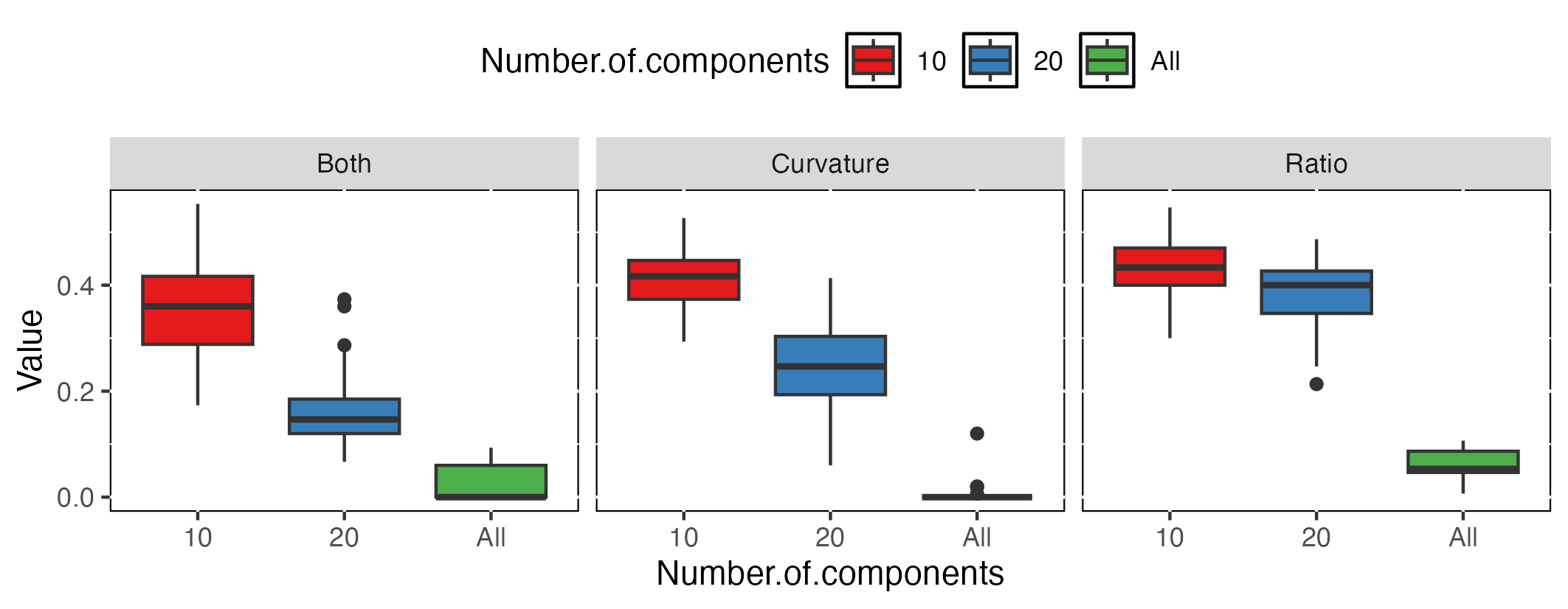}
    \end{minipage}
        \begin{minipage}{0.03\linewidth}\centering
        \rotatebox[origin=center]{90}{100 Realisations}
    \end{minipage}
    \begin{minipage}{0.93\linewidth}\centering
        \includegraphics[height=5.5cm, width=12.3cm]{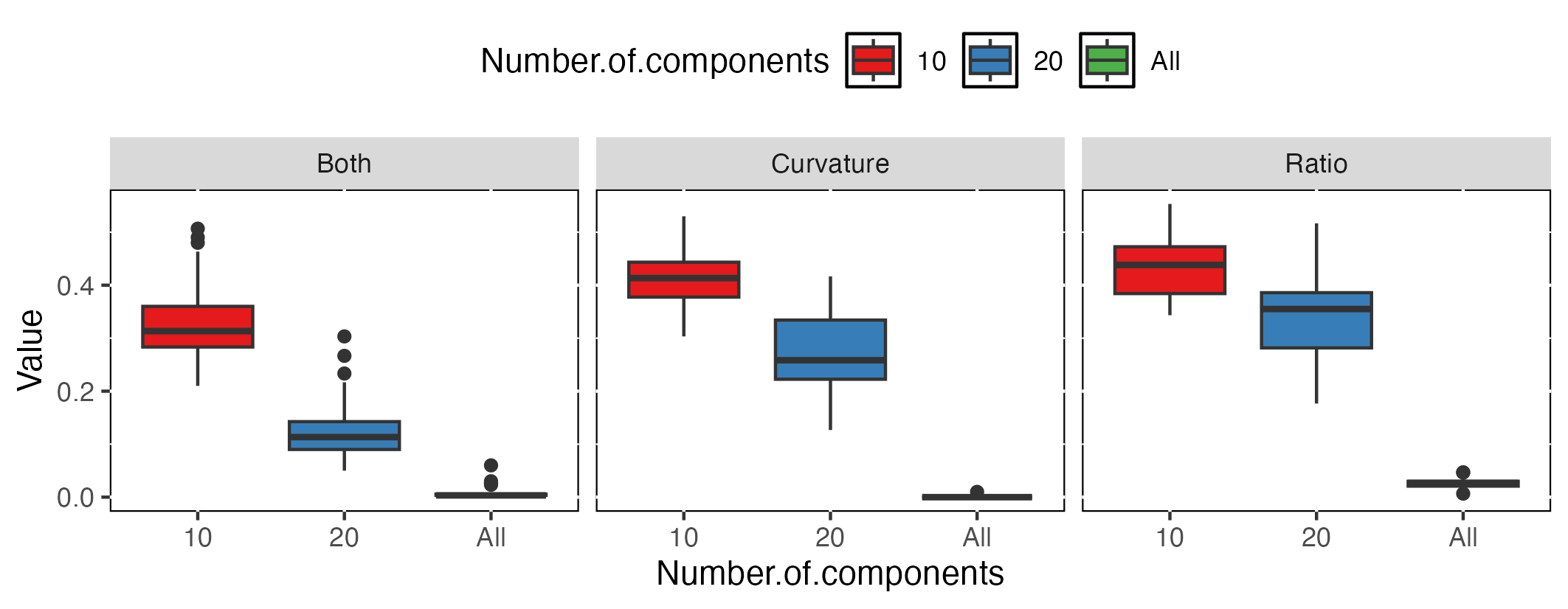}
    \end{minipage}
    \caption{Boxplots of misclassification rate for 50 runs of hierarchical clustering algorithm when considering samples of 20 (top), 50 (middle) and 100 (bottom) realisations using both ratio and curvature, only the curvature and only the ratio for discrimination, respectively. For each setting, misclassification rates for different number of components considered (namely 10, 20 and 'All') are shown. Note that the characteristics were obtained using an osculating disc of radius $r=5$ on the simulated data.}
    \label{fig:hc_box_5_red}
\end{figure}

As for the previous cases, each setting is run 50 times in order to obtain boxplots of the misclassification rate. The results when considering 20 and 50 realisations shown in Figure \ref{fig:hc_box_5_red} suggest that the classifier again does not become substantially more accurate when provided with additional data.

\section{Application}

\begin{figure}[]
\centering
Sample 'MP1'\\ \includegraphics[height=1.8cm, width = 13cm]{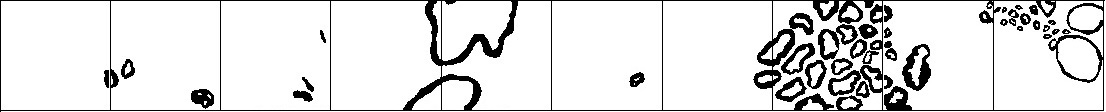}\\
Sample 'MP2'\\ \includegraphics[height=1.8cm, width = 13cm]{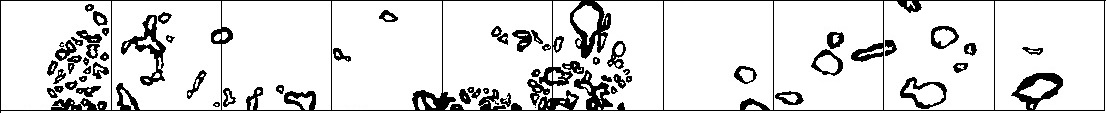}\\
Sample 'MP3'\\ \includegraphics[height=1.8cm, width = 13cm]{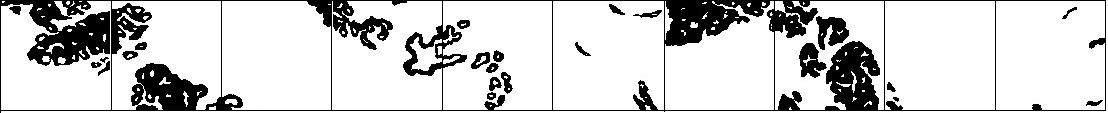}\\
Sample 'MP4'\\ \includegraphics[height=1.8cm, width = 13cm]{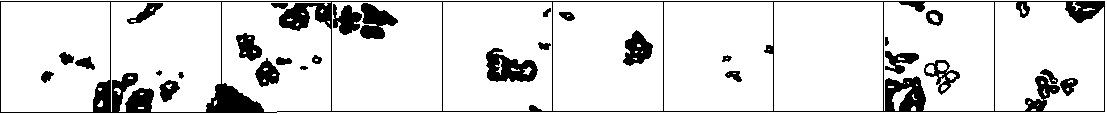}\\
Sample 'MP5'\\ \includegraphics[height=1.8cm, width = 13cm]{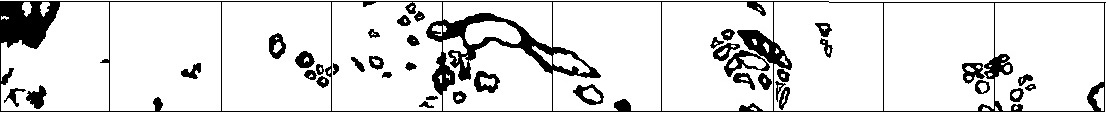}\\
Sample 'MP6'\\ \includegraphics[height=1.8cm, width = 13cm]{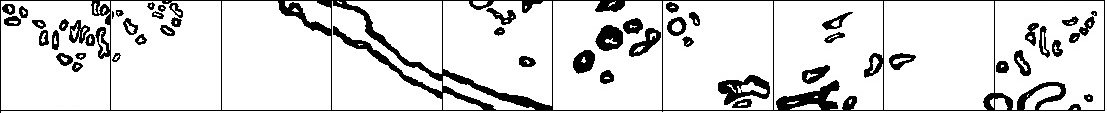}\\
Sample 'MP7'\\ \includegraphics[height=1.8cm, width = 13cm]{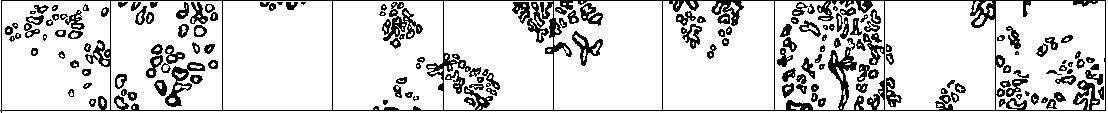}\\
Sample 'MP8'\\ \includegraphics[height=1.8cm, width = 13cm]{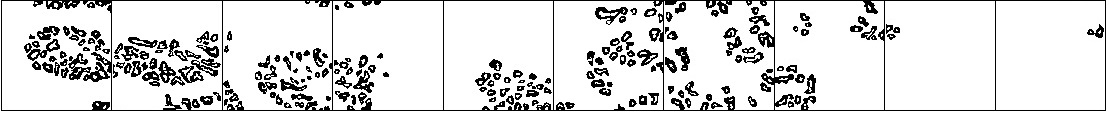}\\
\caption{Samples of mastopathic breast tissue \cite{mrkvicka:2011}, \cite{gotovac:2019}}
\label{fig:masto}
\end{figure}

\begin{figure}[]
 \centering
Sample 'MC1'\\ \includegraphics[height=1.8cm, width = 13cm]{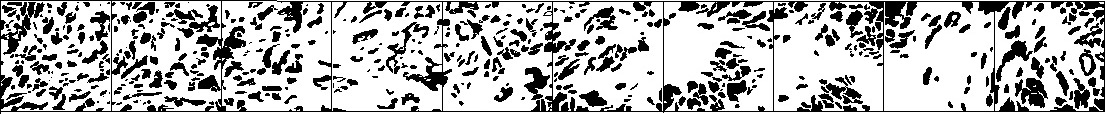}\\
Sample 'MC2'\\ \includegraphics[height=1.8cm, width = 13cm]{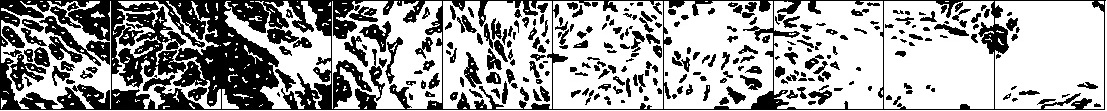}\\
Sample 'MC3'\\ \includegraphics[height=1.8cm, width = 13cm]{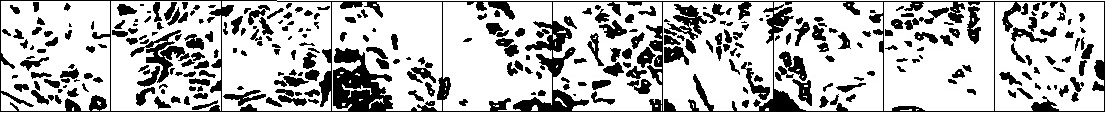}\\
Sample 'MC4'\\ \includegraphics[height=1.8cm, width = 13cm]{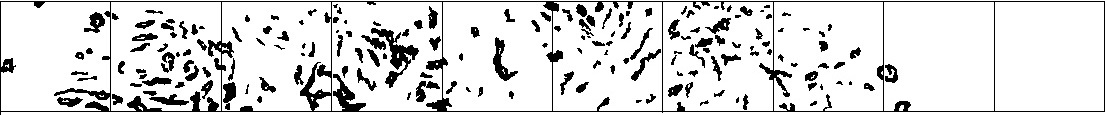}\\
Sample 'MC5'\\ \includegraphics[height=1.8cm, width = 13cm]{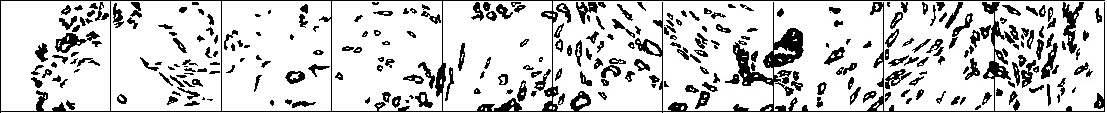}\\
Sample 'MC6'\\ \includegraphics[height=1.8cm, width = 13cm]{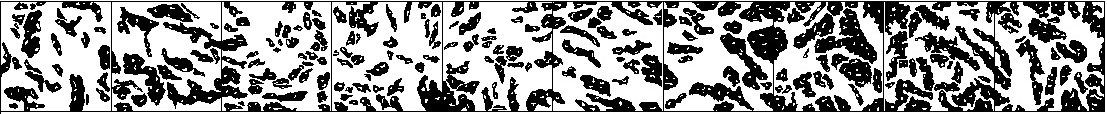}\\
Sample 'MC7'\\ \includegraphics[height=1.8cm, width = 13cm]{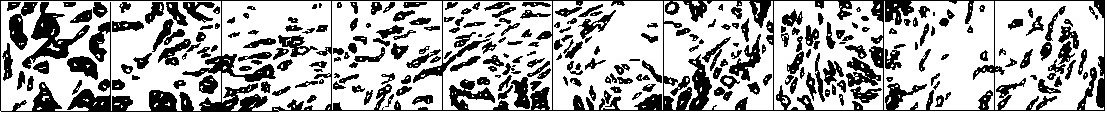}\\
Sample 'MC8'\\ \includegraphics[height=1.8cm, width = 13cm]{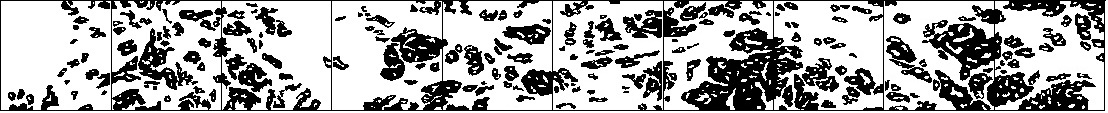}
\caption{Samples of mammary cancer \cite{mrkvicka:2011}, \cite{gotovac:2019}}
\label{fig:mamca}
\end{figure}

Once we have shown that the classifier is able to distinguish between simulated random processes, we will apply it to the real data. Different types of benign or malignant changes can be indicated by the morphology of the tissue located between the lactiferous duct system and the mammary glands \cite{mrkvicka:2011}. In our study, we will consider two types of mammary tissue - mastopathic (referred to~as~Masto or 'MP' only from now on) and mammary cancer tissue (referred to~as~Mamca or 'MC' only). Note that this data has already been studied in \cite{mrkvicka:2011}, \cite{gotovac:2019} and~\cite{gotovac:2021b}. The samples (in the form of binary images containing 10 subsamples of size $512 \times 512$ representing cross-sections of the duct system), which are used in our study, are shown in~Figure \ref{fig:masto} and Figure \ref{fig:mamca}, with black areas representing the aforementioned tissue. The data of mammary cancer and mastopathic tissue were kindly provided by the authors of~\cite{mrkvicka:2011}~and modified by merging subsamples by the author of~\cite{gotovac:2019}.

Initially, we identified the components conventionally and then calculated the corresponding $C$-functions and $P/A$-ratios for both values of $r$. 
Since we were provided with only 8 images
of size 512 $\times$ 5120 pixels
of each tissue, for better learning, we had to augment our data. For mastopathic tissue, we merged the first four images (that is, 'MP1' -- 'MP4') together, while for mammary cancer tissues, we merged the first six images (that is, 'MC1' -- 'MC6') together and randomly sampled a number of components that roughly corresponds to the number of components in the original images (60 for mastopathy and 300 for mammary cancer). The procedure was repeated 200 times, and in this way we obtained 200 realisations that would be used for the training stage. Similarly, we merged the last two images (that is, 'MP5' and 'MP6' for 'MP', and 'MC7' and 'MC8' for 'MC' tissue) together and sampled the components in the same way as for the training data. This was repeated 50 times, so in~the end we had 50 realisations that would be used for the testing phase. The images 'MP7' and 'MP8' were excluded from selection because the results obtained for them in \cite{gotovac:2021b} were not satisfactory in the sense that they were assessed as dissimilar to the remaining 'MP'~observations.

To assess the classification problem, we follow the same procedure as that used for the simulated data. 

\subsection{Supervised classification}

In the supervised case, the data are divided into train set and test set with a 3:1 ratio (which means that 75\% of the realisations is used for training, while 25\% is used for testing the performance of the classifier). Since we wanted to test how fast the classifier learns and how much the amount of data at our disposal affects its performance, we again used three settings: 
\begin{itemize}
\itemsep0em
    \item in the first setting we used a sample of 20 randomly chosen realisations from each type of~mammary tissue (further 'class'), mastopathic (class 'MP') and cancerous (class~'MC'), meaning that in the training set we have 30 realisations (15 of each class, 'MP'~and~'MC') in the training set and 10 realisations for testing purpose (5~of~each~class)
    \vspace{-2pt}
    \item in the second setting we used a sample of 50 randomly chosen realisations from each class, meaning that in the training set we have 74 realisations (37 of each class) in the training set and 26 realisations for testing purpose (13 of each class)
    \vspace{-2pt}
    \item in the third setting we used a sample of 100 randomly chosen realisations from each class, meaning that we have 150 realisations (75 of each class) in the training set and 50 realisations for testing purpose (25 of each class).
\end{itemize}

Each of the above-mentioned settings is then split into three subsettings according to the characteristic which is used for discrimination, namely 'ratio', 'curvature', and 'both'. After that, the classifier is learnt three times for different numbers of components which we~use for calculating the $\mathcal N$-distance (i.e. 10, 20 and 'All'). After the learning stage, we use the test set and predict the labels using the posterior probabilities calculated for each class. 

\begin{figure}[!ht]
    \centering
    \begin{minipage}{0.03\linewidth}\centering
        \rotatebox[origin=center]{90}{Ratio}
    \end{minipage}
    \begin{minipage}{0.93\linewidth}\centering
        \includegraphics[height=5.5cm, width=12.5cm]{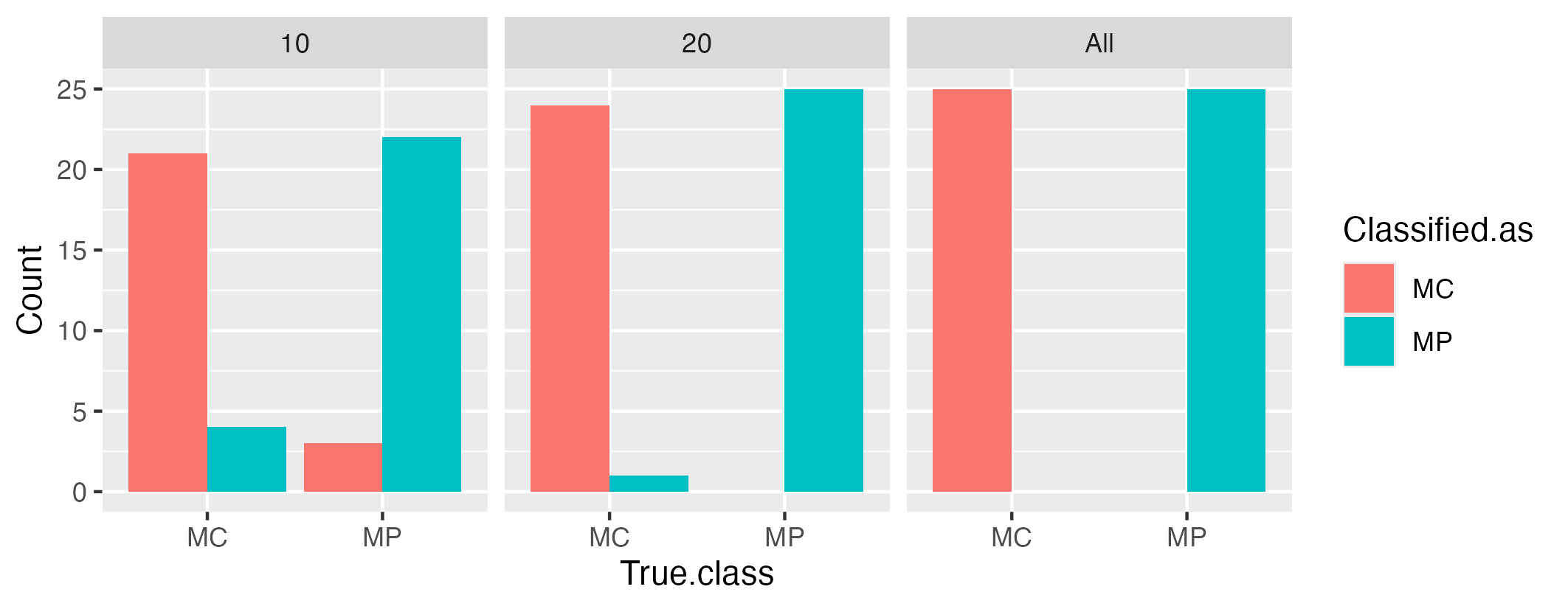}
    \end{minipage}
    \begin{minipage}{0.03\linewidth}\centering
        \rotatebox[origin=center]{90}{Curvature}
    \end{minipage}
    \begin{minipage}{0.93\linewidth}\centering
        \includegraphics[height=5.5cm, width=12.5cm]{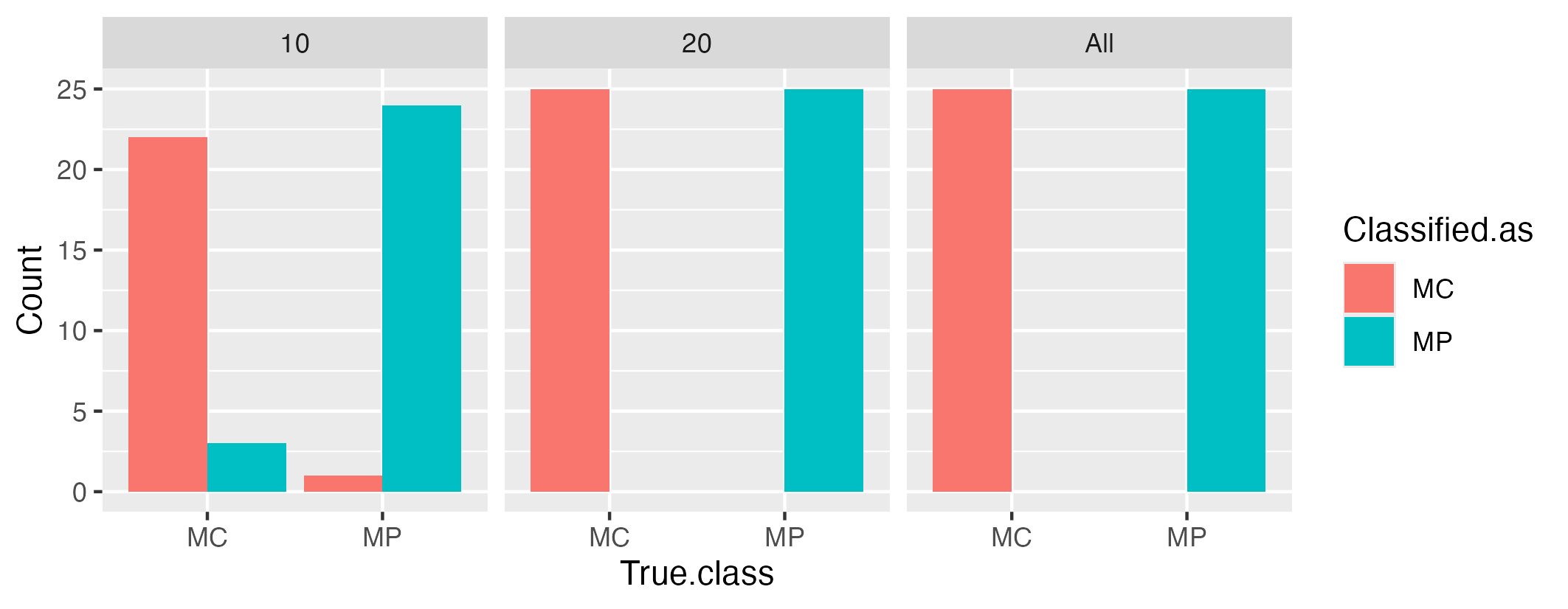}
    \end{minipage}
    \begin{minipage}{0.03\linewidth}\centering
        \rotatebox[origin=center]{90}{Both}
    \end{minipage}
    \begin{minipage}{0.93\linewidth}\centering
        \includegraphics[height=5.5cm, width=12.5cm]{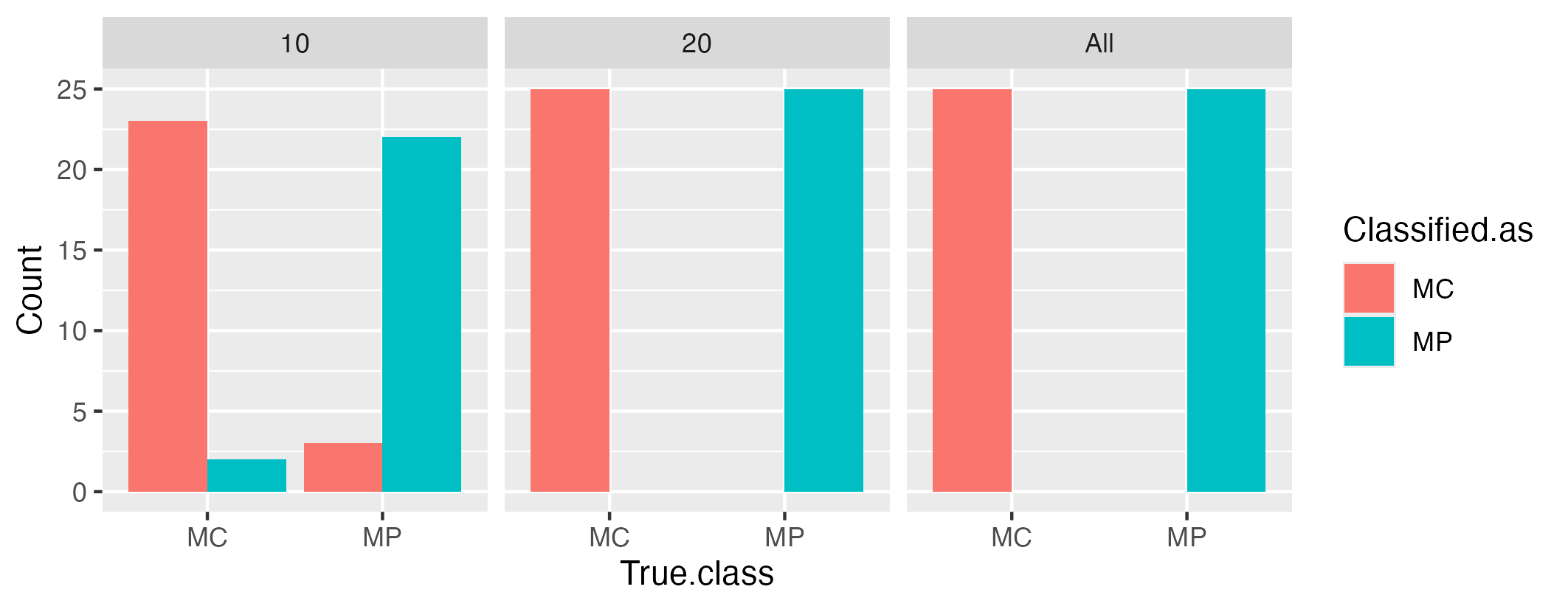}
    \end{minipage}
    \caption{Histograms of $k$-nearest neighbours classification accuracy using only the ratio, only the curvature and both ratio and curvature for discrimination when using a sample of 10, 20, and 'All' components, respectively. Misclassification rates are 4.7\%, 0.7\% and 0\% for 10, 20 and 'All' components, respectively, when using only the ratio, 2.7\%, 0\% and 0\% when using only the curvature and 3.3\%, 0\% and 0\% when using both characteristics for a sample of 100 realisations that were osculated by a disc of radius $r=5$.}
    \label{fig:knn_100_best_tissues_5}
\end{figure}

\begin{figure}[!ht]
    \centering
    \begin{minipage}{0.03\linewidth}\centering
        \rotatebox[origin=center]{90}{20 Realisations}
    \end{minipage}
    \begin{minipage}{0.93\linewidth}\centering
        \includegraphics[height=5.5cm, width=12.3cm]{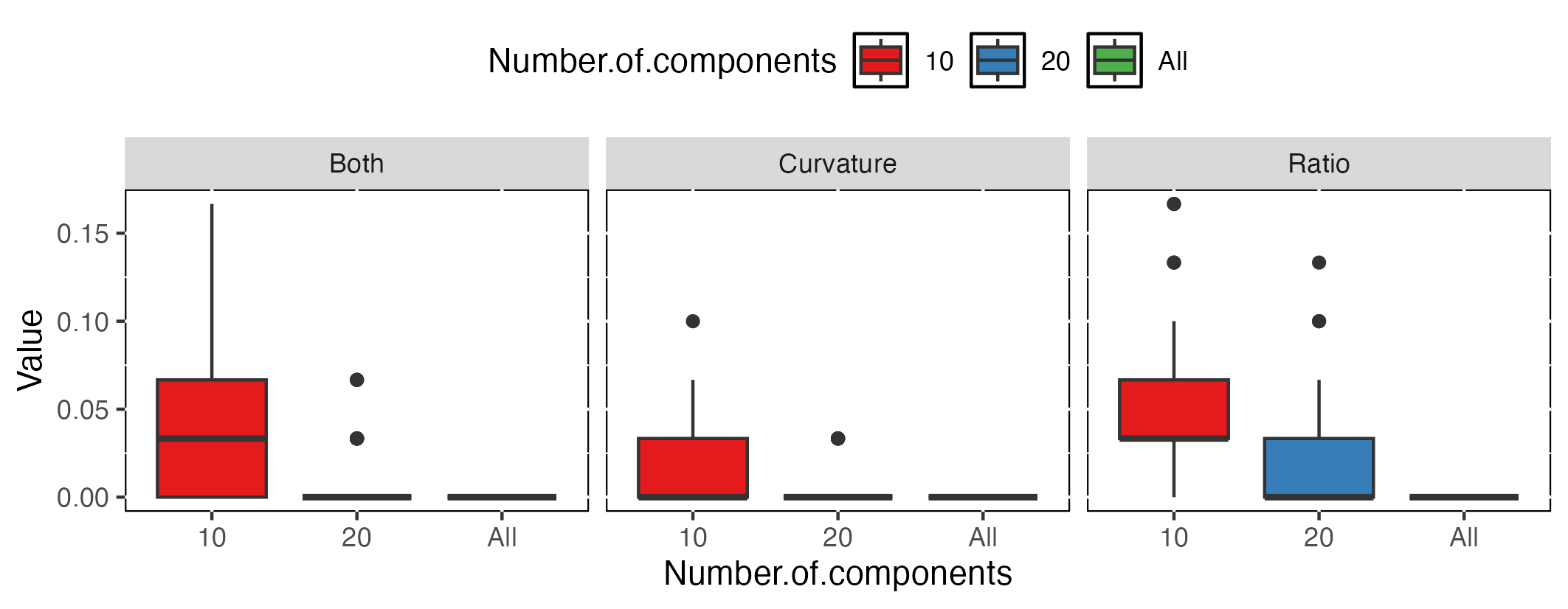}
    \end{minipage}
    \begin{minipage}{0.03\linewidth}\centering
        \rotatebox[origin=center]{90}{50 Realisations}
    \end{minipage}
    \begin{minipage}{0.93\linewidth}\centering
        \includegraphics[height=5.5cm, width=12.3cm]{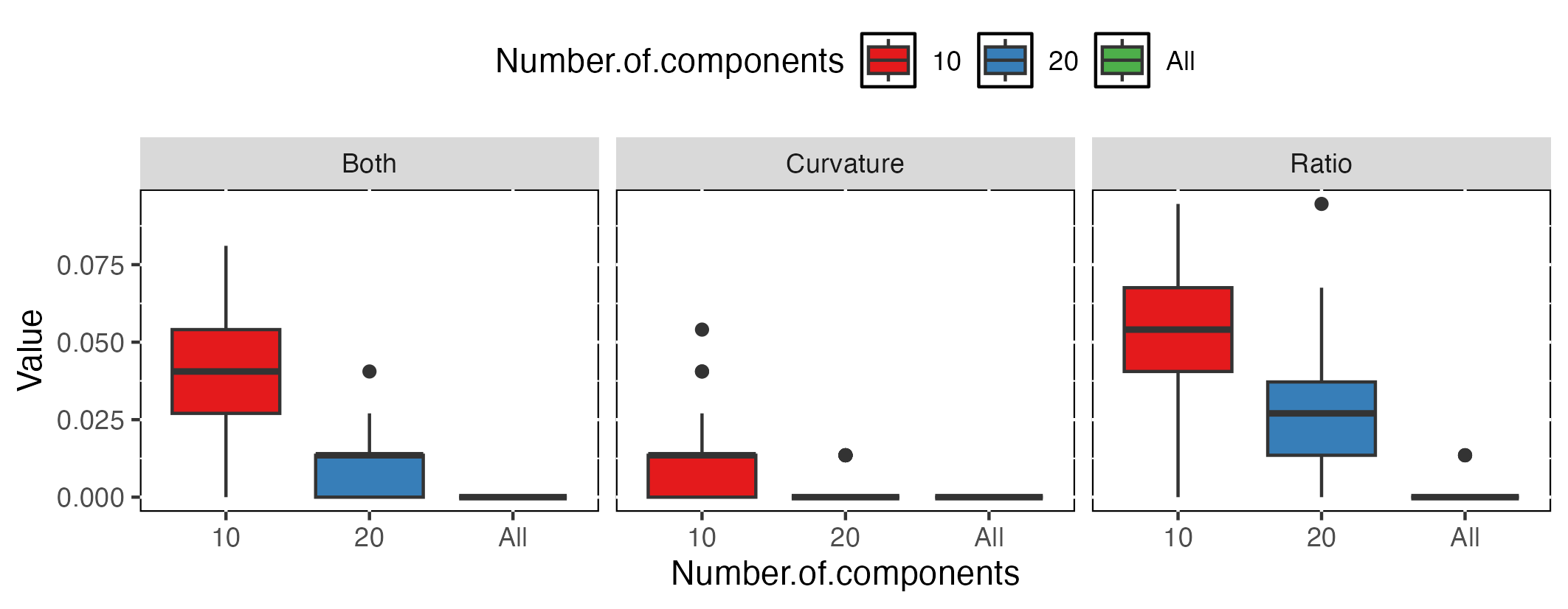}
    \end{minipage}
    \begin{minipage}{0.03\linewidth}\centering
        \rotatebox[origin=center]{90}{100 Realisations}
    \end{minipage}
    \begin{minipage}{0.93\linewidth}\centering
        \includegraphics[height=5.5cm, width=12.3cm]{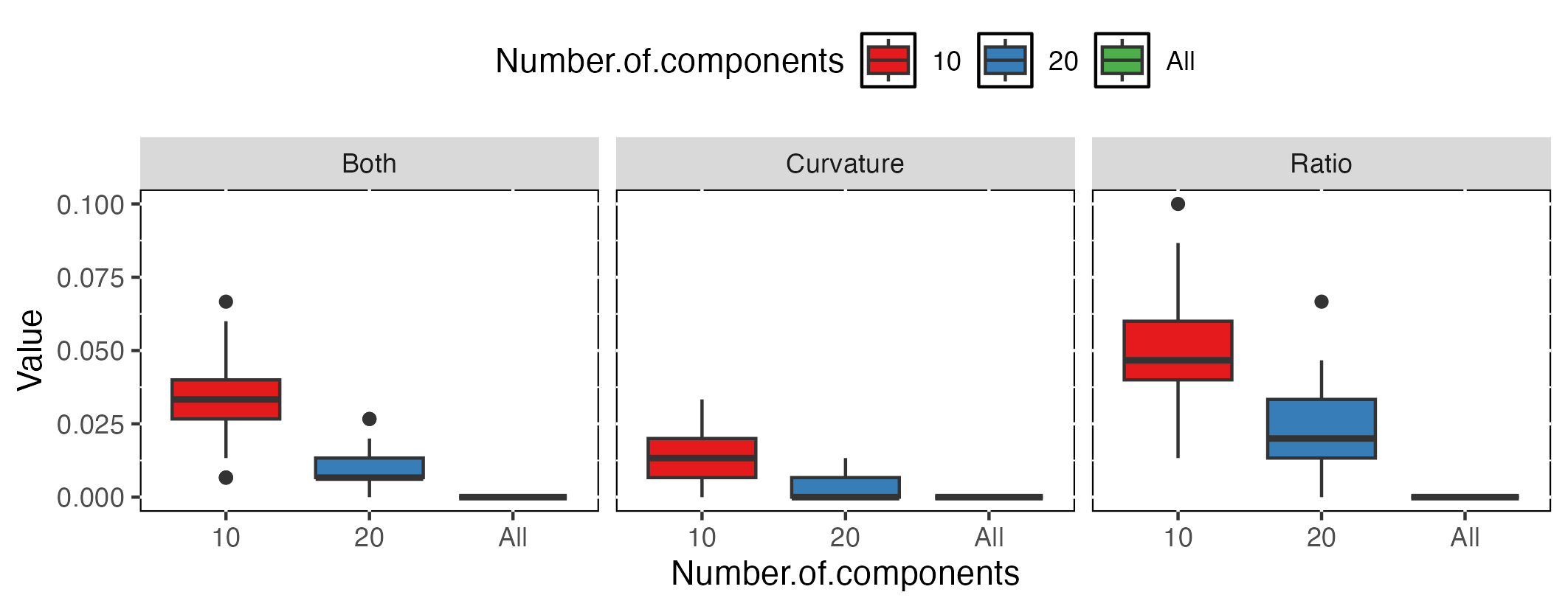}
    \end{minipage}
    \caption{Boxplots of misclassification rate for 50 runs of $k$-nearest neighbours algorithm when considering samples of 20 (top), 50 (middle) and 100 (bottom) realisations using both ratio and curvature, only the curvature and only the ratio for discrimination, respectively. For each setting, misclassification rates for different number of components considered (namely 10, 20, and 'All') are shown. Note that the characteristics were obtained using an osculating disc of radius $r=5$.}
    \label{fig:box_knn_tissues_5_red}
\end{figure}

The classification results for the data obtained using the osculating circle with radius $r=5$ and 100 realisations are shown in Figure \ref{fig:knn_100_best_tissues_5}. We can see that after the initial run, the classification precision follows the pattern observed for the simulated data -- it increases with the growing sample size (i.e., it is the lowest when only 10~components are used and the highest when 'All' components are used) for all settings (i.e.,~for different number of realisations considered). The results for 20 and 50 realisations, as well as for $r=3$ are presented in the Supplementary Material. After the initial run, we repeat the procedure 50 times to obtain boxplots of misclassification rates. The results are shown in Figure \ref{fig:box_knn_tissues_5_red}, and we observe that, as in the initial run, the classification is most precise when using 'All' components in all settings.

\subsection{Unsupervised classification}
Following the same procedure as for the simulated data, we test our classifiers on medical data. The results for the data obtained using the osculating circle with radius $r=3$ are presented in Supplementary Material.

\subsubsection{Non-hierarchical clustering}

\begin{figure}[!ht]
    \centering
    \begin{minipage}{0.03\linewidth}\centering
        \rotatebox[origin=center]{90}{Ratio}
    \end{minipage}
    \begin{minipage}{0.93\linewidth}\centering
        \includegraphics[height=5.5cm, width=12.5cm]{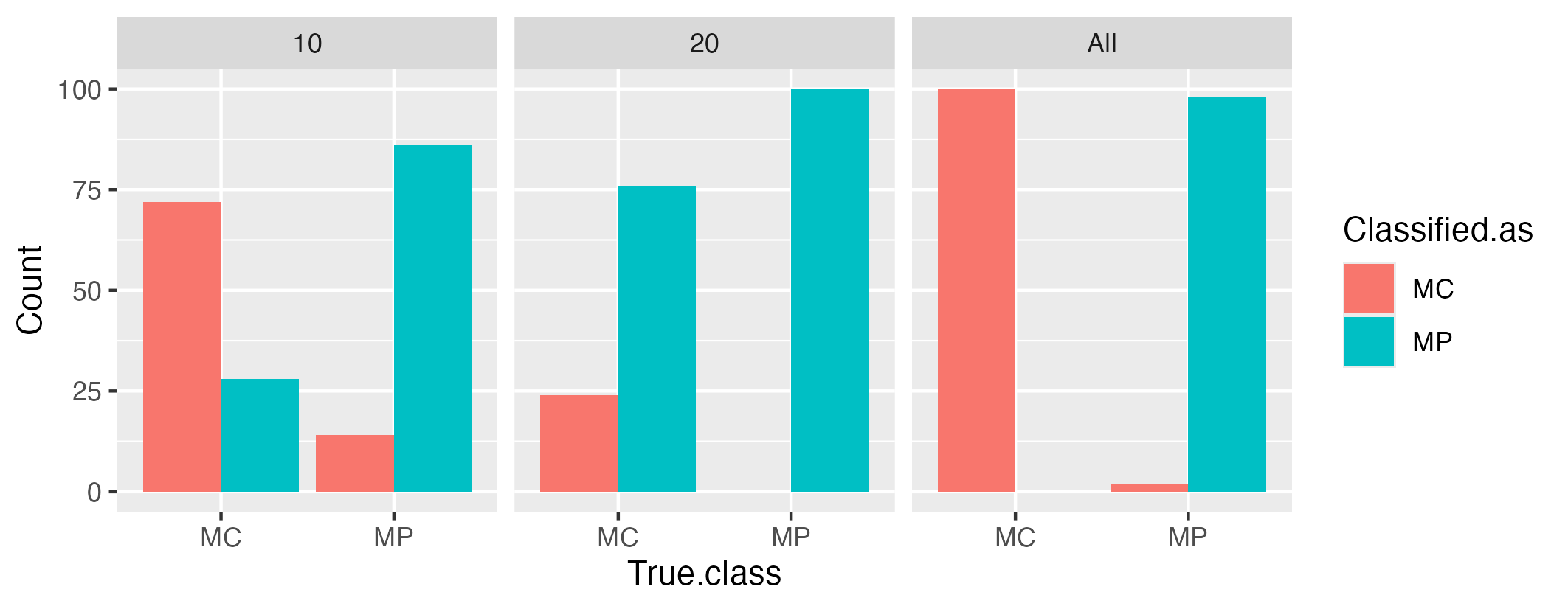}
    \end{minipage}
    \begin{minipage}{0.03\linewidth}\centering
        \rotatebox[origin=center]{90}{Curvature}
    \end{minipage}
    \begin{minipage}{0.93\linewidth}\centering
        \includegraphics[height=5.5cm, width=12.5cm]{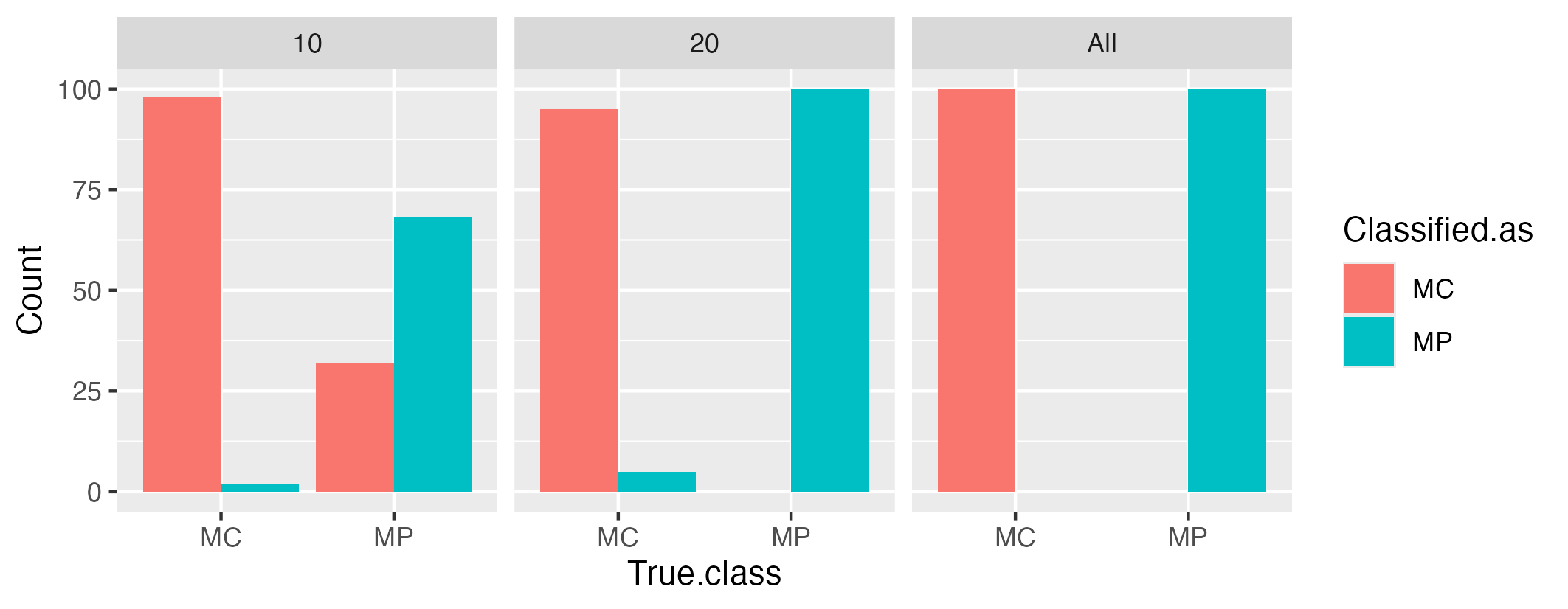}
    \end{minipage}
    \begin{minipage}{0.03\linewidth}\centering
        \rotatebox[origin=center]{90}{Both}
    \end{minipage}
    \begin{minipage}{0.93\linewidth}\centering
        \includegraphics[height=5.5cm, width=12.5cm]{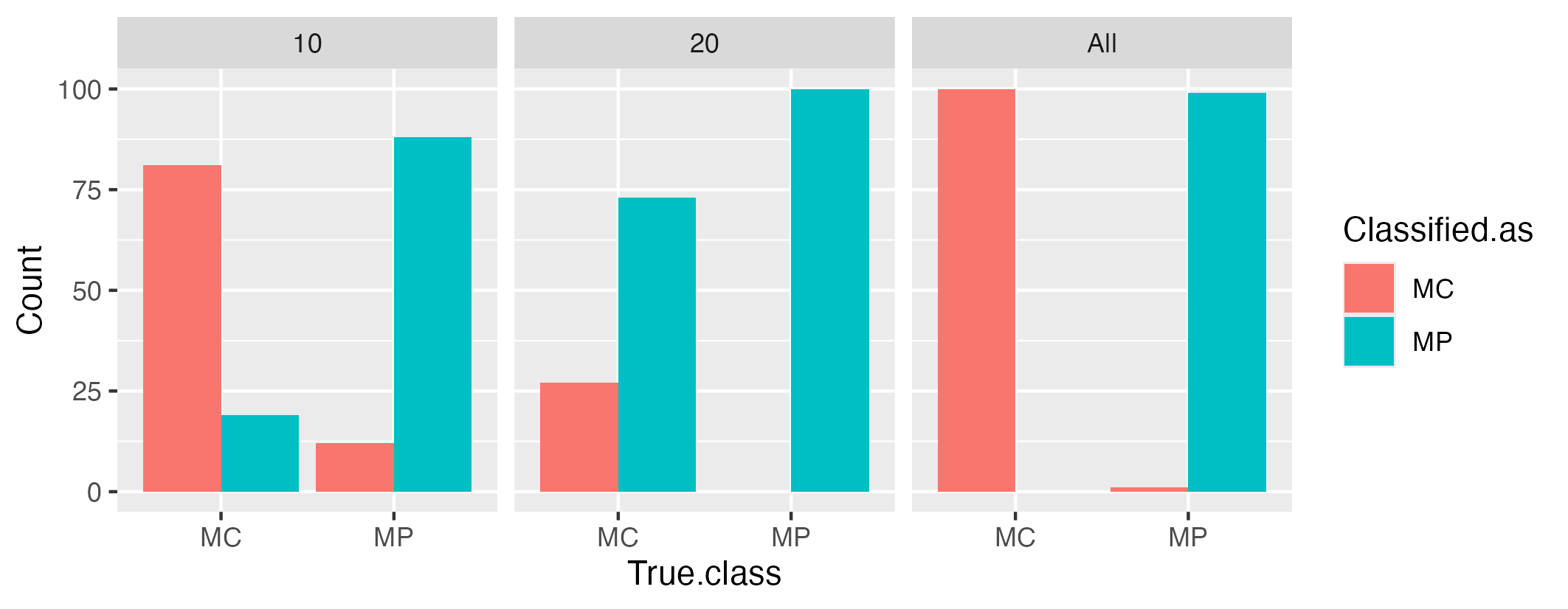}
    \end{minipage}
    \caption{Histograms of $k$-medoids classification accuracy using only the ratio, only the curvature and both ratio and curvature for discrimination when using a sample of 10, 20, and 'All' components, respectively. Misclassification rates are 21\%, 38\% and 1\% for 10, 20 and 'All' components, respectively, when using only the ratio, 17\%, 2.5\% and 0\% when using only the curvature and 15.5\%, 36.5\% and 0.5\% when using both characteristics for a sample of 100 realisations that were osculated by a disc of radius $r=5$.}
    \label{fig:kmed_100_best_tissues_5}
\end{figure}

The classification results for the data obtained using the osculating circle with radius $r=5$ are shown in Figure \ref{fig:kmed_100_best_tissues_5} (100~realisations). We can see that after the initial run, the classification precision follows the same trend observed for the simulated data -- it improves with the increasing sample size. Specifically, it is the lowest when only 10~components are used and the highest when 'All' components are used. This holds consistently across all settings. The results for the case when 20 and 50 realisations are considered are presented in the Supplementary Material.

\begin{figure}[!ht]
    \centering
    \begin{minipage}{0.03\linewidth}\centering
        \rotatebox[origin=center]{90}{20 Realisations}
    \end{minipage}
    \begin{minipage}{0.93\linewidth}\centering
        \includegraphics[height=5.5cm, width=12.3cm]{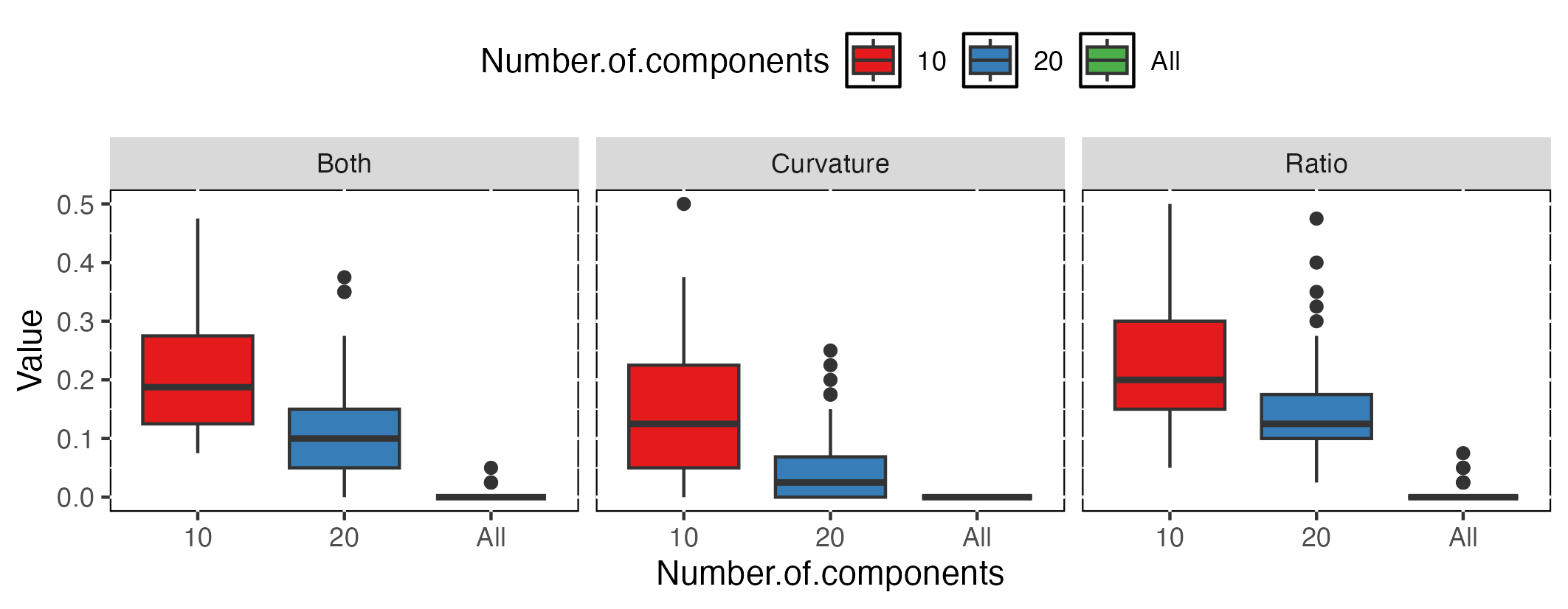}
    \end{minipage}
    \begin{minipage}{0.03\linewidth}\centering
        \rotatebox[origin=center]{90}{50 Realisations}
    \end{minipage}
    \begin{minipage}{0.93\linewidth}\centering
        \includegraphics[height=5.5cm, width=12.3cm]{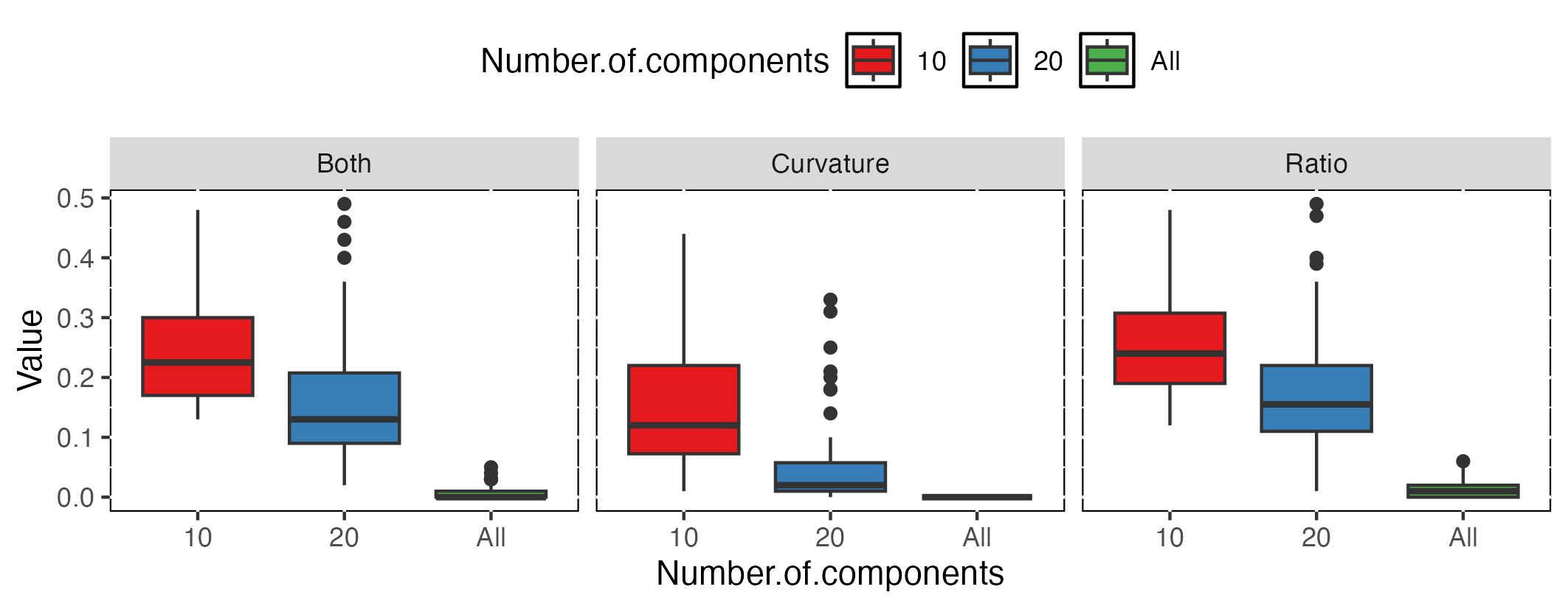}
    \end{minipage}
    \begin{minipage}{0.03\linewidth}\centering
        \rotatebox[origin=center]{90}{100 Realisations}
    \end{minipage}
    \begin{minipage}{0.93\linewidth}\centering
        \includegraphics[height=5.5cm, width=12.3cm]{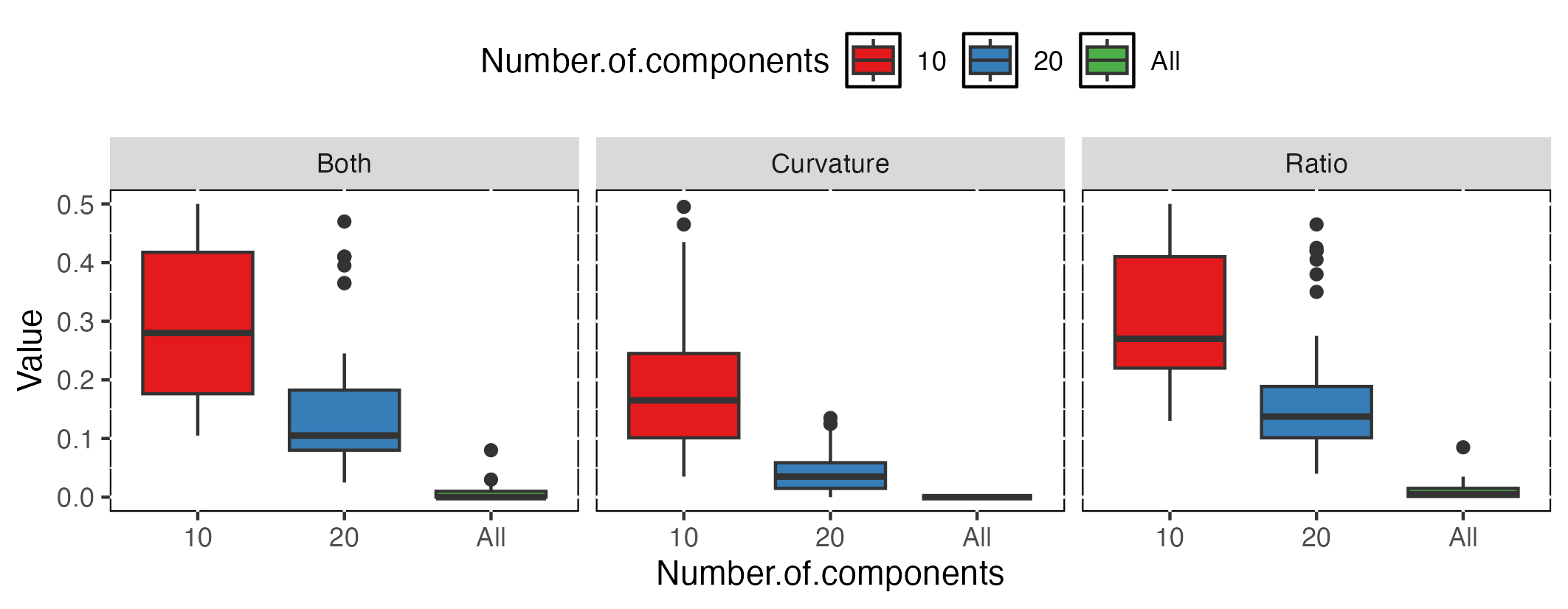}
    \end{minipage}
    \caption{Boxplots of misclassification rate for 50 runs of $k$-medoids algorithm when considering samples of 20 (top), 50 (middle) and 100 (bottom) realisations using both ratio and curvature, only the curvature and only the ratio for discrimination, respectively. For each setting, misclassification rates for different number of components considered (namely 10, 20, and 'All') are shown. Note that the characteristics were obtained using an osculating disc of radius $r=5$.}
    \label{fig:box_kmed_tissues_5}
\end{figure}

 The results for 50 runs, shown in Figure \ref{fig:box_kmed_tissues_5}, mirror those of the initial run, suggesting that the classification performs with the highest precision when 'All' components are used across all settings. Comparing boxplots when 20 and 100 realisations are used, we observe that the classifier does not become more accurate when more realisations are at its disposal.

\subsubsection{Hierarchical clustering}

The classification results for 100~realisations are presented in Figure \ref{fig:hc_100_best_tissues_5}, while the results for 20 and 50 are presented in the Supplementary Material. We find that after the initial run, the classification precision follows the same pattern as in the simulated data -- it improves with the increasing sample size. In particular, it is the lowest when only 10~components are considered and the highest when 'All' components are used).

\begin{figure}[!ht]
    \centering
\begin{minipage}{0.03\linewidth}\centering
        \rotatebox[origin=center]{90}{20 Realisations}
    \end{minipage}
    \begin{minipage}{0.93\linewidth}\centering
        \includegraphics[height=5.5cm, width=12.3cm]{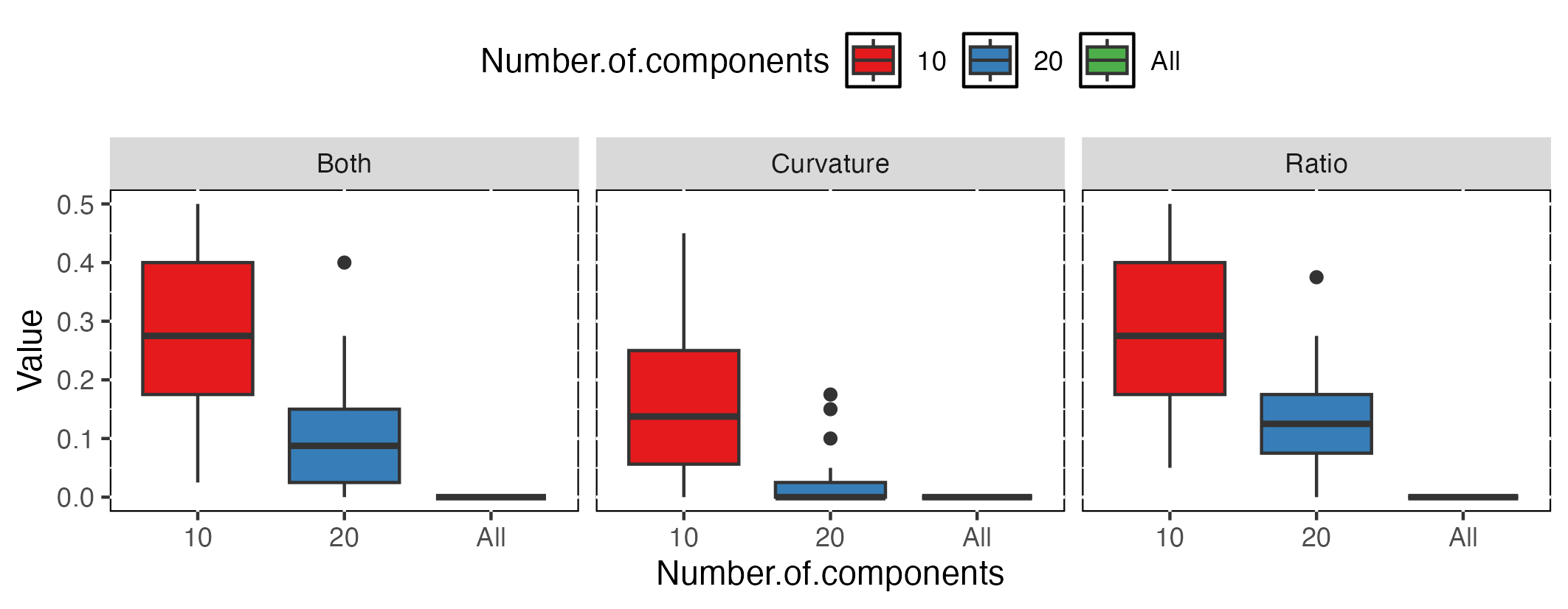}
    \end{minipage}
    \begin{minipage}{0.03\linewidth}\centering
        \rotatebox[origin=center]{90}{50 Realisations}
    \end{minipage}
    \begin{minipage}{0.93\linewidth}\centering
        \includegraphics[height=5.5cm, width=12.3cm]{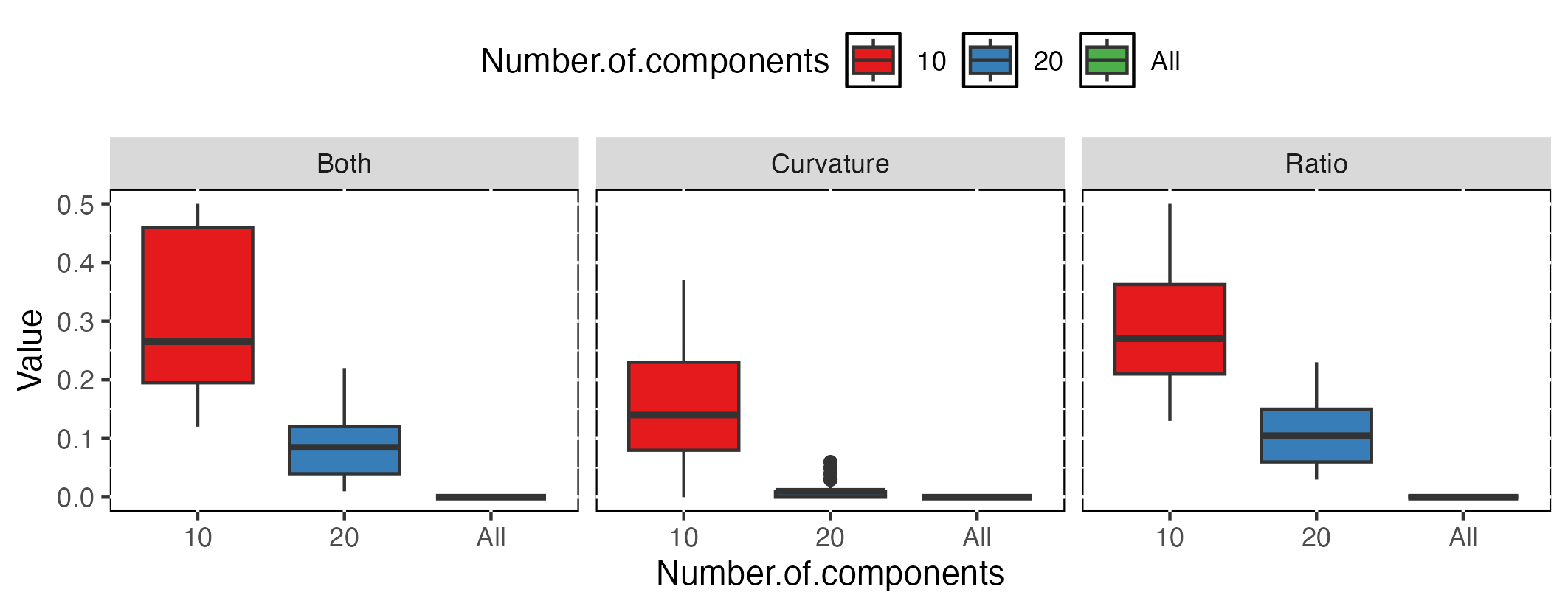}
    \end{minipage}
    \begin{minipage}{0.03\linewidth}\centering
        \rotatebox[origin=center]{90}{100 Realisations}
    \end{minipage}
    \begin{minipage}{0.93\linewidth}\centering
        \includegraphics[height=5.5cm, width=12.3cm]{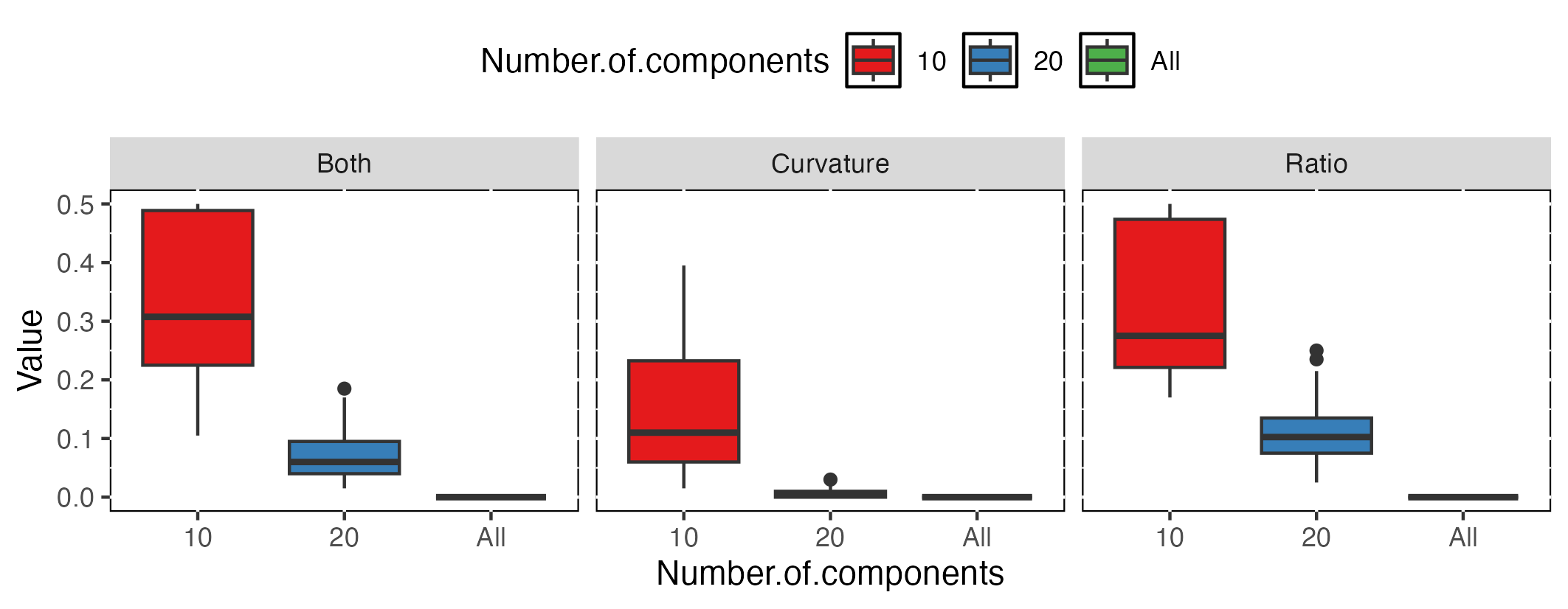}
    \end{minipage}
    \caption{Boxplots of misclassification rate for 50 runs of hierarchical clustering algorithm when considering samples of 20 (top), 50 (middle) and 100 (bottom) realisations using both ratio and curvature, only the curvature and only the ratio for discrimination, respectively. For each setting, misclassification rates for different number of components considered (namely 10, 20, and 'All') are shown. Note that the characteristics were obtained using an osculating disc of radius $r=5$.}
    \label{fig:box_hc_tissues_5}
\end{figure}

 The results for 50 runs are shown in Figure \ref{fig:box_hc_tissues_5}.
 We can see that the values reflect the ones in the initial run, meaning that the classification is most precise when using 'All' components, consistently across all settings. Again, comparing boxplots for 20 and 100 realisations, we observe that the classifier does not become more accurate when fed with more data. 

\begin{figure}[!ht]
    \centering
    \begin{minipage}{0.03\linewidth}\centering
        \rotatebox[origin=center]{90}{Ratio}
    \end{minipage}
    \begin{minipage}{0.93\linewidth}\centering
        \includegraphics[height=5.5cm, width=12.5cm]{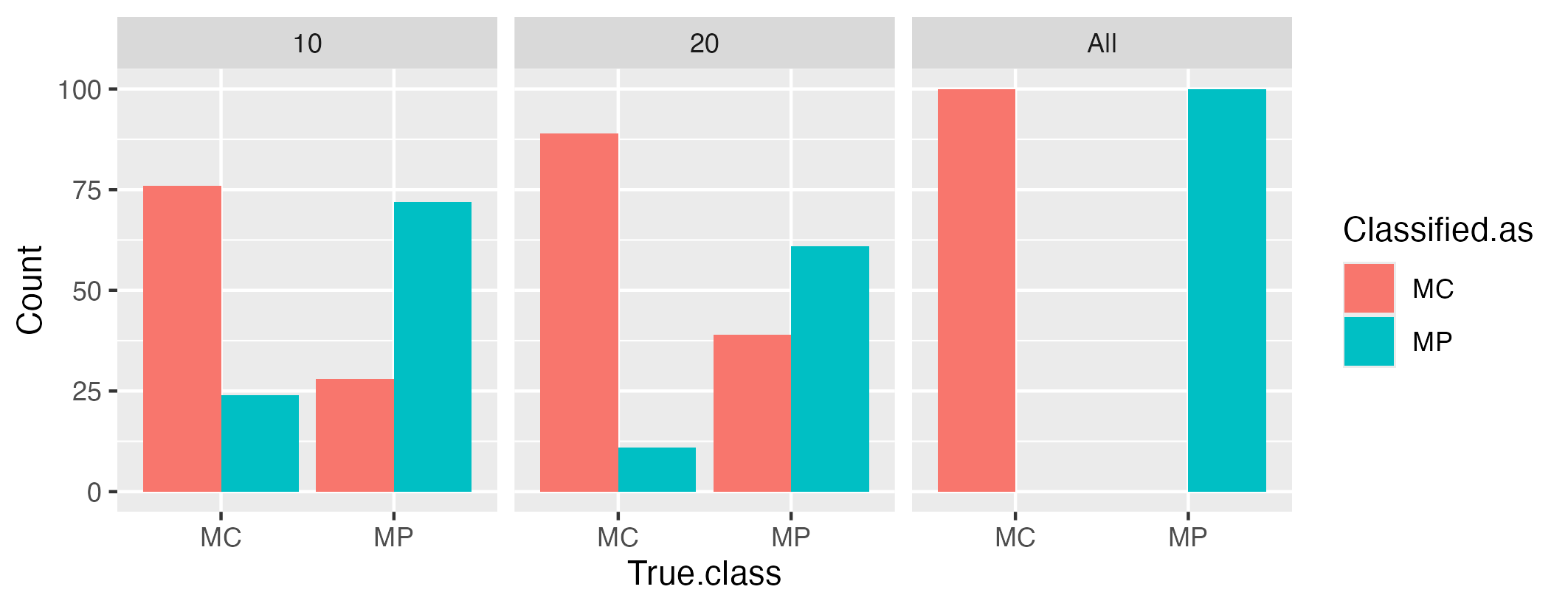}
    \end{minipage}
    \begin{minipage}{0.03\linewidth}\centering
        \rotatebox[origin=center]{90}{Curvature}
    \end{minipage}
    \begin{minipage}{0.93\linewidth}\centering
        \includegraphics[height=5.5cm, width=12.5cm]{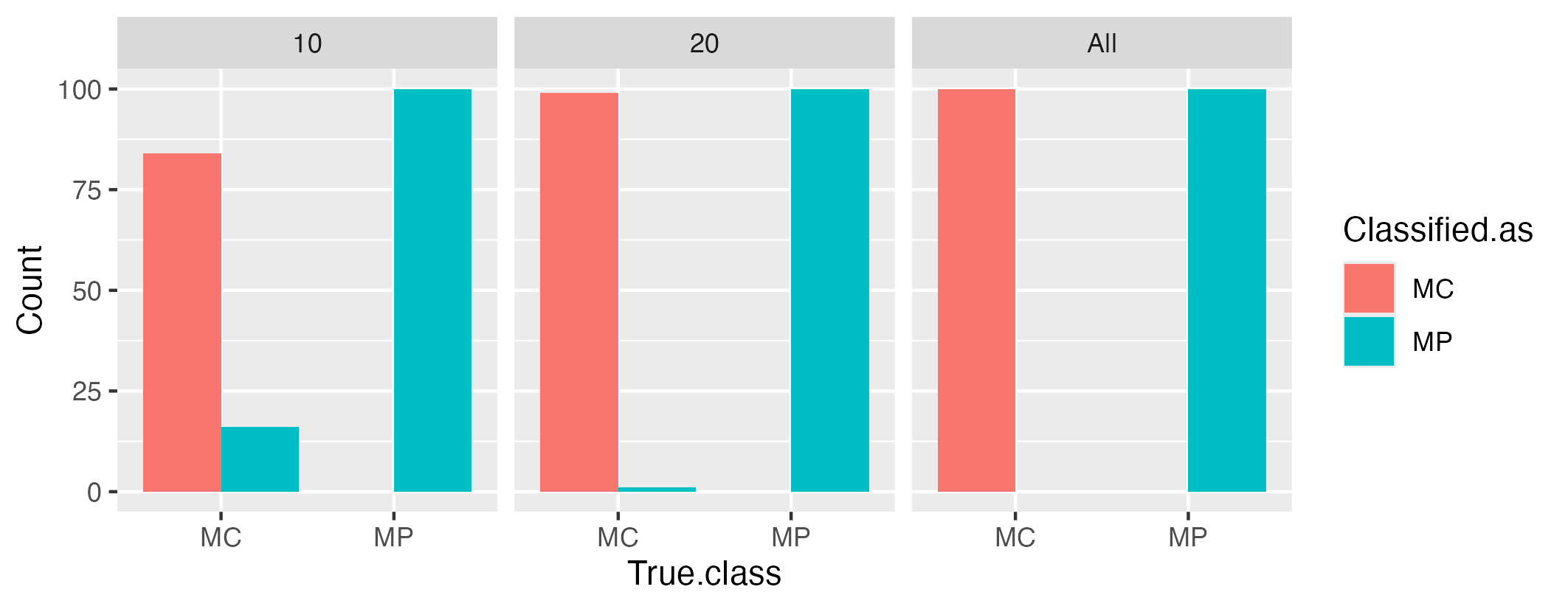}
    \end{minipage}
    \begin{minipage}{0.03\linewidth}\centering
        \rotatebox[origin=center]{90}{Both}
    \end{minipage}
    \begin{minipage}{0.93\linewidth}\centering
        \includegraphics[height=5.5cm, width=12.5cm]{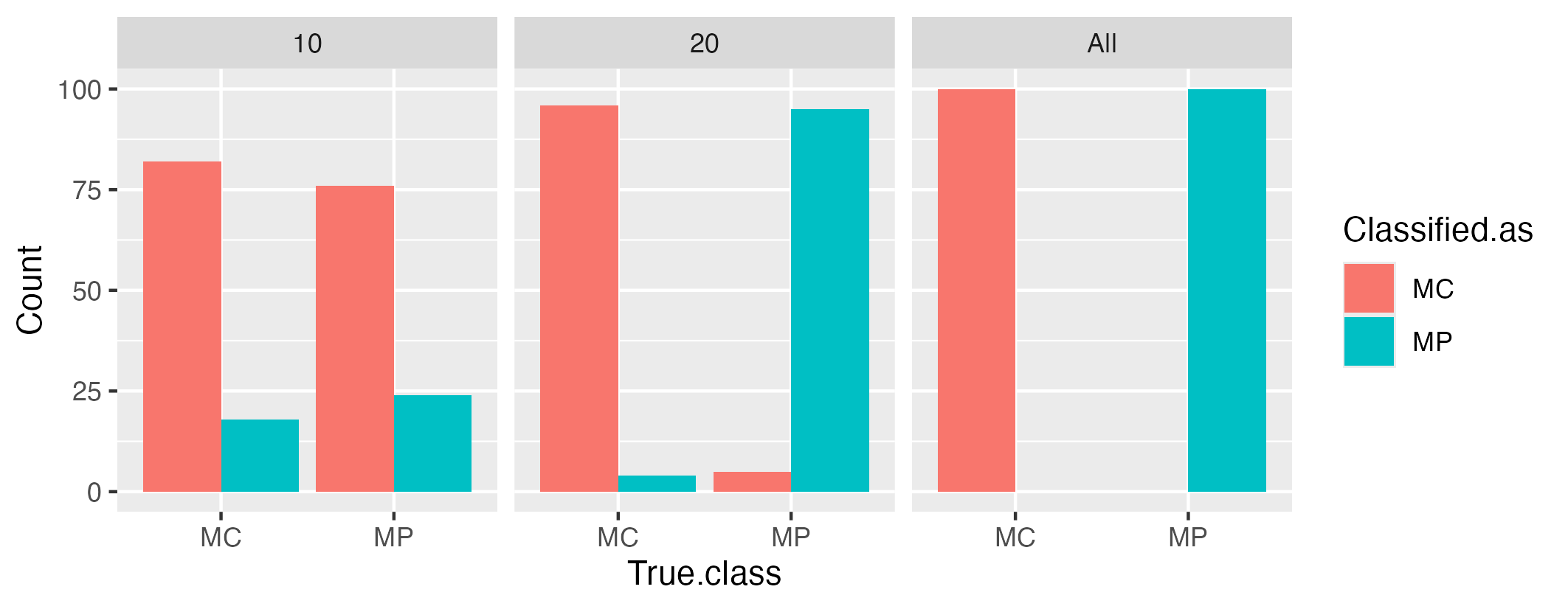}
    \end{minipage}
    \caption{Histograms of hierarchical clustering classification accuracy using only the ratio, only the curvature and both ratio and curvature for discrimination when using a sample of 10, 20, and 'All' components, respectively. Misclassification rates are 26\%, 25\% and 0\% for 10, 20 and 'All' components, respectively, when using only the ratio, 8\%, 0.5\% and 0\% when using only the curvature and 47\%, 4.5\% and 0\% when using both characteristics for a sample of 100 realisations that were osculated by a disc of radius $r=5$.}
    \label{fig:hc_100_best_tissues_5}
\end{figure}

\section{Discussion}
We have proposed both supervised and unsupervised methods for classification of random set realisations. The methods rely on the similarity measure in the form of the convex combination of $\mathcal N$-distances between the distributions of the $C$-function describing the curvature of the boundary of the connected components and $\mathcal N$-distance between distributions of $P/A$-ratios  of the connected components in the realisations. The obtained similarity between realisations was an input for supervised clustering based on the highest estimated posterior probability, and $k$-medoids and hierarchical clustering algorithms. The simulation study performed on the random set models showed satisfactory misclassification rates, which are lowest when using the maximum number of connected components sampled from the realisations.

The maximum misclassification rates for each characteristic separately is higher, while it decreases when considering both characteristics, confirming that combining curvature and perimeter–area information leads to a more reliable classification of random set realisations. The simulation results indicate that the proposed similarity-based classifiers perform robustly across different random set models, with accuracy improving as the number of connected components increases. Moreover, the application to histological images demonstrates that the approach can effectively distinguish between mastopathy and mammary cancer tissue, highlighting its potential for use in medical image analysis.

Overall, the study establishes a general framework for classifying realisations of random sets using functional characteristics and $\mathcal N$-distances as similarity measures. The combination of supervised and unsupervised approaches provides flexibility for both labelled and unlabelled data. Future work may focus on extending the methodology to higher-dimensional random sets, exploring additional geometric or topological features, and investigating the theoretical properties of the proposed classifiers, including their asymptotic behaviour and computational optimisation for large-scale~datasets.

\bmhead{Acknowledgements}

The research of B. Radović and K. Helisov\'a was supported by the Grant Agency of the Czech Technical University in Prague, grant No. SGS25/050/OHK3/1T/13.

Computational resources were provided by the e-INFRA CZ project (ID:90254), supported by the Ministry of Education, Youth and Sports of the Czech Republic.

\section*{Declarations}

The authors certify that they have no affiliation with or involvement in any organization or entity that has a financial or non-financial interest in the topics or materials discussed in this manuscript.

\section*{Supplementary Material}
Supplementary Material provides additional results for both simulated and real medical data osculated using a circle with radius $r=3$, as well as further results for $r=5$.


\end{document}


\title[Classification of Realisations of Random Sets]{Supplementary Material}
\title{\centering Supplementary Material \\Classification of Realisations of Random Sets}

\author*[1]{\fnm{Bogdan} \sur{Radovi\'c}}\email{
radovbog@fel.cvut.cz}

\author[2]{\fnm{Vesna} \sur{Gotovac \DJ oga\v{s}}}\email{vgotovac@pmfst.hr}

\author[1]{\fnm{Kate\v{r}ina} \sur{Helisov\'a}}\email{heliskat@fel.cvut.cz}

\affil*[1]{\orgdiv{Department of Mathematics}, \orgname{Faculty of Electrical Engineering, Czech Technical University in Prague}, \orgaddress{\street{Technick\'{a} 2}, \city{Prague}, \postcode{166 27}, \country{Czech~Republic}}}

\affil[2]{\orgdiv{Department of Mathematics}, \orgname{Faculty of Science, University of Split}, \orgaddress{\street{Ru\dj{}era Bo\v{s}kovi\'{c}a 33}, \city{Split}, \postcode{21000}, \country{Croatia}}}

\maketitle

\section{Simulation study}
This section presents additional classification results for the simulated data. 

\subsection{Supervised classification}

\begin{figure}[ht]
    \centering
    \begin{minipage}{0.03\linewidth}\centering
        \rotatebox[origin=center]{90}{Ratio}
    \end{minipage}
    \begin{minipage}{0.93\linewidth}\centering
        \includegraphics[height=5.5cm, width=12.5cm]{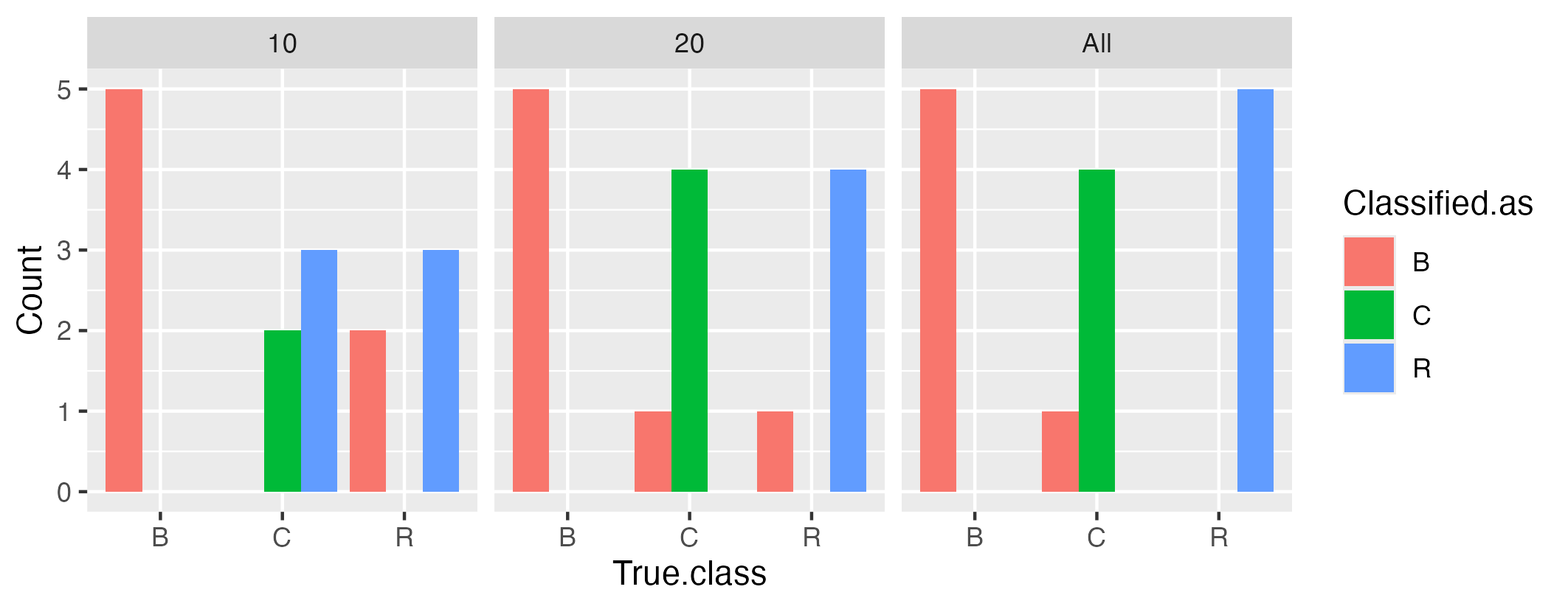}
    \end{minipage}
    \begin{minipage}{0.03\linewidth}\centering
        \rotatebox[origin=center]{90}{Curvature}
    \end{minipage}
    \begin{minipage}{0.93\linewidth}\centering
        \includegraphics[height=5.5cm, width=12.5cm]{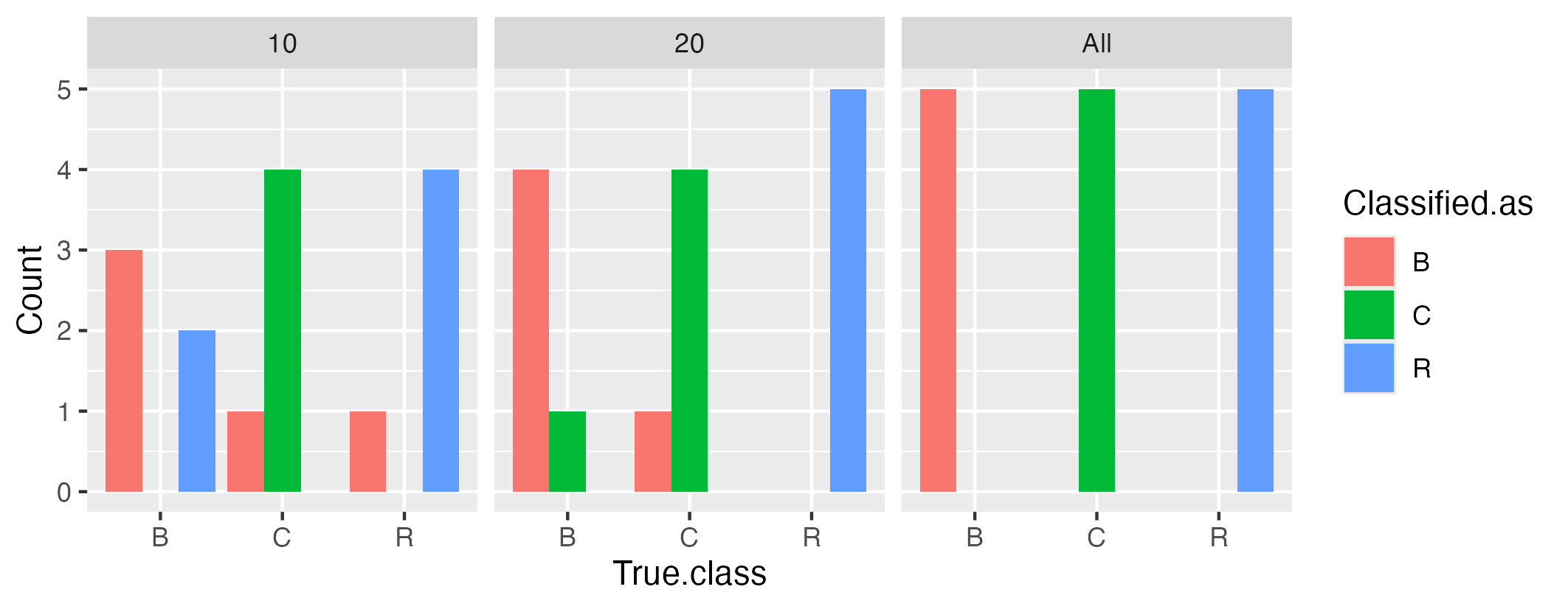}
    \end{minipage}
    \begin{minipage}{0.03\linewidth}\centering
        \rotatebox[origin=center]{90}{Both}
    \end{minipage}
    \begin{minipage}{0.93\linewidth}\centering
        \includegraphics[height=5.5cm, width=12.5cm]{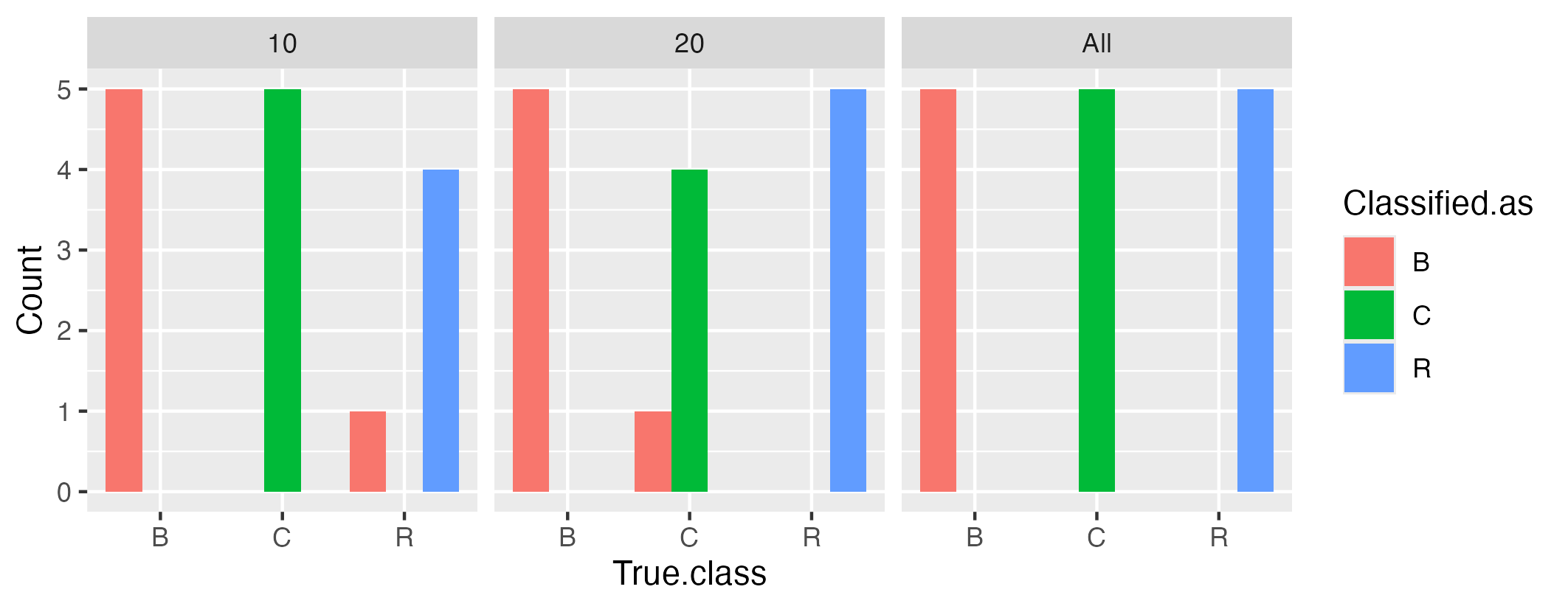}
    \end{minipage}
    \caption{Histograms of $k$-nearest neighbours classification accuracy using only the ratio, only the curvature and both ratio and curvature for discrimination when using a sample of 10, 20, and 'All' components, respectively. Misclassification rates are 11.1\%, 4.4\% and 2.2\% for 10, 20 and 'All' components, respectively, when using only the ratio, 8.9\%, 4.4\% and 0\% when using only the curvature, and 2.2\%, 2.2\% and 0\% when using both characteristics for a sample of 20 realisations that were osculated by a disc of radius $r=5$.}
    \label{fig:knn_20_best_5}
\end{figure}

\begin{figure}[!ht]
    \centering
    \begin{minipage}{0.03\textwidth}\centering
        \rotatebox[origin=center]{90}{Ratio}
    \end{minipage}
    \begin{minipage}{0.93\textwidth}\centering
        \includegraphics[height=5.5cm, width=12.5cm]{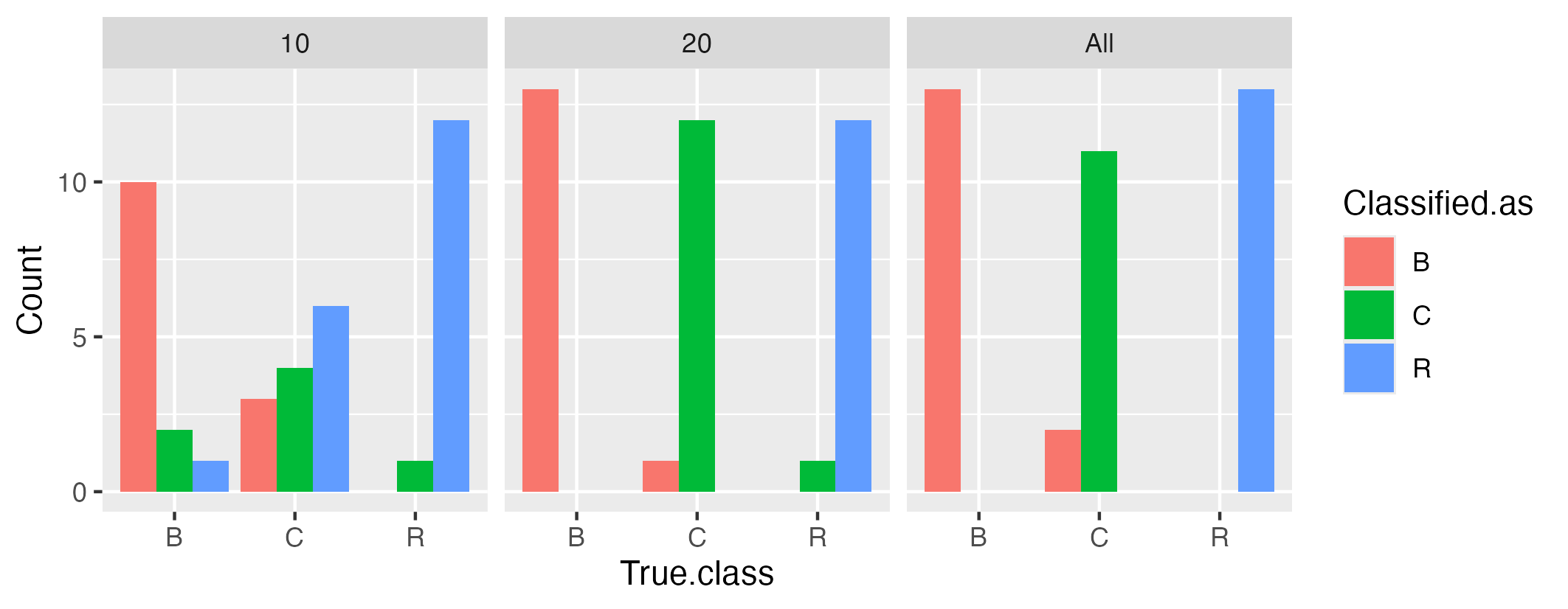}
    \end{minipage}
    \begin{minipage}{0.03\linewidth}\centering
        \rotatebox[origin=center]{90}{Curvature}
    \end{minipage}
    \begin{minipage}{0.93\linewidth}\centering
        \includegraphics[height=5.5cm, width=12.5cm]{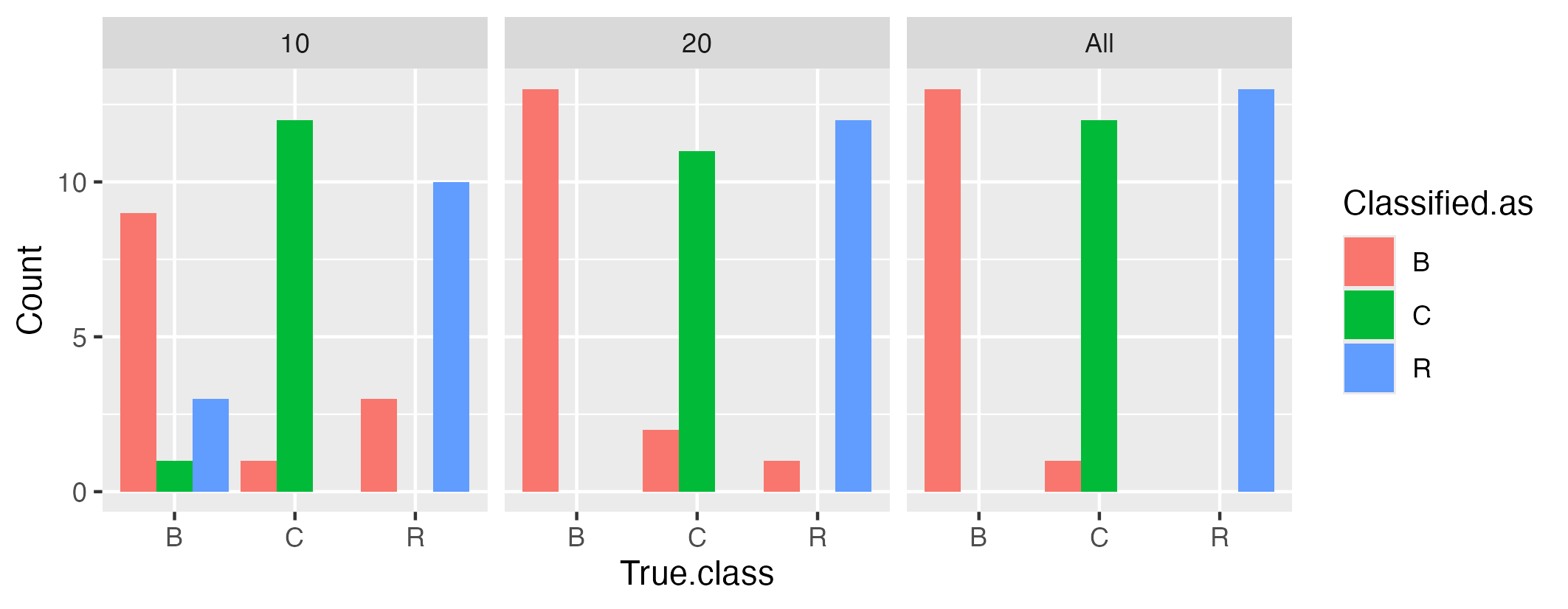}
    \end{minipage}
    \begin{minipage}{0.03\linewidth}\centering
        \rotatebox[origin=center]{90}{Both}
    \end{minipage}
    \begin{minipage}{0.93\linewidth}\centering
        \includegraphics[height=5.5cm, width=12.5cm]{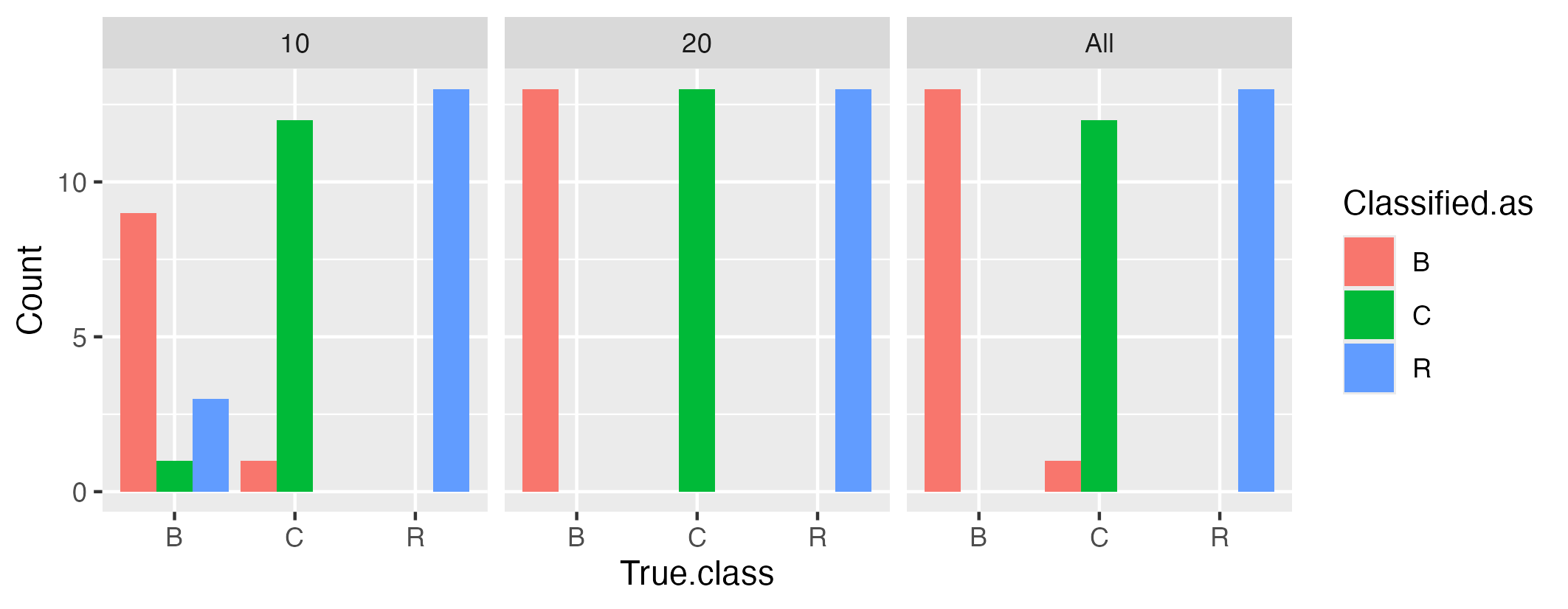}
    \end{minipage}
    \caption{Histograms of $k$-nearest neighbours classification accuracy using only the ratio, only the curvature and both ratio and curvature for discrimination when using a sample of 10, 20, and 'All' components, respectively. Misclassification rates are 11.7\%, 1.8\% and 1.8\% for 10, 20 and 'All' components, respectively, when using only the ratio, 7.2\%, 2.7\% and 0.9\% when using only the curvature, and 4.5\%, 0\% and 0.9\% when using both characteristics for a sample of 50 realisations that were osculated by a disc of radius $r=5$.}
    \label{fig:knn_50_best_5}
\end{figure}

Histograms of classification accuracy using $k$-nearest neighbours algorithm when 20 and 50 realisations are considered, on the data obtained using the osculating circle with radius $r=5$ are shown in Figures \ref{fig:knn_20_best_5} and \ref{fig:knn_50_best_5}, respectively.

\begin{figure}[!ht]
    \centering
    \begin{minipage}{0.03\linewidth}\centering
        \rotatebox[origin=center]{90}{Ratio}
    \end{minipage}
    \begin{minipage}{0.93\linewidth}\centering
        \includegraphics[height=5.5cm, width=12.5cm]{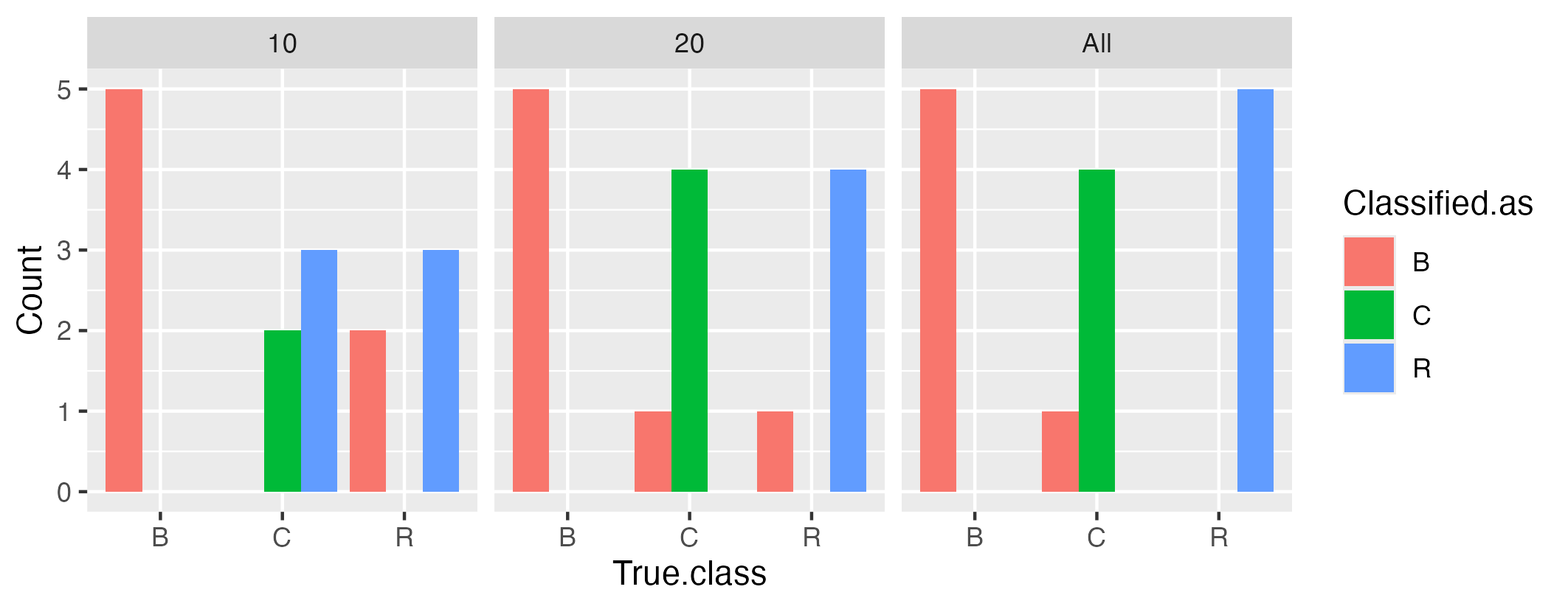}
    \end{minipage}
    \begin{minipage}{0.03\linewidth}\centering
        \rotatebox[origin=center]{90}{Curvature}
    \end{minipage}
    \begin{minipage}{0.93\linewidth}\centering
        \includegraphics[height=5.5cm, width=12.5cm]{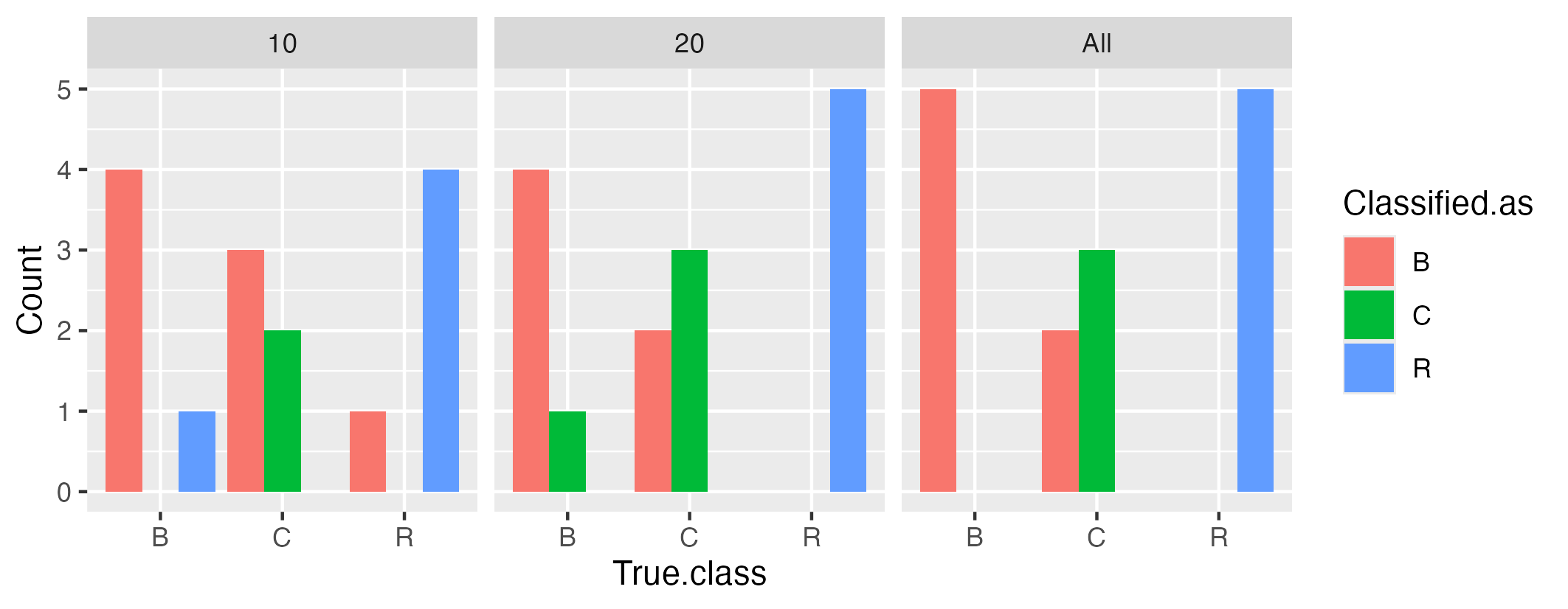}
    \end{minipage}
    \begin{minipage}{0.03\linewidth}\centering
        \rotatebox[origin=center]{90}{Both}
    \end{minipage}
    \begin{minipage}{0.93\linewidth}\centering
        \includegraphics[height=5.5cm, width=12.5cm]{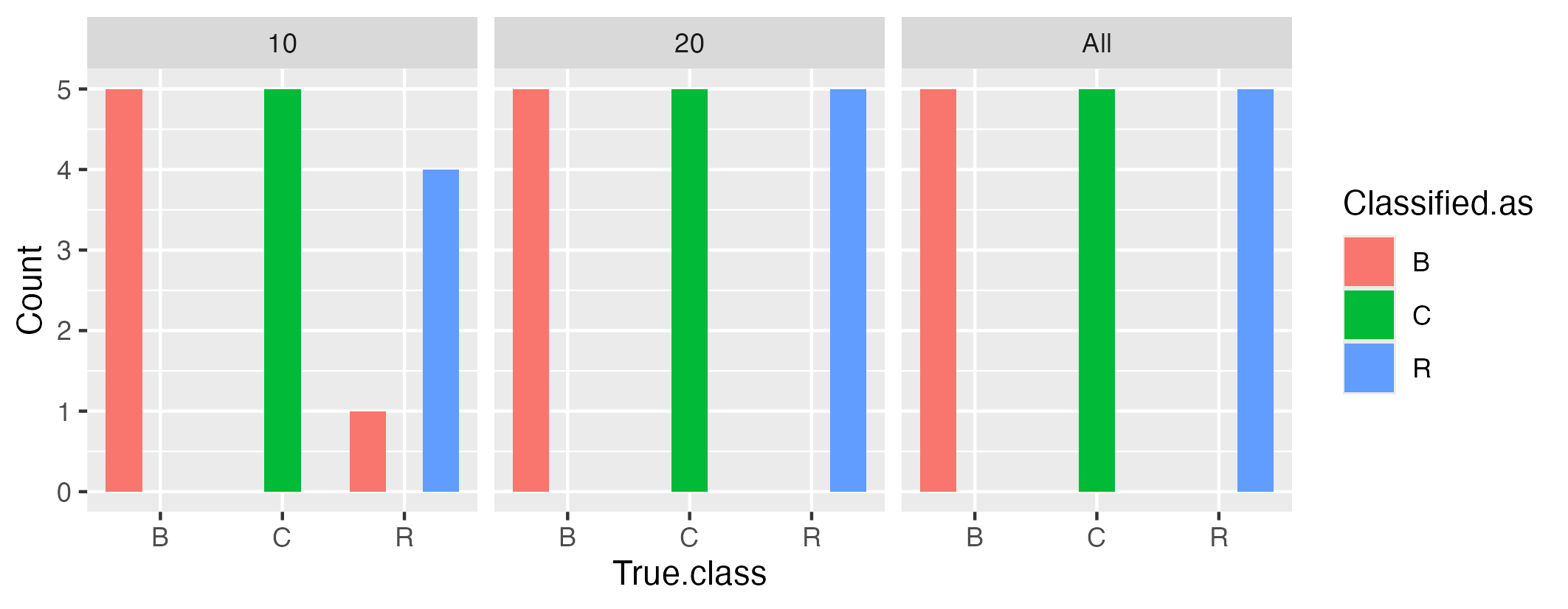}
    \end{minipage}
    \caption{Histograms of $k$-nearest neighbours classification accuracy using only the ratio, only the curvature and both ratio and curvature for discrimination when using a sample of 10, 20, and 'All' components, respectively. Misclassification rates are 11.1\%, 4.4\% and 2.2\% for 10, 20 and 'All' components, respectively, when using only the ratio, 11.1\%, 6.7\% and 4.4\% when using only the curvature, and 2.2\%, 0\% and 0\% when using both characteristics for a sample of 20 realisations that were osculated by a disc of radius $r=3$.}
    \label{fig:knn_20_best_3}
\end{figure}
\begin{figure}[!ht]
    \centering
    \begin{minipage}{0.03\linewidth}\centering
        \rotatebox[origin=center]{90}{Ratio}
    \end{minipage}
    \begin{minipage}{0.93\linewidth}\centering
        \includegraphics[height=5.5cm, width=12.5cm]{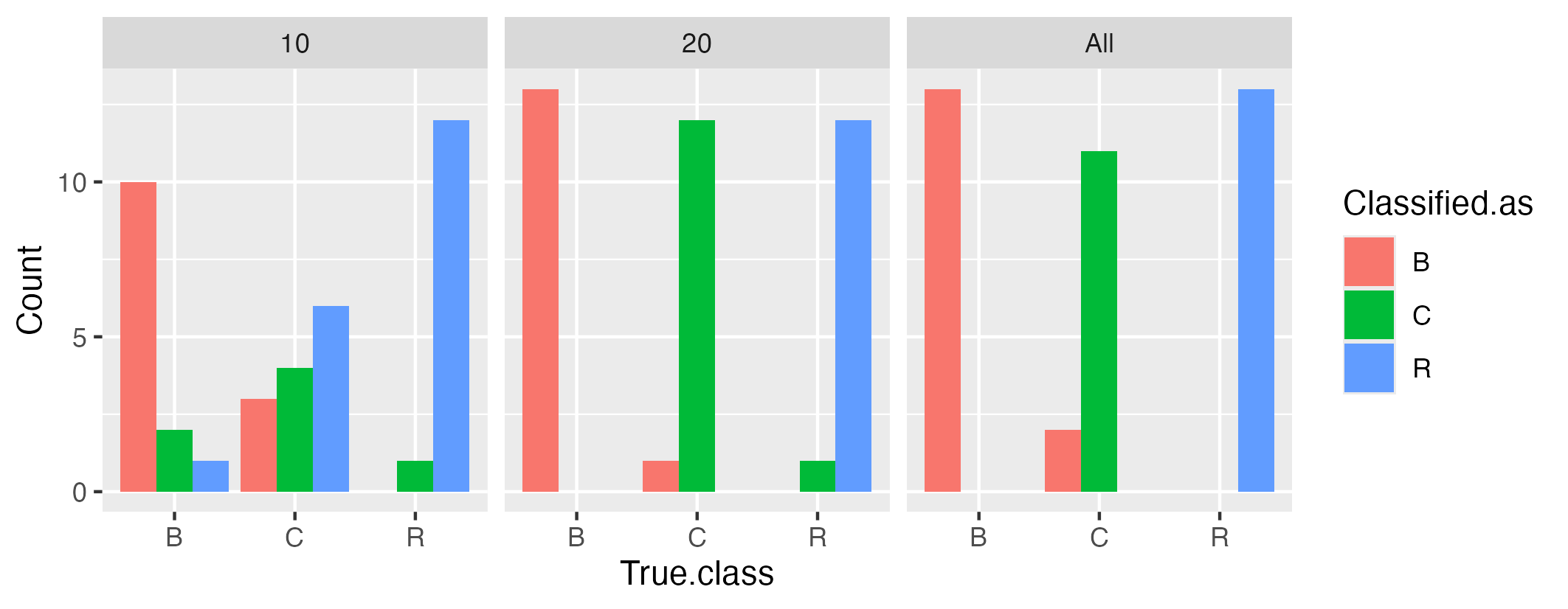}
    \end{minipage}
    \begin{minipage}{0.03\linewidth}\centering
        \rotatebox[origin=center]{90}{Curvature}
    \end{minipage}
    \begin{minipage}{0.93\linewidth}\centering
        \includegraphics[height=5.5cm, width=12.5cm]{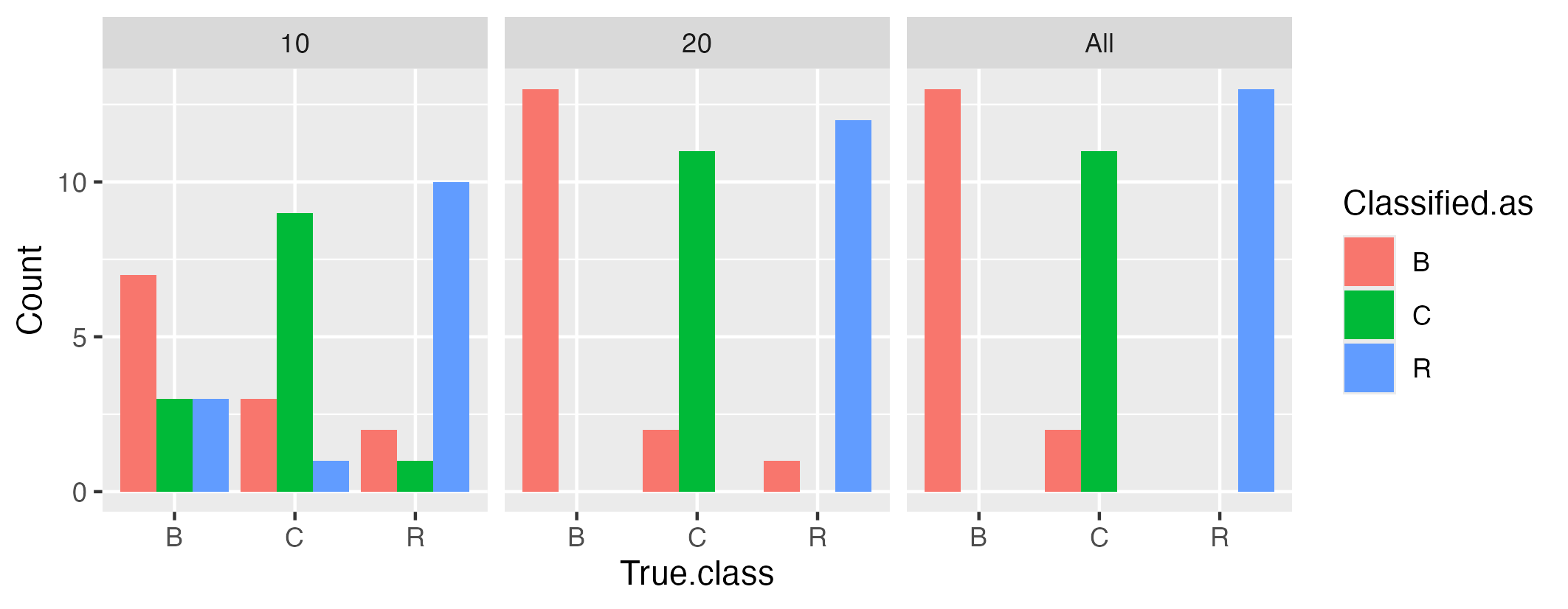}
    \end{minipage}
    \begin{minipage}{0.03\linewidth}\centering
        \rotatebox[origin=center]{90}{Both}
    \end{minipage}
    \begin{minipage}{0.93\linewidth}\centering
        \includegraphics[height=5.5cm, width=12.5cm]{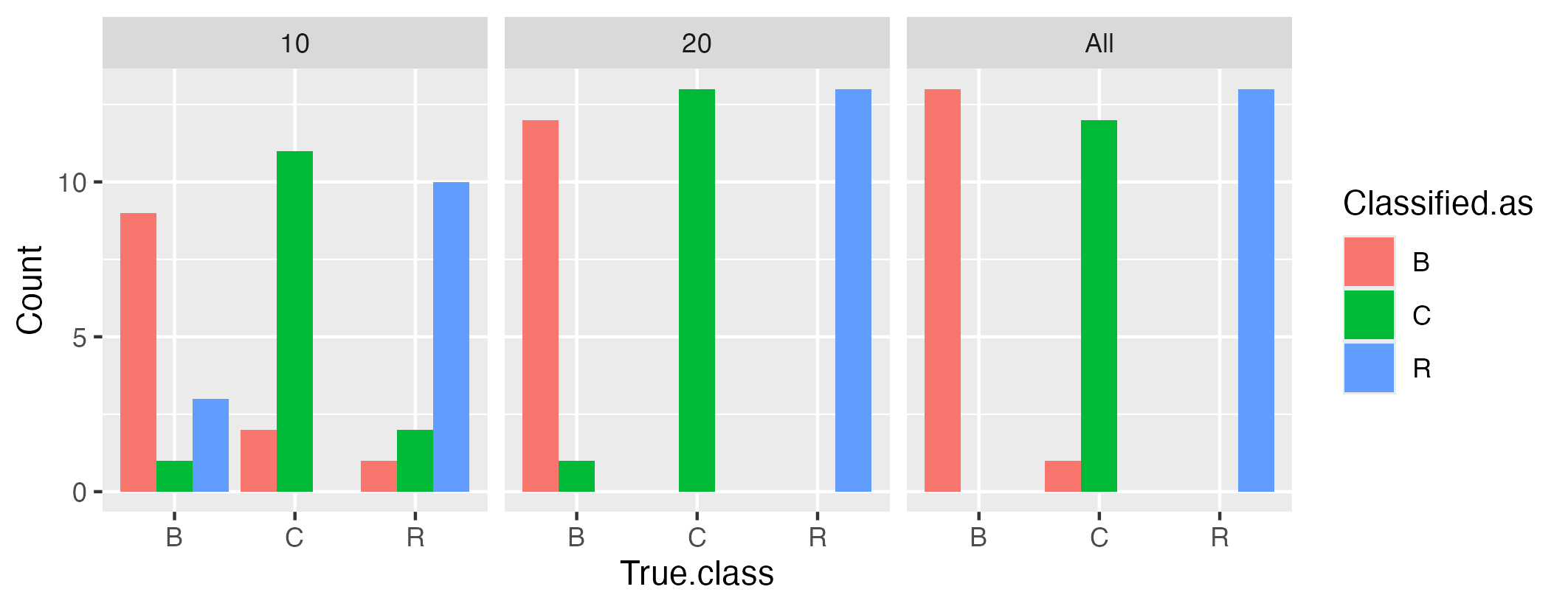}
    \end{minipage}
    \caption{Histograms of $k$-nearest neighbours classification accuracy using only the ratio, only the curvature and both ratio and curvature for discrimination when using a sample of 10, 20, and 'All' components, respectively. Misclassification rates are 11.7\%, 1.8\% and 1.8\% for 10, 20 and 'All' components, respectively, when using only the ratio, 11.7\%, 2.7\% and 1.8\% when using only the curvature, and 8.1\%, 0.9\% and 0.9\% when using both characteristics for a sample of 50 realisations that were osculated by a disc of radius $r=3$.}
    \label{fig:knn_50_best_3}
\end{figure}
\begin{figure}[!ht]
    \centering
    \begin{minipage}{0.03\linewidth}\centering
        \rotatebox[origin=center]{90}{Ratio}
    \end{minipage}
    \begin{minipage}{0.93\linewidth}\centering
        \includegraphics[height=5.5cm, width=12.5cm]{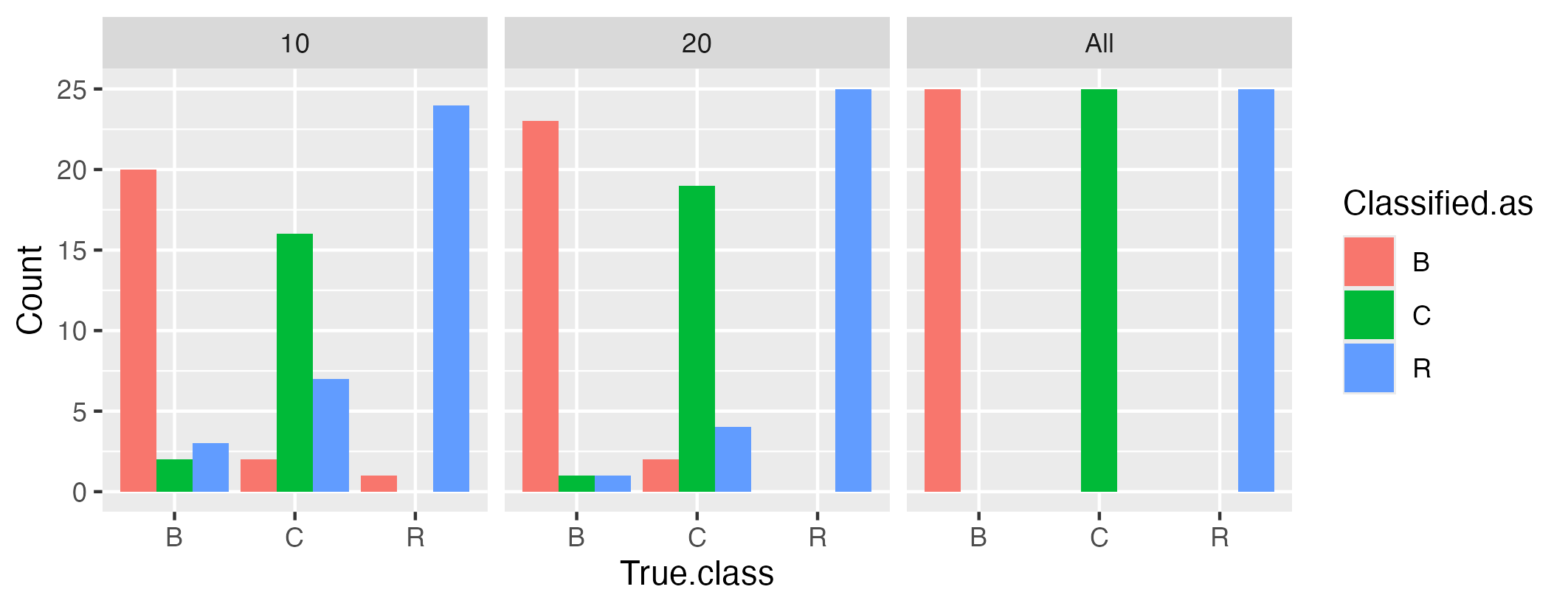}
    \end{minipage}
    \begin{minipage}{0.03\linewidth}\centering
        \rotatebox[origin=center]{90}{Curvature}
    \end{minipage}
    \begin{minipage}{0.93\linewidth}\centering
        \includegraphics[height=5.5cm, width=12.5cm]{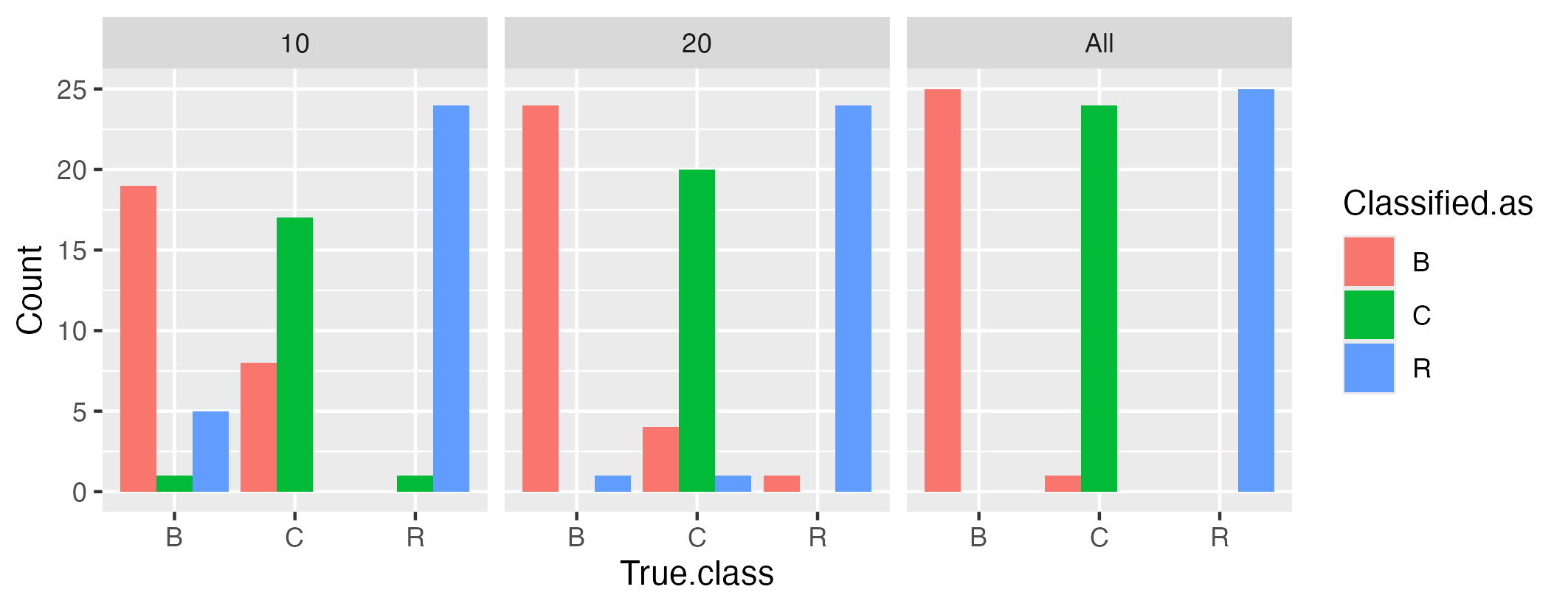}
    \end{minipage}
    \begin{minipage}{0.03\linewidth}\centering
        \rotatebox[origin=center]{90}{Both}
    \end{minipage}
    \begin{minipage}{0.93\linewidth}\centering
        \includegraphics[height=5.5cm, width=12.5cm]{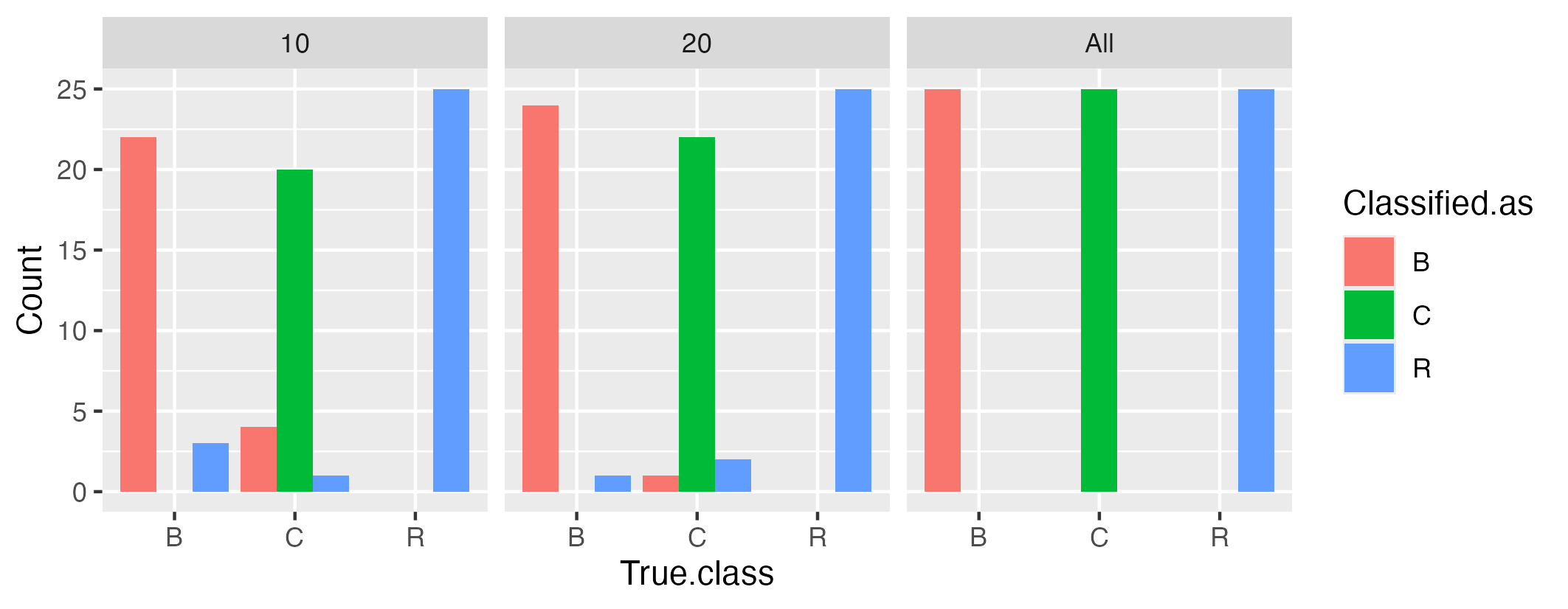}
    \end{minipage}
    \caption{Histograms of $k$-nearest neighbours classification accuracy using only the ratio, only the curvature and both ratio and curvature for discrimination when using a sample of 10, 20, and 'All' components, respectively. Misclassification rates are 6.7\%, 3.6\% and 0\% for 10, 20 and 'All' components, respectively, when using only the ratio, 6.7\%, 3.1\% and 0.4\% when using only the curvature, and 3.6\%, 1.8\% and 0\% when using both characteristics for a sample of 100 realisations that were osculated by a disc of radius $r=3$.}
    \label{fig:knn_100_best_3}
\end{figure}
\begin{figure}[!ht]
    \centering
    \begin{minipage}{0.03\linewidth}\centering
        \rotatebox[origin=center]{90}{20 Realisations}
    \end{minipage}
    \begin{minipage}{0.93\linewidth}\centering
        \includegraphics[height=5.5cm, width=12.3cm]{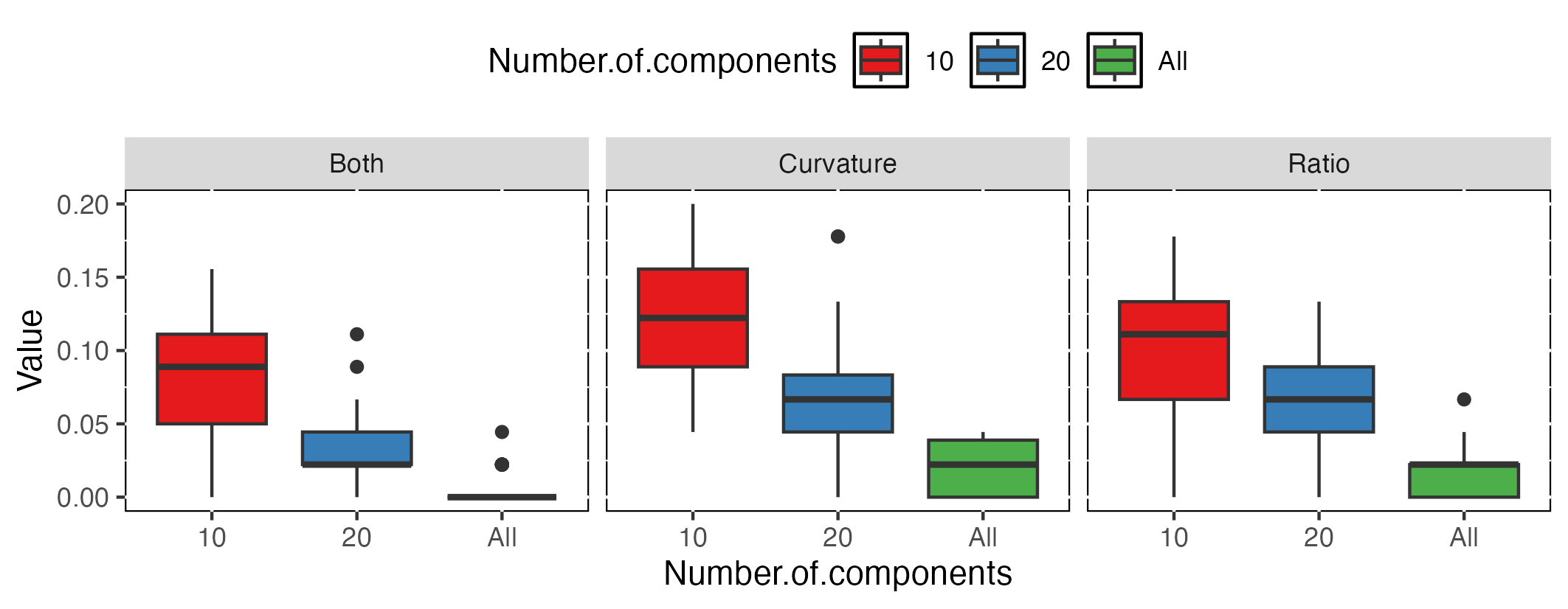}
    \end{minipage}
    \begin{minipage}{0.03\linewidth}\centering
        \rotatebox[origin=center]{90}{50 Realisations}
    \end{minipage}
    \begin{minipage}{0.93\linewidth}\centering
        \includegraphics[height=5.5cm, width=12.3cm]{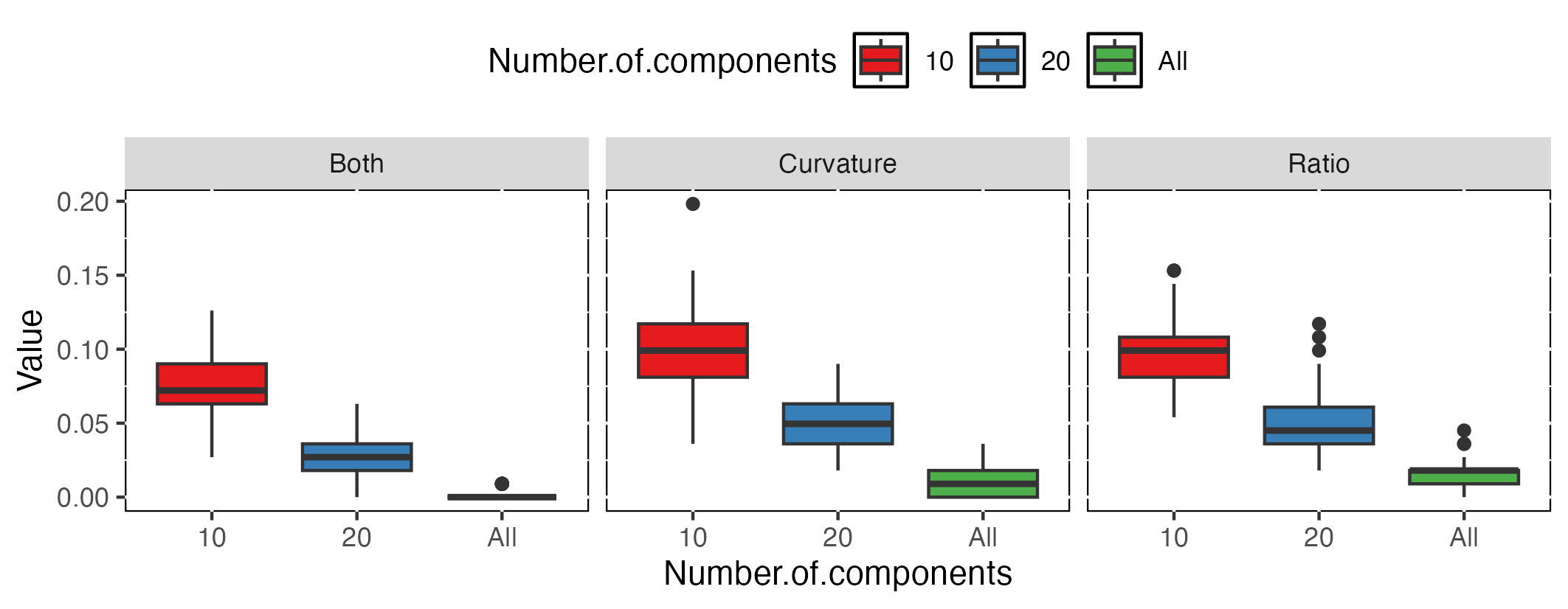}
    \end{minipage}
    \begin{minipage}{0.03\linewidth}\centering
        \rotatebox[origin=center]{90}{100 Realisations}
    \end{minipage}
    \begin{minipage}{0.93\linewidth}\centering
        \includegraphics[height=5.5cm, width=12.3cm]{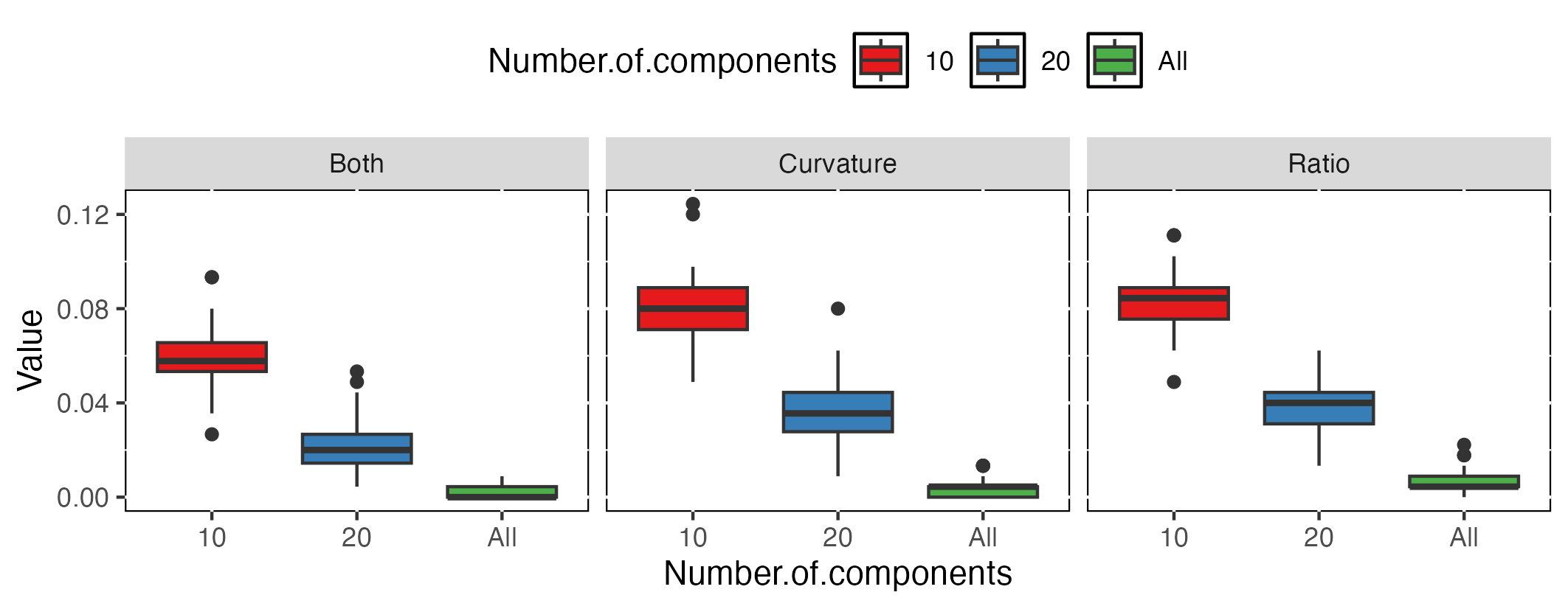}
    \end{minipage}
    \caption{Boxplots of misclassification rate for 50 runs of $k$-nearest neighbours algorithm when considering samples of 20 (top), 50 (middle) and 100 (bottom) realisations using both ratio and curvature, only the curvature and only the ratio for discrimination, respectively. For each setting, misclassification rates for different number of components considered (namely 10, 20 and 'All') are shown. Note that the characteristics were obtained using an osculating disc of radius $r=3$ on the simulated data.}
    \label{fig:knn_box_3}
\end{figure}

The results presented in Figures \ref{fig:knn_20_best_3}, \ref{fig:knn_50_best_3} and \ref{fig:knn_100_best_3} represent the histograms of classification accuracy for supervised classification using $k$-nearest neighbours algorithm when 20, 50 and 100 realisations are considered, respectively, on the data obtained using the osculating circle with radius $r=3$. The corresponding boxplot of misclassification rate, obtained after 50 runs of the algorithm, can be seen in Figure \ref{fig:knn_box_3}.

\subsection{Unupervised classification}
\subsubsection{Non-hierarchical clustering}

\begin{figure}[!ht]
    \centering
    \begin{minipage}{0.03\linewidth}\centering
        \rotatebox[origin=center]{90}{Ratio}
    \end{minipage}
    \begin{minipage}{0.93\linewidth}\centering
        \includegraphics[height=5.5cm, width=12.5cm]{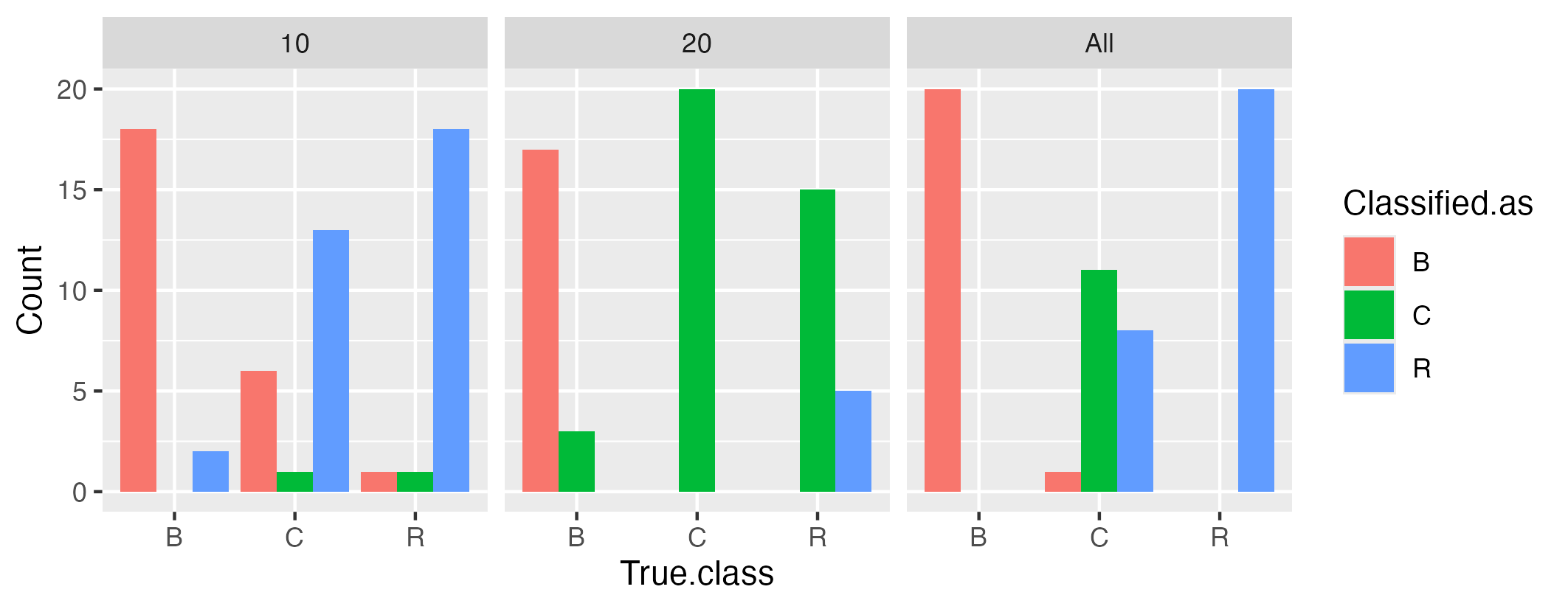}
    \end{minipage}
    \begin{minipage}{0.03\linewidth}\centering
        \rotatebox[origin=center]{90}{Curvature}
    \end{minipage}
    \begin{minipage}{0.93\linewidth}\centering
        \includegraphics[height=5.5cm, width=12.5cm]{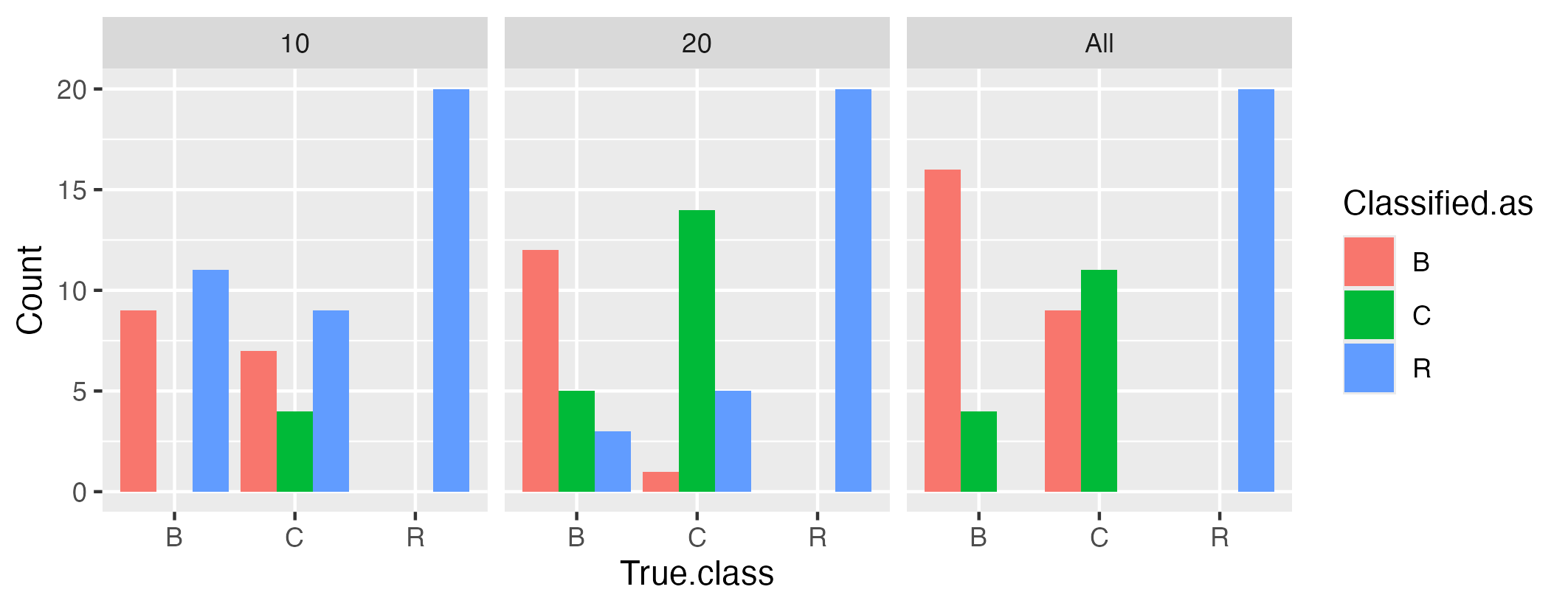}
    \end{minipage}
    \begin{minipage}{0.03\linewidth}\centering
        \rotatebox[origin=center]{90}{Both}
    \end{minipage}
    \begin{minipage}{0.93\linewidth}\centering
        \includegraphics[height=5.5cm, width=12.5cm]{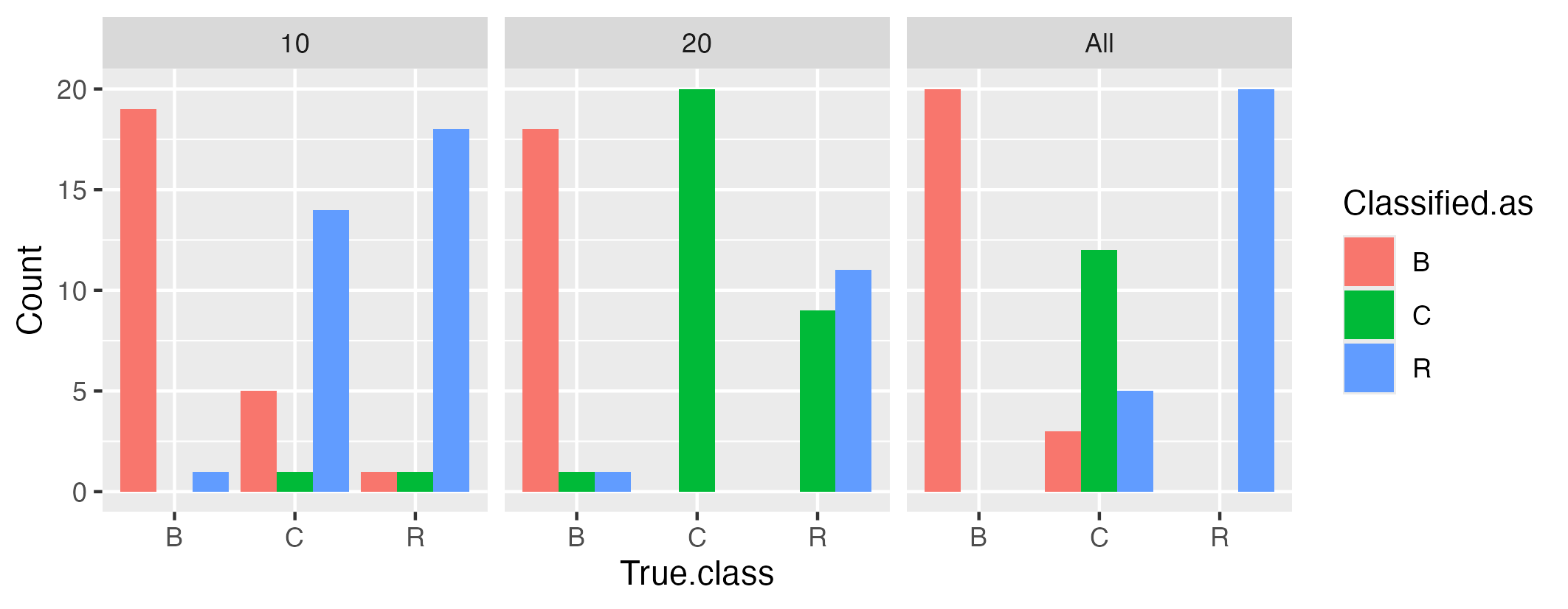}
    \end{minipage}
    \caption{Histograms of $k$-medoids classification accuracy using only the ratio, only the curvature and both ratio and curvature for discrimination when using a sample of 10, 20, and 'All' components, respectively. Misclassification rates are 38.3\%, 30\% and 15\% for 10, 20 and 'All' components, respectively, when using only the ratio, 45\%, 23.3\% and 21.7\% when using only the curvature, and 36.7\%, 18.3\% and 13.3\% when using both characteristics for a sample of 20 realisations that were osculated by a disc of radius $r=5$.}
    \label{fig:kmed_20_best_5}
\end{figure}

\begin{figure}[!ht]
    \centering
    \begin{minipage}{0.03\linewidth}\centering
        \rotatebox[origin=center]{90}{Ratio}
    \end{minipage}
    \begin{minipage}{0.93\linewidth}\centering
        \includegraphics[height=5.5cm, width=12.5cm]{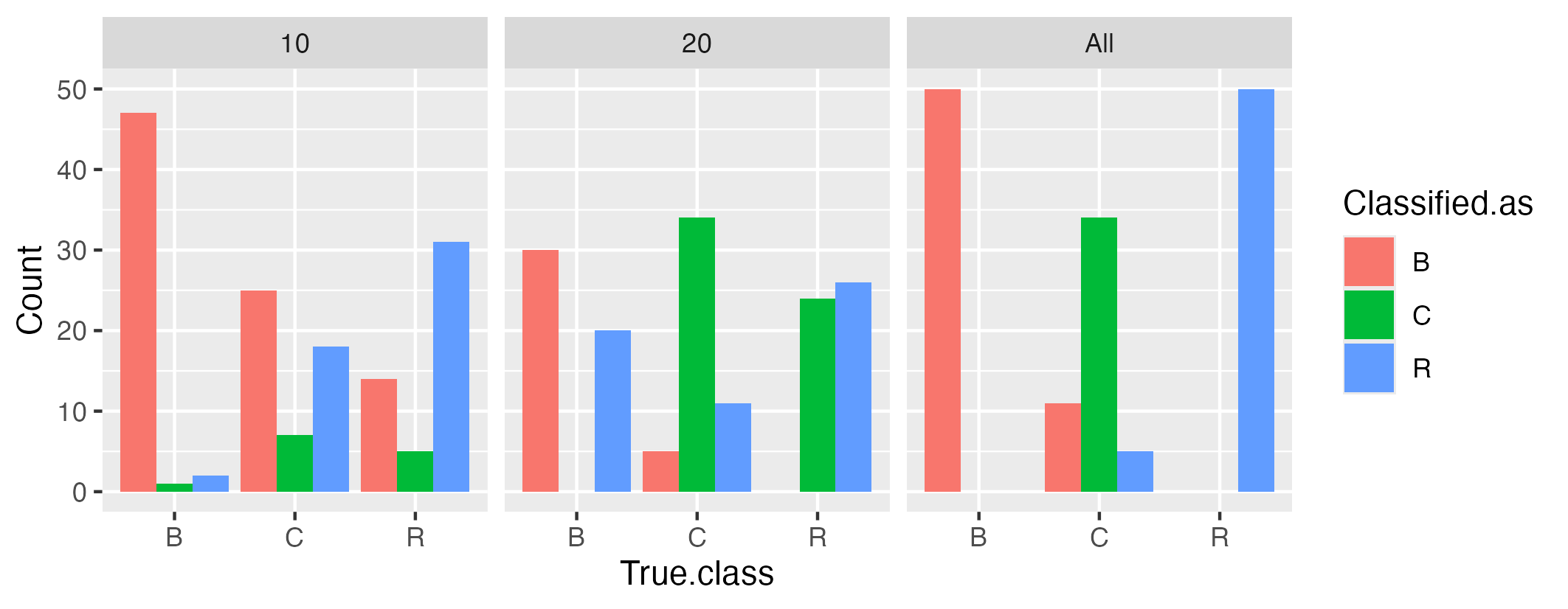}
    \end{minipage}
    \begin{minipage}{0.03\linewidth}\centering
        \rotatebox[origin=center]{90}{Curvature}
    \end{minipage}
    \begin{minipage}{0.93\linewidth}\centering
        \includegraphics[height=5.5cm, width=12.5cm]{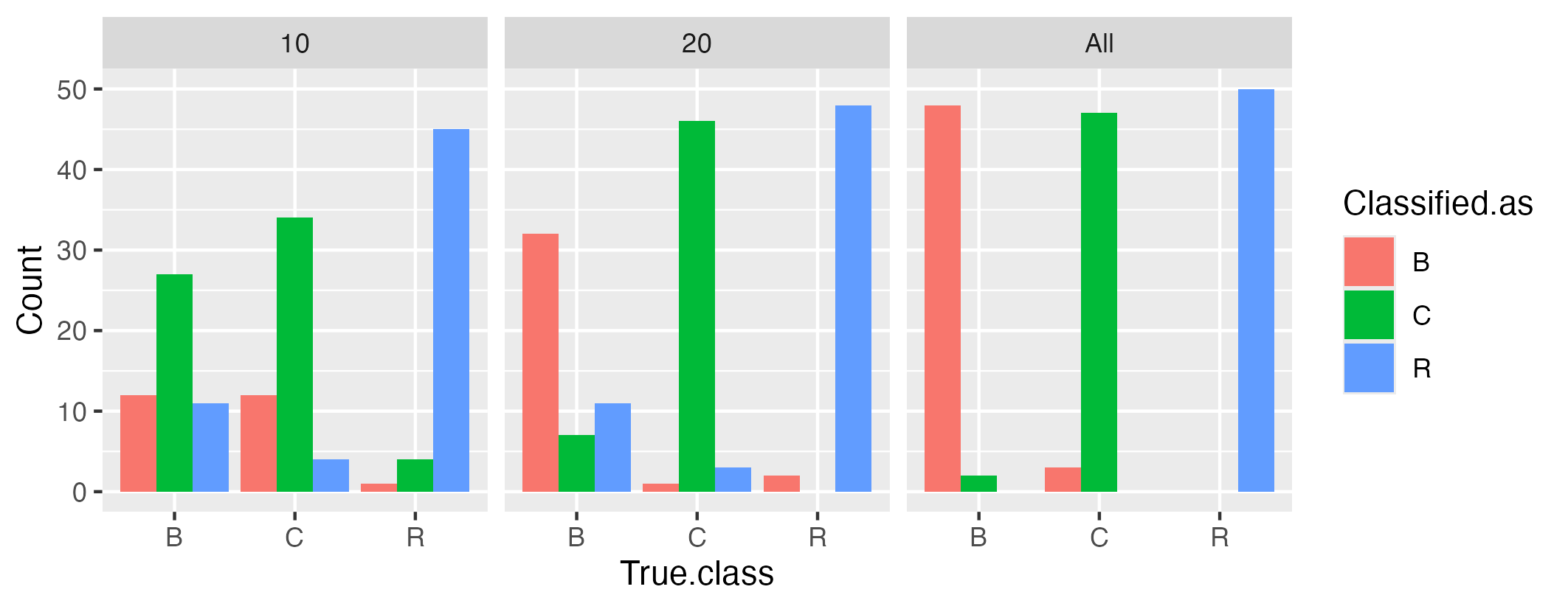}
    \end{minipage}
    \begin{minipage}{0.03\linewidth}\centering
        \rotatebox[origin=center]{90}{Both}
    \end{minipage}
    \begin{minipage}{0.93\linewidth}\centering
        \includegraphics[height=5.5cm, width=12.5cm]{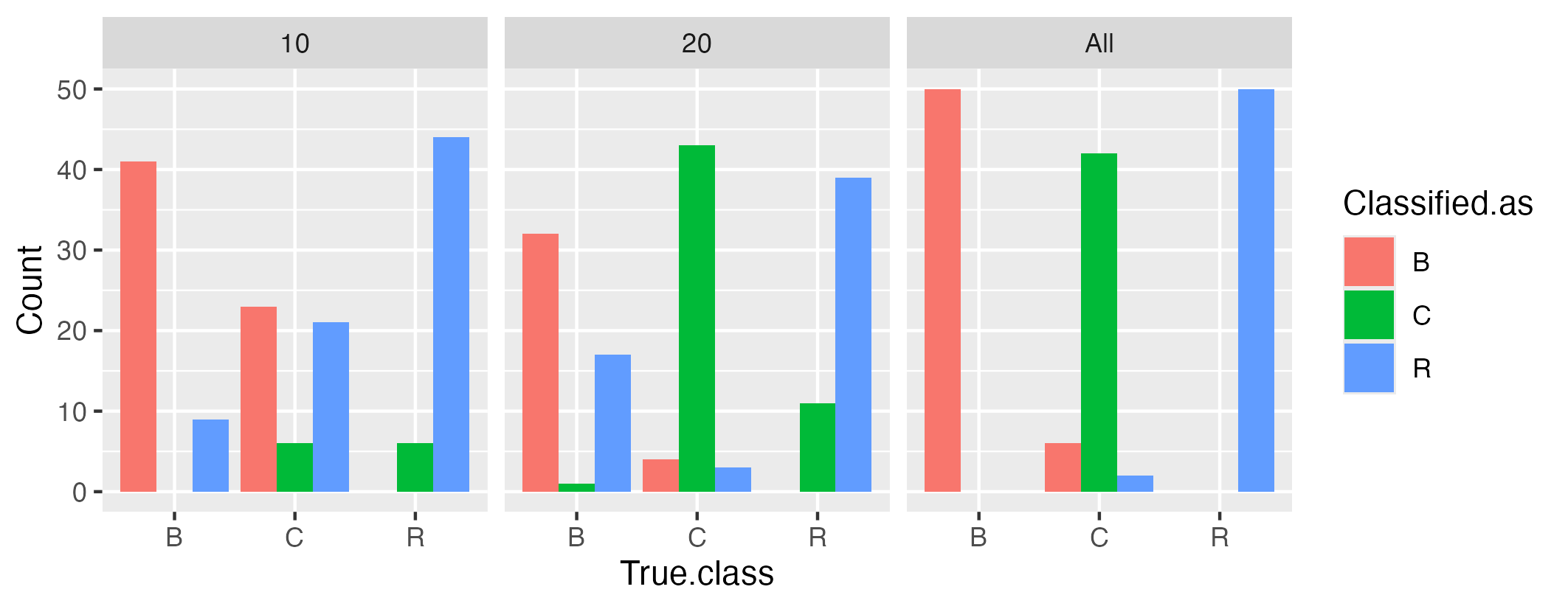}
    \end{minipage}
    \caption{Histograms of $k$-medoids classification accuracy using only the ratio, only the curvature and both ratio and curvature for discrimination when using a sample of 10, 20, and 'All' components, respectively. Misclassification rates are 43.3\%, 40\% and 10.7\% for 10, 20 and 'All' components, respectively, when using only the ratio, 39.3\%, 16\% and 3.3\% when using only the curvature, and 39.3\%, 24\% and 5.3\% when using both characteristics for a sample of 50 realisations that were osculated by a disc of radius $r=5$.}
    \label{fig:kmed_50_best_5}
\end{figure}

Histograms of classification accuracy using $k$-medoids algorithm when 20 and 50 realisations are considered, on the data obtained using the osculating circle with radius $r=5$ are shown in Figures \ref{fig:kmed_20_best_5} and \ref{fig:kmed_50_best_5}, respectively.

\begin{figure}[!ht]
    \centering
    \begin{minipage}{0.03\linewidth}\centering
        \rotatebox[origin=center]{90}{Ratio}
    \end{minipage}
    \begin{minipage}{0.93\linewidth}\centering
        \includegraphics[height=5.5cm, width=12.5cm]{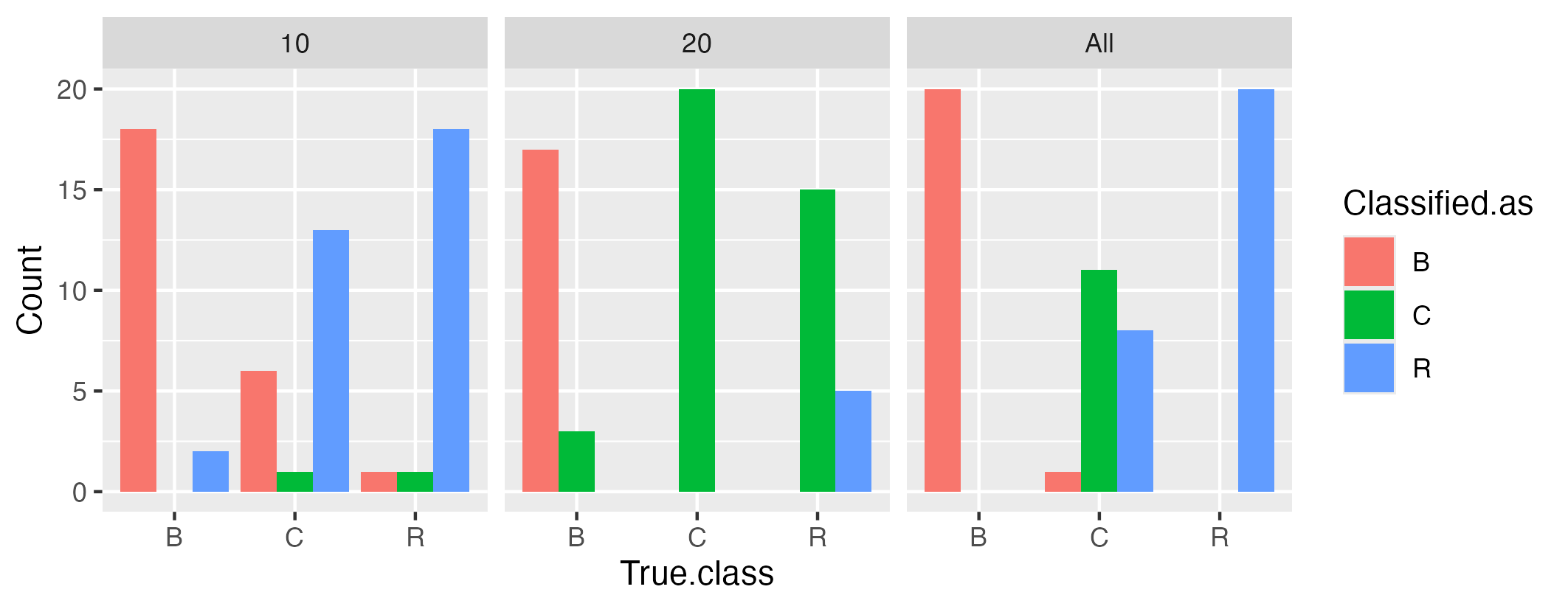}
    \end{minipage}
    \begin{minipage}{0.03\linewidth}\centering
        \rotatebox[origin=center]{90}{Curvature}
    \end{minipage}
    \begin{minipage}{0.93\linewidth}\centering
        \includegraphics[height=5.5cm, width=12.5cm]{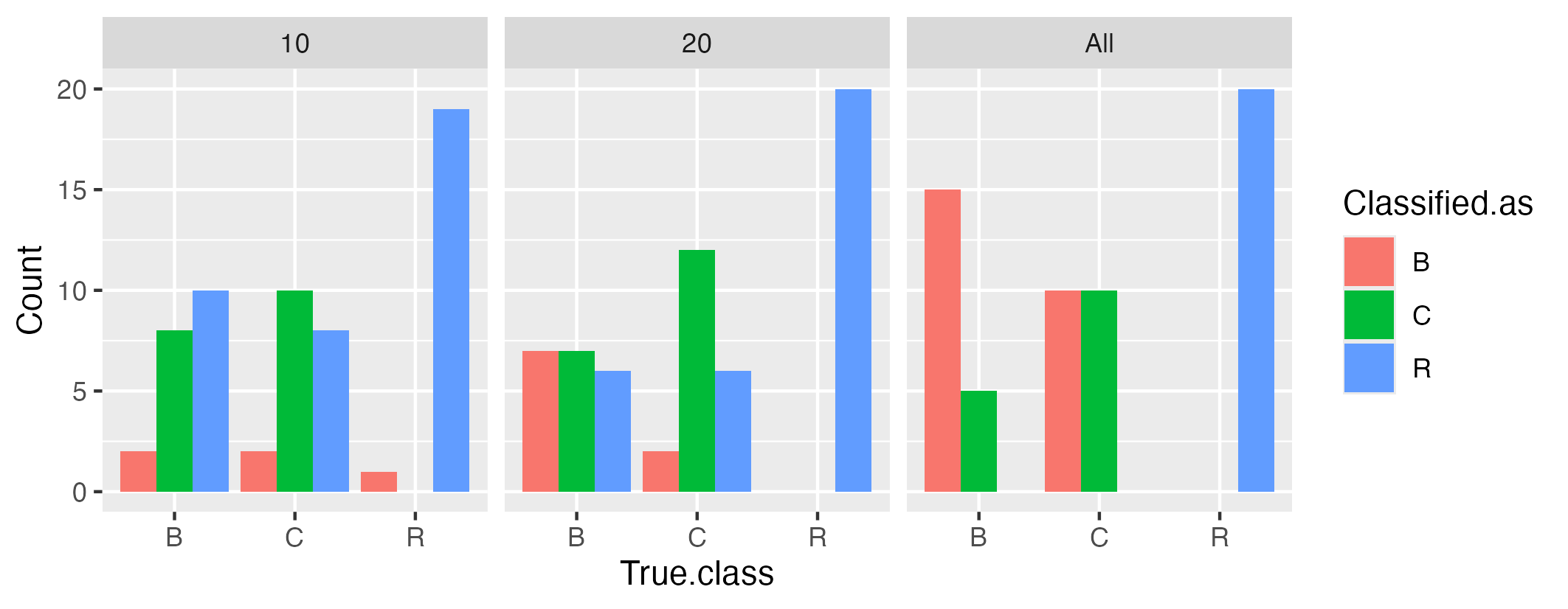}
    \end{minipage}
    \begin{minipage}{0.03\linewidth}\centering
        \rotatebox[origin=center]{90}{Both}
    \end{minipage}
    \begin{minipage}{0.93\linewidth}\centering
        \includegraphics[height=5.5cm, width=12.5cm]{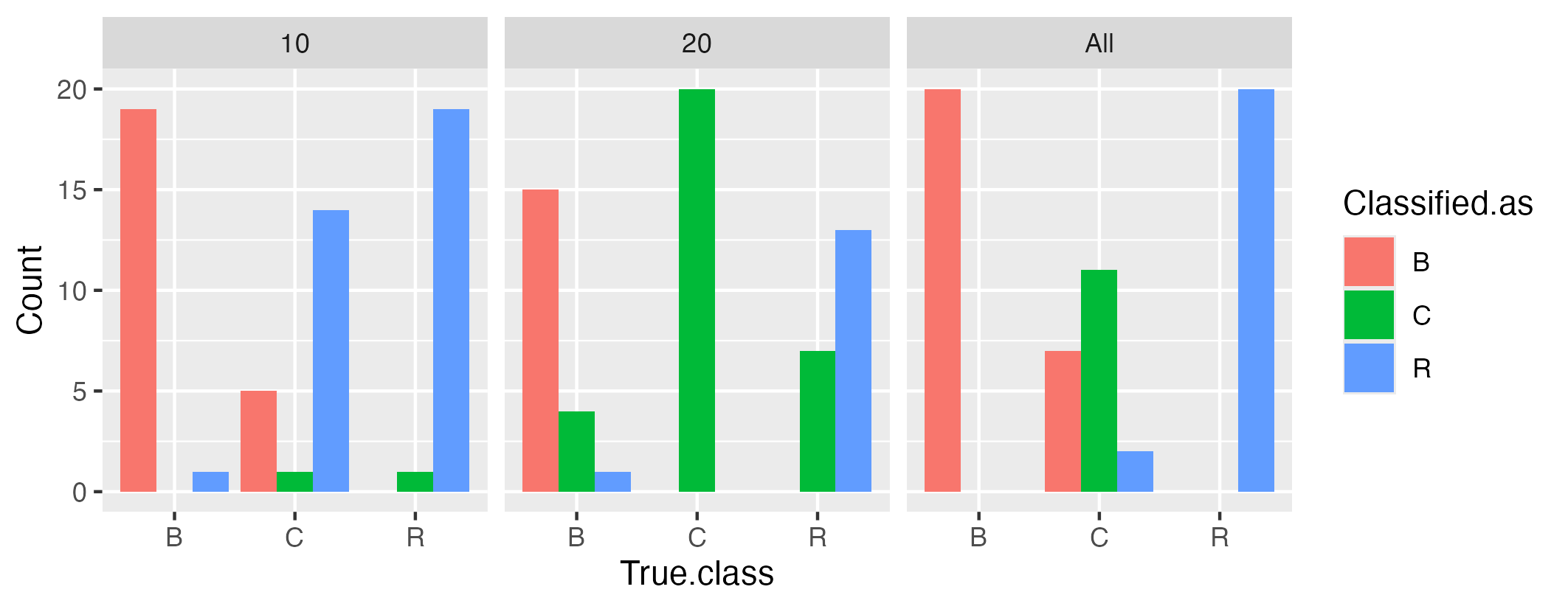}
    \end{minipage}
    \caption{Histograms of $k$-medoids classification accuracy using only the ratio, only the curvature and both ratio and curvature for discrimination when using a sample of 10, 20, and 'All' components, respectively. Misclassification rates are 38.3\%, 30\% and 15\% for 10, 20 and 'All' components, respectively, when using only the ratio, 48.3\%, 35\% and 25\% when using only the curvature, and 35\%, 20\% and 15\% when using both characteristics for a sample of 20 realisations that were osculated by a disc of radius $r=3$.}
    \label{fig:kmed_20_best_3}
\end{figure}

\begin{figure}[!ht]
    \centering
    \begin{minipage}{0.03\linewidth}\centering
        \rotatebox[origin=center]{90}{Ratio}
    \end{minipage}
    \begin{minipage}{0.93\linewidth}\centering
        \includegraphics[height=5.5cm, width=12.5cm]{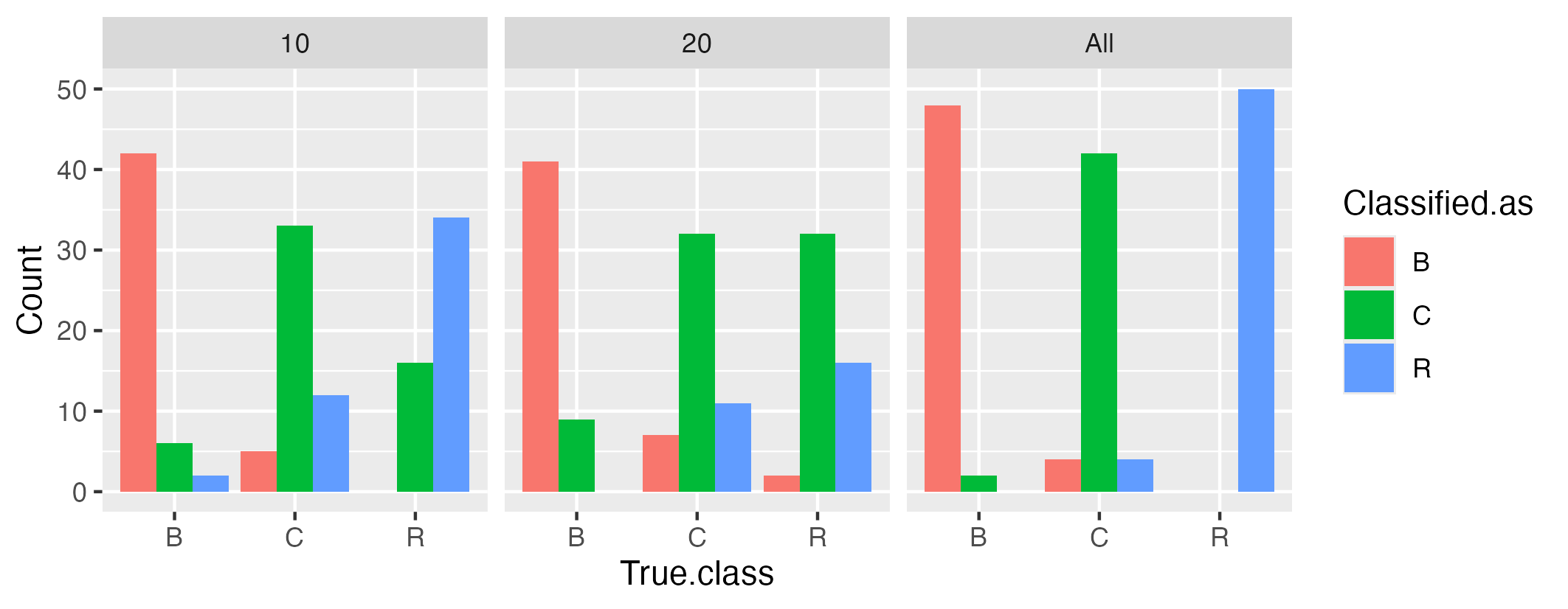}
    \end{minipage}
    \begin{minipage}{0.03\linewidth}\centering
        \rotatebox[origin=center]{90}{Curvature}
    \end{minipage}
    \begin{minipage}{0.93\linewidth}\centering
        \includegraphics[height=5.5cm, width=12.5cm]{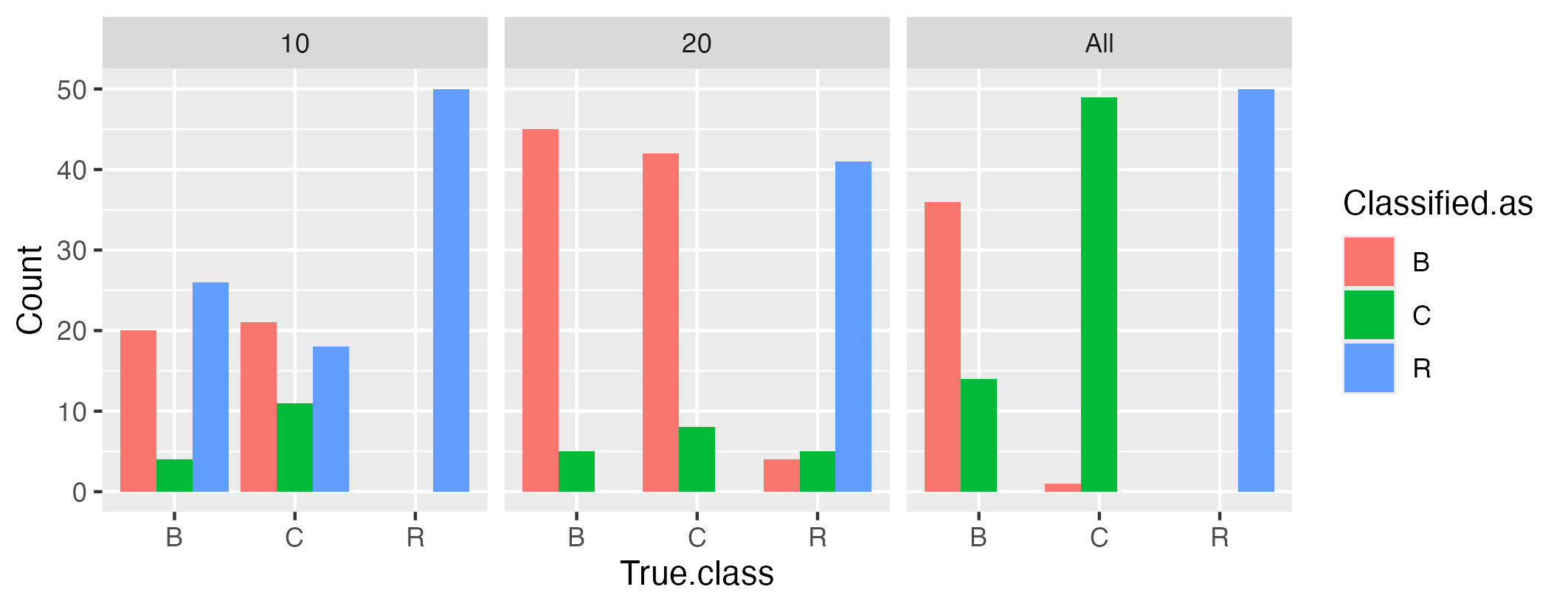}
    \end{minipage}
    \begin{minipage}{0.03\linewidth}\centering
        \rotatebox[origin=center]{90}{Both}
    \end{minipage}
    \begin{minipage}{0.93\linewidth}\centering
        \includegraphics[height=5.5cm, width=12.5cm]{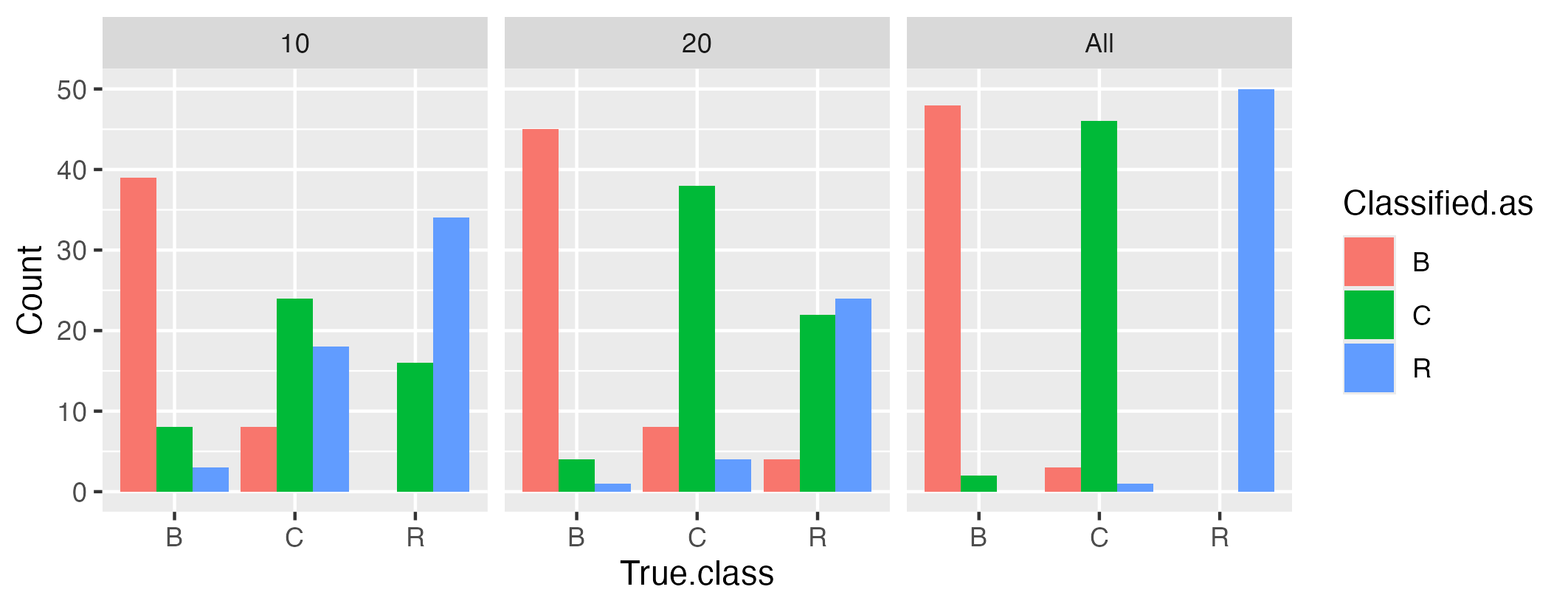}
    \end{minipage}
    \caption{Histograms of $k$-medoids classification accuracy using only the ratio, only the curvature and both ratio and curvature for discrimination when using a sample of 10, 20, and 'All' components, respectively. Misclassification rates are 27.3\%, 40.7\% and 6.7\% for 10, 20 and 'All' components, respectively, when using only the ratio, 46\%, 37.3\% and 10\% when using only the curvature, and 35.3\%, 28.7\% and 4\% when using both characteristics for a sample of 50 realisations that were osculated by a disc of radius $r=3$.}
    \label{fig:kmed_50_best_3}
\end{figure}

\begin{figure}[!ht]
    \centering
    \begin{minipage}{0.03\linewidth}\centering
        \rotatebox[origin=center]{90}{Ratio}
    \end{minipage}
    \begin{minipage}{0.93\linewidth}\centering
        \includegraphics[height=5.5cm, width=12.5cm]{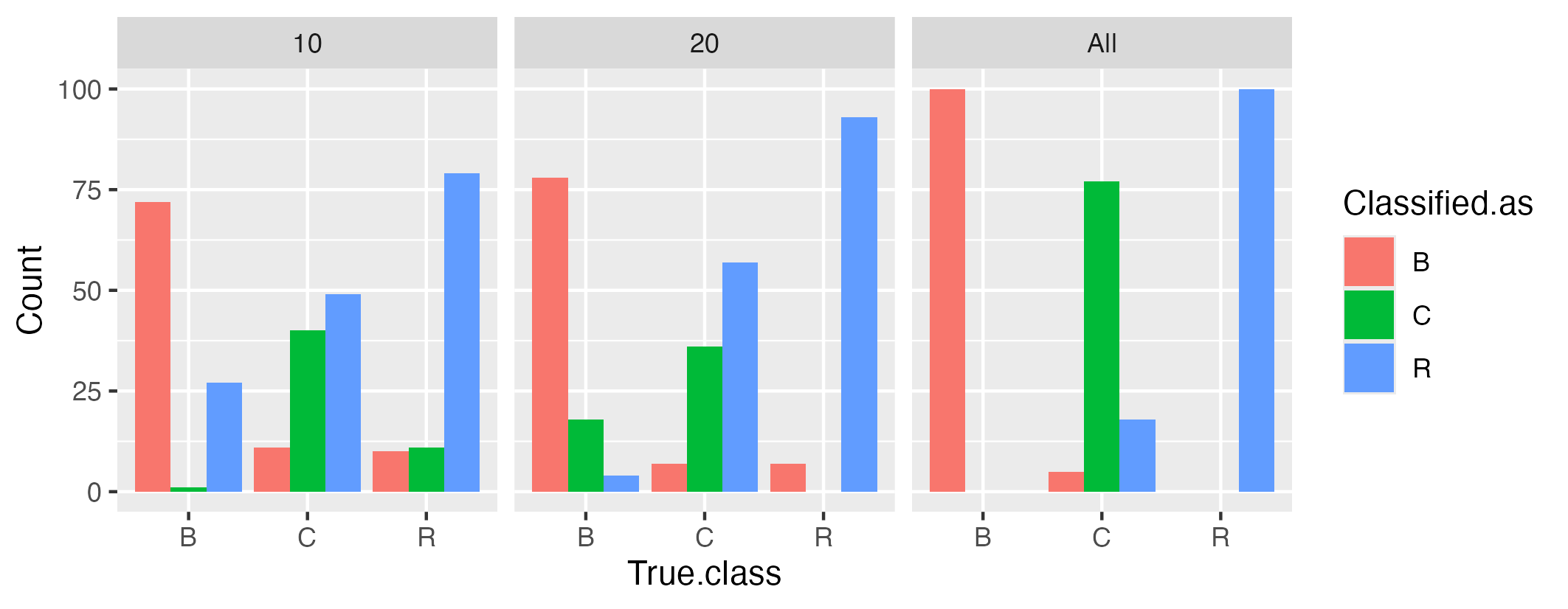}
    \end{minipage}
    \begin{minipage}{0.03\linewidth}\centering
        \rotatebox[origin=center]{90}{Curvature}
    \end{minipage}
    \begin{minipage}{0.93\linewidth}\centering
        \includegraphics[height=5.5cm, width=12.5cm]{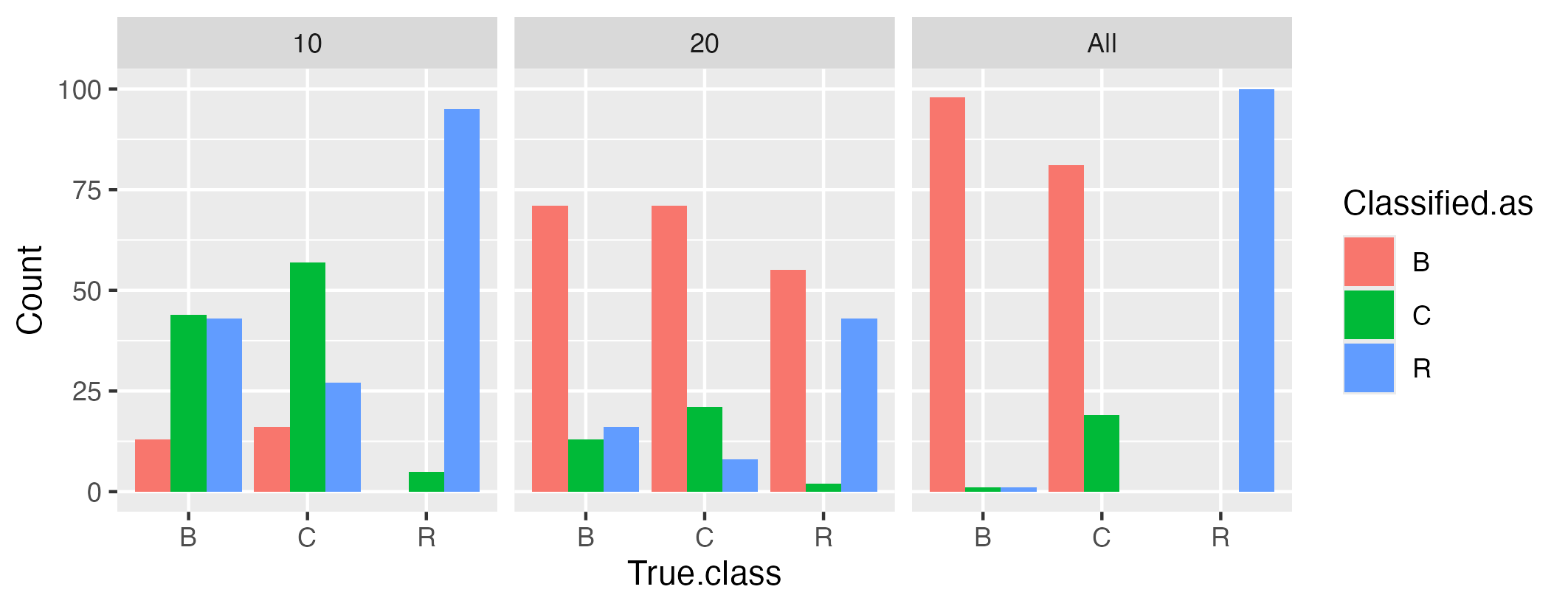}
    \end{minipage}
    \begin{minipage}{0.03\linewidth}\centering
        \rotatebox[origin=center]{90}{Both}
    \end{minipage}
    \begin{minipage}{0.93\linewidth}\centering
        \includegraphics[height=5.5cm, width=12.5cm]{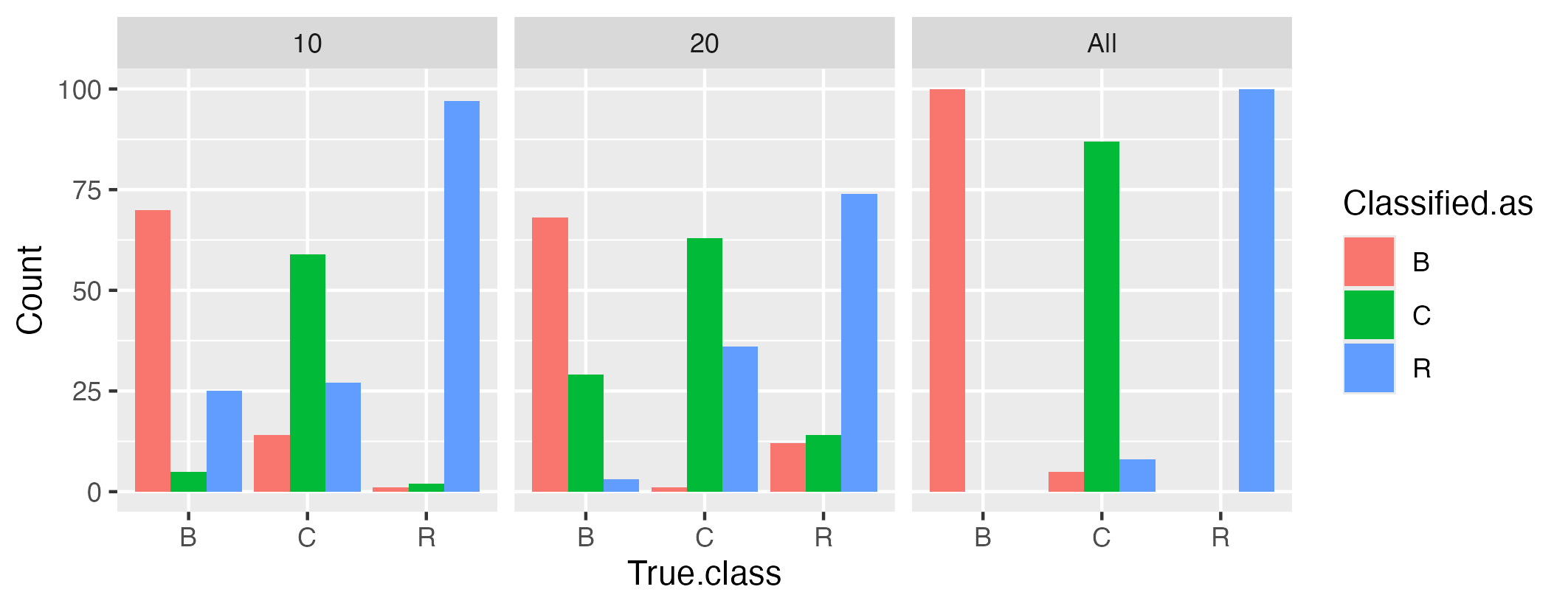}
    \end{minipage}
    \caption{Histograms of $k$-medoids classification accuracy using only the ratio, only the curvature and both ratio and curvature for discrimination when using a sample of 10, 20, and 'All' components, respectively. Misclassification rates are 36.3\%, 31\% and 7.7\% for 10, 20 and 'All' components, respectively, when using only the ratio, 45\%, 55\% and 27.7\% when using only the curvature, and 24.7\%, 31.7\% and 4.3\% when using both characteristics for a sample of 100 realisations that were osculated by a disc of radius $r=3$.}
    \label{fig:kmed_100_best_3}
\end{figure}

\begin{figure}[!ht]
    \centering
    \begin{minipage}{0.03\linewidth}\centering
        \rotatebox[origin=center]{90}{20 Realisations}
    \end{minipage}
    \begin{minipage}{0.93\linewidth}\centering
        \includegraphics[height=5.5cm, width=12.3cm]{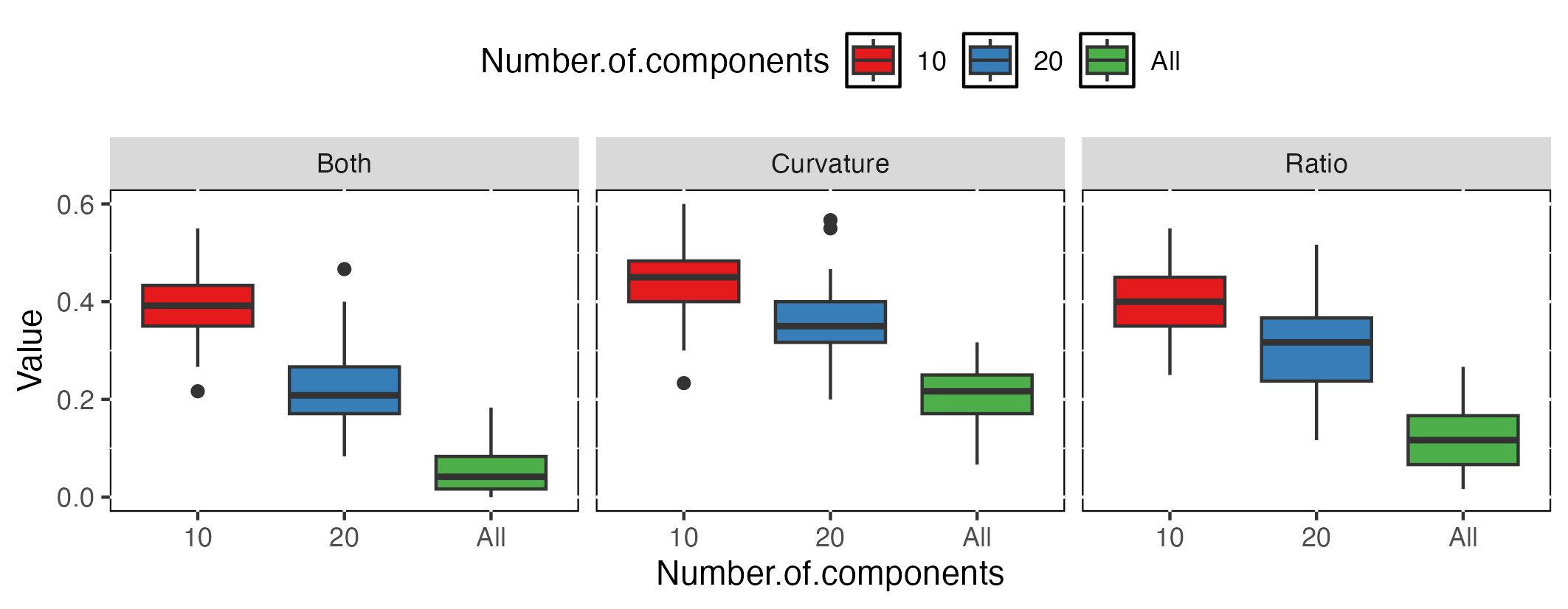}
    \end{minipage}
    \begin{minipage}{0.03\linewidth}\centering
        \rotatebox[origin=center]{90}{50 Realisations}
    \end{minipage}
    \begin{minipage}{0.93\linewidth}\centering
        \includegraphics[height=5.5cm, width=12.3cm]{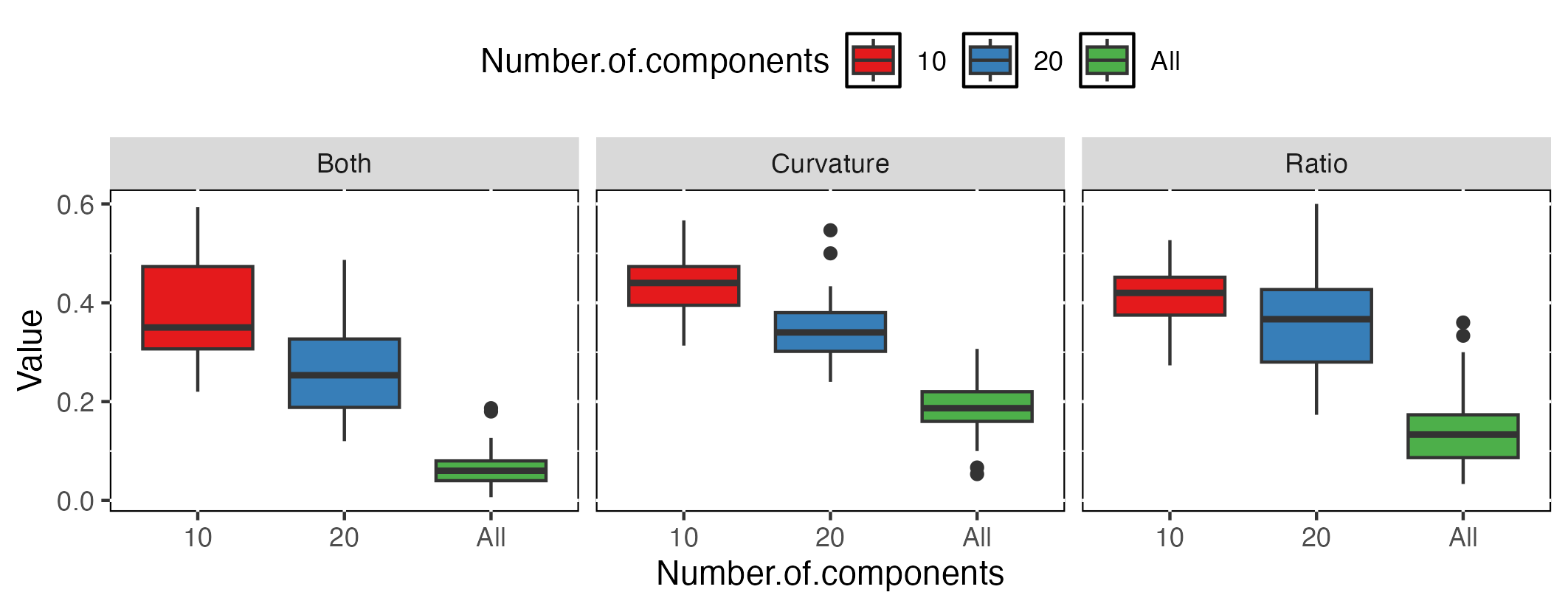}
    \end{minipage}
    \begin{minipage}{0.03\linewidth}\centering
        \rotatebox[origin=center]{90}{100 Realisations}
    \end{minipage}
    \begin{minipage}{0.93\linewidth}\centering
        \includegraphics[height=5.5cm, width=12.3cm]{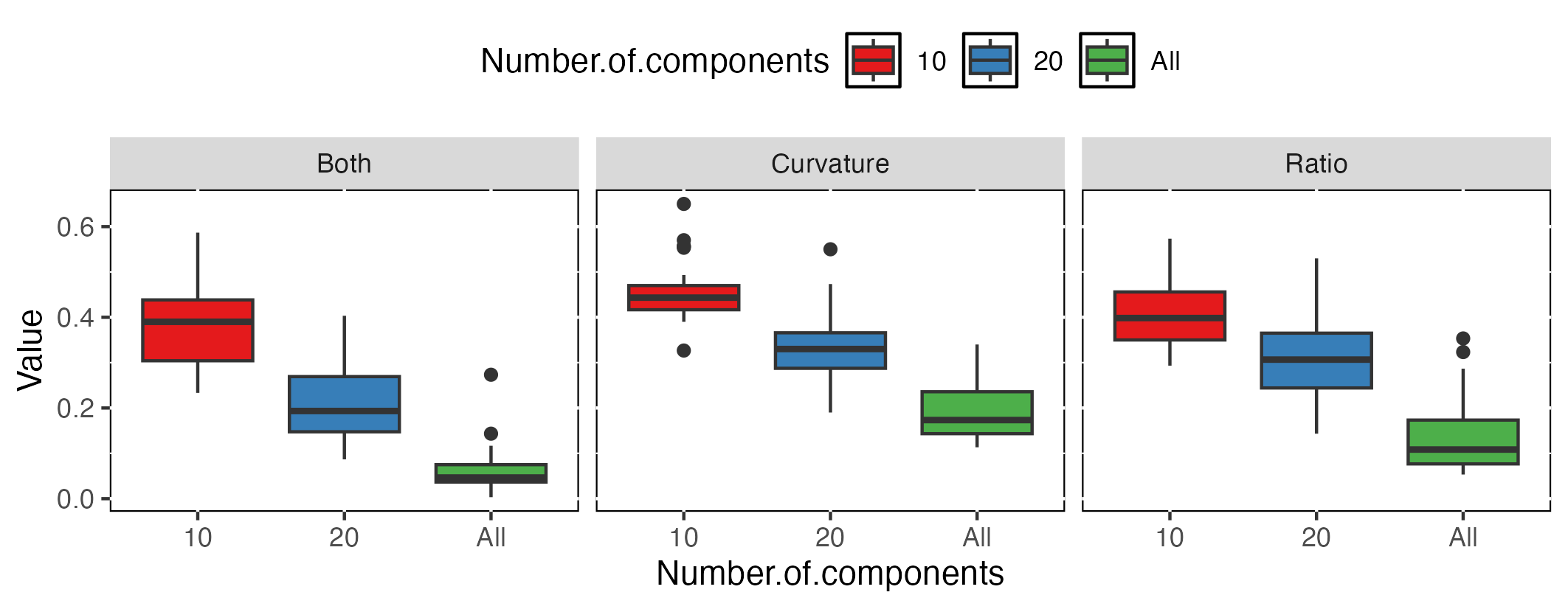}
    \end{minipage}
    \caption{Boxplots of misclassification rate for 50 runs of $k$-medoids algorithm when considering samples of 20 (top), 50 (middle) and 100 (bottom) realisations using both ratio and curvature, only the curvature and only the ratio for discrimination, respectively. For each setting, misclassification rates for different number of components considered (namely 10, 20 and 'All') are shown. Note that the characteristics were obtained using an osculating disc of radius $r=3$ on the simulated data.}
    \label{fig:kmed_box_3}
\end{figure}

Figures \ref{fig:kmed_20_best_3}, \ref{fig:kmed_50_best_3} and \ref{fig:kmed_100_best_3} represent the histograms of classification accuracy for non-hierarchical unsupervised clustering based on $k$-medoids algorithm when 20, 50 and 100 realisations are considered, respectively, on the data obtained using the osculating circle with radius $r=3$. The corresponding boxplot, obtained after 50 runs of the alhorithm, can be seen in Figure \ref{fig:kmed_box_3}.


\subsubsection{Hierarchical clustering}

\begin{figure}[!ht]
    \centering
    \begin{minipage}{0.03\linewidth}\centering
        \rotatebox[origin=center]{90}{Ratio}
    \end{minipage}
    \begin{minipage}{0.93\linewidth}\centering
        \includegraphics[height=5.5cm, width=12.5cm]{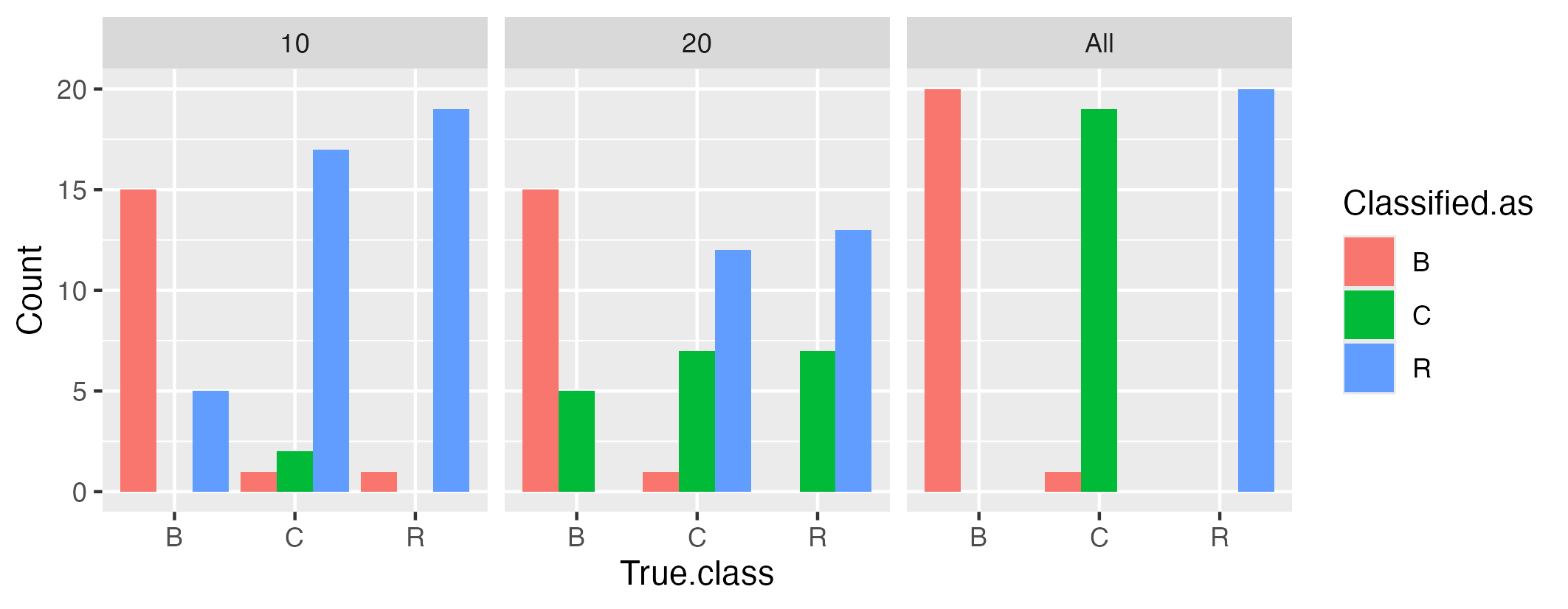}
    \end{minipage}
    \begin{minipage}{0.03\linewidth}\centering
        \rotatebox[origin=center]{90}{Curvature}
    \end{minipage}
    \begin{minipage}{0.93\linewidth}\centering
        \includegraphics[height=5.5cm, width=12.5cm]{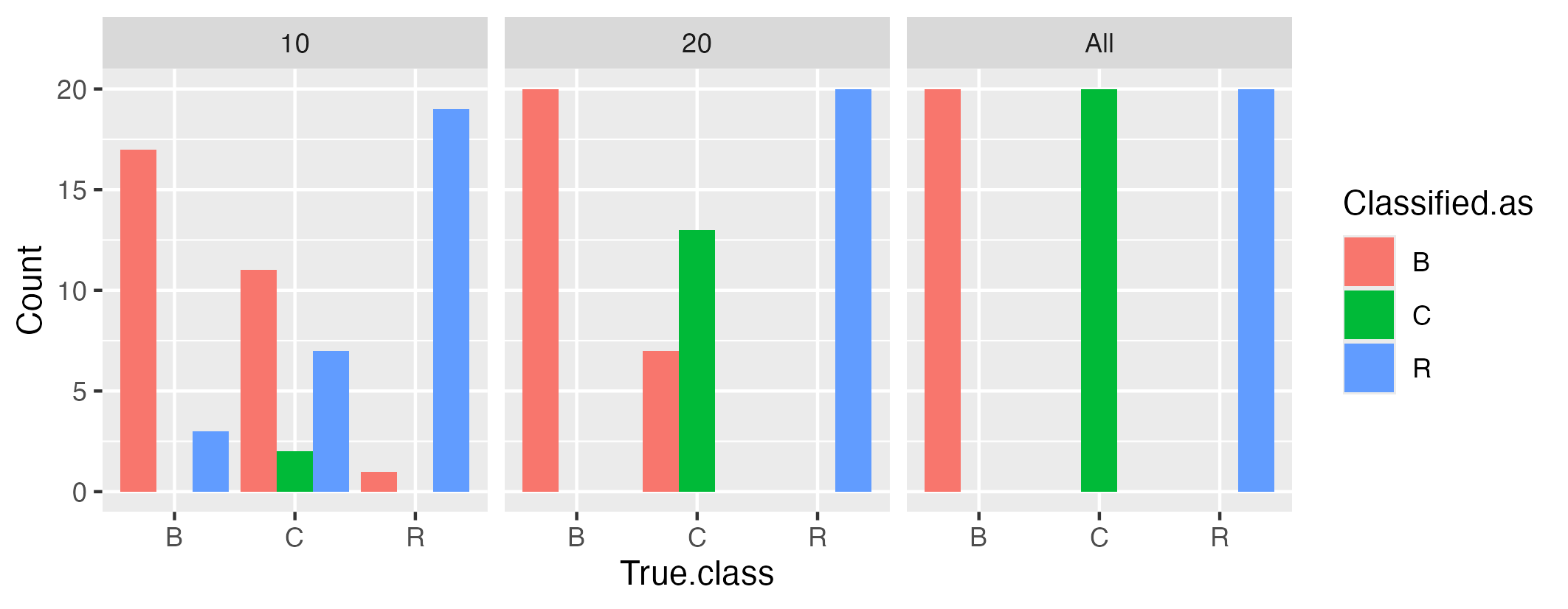}
    \end{minipage}
    \begin{minipage}{0.03\linewidth}\centering
        \rotatebox[origin=center]{90}{Both}
    \end{minipage}
    \begin{minipage}{0.93\linewidth}\centering
        \includegraphics[height=5.5cm, width=12.5cm]{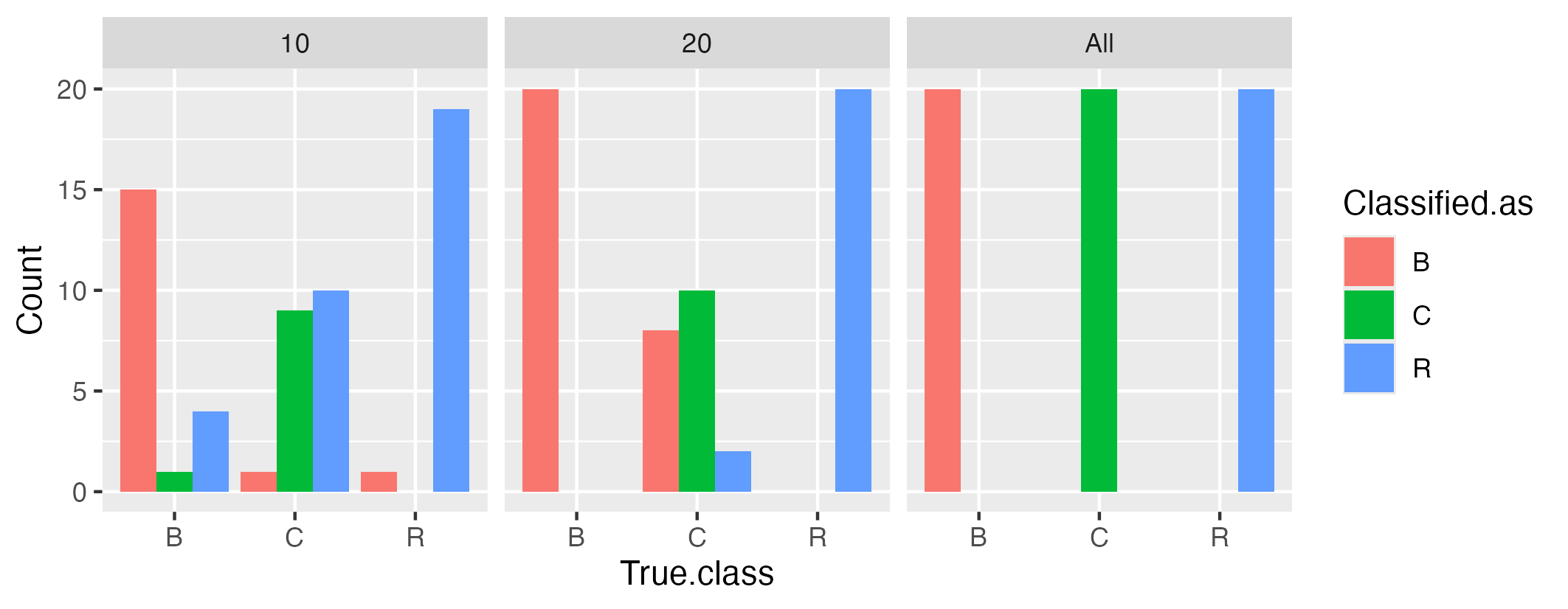}
    \end{minipage}
    \caption{Histograms of hierarchical clustering classification accuracy using only the ratio, only the curvature and both ratio and curvature for discrimination when using a sample of 10, 20, and 'All' components, respectively. Misclassification rates are 40\%, 41.7\% and 1.7\% for 10, 20 and 'All' components, respectively, when using only the ratio, 36.7\%, 11.7\% and 0\% when using only the curvature, and 28.3\%, 16.7\% and 0\% when using both characteristics for a sample of 20 realisations that were osculated by a disc of radius $r=5$.}
    \label{fig:hc_20_best_5}
\end{figure}

\begin{figure}[!ht]
    \centering
    \begin{minipage}{0.03\linewidth}\centering
        \rotatebox[origin=center]{90}{Ratio}
    \end{minipage}
    \begin{minipage}{0.93\linewidth}\centering
        \includegraphics[height=5.5cm, width=12.5cm]{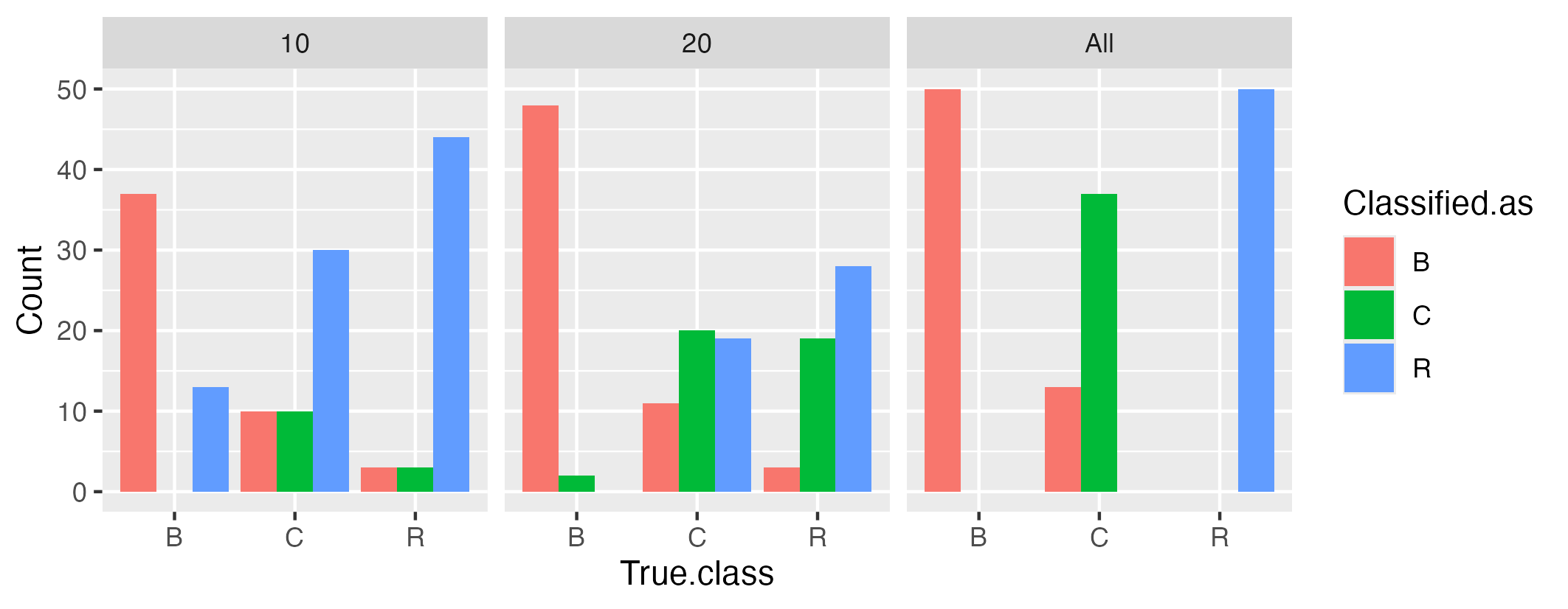}
    \end{minipage}
    \begin{minipage}{0.03\linewidth}\centering
        \rotatebox[origin=center]{90}{Curvature}
    \end{minipage}
    \begin{minipage}{0.93\linewidth}\centering
        \includegraphics[height=5.5cm, width=12.5cm]{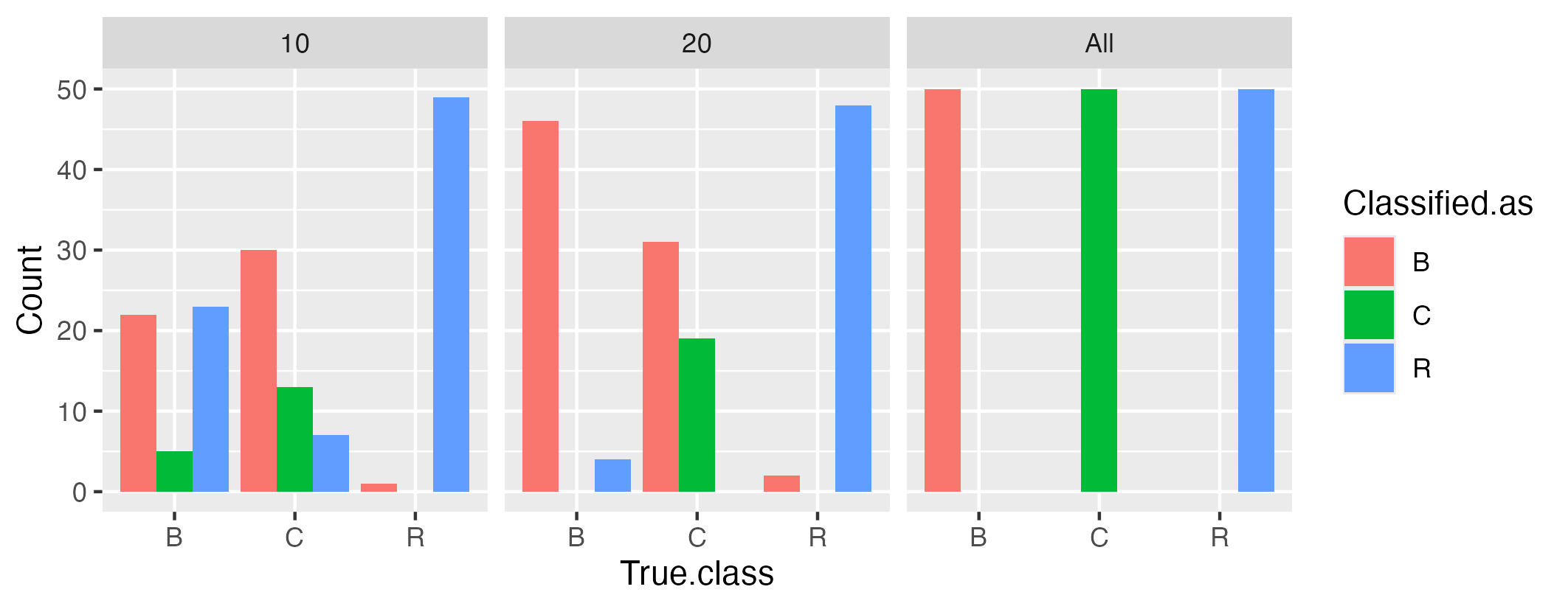}
    \end{minipage}
    \begin{minipage}{0.03\linewidth}\centering
        \rotatebox[origin=center]{90}{Both}
    \end{minipage}
    \begin{minipage}{0.93\linewidth}\centering
        \includegraphics[height=5.5cm, width=12.5cm]{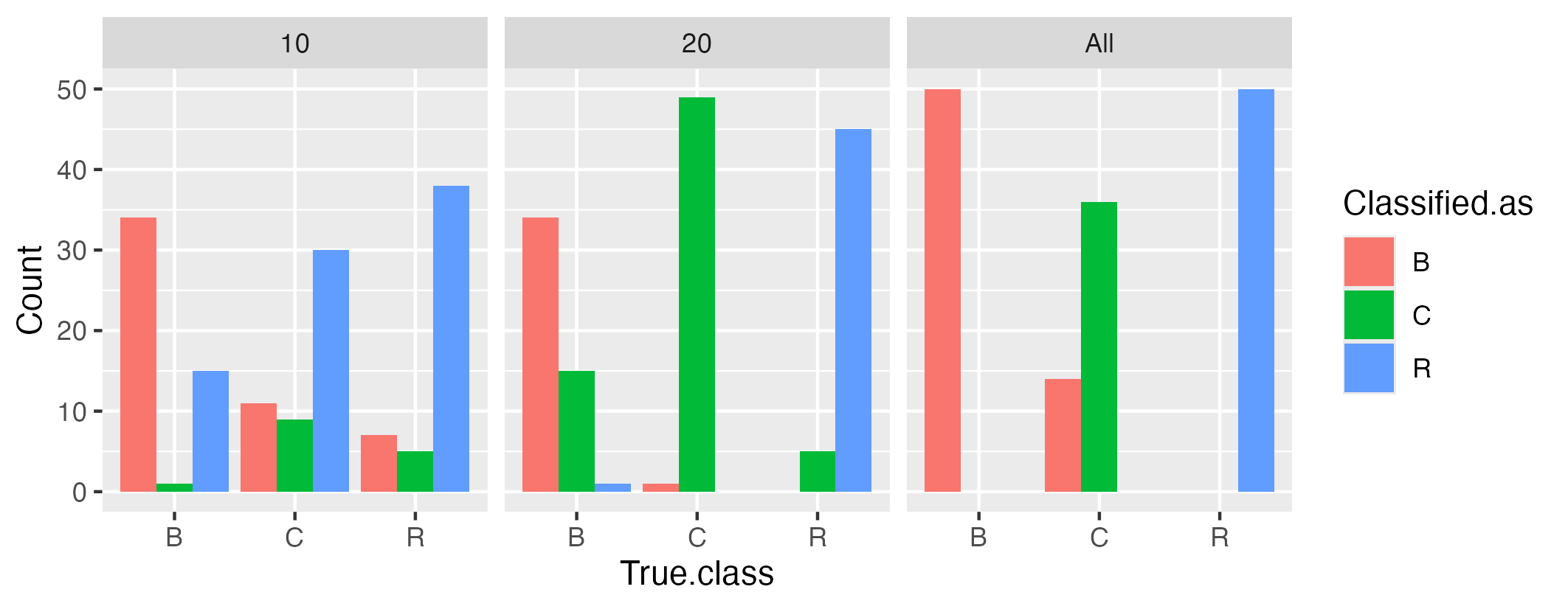}
    \end{minipage}
    \caption{Histograms of hierarchical clustering classification accuracy using only the ratio, only the curvature and both ratio and curvature for discrimination when using a sample of 10, 20, and 'All' components, respectively. Misclassification rates are 39.3\%, 36\% and 8.7\% for 10, 20 and 'All' components, respectively, when using only the ratio, 44\%, 24.7\% and 0\% when using only the curvature, and 46\%, 14.7\% and 9.3\% when using both characteristics for a sample of 50 realisations that were osculated by a disc of radius $r=5$.}
    \label{fig:hc_50_best_5}
\end{figure}

Histograms of classification accuracy using Ward's agglomerative algorithm when 20 and 50 realisations are considered, on the data obtained using the osculating circle with radius $r=5$ are shown in Figures \ref{fig:hc_20_best_5} and \ref{fig:hc_50_best_5}, respectively.

\begin{figure}[!ht]
    \centering
    \begin{minipage}{0.03\linewidth}\centering
        \rotatebox[origin=center]{90}{Ratio}
    \end{minipage}
    \begin{minipage}{0.93\linewidth}\centering
        \includegraphics[height=5.5cm, width=12.5cm]{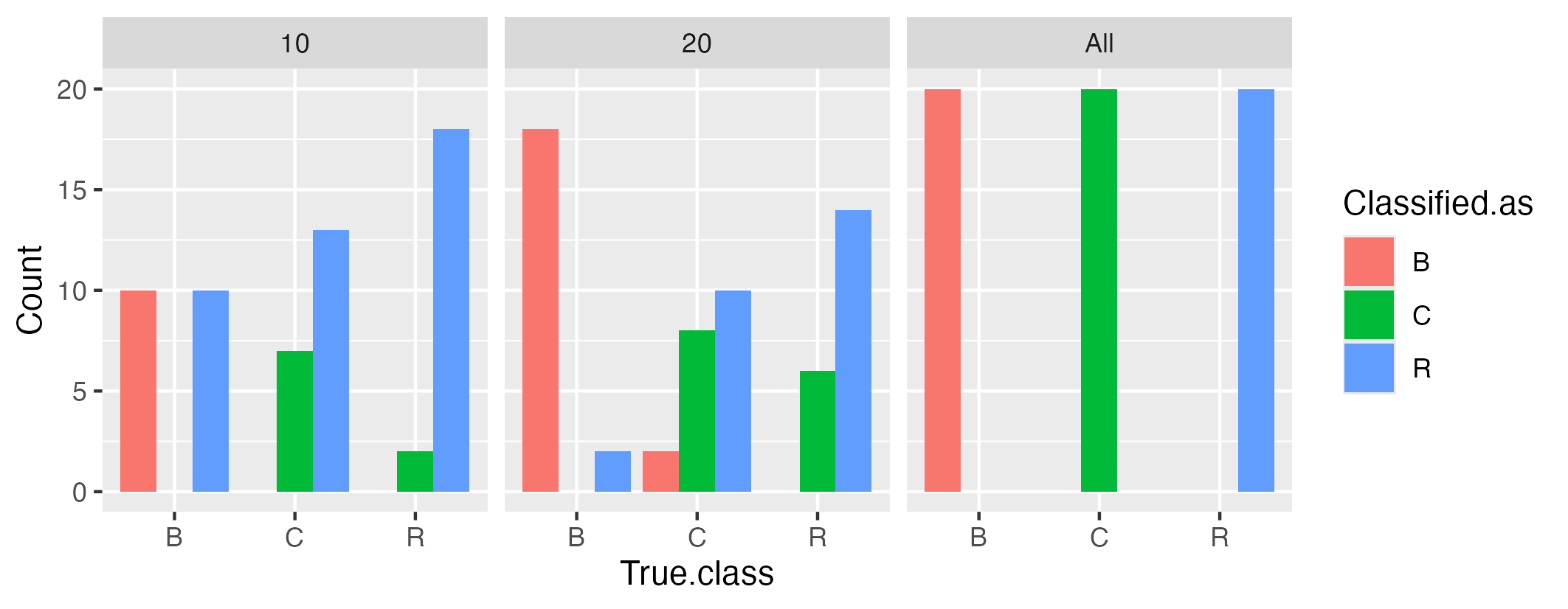}
    \end{minipage}
    \begin{minipage}{0.03\linewidth}\centering
        \rotatebox[origin=center]{90}{Curvature}
    \end{minipage}
    \begin{minipage}{0.93\linewidth}\centering
        \includegraphics[height=5.5cm, width=12.5cm]{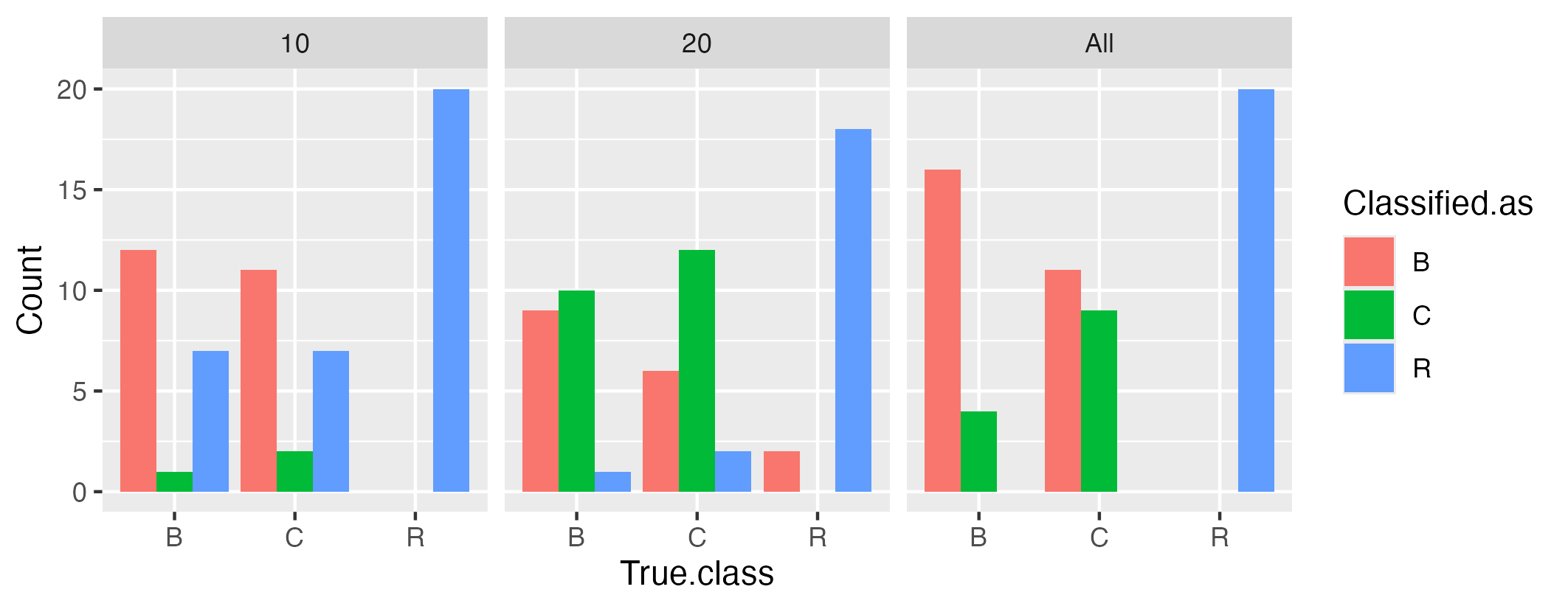}
    \end{minipage}
    \begin{minipage}{0.03\linewidth}\centering
        \rotatebox[origin=center]{90}{Both}
    \end{minipage}
    \begin{minipage}{0.93\linewidth}\centering
        \includegraphics[height=5.5cm, width=12.5cm]{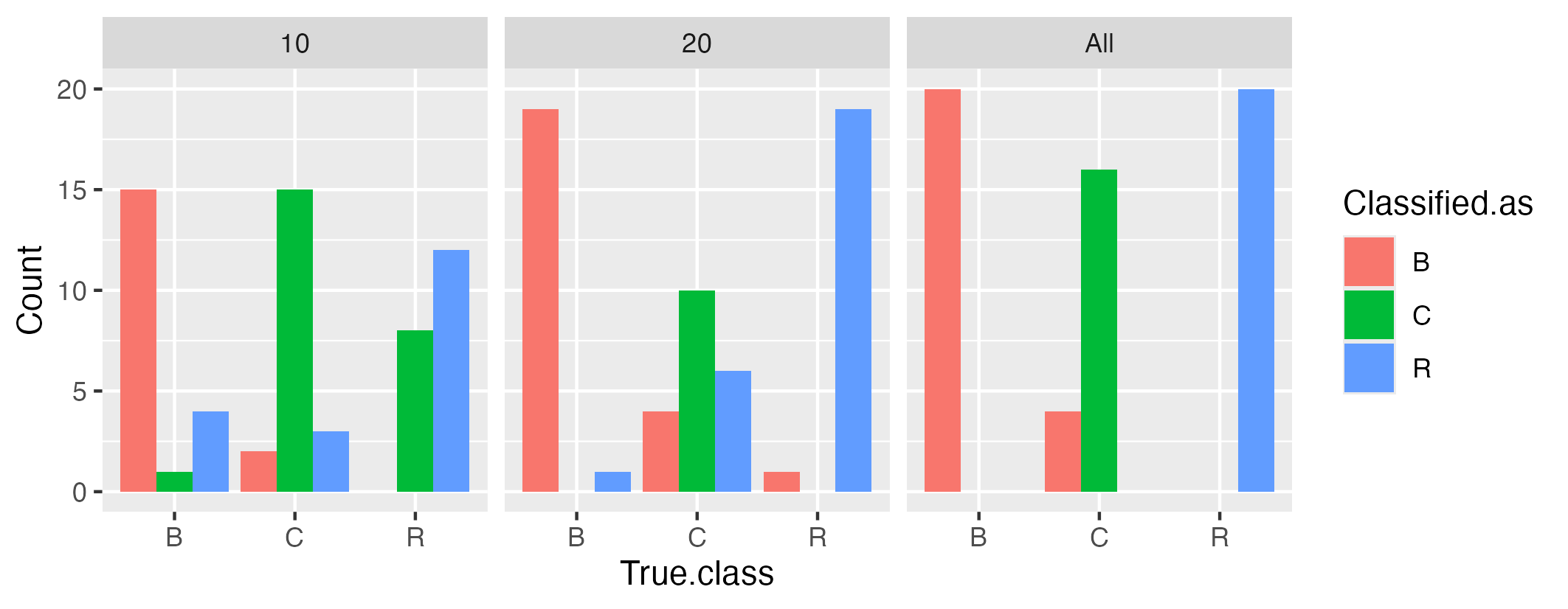}
    \end{minipage}
    \caption{Histograms of hierarchical clustering classification accuracy using only the ratio, only the curvature and both ratio and curvature for discrimination when using a sample of 10, 20, and 'All' components, respectively. Misclassification rates are 41.7\%, 33.3\% and 0\% for 10, 20 and 'All' components, respectively, when using only the ratio, 43.3\%, 35\% and 25\% when using only the curvature, and 30\%, 20\% and 6.7\% when using both characteristics for a sample of 20 realisations that were osculated by a disc of radius $r=3$.}
    \label{fig:hc_20_best_3}
\end{figure}

\begin{figure}[!ht]
    \centering
    \begin{minipage}{0.03\linewidth}\centering
        \rotatebox[origin=center]{90}{Ratio}
    \end{minipage}
    \begin{minipage}{0.93\linewidth}\centering
        \includegraphics[height=5.5cm, width=12.5cm]{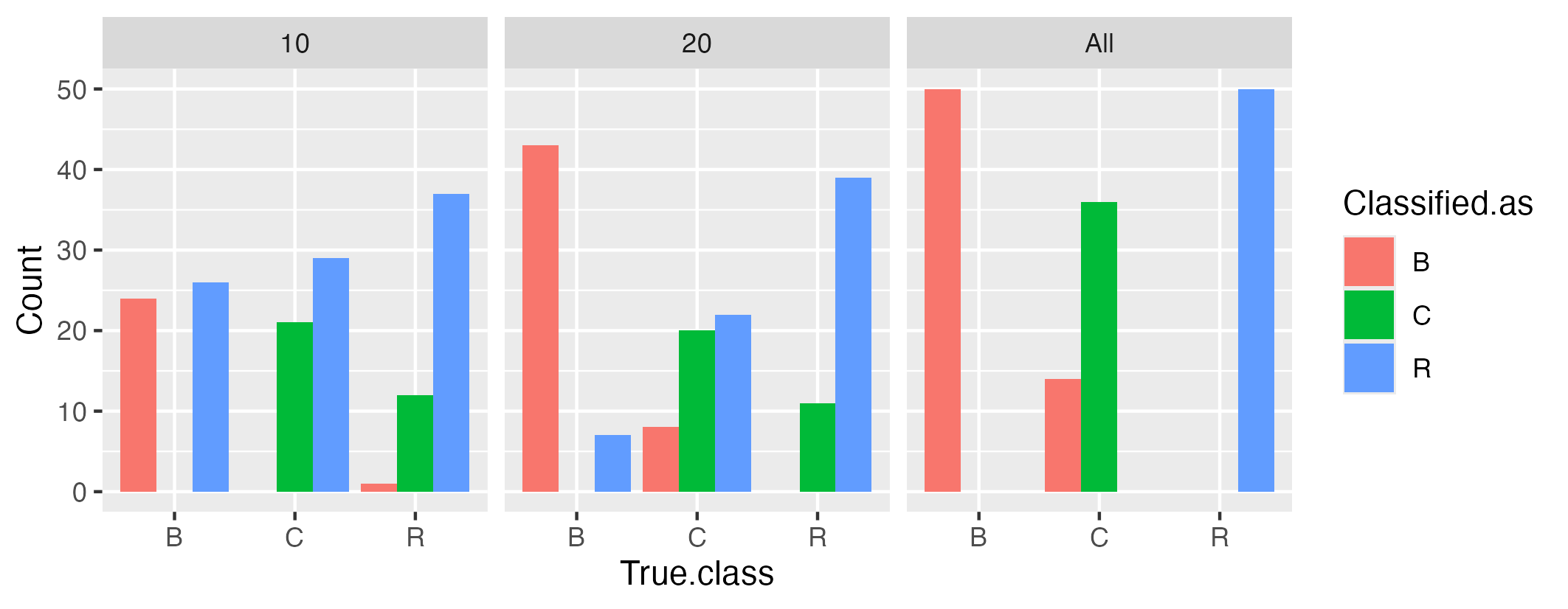}
    \end{minipage}
    \begin{minipage}{0.03\linewidth}\centering
        \rotatebox[origin=center]{90}{Curvature}
    \end{minipage}
    \begin{minipage}{0.93\linewidth}\centering
        \includegraphics[height=5.5cm, width=12.5cm]{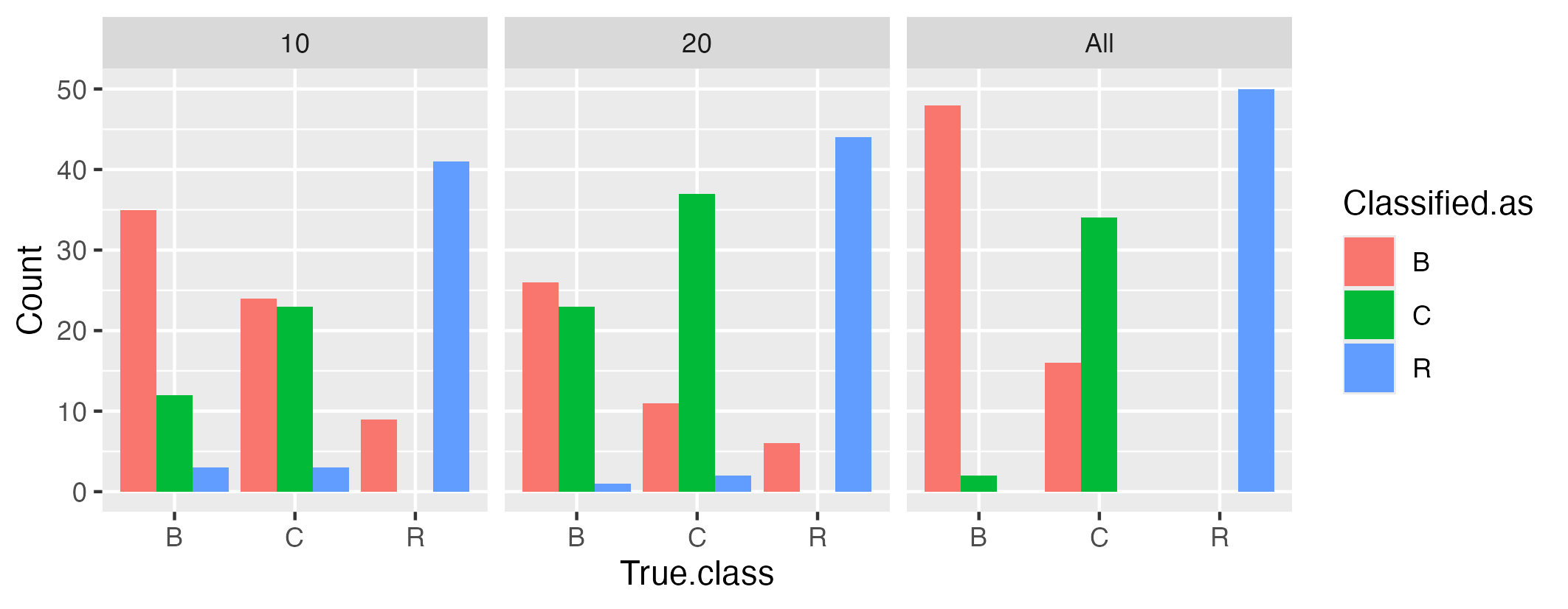}
    \end{minipage}
    \begin{minipage}{0.03\linewidth}\centering
        \rotatebox[origin=center]{90}{Both}
    \end{minipage}
    \begin{minipage}{0.93\linewidth}\centering
        \includegraphics[height=5.5cm, width=12.5cm]{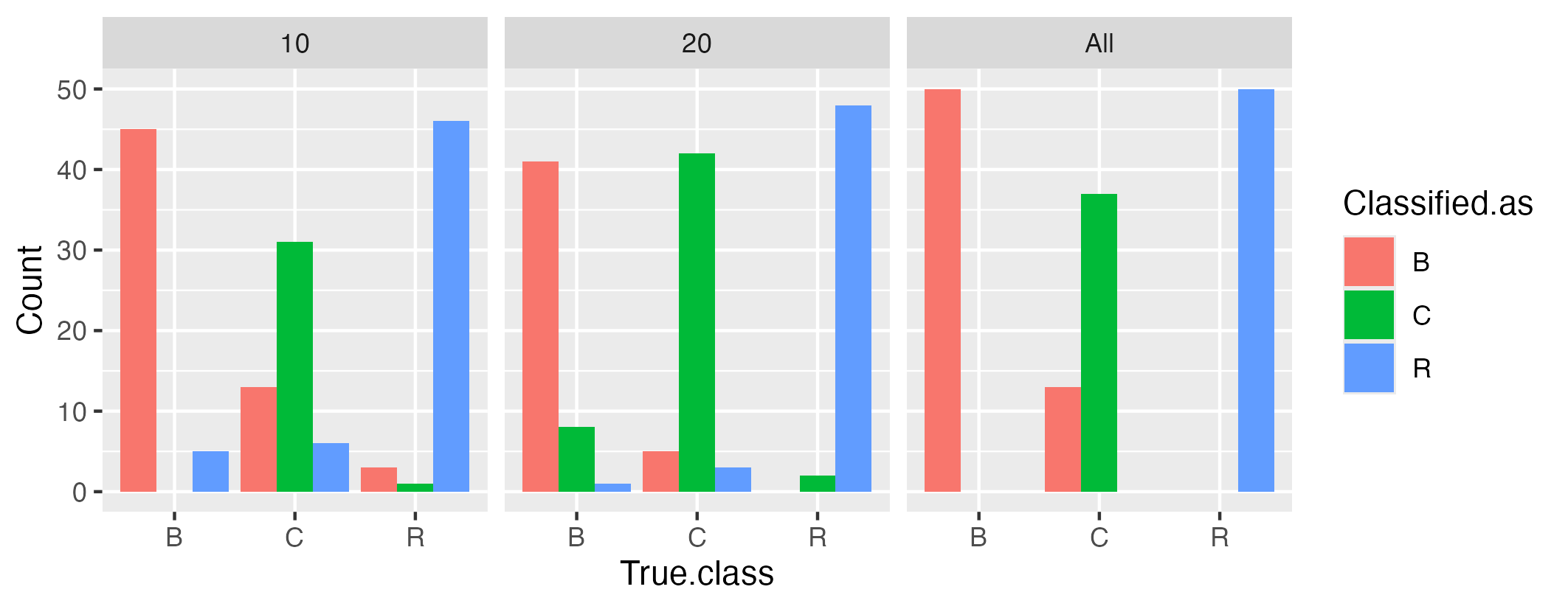}
    \end{minipage}
    \caption{Histograms of hierarchical clustering classification accuracy using only the ratio, only the curvature and both ratio and curvature for discrimination when using a sample of 10, 20, and 'All' components, respectively. Misclassification rates are 45.3\%, 32\% and 9.3\% for 10, 20 and 'All' components, respectively, when using only the ratio, 34\%, 28.7\% and 12\% when using only the curvature, and 18.7\%, 12.7\% and 8.7\% when using both characteristics for a sample of 50 realisations that were osculated by a disc of radius $r=3$.}
    \label{fig:hc_50_best_3}
\end{figure}

\begin{figure}[!ht]
    \centering
    \begin{minipage}{0.03\linewidth}\centering
        \rotatebox[origin=center]{90}{Ratio}
    \end{minipage}
    \begin{minipage}{0.93\linewidth}\centering
        \includegraphics[height=5.5cm, width=12.5cm]{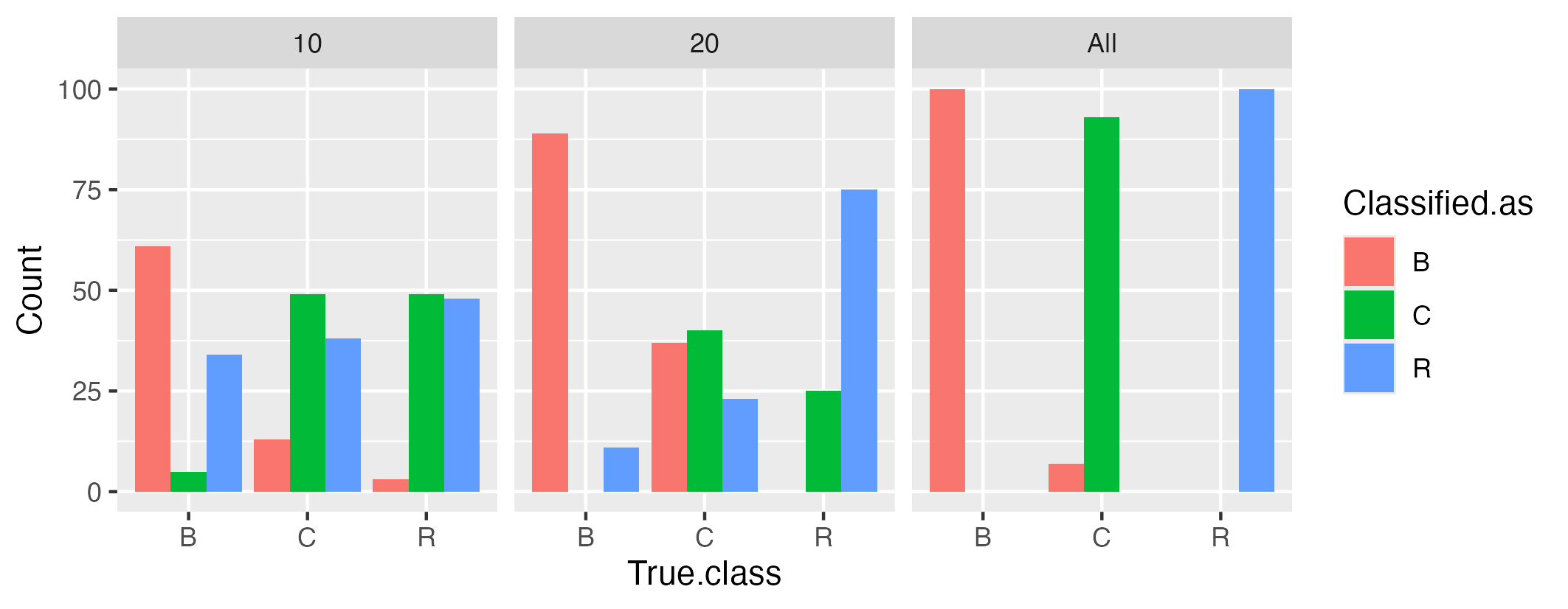}
    \end{minipage}
    \begin{minipage}{0.03\linewidth}\centering
        \rotatebox[origin=center]{90}{Curvature}
    \end{minipage}
    \begin{minipage}{0.93\linewidth}\centering
        \includegraphics[height=5.5cm, width=12.5cm]{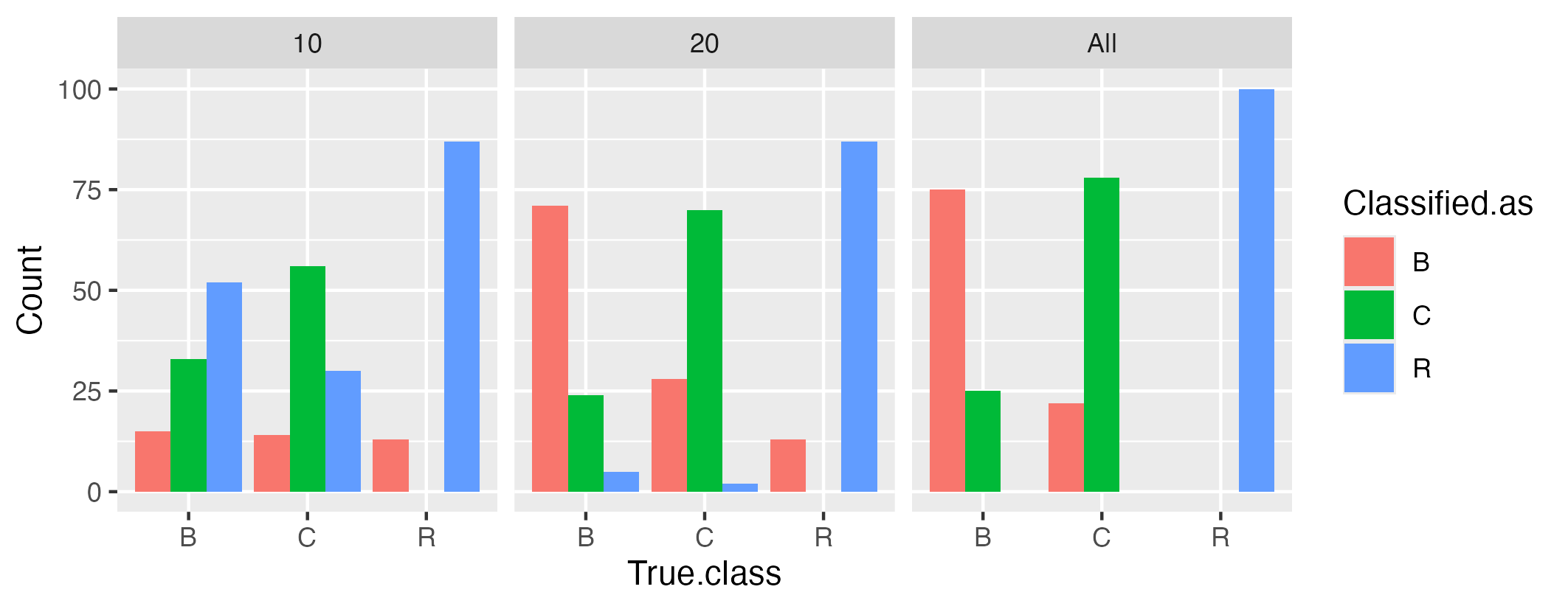}
    \end{minipage}
    \begin{minipage}{0.03\linewidth}\centering
        \rotatebox[origin=center]{90}{Both}
    \end{minipage}
    \begin{minipage}{0.93\linewidth}\centering
        \includegraphics[height=5.5cm, width=12.5cm]{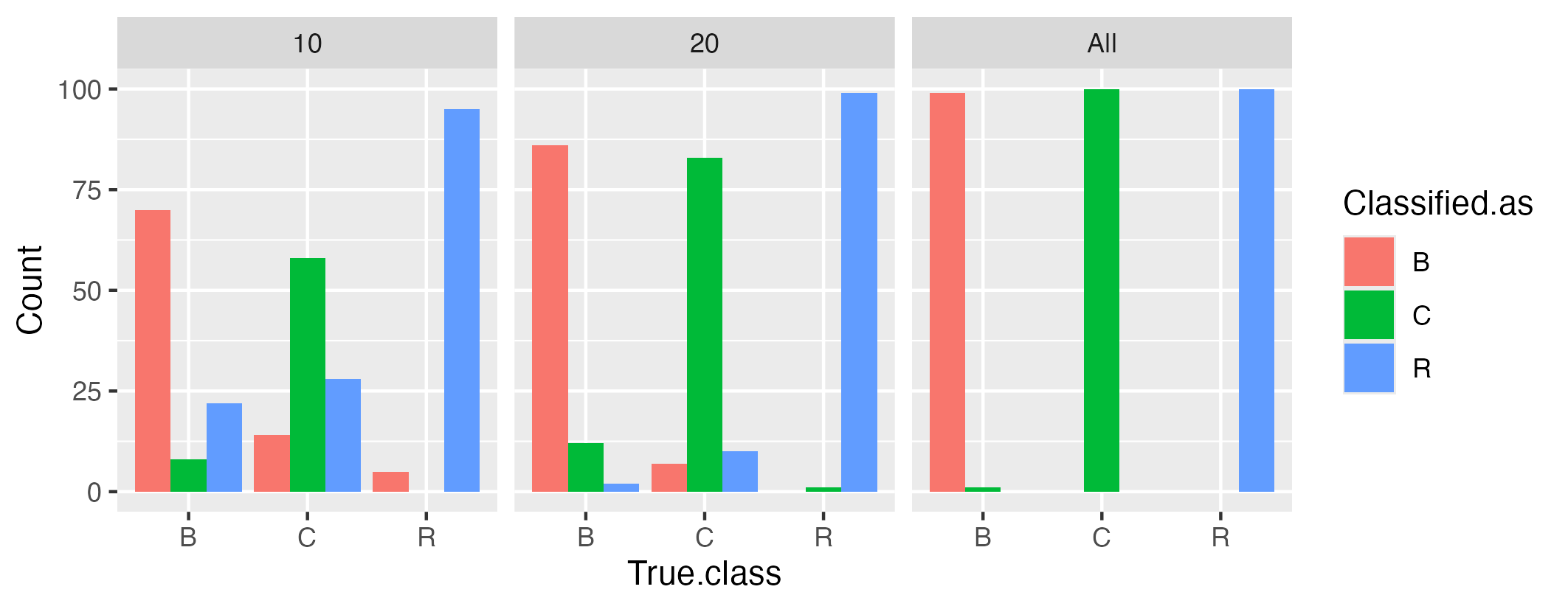}
    \end{minipage}
    \caption{Histograms of hierarchical clustering classification accuracy using only the ratio, only the curvature and both ratio and curvature for discrimination when using a sample of 10, 20, and 'All' components, respectively. Misclassification rates are 47.3\%, 32\% and 2.3\% for 10, 20 and 'All' components, respectively, when using only the ratio, 47.3\%, 24\% and 15.7\% when using only the curvature, and 25.6\%, 10.7\% and 0.3\% when using both characteristics for a sample of 100 realisations that were osculated by a disc of radius $r=3$.}
    \label{fig:hc_100_best_3}
\end{figure}

\begin{figure}[!ht]
    \centering
    \begin{minipage}{0.03\linewidth}\centering
        \rotatebox[origin=center]{90}{20 Realisations}
    \end{minipage}
    \begin{minipage}{0.93\linewidth}\centering
        \includegraphics[height=5.5cm, width=12.3cm]{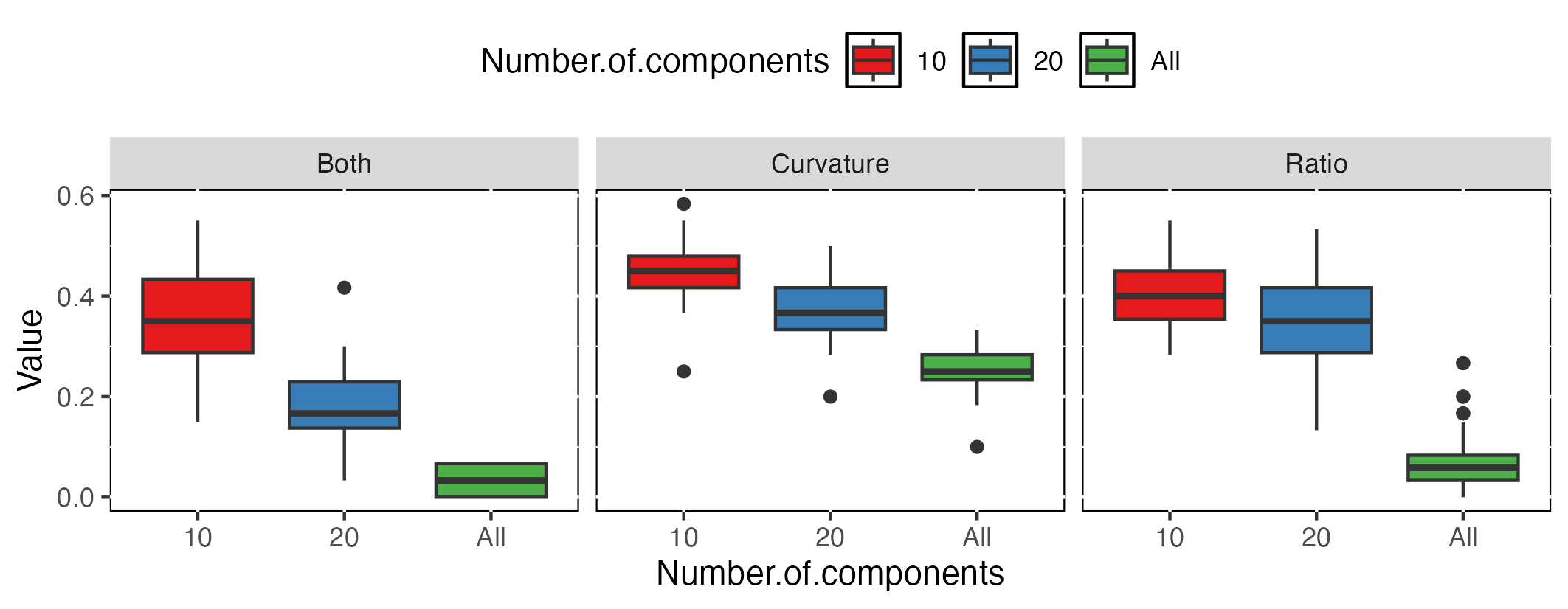}
    \end{minipage}
    \begin{minipage}{0.03\linewidth}\centering
        \rotatebox[origin=center]{90}{50 Realisations}
    \end{minipage}
    \begin{minipage}{0.93\linewidth}\centering
        \includegraphics[height=5.5cm, width=12.3cm]{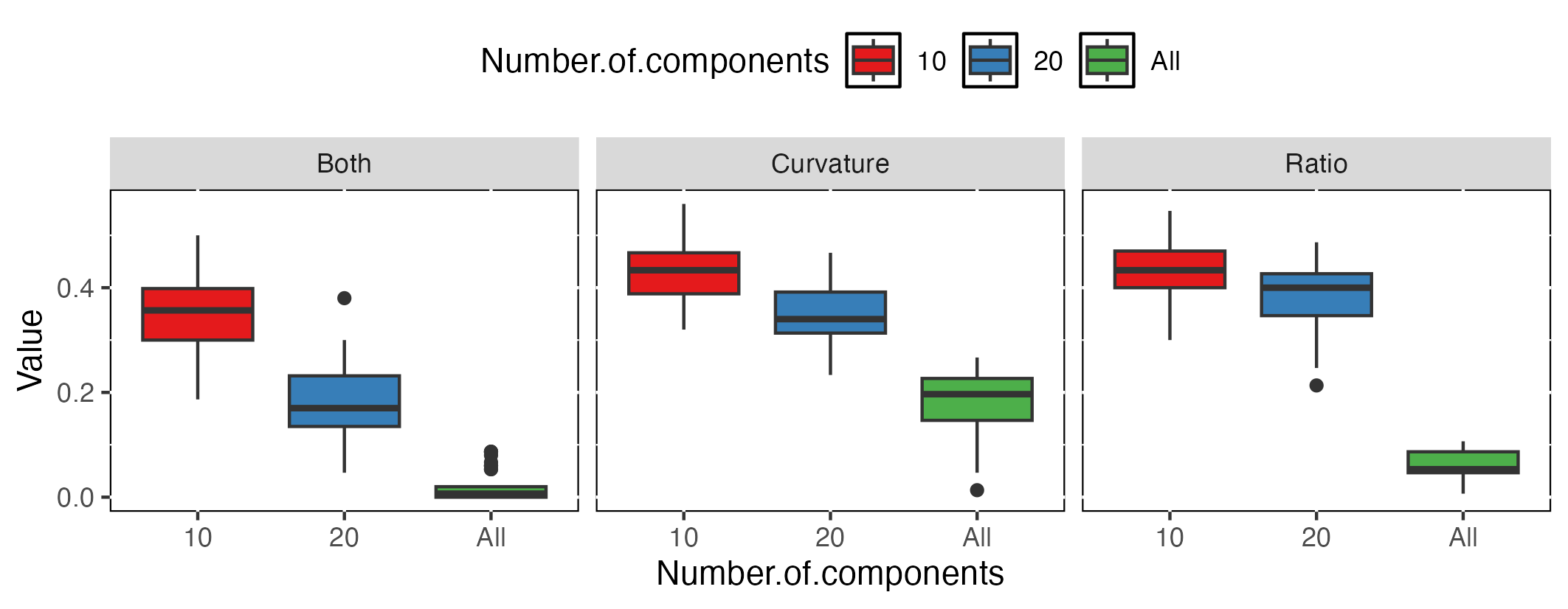}
    \end{minipage}
    \begin{minipage}{0.03\linewidth}\centering
        \rotatebox[origin=center]{90}{100 Realisations}
    \end{minipage}
    \begin{minipage}{0.93\linewidth}\centering
        \includegraphics[height=5.5cm, width=12.3cm]{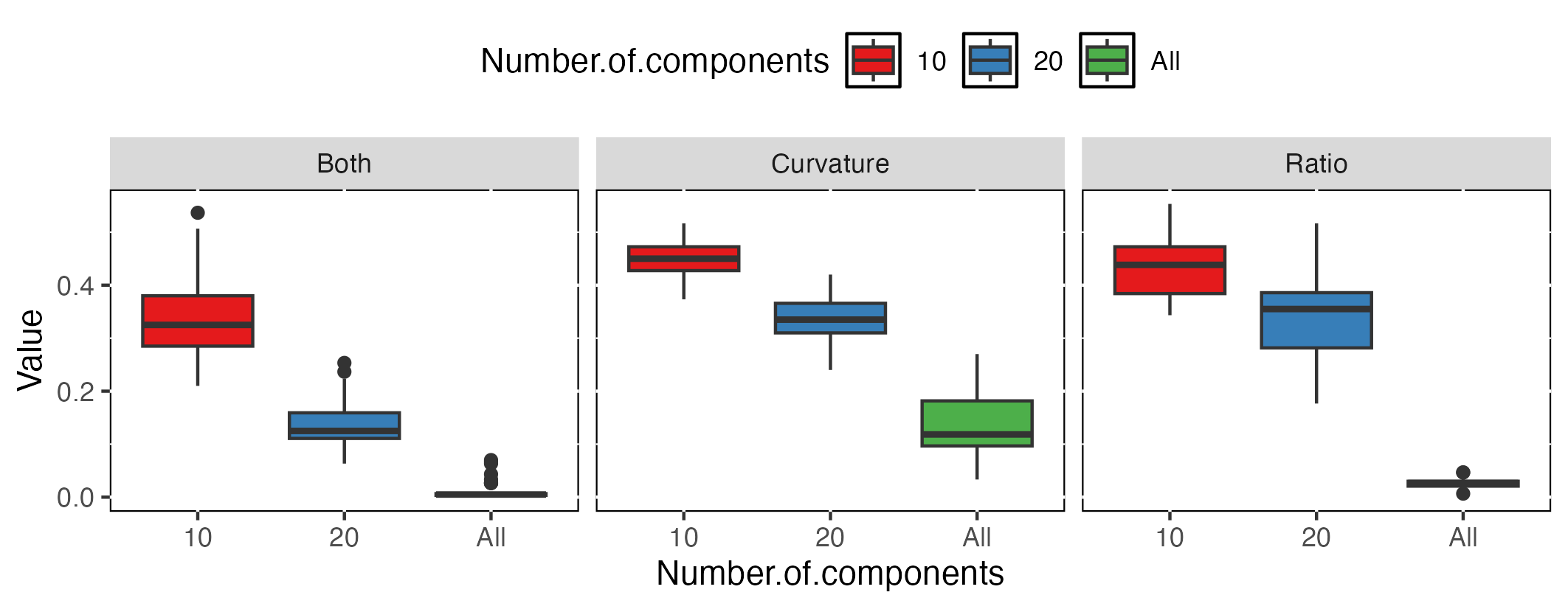}
    \end{minipage}
    \caption{Boxplots of misclassification rate for 50 runs of hierarchical clustering algorithm when considering samples of 20 (top), 50 (middle) and 100 (bottom) realisations using both ratio and curvature, only the curvature and only the ratio for discrimination, respectively. For each setting, misclassification rates for different number of components considered (namely 10, 20 and 'All') are shown. Note that the characteristics were obtained using an osculating disc of radius $r=3$ on the simulated data.}
    \label{fig:hc_box_3}
\end{figure}

Figures \ref{fig:hc_20_best_3}, \ref{fig:hc_50_best_3} and \ref{fig:hc_100_best_3} show the histograms of classification accuracy for hierarchical unsupervised clustering based on Ward's algorithm when 20, 50 and 100 realisations are considered, respectively, on the data obtained using the osculating circle with radius $r=3$. The corresponding boxplot of misclassification rate, obtained after 50 runs of the algorithm, can be seen in Figure \ref{fig:hc_box_3}.



\section{Application}

This section presents additional classification results for the real medical data.

\subsection{Supervised classification}

\begin{figure}[!ht]
    \centering
    \begin{minipage}{0.03\linewidth}\centering
        \rotatebox[origin=center]{90}{Ratio}
    \end{minipage}
    \begin{minipage}{0.93\linewidth}\centering
        \includegraphics[height=5.5cm, width=12.5cm]{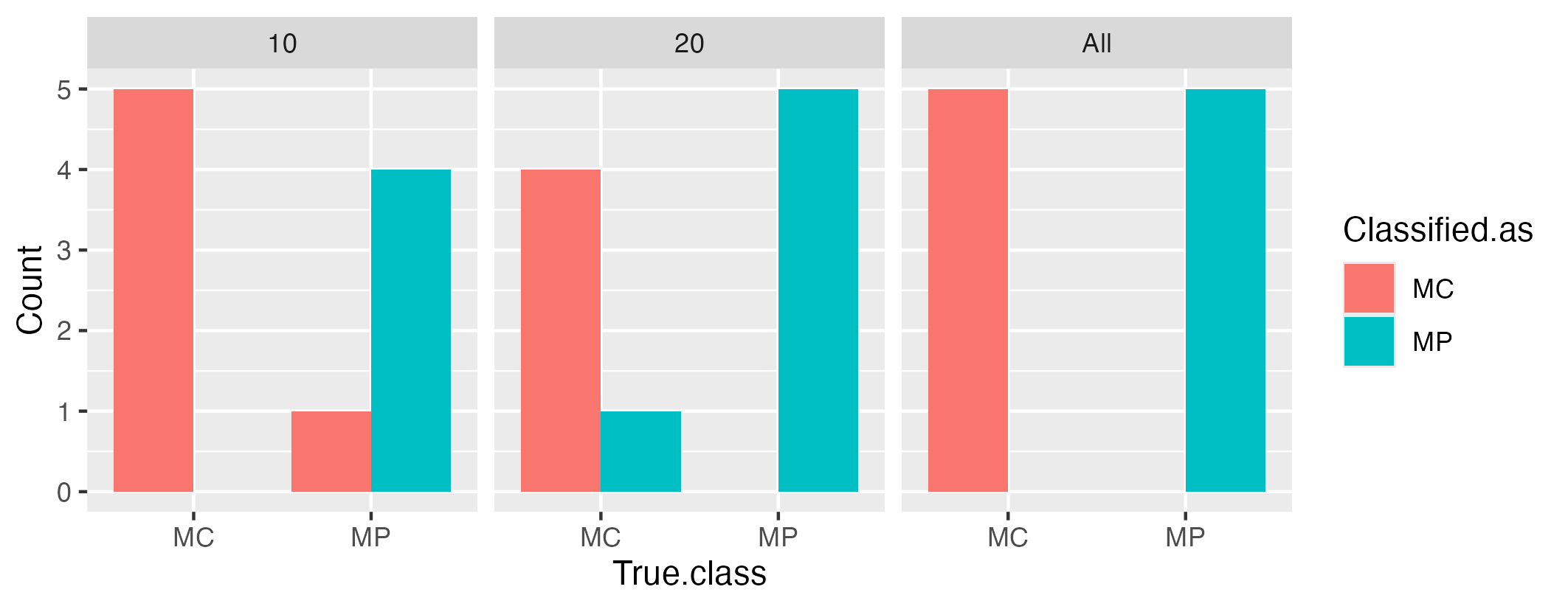}
    \end{minipage}
    \begin{minipage}{0.03\linewidth}\centering
        \rotatebox[origin=center]{90}{Curvature}
    \end{minipage}
    \begin{minipage}{0.93\linewidth}\centering
        \includegraphics[height=5.5cm, width=12.5cm]{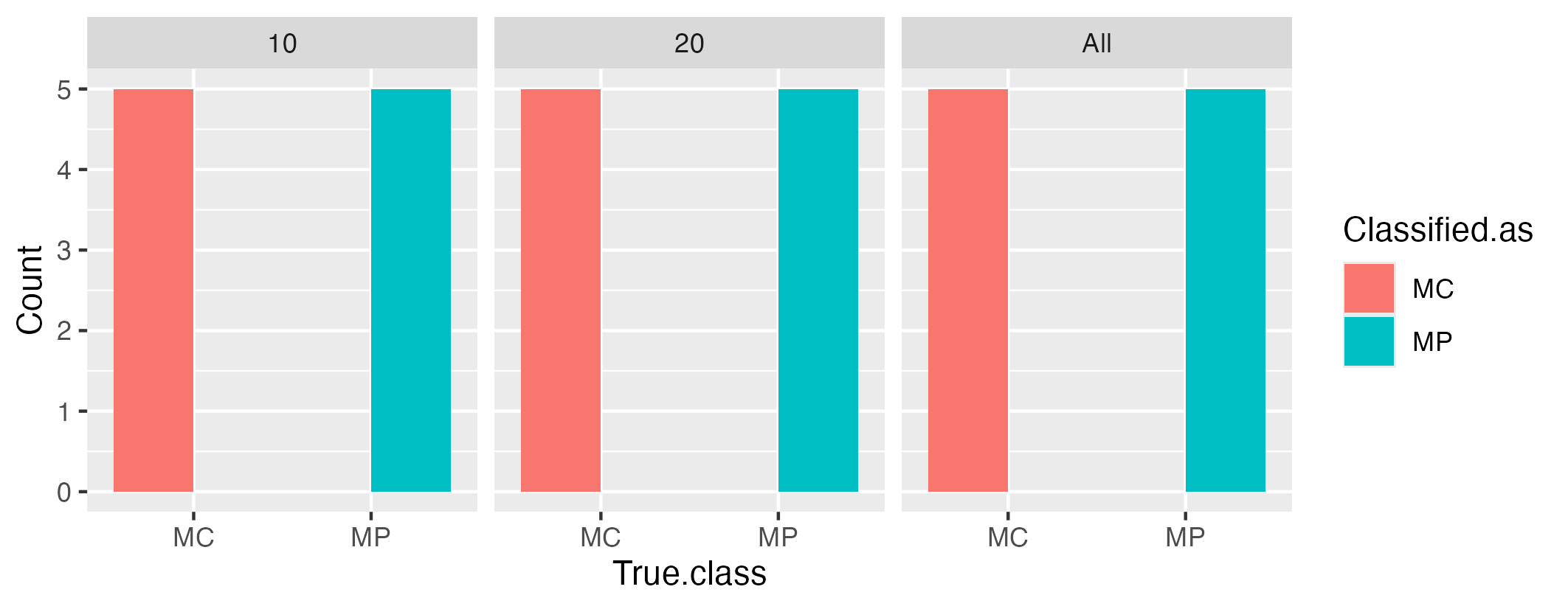}
    \end{minipage}
    \begin{minipage}{0.03\linewidth}\centering
        \rotatebox[origin=center]{90}{Both}
    \end{minipage}
    \begin{minipage}{0.93\linewidth}\centering
        \includegraphics[height=5.5cm, width=12.5cm]{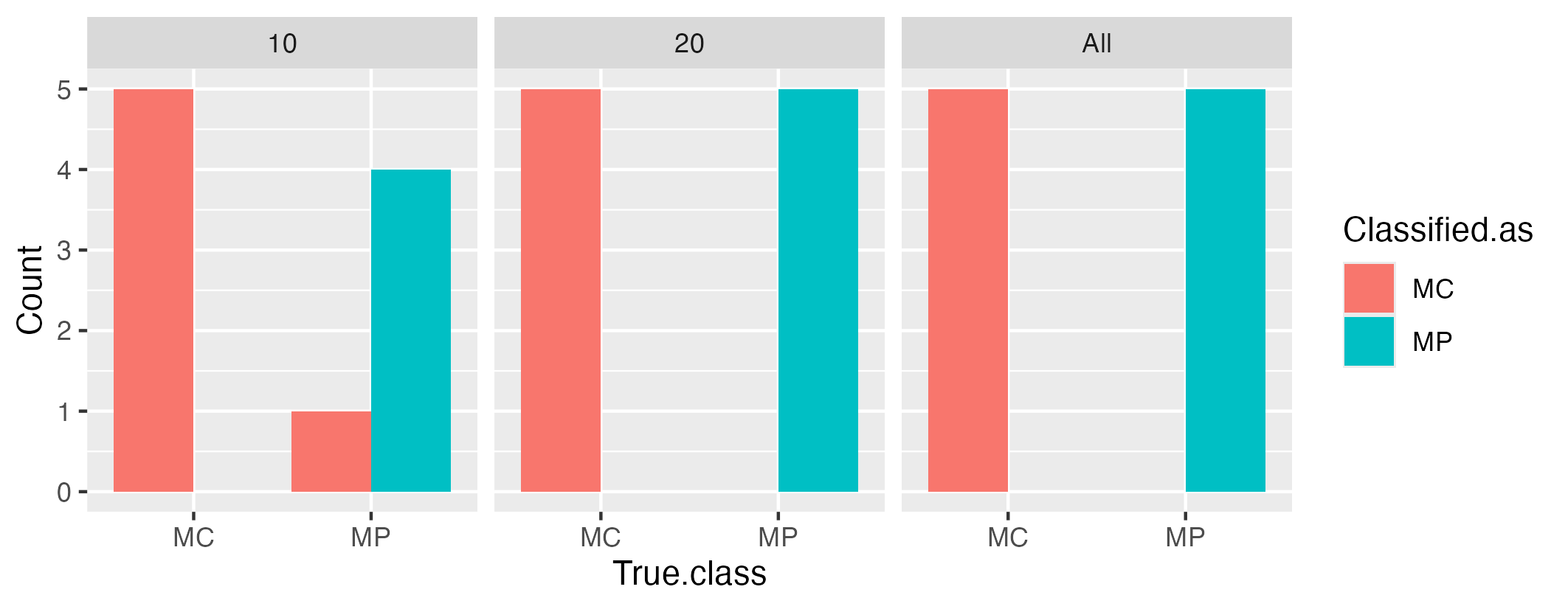}
    \end{minipage}
    \caption{Histograms of $k$-nearest neighbours classification accuracy using only the ratio, only the curvature and both ratio and curvature for discrimination when using a sample of 10, 20, and 'All' components, respectively. Misclassification rates are 3.3\%, 3.3\% and 0\% for 10, 20 and 'All' components, respectively, when using only the ratio, 0\%, 0\% and 0\% when using only the curvature and 3.3\%, 0\% and 0\% when using both characteristics for a sample of 20 realisations that were osculated by a disc of radius $r=5$.}
    \label{fig:knn_20_best_tissues_5}
\end{figure}
\begin{figure}[!ht]
    \centering
    \begin{minipage}{0.03\linewidth}\centering
        \rotatebox[origin=center]{90}{Ratio}
    \end{minipage}
    \begin{minipage}{0.93\linewidth}\centering
        \includegraphics[height=5.5cm, width=12.5cm]{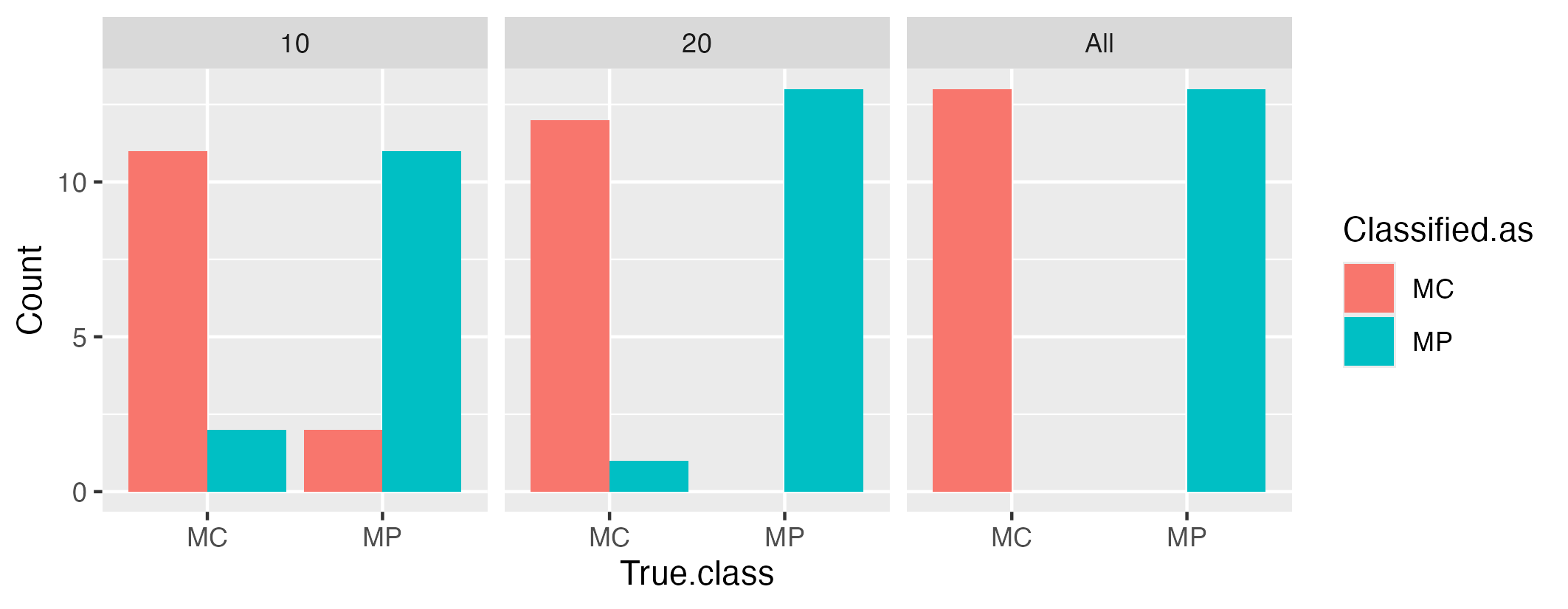}
    \end{minipage}
    \begin{minipage}{0.03\linewidth}\centering
        \rotatebox[origin=center]{90}{Curvature}
    \end{minipage}
    \begin{minipage}{0.93\linewidth}\centering
        \includegraphics[height=5.5cm, width=12.5cm]{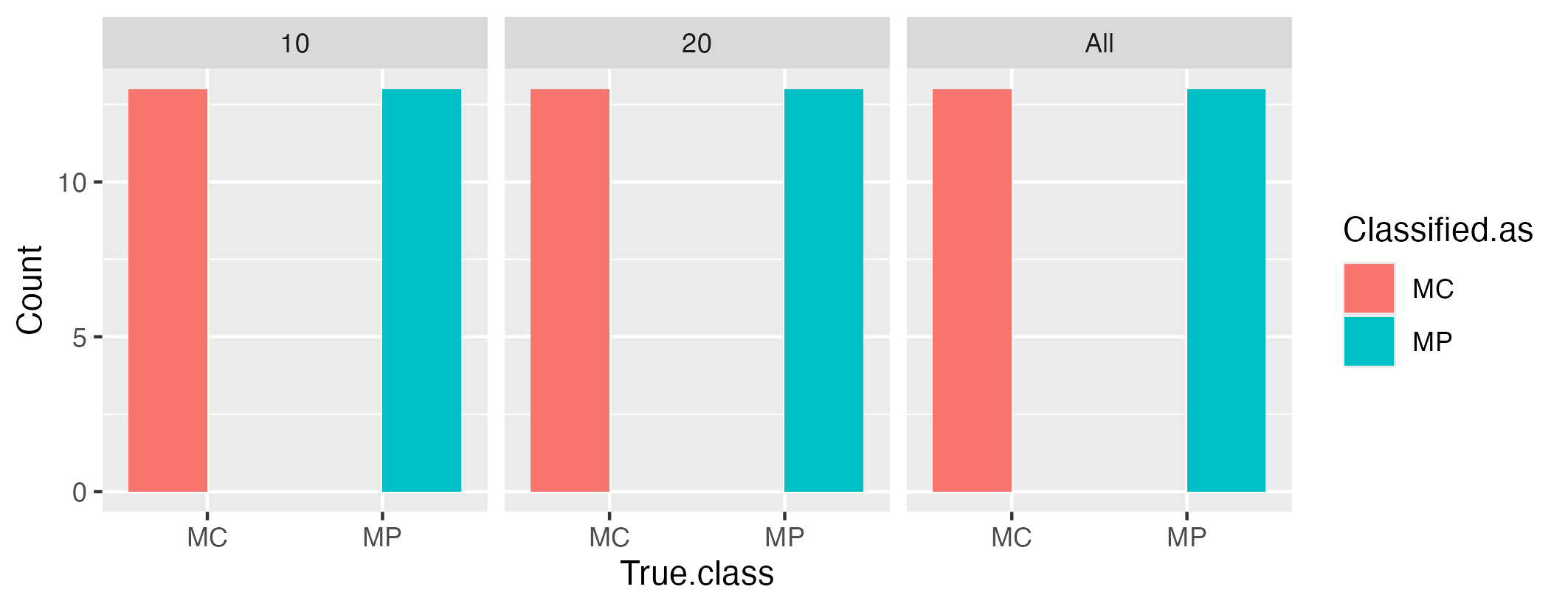}
    \end{minipage}
    \begin{minipage}{0.03\linewidth}\centering
        \rotatebox[origin=center]{90}{Both}
    \end{minipage}
    \begin{minipage}{0.93\linewidth}\centering
        \includegraphics[height=5.5cm, width=12.5cm]{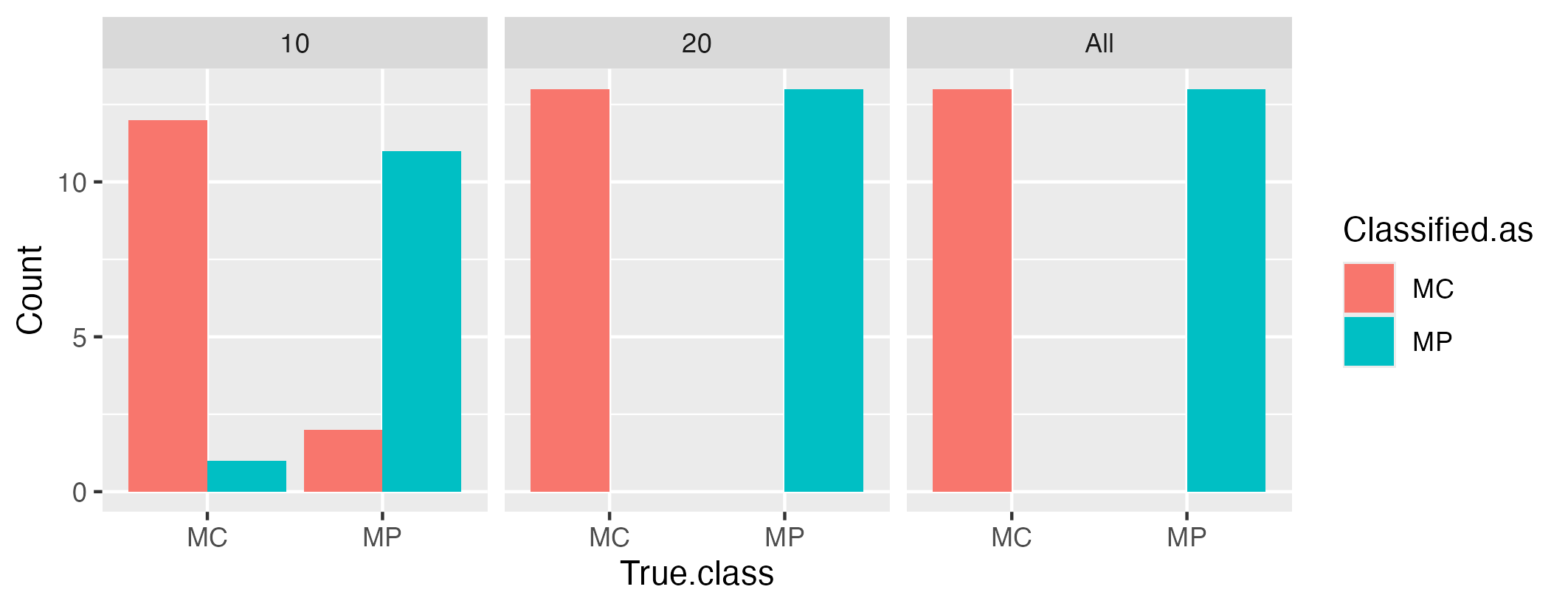}
    \end{minipage}
    \caption{Histograms of $k$-nearest neighbours classification accuracy using only the ratio, only the curvature and both ratio and curvature for discrimination when using a sample of 10, 20, and 'All' components, respectively. Misclassification rates are 5.4\%, 1.4\% and 0\% for 10, 20 and 'All' components, respectively, when using only the ratio, 0\%, 0\% and 0\% when using only the curvature and 4.1\%, 0\% and 0\% when using both characteristics for a sample of 50 realisations that were osculated by a disc of radius $r=5$.}
    \label{fig:knn_50_best_tissues_5}
\end{figure}

Histograms of classification accuracy using $k$-nearest neighbours algorithm on data obtained with osculating circle with radius $r=5$, when 20 and 50 realisations are considered, are shown in Figures \ref{fig:knn_20_best_tissues_5} and \ref{fig:knn_50_best_tissues_5}, respectively.

\begin{figure}[!ht]
    \centering
    \begin{minipage}{0.03\linewidth}\centering
        \rotatebox[origin=center]{90}{Ratio}
    \end{minipage}
    \begin{minipage}{0.93\linewidth}\centering
        \includegraphics[height=5.5cm, width=12.5cm]{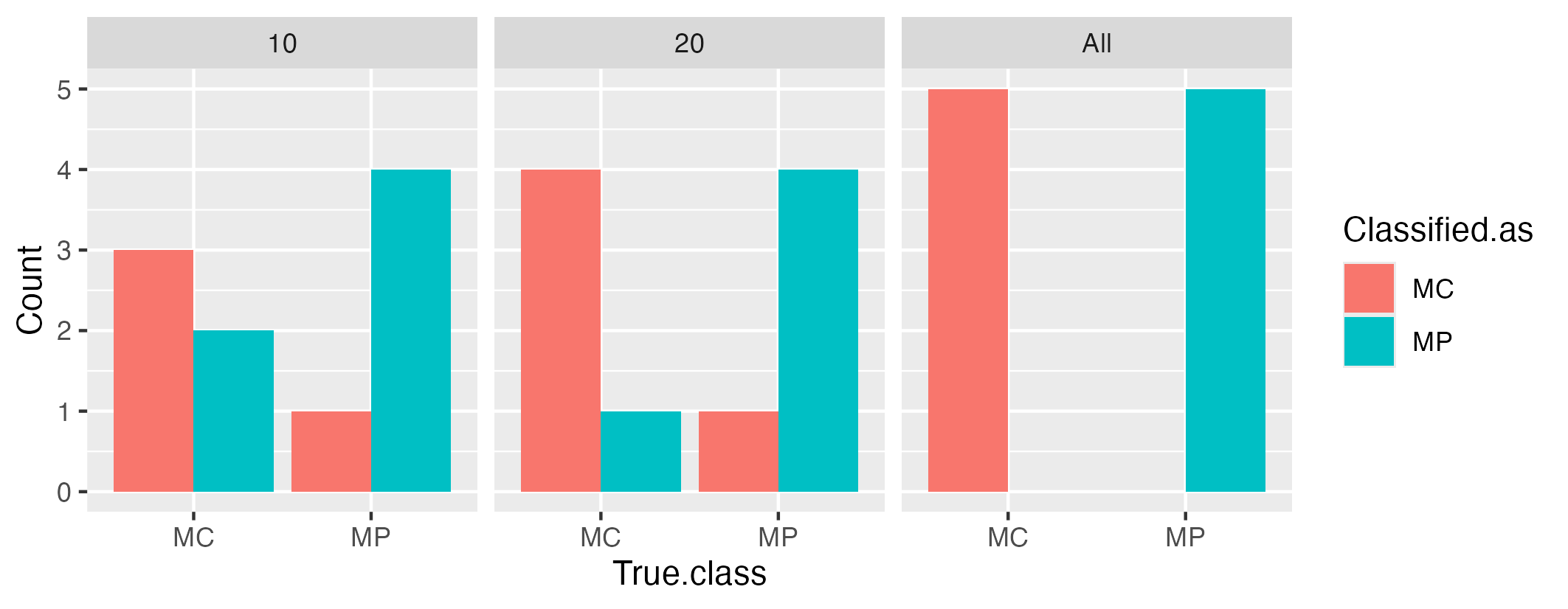}
    \end{minipage}
    \begin{minipage}{0.03\linewidth}\centering
        \rotatebox[origin=center]{90}{Curvature}
    \end{minipage}
    \begin{minipage}{0.93\linewidth}\centering
        \includegraphics[height=5.5cm, width=12.5cm]{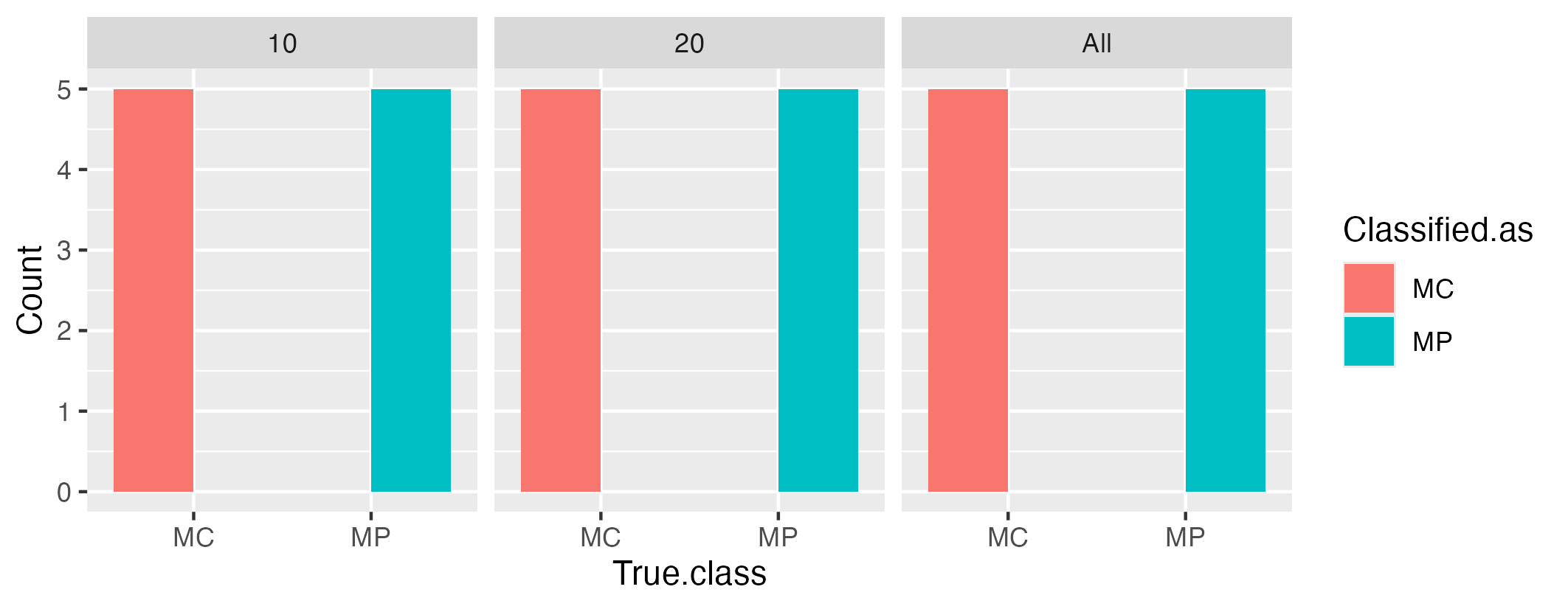}
    \end{minipage}
    \begin{minipage}{0.03\linewidth}\centering
        \rotatebox[origin=center]{90}{Both}
    \end{minipage}
    \begin{minipage}{0.93\linewidth}\centering
        \includegraphics[height=5.5cm, width=12.5cm]{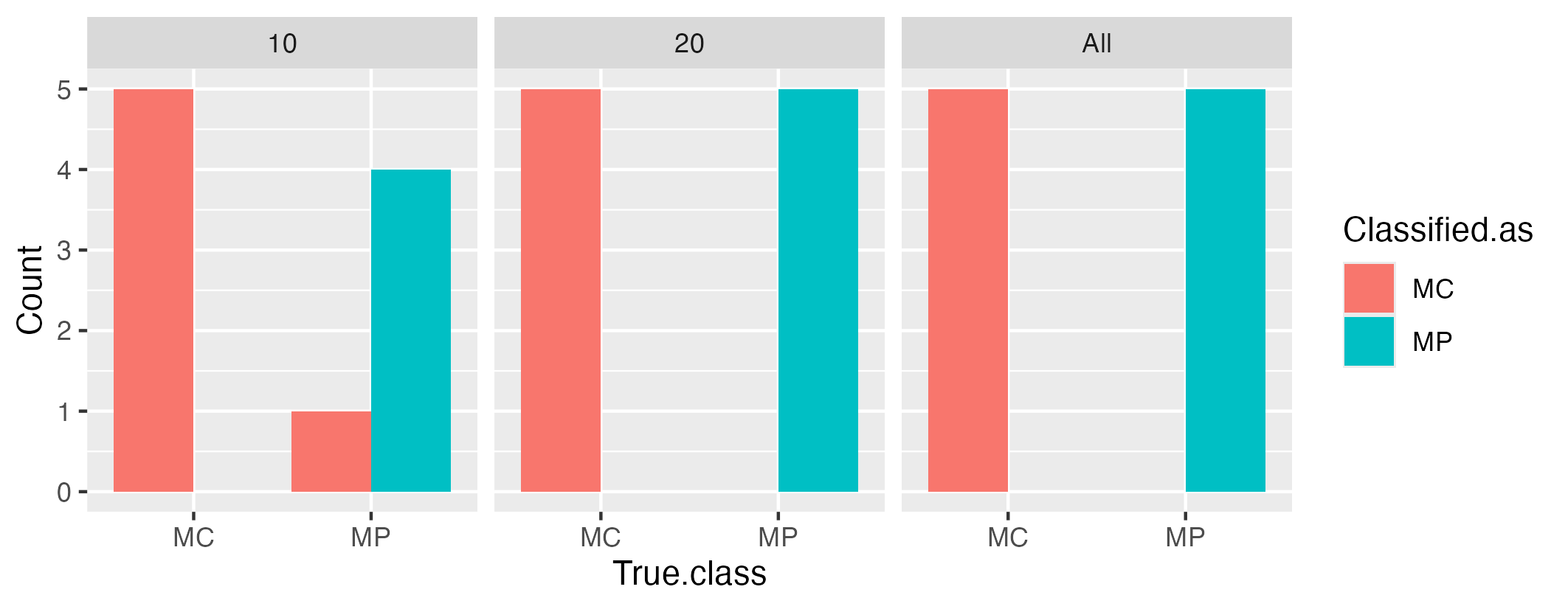}
    \end{minipage}
    \caption{Histograms of $k$-nearest neighbours classification accuracy using only the ratio, only the curvature and both ratio and curvature for discrimination when using a sample of 10, 20, and 'All' components, respectively. Misclassification rates are 10\%, 6.7\% and 0\% for 10, 20 and 'All' components, respectively, when using only the ratio, 0\%, 0\% and 0\% when using only the curvature and 3.3\%, 0\% and 0\% when using both characteristics for a sample of 20 realisations that were osculated by a disc of radius $r=3$.}
    \label{fig:knn_20_best_tissues_3}
\end{figure}

\begin{figure}[!ht]
    \centering
    \begin{minipage}{0.03\linewidth}\centering
        \rotatebox[origin=center]{90}{Ratio}
    \end{minipage}
    \begin{minipage}{0.93\linewidth}\centering
        \includegraphics[height=5.5cm, width=12.5cm]{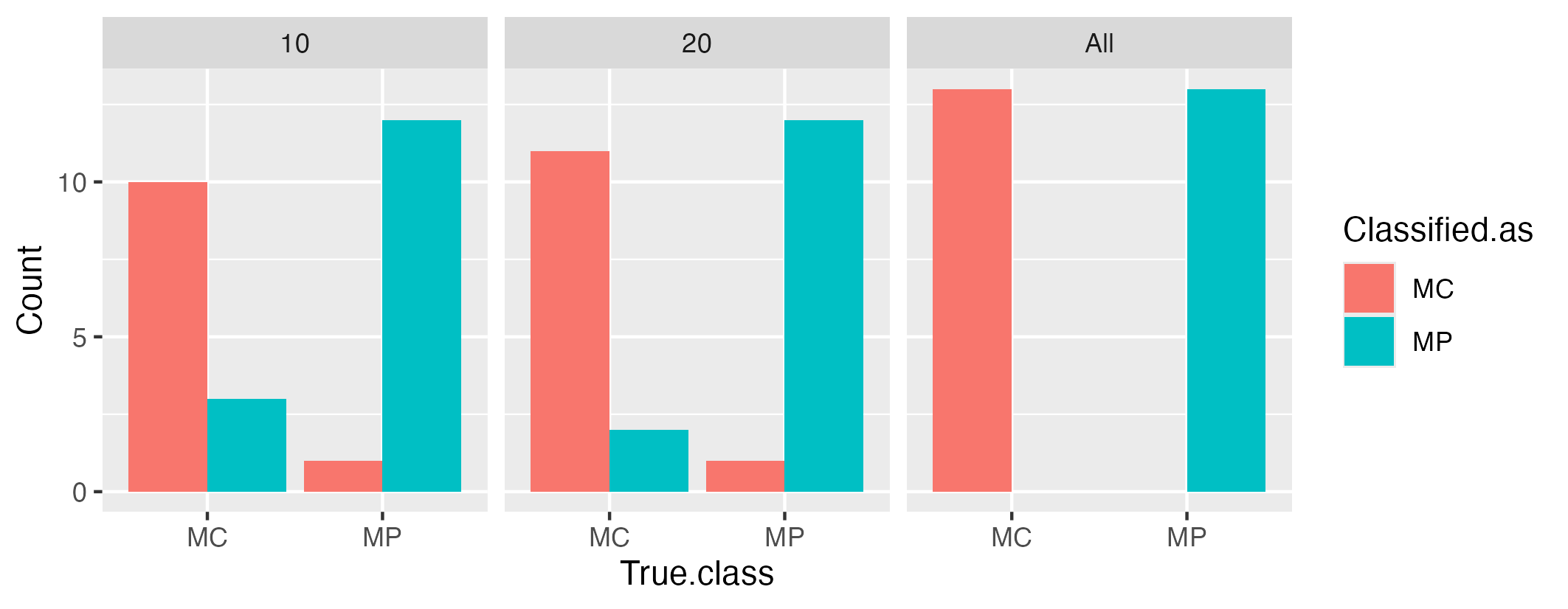}
    \end{minipage}
    \begin{minipage}{0.03\linewidth}\centering
        \rotatebox[origin=center]{90}{Curvature}
    \end{minipage}
    \begin{minipage}{0.93\linewidth}\centering
        \includegraphics[height=5.5cm, width=12.5cm]{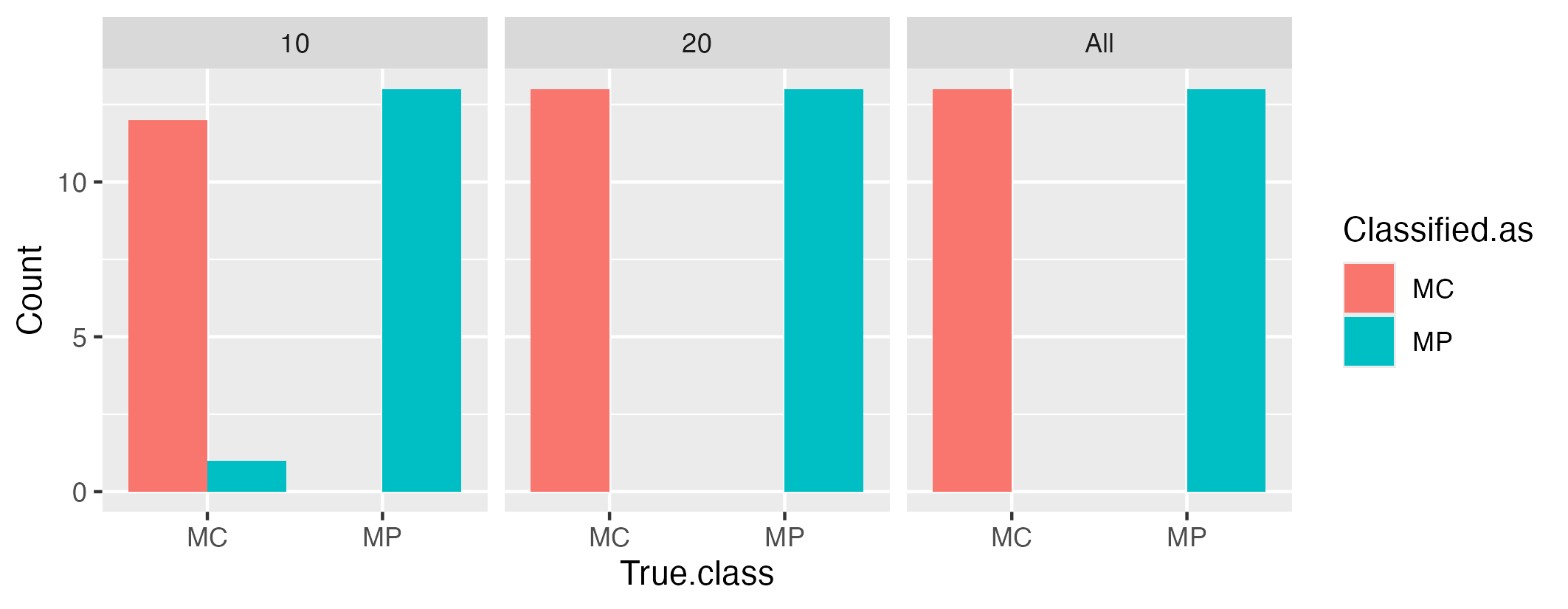}
    \end{minipage}
    \begin{minipage}{0.03\linewidth}\centering
        \rotatebox[origin=center]{90}{Both}
    \end{minipage}
    \begin{minipage}{0.93\linewidth}\centering
        \includegraphics[height=5.5cm, width=12.5cm]{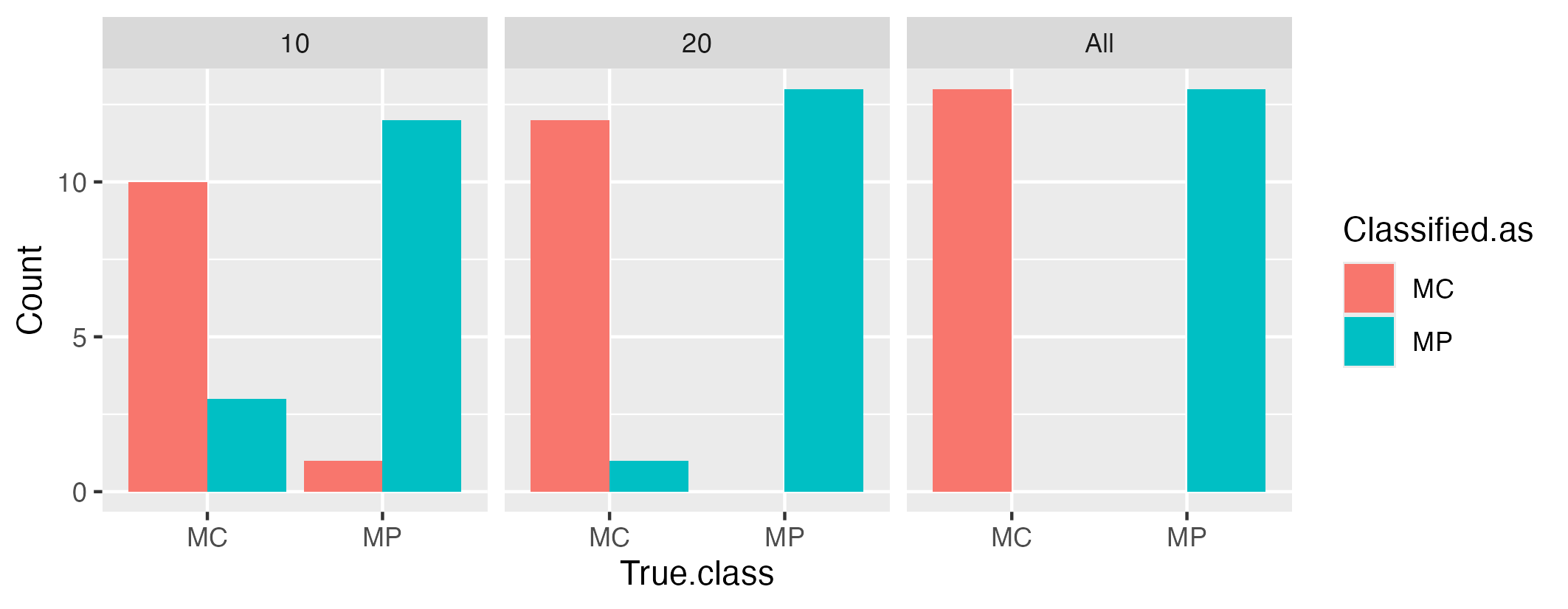}
    \end{minipage}
    \caption{Histograms of $k$-nearest neighbours classification accuracy using only the ratio, only the curvature and both ratio and curvature for discrimination when using a sample of 10, 20, and 'All' components, respectively. Misclassification rates are 5.4\%, 4.1\% and 0\% for 10, 20 and 'All' components, respectively, when using only the ratio, 1.4\%, 0\% and 0\% when using only the curvature and 5.4\%, 1.4\% and 0\% when using both characteristics for a sample of 50 realisations that were osculated by a disc of radius $r=3$.}
    \label{fig:knn_50_best_tissues_3}
\end{figure}

\begin{figure}[!ht]
    \centering
    \begin{minipage}{0.03\linewidth}\centering
        \rotatebox[origin=center]{90}{Ratio}
    \end{minipage}
    \begin{minipage}{0.93\linewidth}\centering
        \includegraphics[height=5.5cm, width=12.5cm]{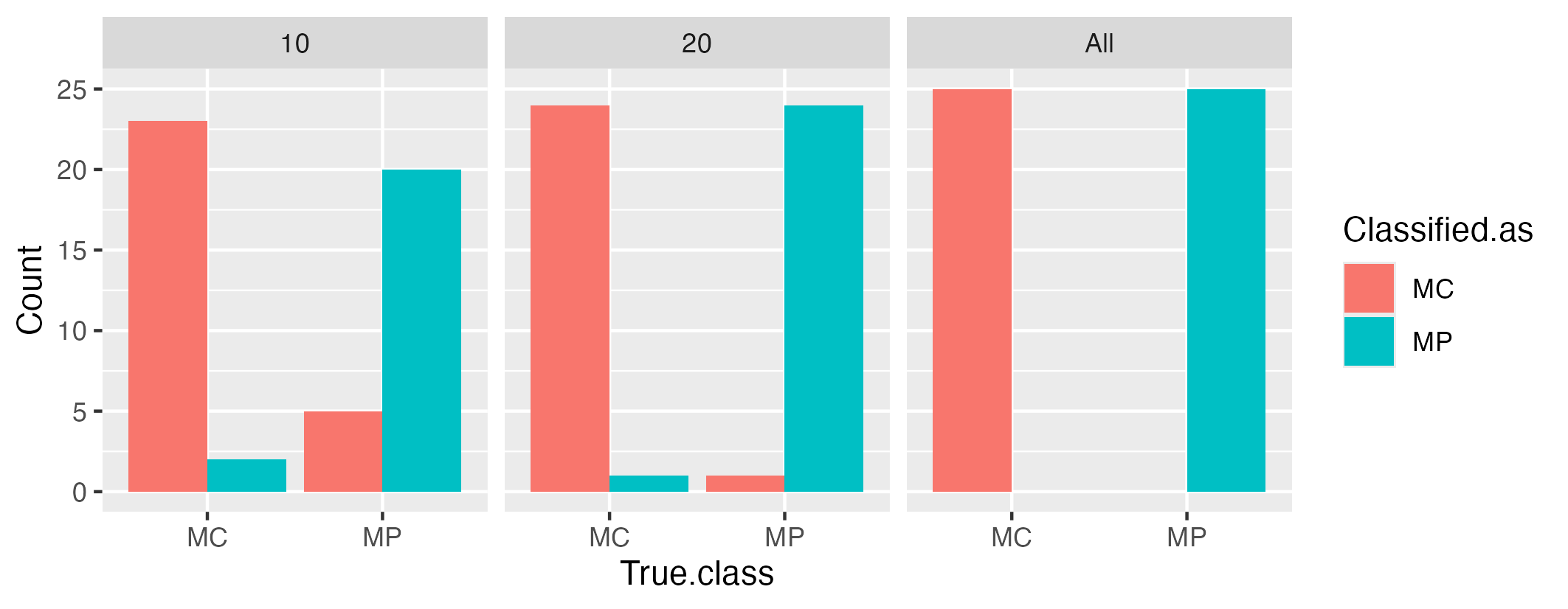}
    \end{minipage}
    \begin{minipage}{0.03\linewidth}\centering
        \rotatebox[origin=center]{90}{Curvature}
    \end{minipage}
    \begin{minipage}{0.93\linewidth}\centering
        \includegraphics[height=5.5cm, width=12.5cm]{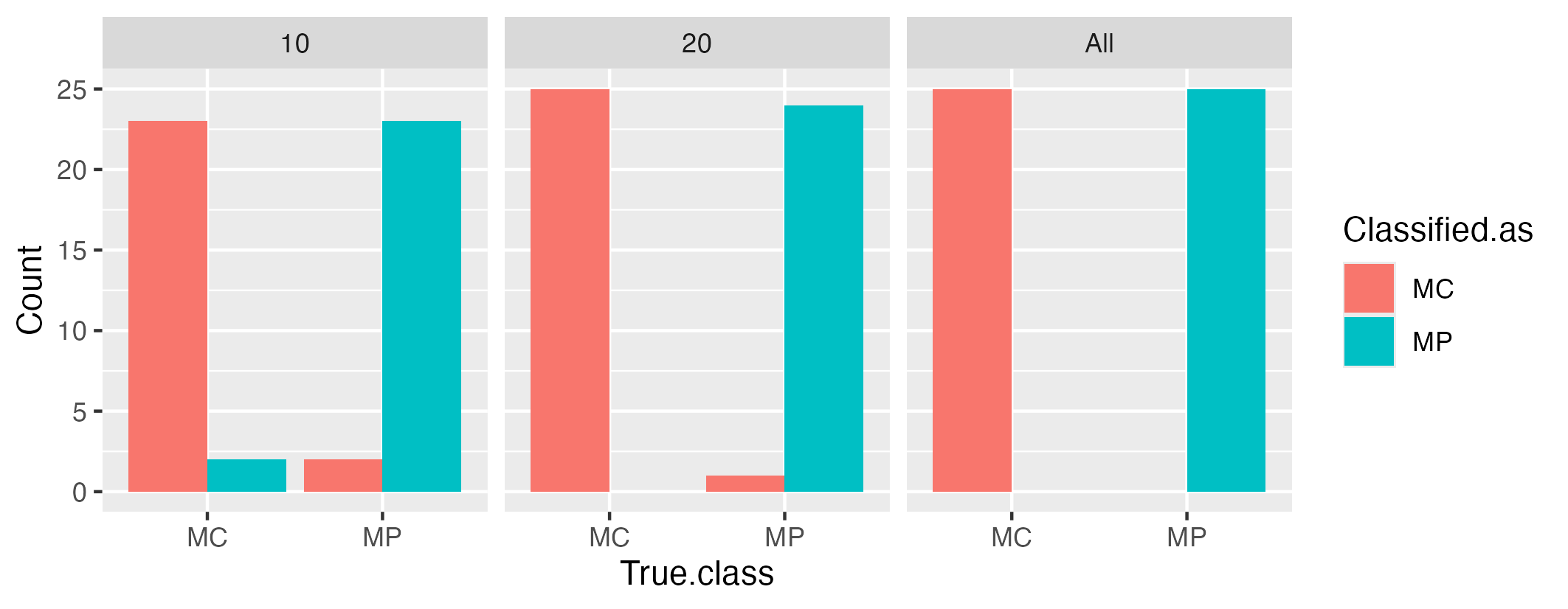}
    \end{minipage}
    \begin{minipage}{0.03\linewidth}\centering
        \rotatebox[origin=center]{90}{Both}
    \end{minipage}
    \begin{minipage}{0.93\linewidth}\centering
        \includegraphics[height=5.5cm, width=12.5cm]{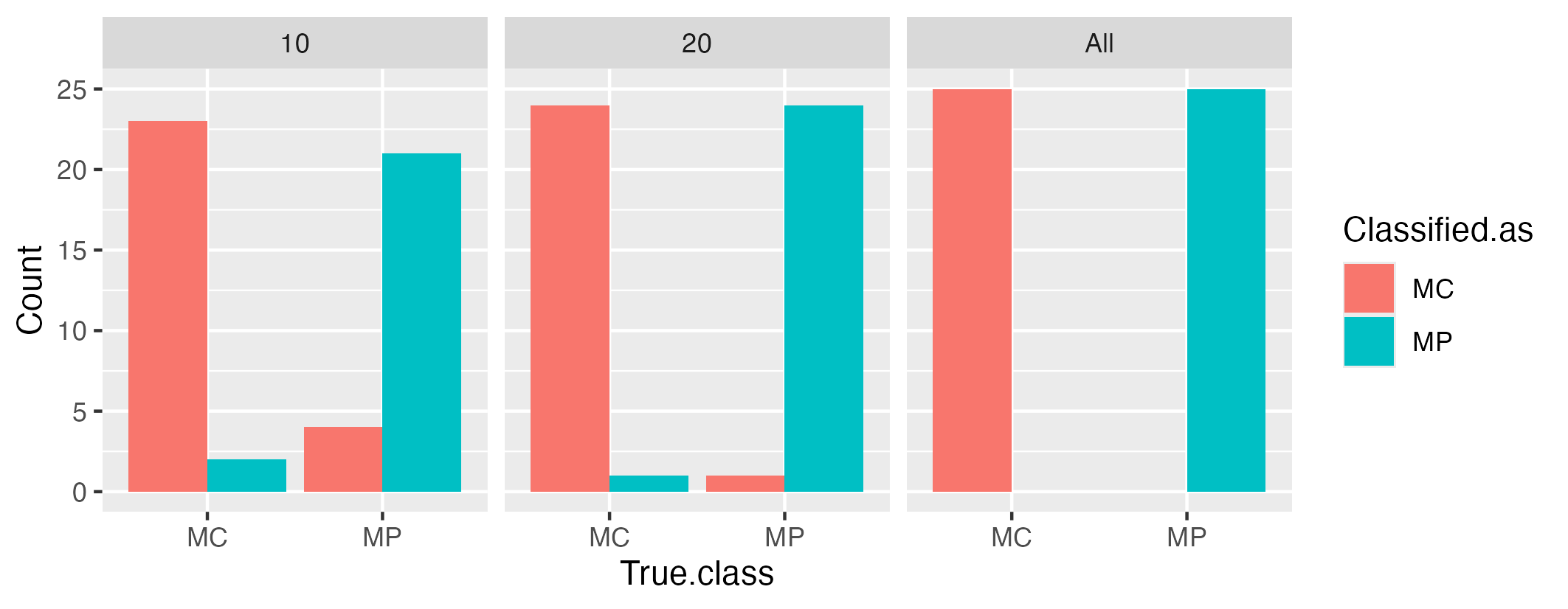}
    \end{minipage}
    \caption{Histograms of $k$-nearest neighbours classification accuracy using only the ratio, only the curvature and both ratio and curvature for discrimination when using a sample of 10, 20, and 'All' components, respectively. Misclassification rates are 4.7\%, 1.3\% and 0\% for 10, 20 and 'All' components, respectively, when using only the ratio, 2.7\%, 0.7\% and 0\% when using only the curvature and 4\%, 1.3\% and 0\% when using both characteristics for a sample of 100 realisations that were osculated by a disc of radius $r=3$.}
    \label{fig:knn_100_best_tissues_3}
\end{figure}

\begin{figure}[!ht]
    \centering
    \begin{minipage}{0.03\linewidth}\centering
        \rotatebox[origin=center]{90}{20 Realisations}
    \end{minipage}
    \begin{minipage}{0.93\linewidth}\centering
        \includegraphics[height=5.5cm, width=12.3cm]{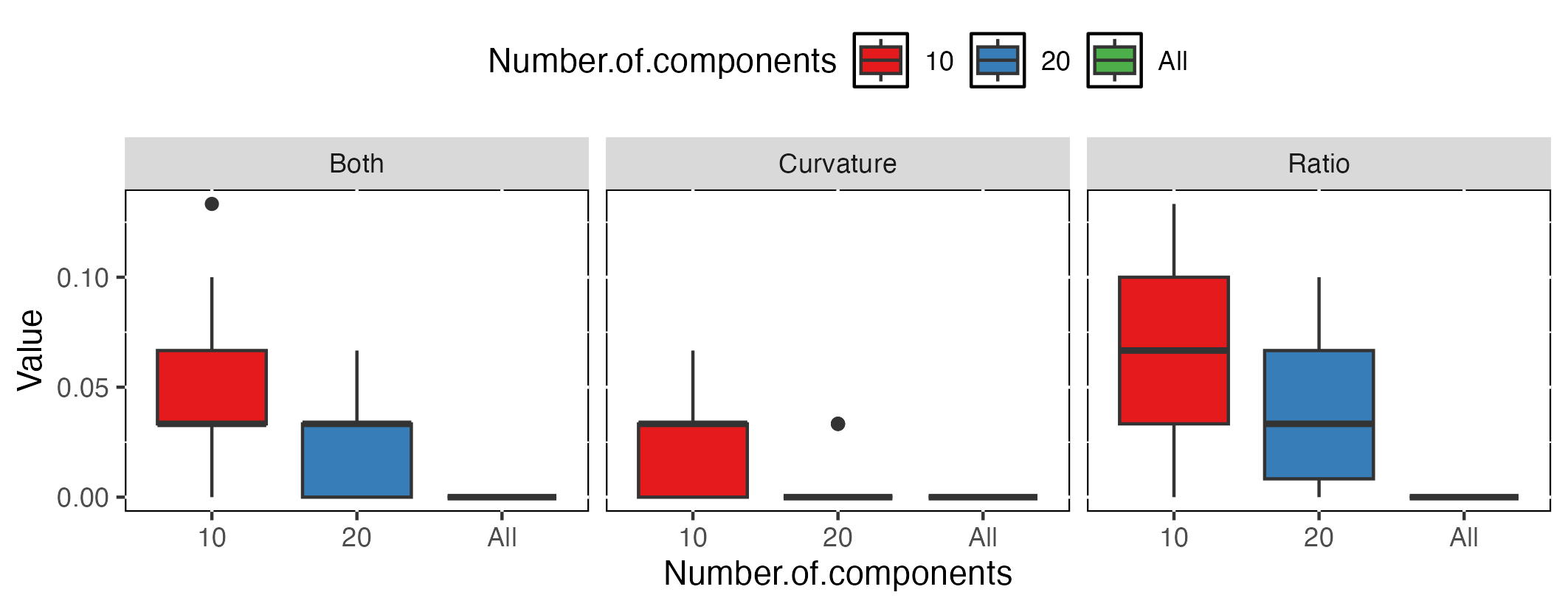}
    \end{minipage}
    \begin{minipage}{0.03\linewidth}\centering
        \rotatebox[origin=center]{90}{50 Realisations}
    \end{minipage}
    \begin{minipage}{0.93\linewidth}\centering
        \includegraphics[height=5.5cm, width=12.3cm]{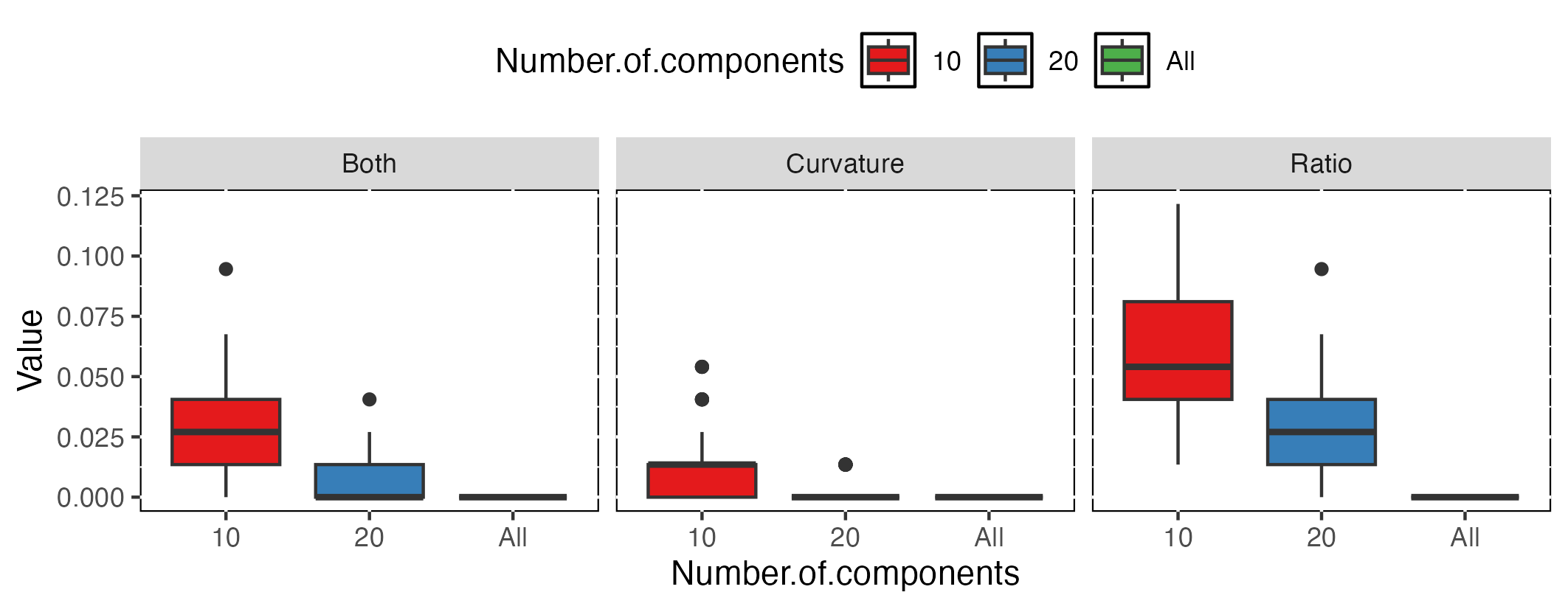}
    \end{minipage}
    \begin{minipage}{0.03\linewidth}\centering
        \rotatebox[origin=center]{90}{100 Realisations}
    \end{minipage}
    \begin{minipage}{0.93\linewidth}\centering
        \includegraphics[height=5.5cm, width=12.3cm]{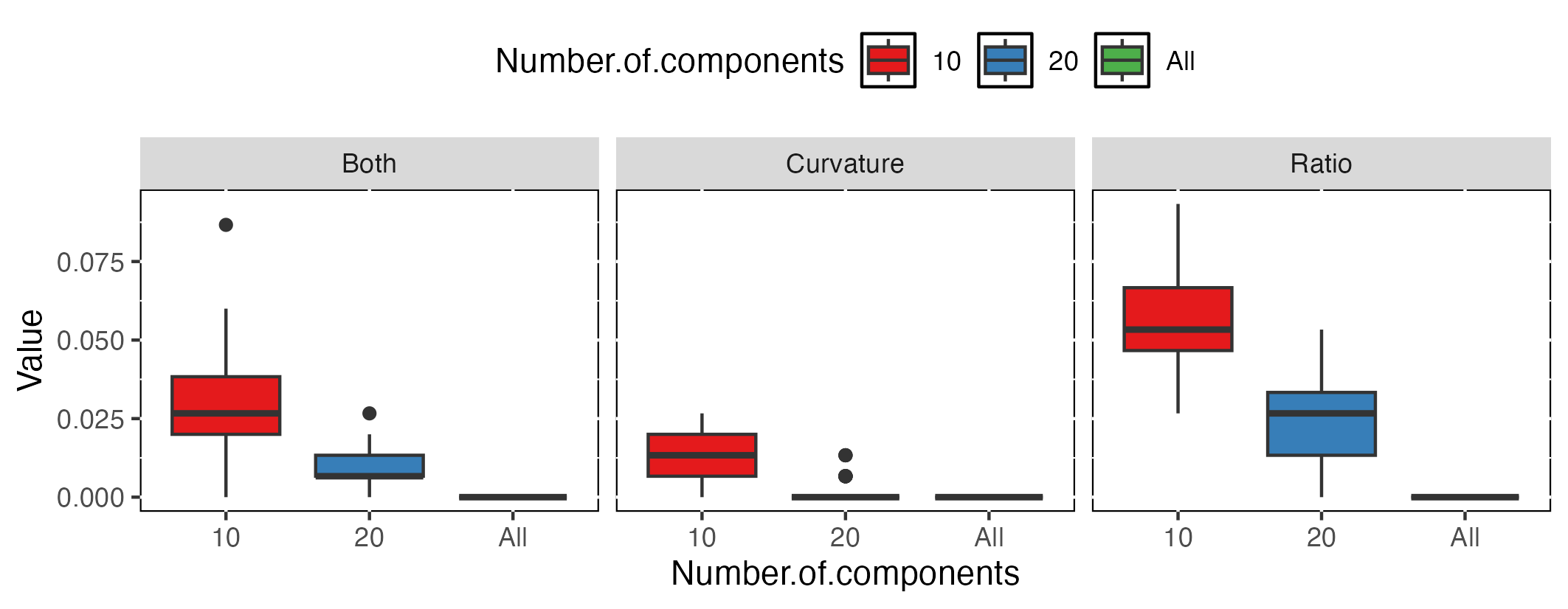}
    \end{minipage}
    \caption{Boxplots of misclassification rate for 50 runs of $k$-nearest neighbours algorithm when considering samples of 20 (top), 50 (middle) and 100 (bottom) realisations using both ratio and curvature, only the curvature and only the ratio for discrimination, respectively. For each setting, misclassification rates for different number of components considered (namely 10, 20, and 'All') are shown. Note that the characteristics were obtained using an osculating disc of radius $r=3$.}
    \label{fig:box_knn_tissues_3}
\end{figure}

Figures \ref{fig:knn_20_best_tissues_3}, \ref{fig:knn_50_best_tissues_3} and \ref{fig:knn_100_best_tissues_3} represent the histograms of classification accuracy for supervised classification using $k$-nearest neighbours algorithm when 20, 50 and 100 realisations are considered, respectively. Corresponding boxplots of misclassification rate are shown in Figure \ref{fig:box_knn_tissues_3}.

\subsection{Unsupervised classification}
\subsubsection{Non-hierarchical clustering}

\begin{figure}[!ht]
    \centering
    \begin{minipage}{0.03\linewidth}\centering
        \rotatebox[origin=center]{90}{Ratio}
    \end{minipage}
    \begin{minipage}{0.93\linewidth}\centering
        \includegraphics[height=5.5cm, width=12.5cm]{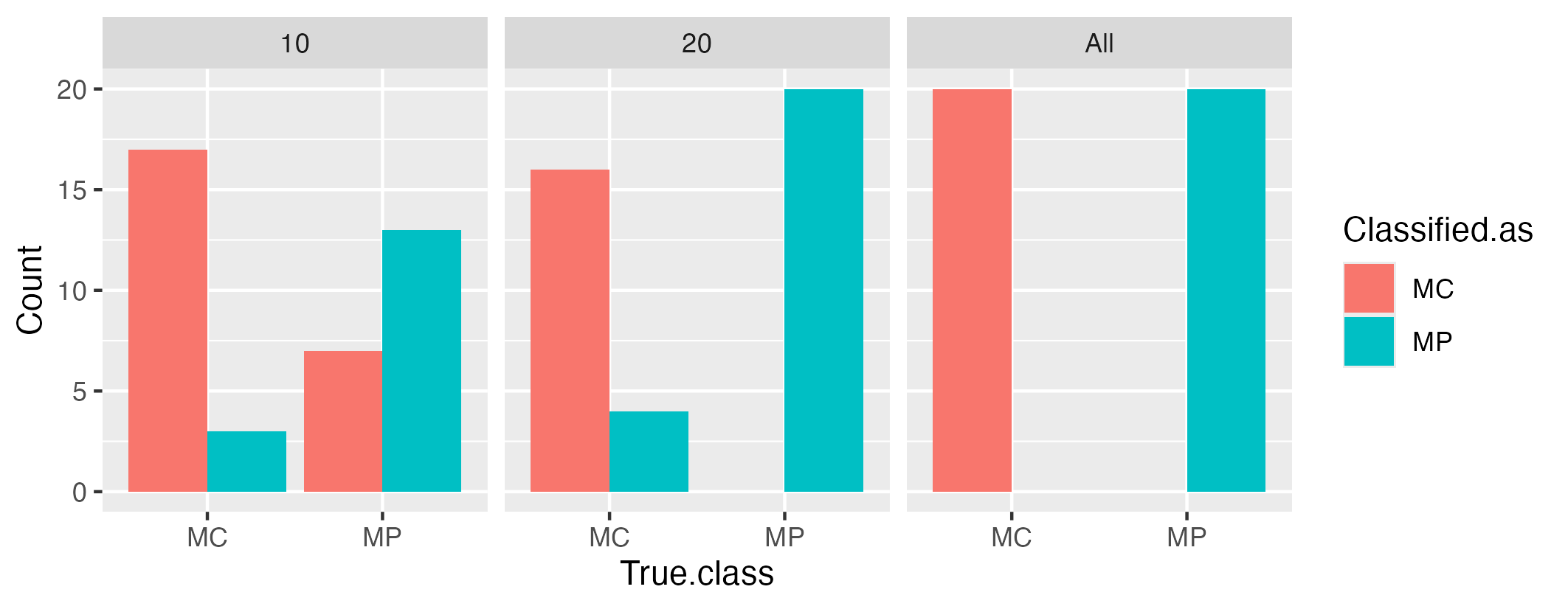}
    \end{minipage}
    \begin{minipage}{0.03\linewidth}\centering
        \rotatebox[origin=center]{90}{Curvature}
    \end{minipage}
    \begin{minipage}{0.93\linewidth}\centering
        \includegraphics[height=5.5cm, width=12.5cm]{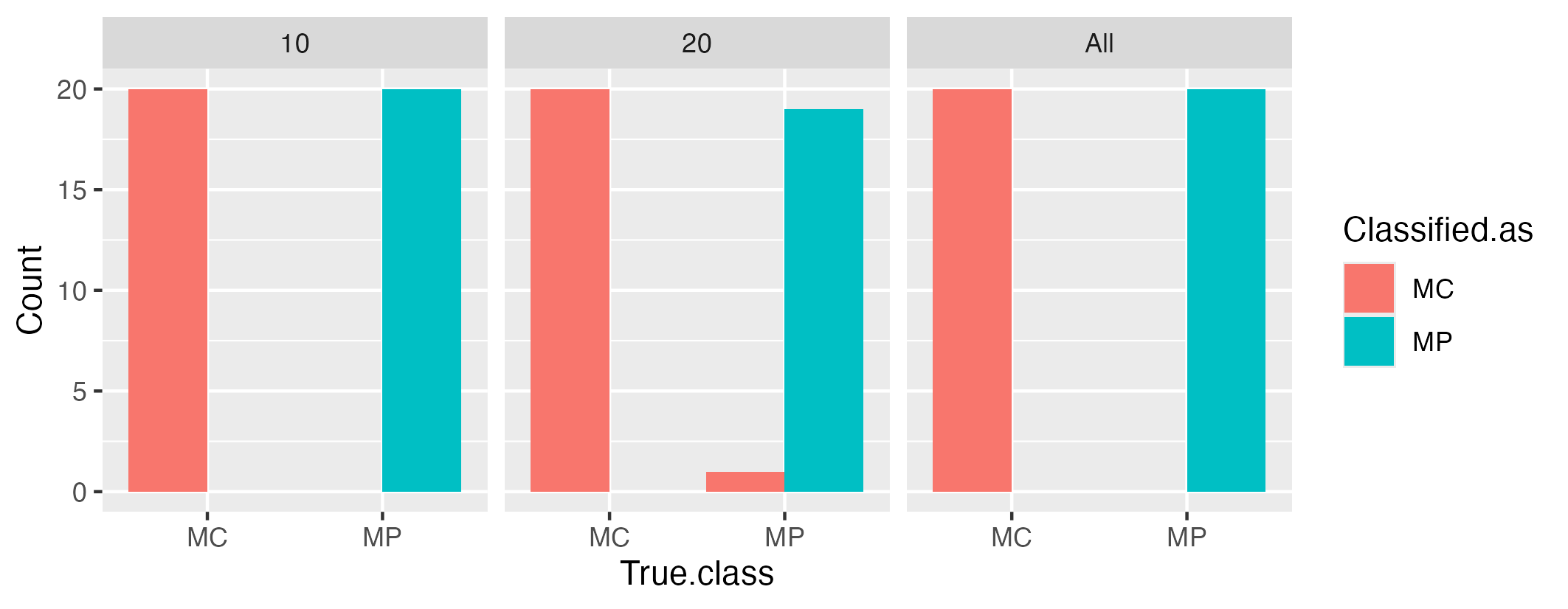}
    \end{minipage}
    \begin{minipage}{0.03\linewidth}\centering
        \rotatebox[origin=center]{90}{Both}
    \end{minipage}
    \begin{minipage}{0.93\linewidth}\centering
        \includegraphics[height=5.5cm, width=12.5cm]{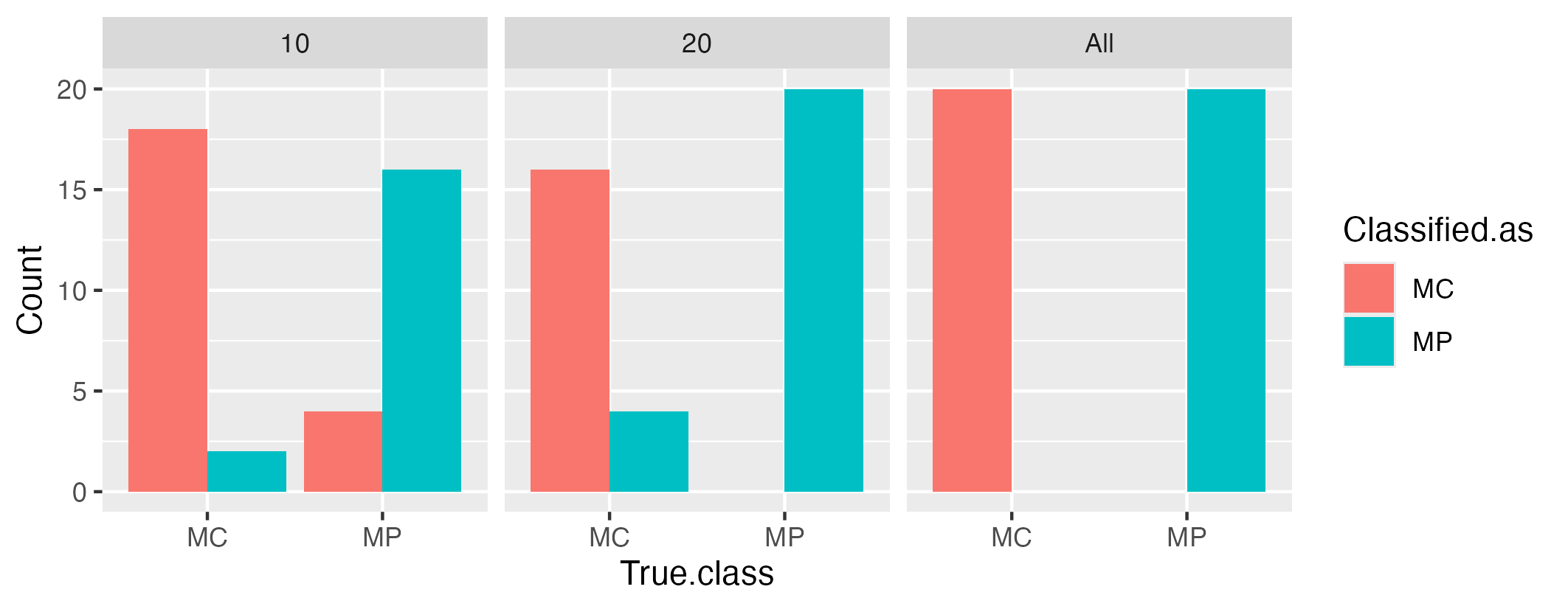}
    \end{minipage}
    \caption{Histograms of $k$-medoids classification accuracy using only the ratio, only the curvature and both ratio and curvature for discrimination when using a sample of 10, 20, and 'All' components, respectively. Misclassification rates are 25\%, 10\% and 0\% for 10, 20 and 'All' components, respectively, when using only the ratio, 0\%, 2.5\% and 0\% when using only the curvature and 15\%, 10\% and 0\% when using both characteristics for a sample of 20 realisations that were osculated by a disc of radius $r=5$.}
    \label{fig:kmed_20_best_tissues_5}
\end{figure}

\begin{figure}[!ht]
    \centering
    \begin{minipage}{0.03\linewidth}\centering
        \rotatebox[origin=center]{90}{Ratio}
    \end{minipage}
    \begin{minipage}{0.93\linewidth}\centering
        \includegraphics[height=5.5cm, width=12.5cm]{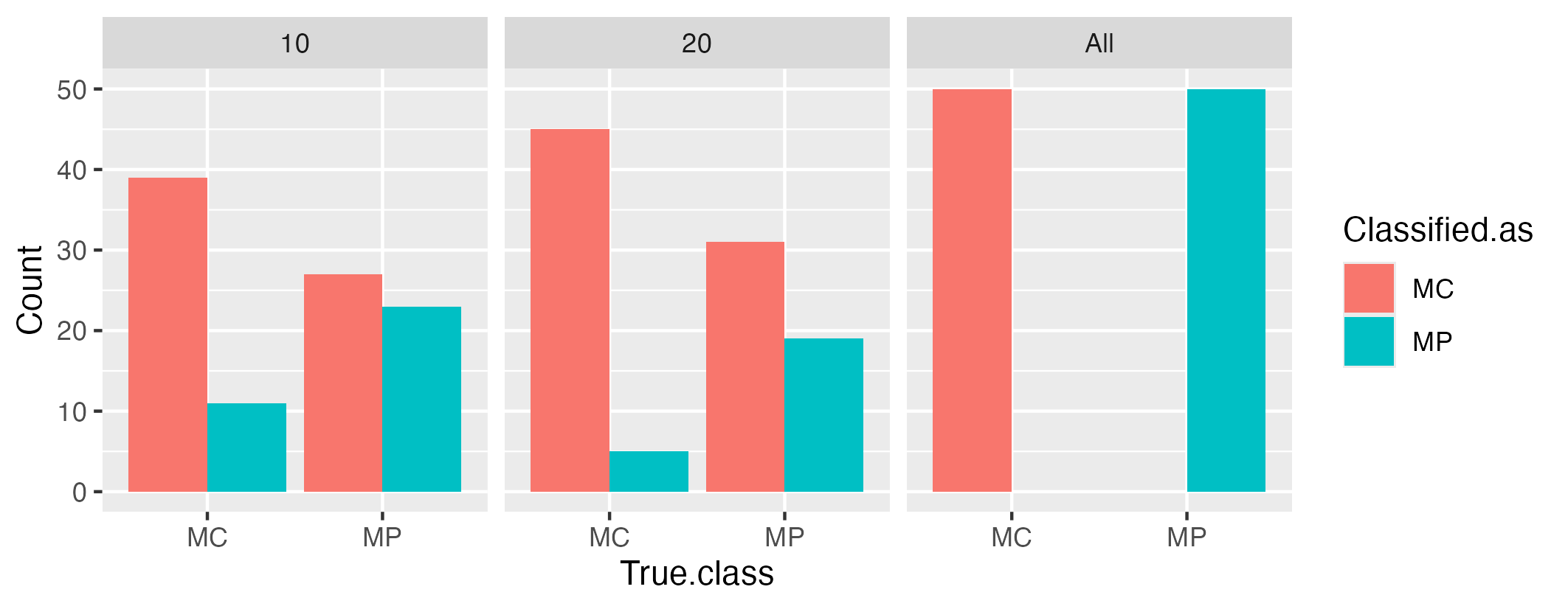}
    \end{minipage}
    \begin{minipage}{0.03\linewidth}\centering
        \rotatebox[origin=center]{90}{Curvature}
    \end{minipage}
    \begin{minipage}{0.93\linewidth}\centering
        \includegraphics[height=5.5cm, width=12.5cm]{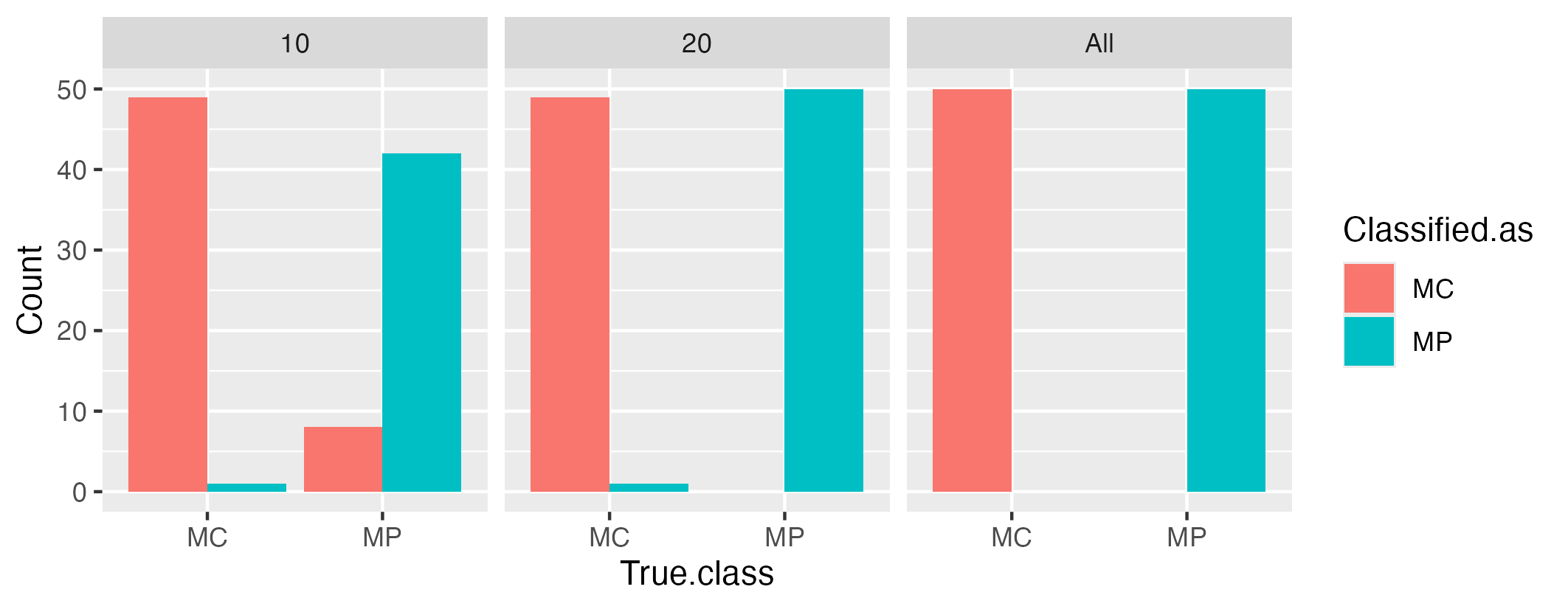}
    \end{minipage}
    \begin{minipage}{0.03\linewidth}\centering
        \rotatebox[origin=center]{90}{Both}
    \end{minipage}
    \begin{minipage}{0.93\linewidth}\centering
        \includegraphics[height=5.5cm, width=12.5cm]{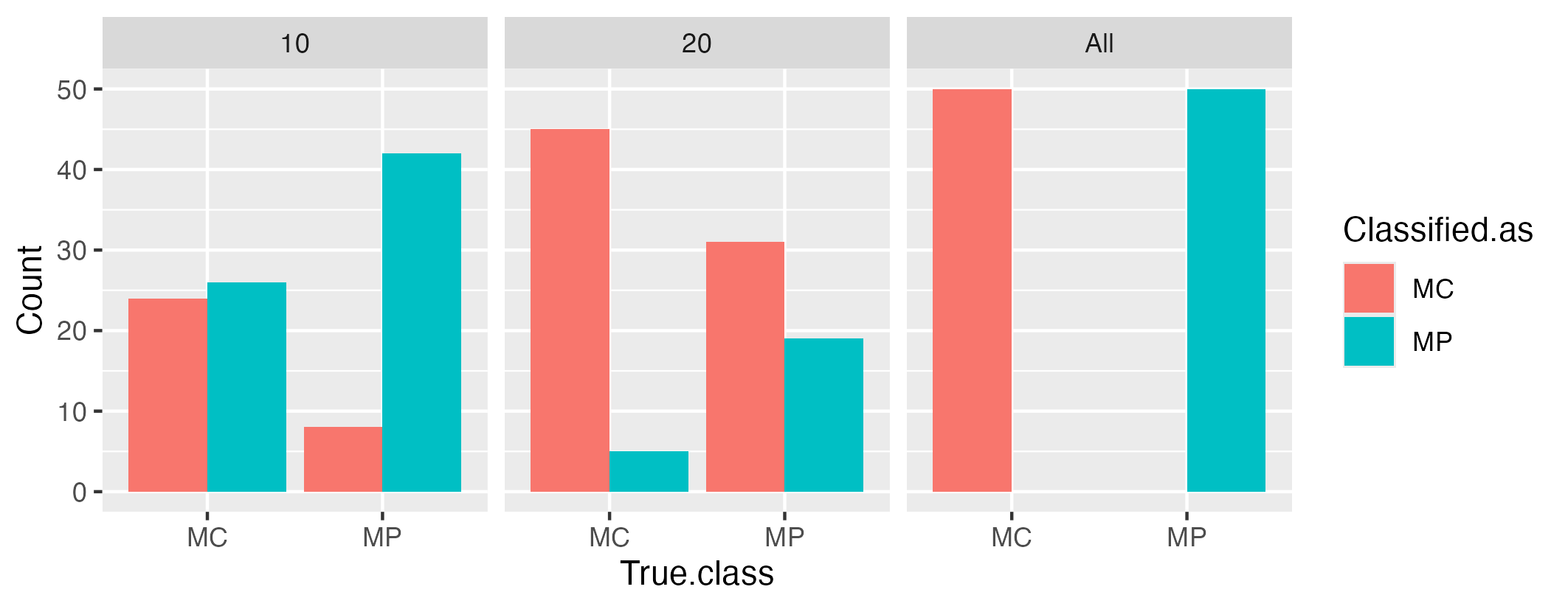}
    \end{minipage}
    \caption{Histograms of $k$-medoids classification accuracy using only the ratio, only the curvature and both ratio and curvature for discrimination when using a sample of 10, 20, and 'All' components, respectively. Misclassification rates are 38\%, 36\% and 0\% for 10, 20 and 'All' components, respectively, when using only the ratio, 9\%, 1\% and 0\% when using only the curvature and 34\%, 36\% and 0\% when using both characteristics for a sample of 50 realisations that were osculated by a disc of radius $r=5$.}
    \label{fig:kmed_50_best_tissues_5}
\end{figure}

Histograms of classification accuracy using $k$-medoids algorithm on data obtained with osculating circle with radius $r=5$, when 20 and 50 realisations are considered, are shown in Figures \ref{fig:kmed_20_best_tissues_5} and \ref{fig:kmed_50_best_tissues_5}, respectively.

\begin{figure}[!ht]
    \centering
    \begin{minipage}{0.03\linewidth}\centering
        \rotatebox[origin=center]{90}{Ratio}
    \end{minipage}
    \begin{minipage}{0.93\linewidth}\centering
        \includegraphics[height=5.5cm, width=12.5cm]{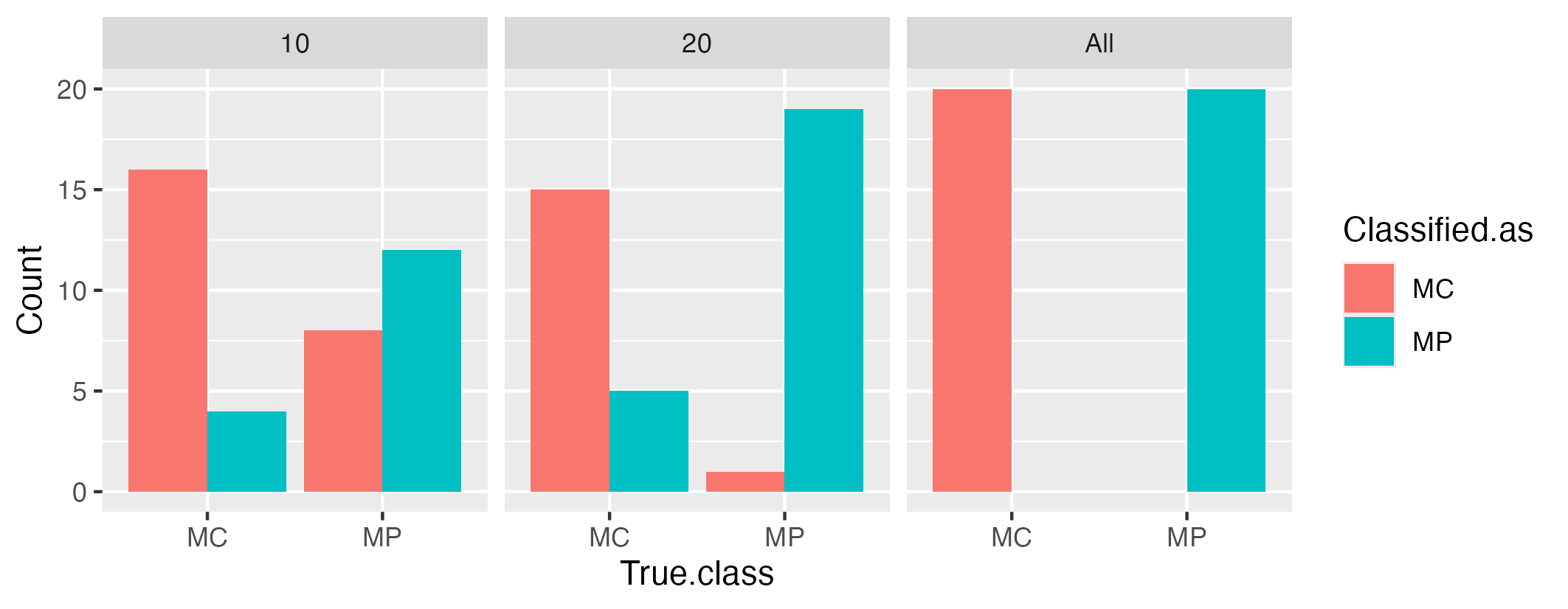}
    \end{minipage}
    \begin{minipage}{0.03\linewidth}\centering
        \rotatebox[origin=center]{90}{Curvature}
    \end{minipage}
    \begin{minipage}{0.93\linewidth}\centering
        \includegraphics[height=5.5cm, width=12.5cm]{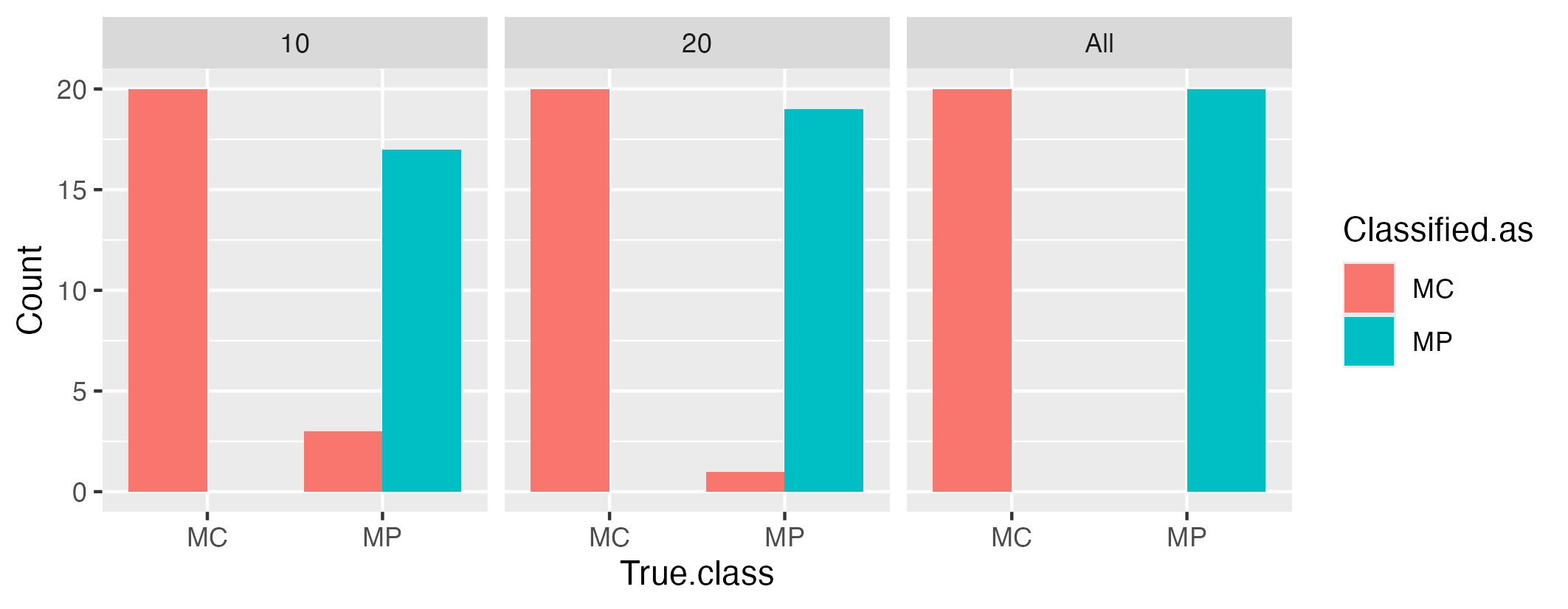}
    \end{minipage}
    \begin{minipage}{0.03\linewidth}\centering
        \rotatebox[origin=center]{90}{Both}
    \end{minipage}
    \begin{minipage}{0.93\linewidth}\centering
        \includegraphics[height=5.5cm, width=12.5cm]{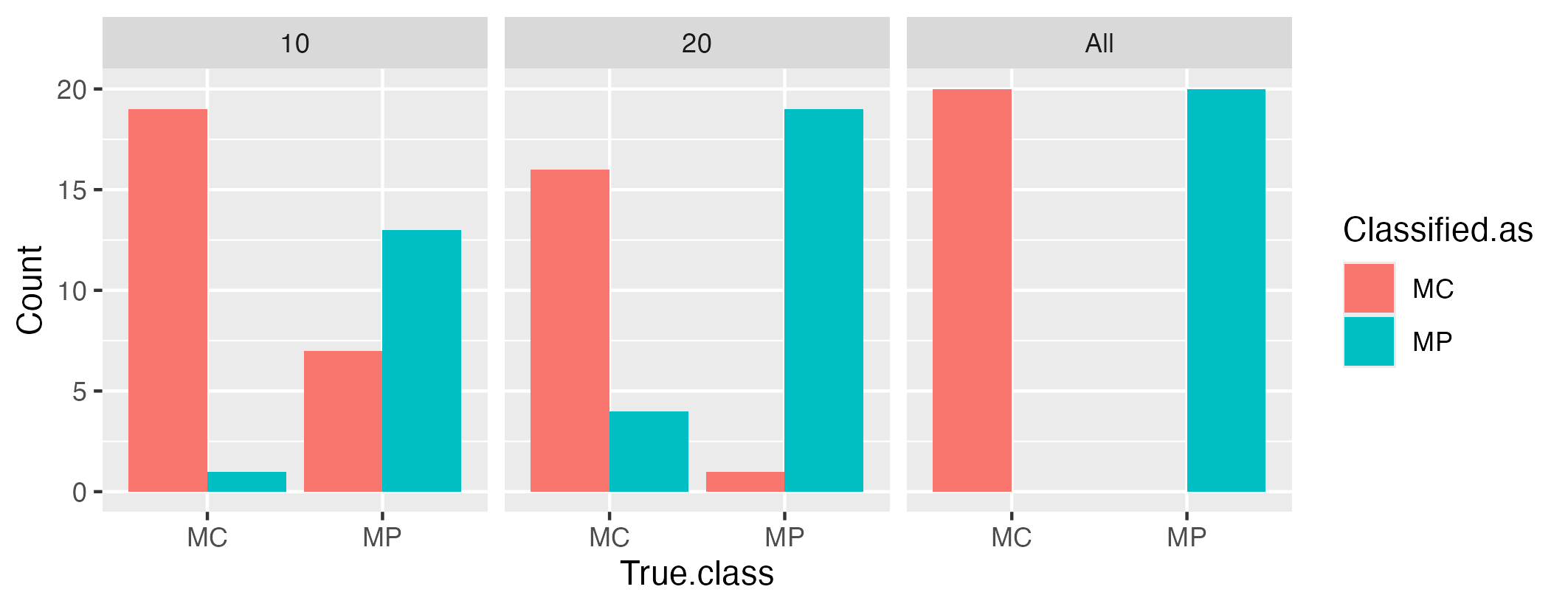}
    \end{minipage}
    \caption{Histograms of $k$-medoids classification accuracy using only the ratio, only the curvature and both ratio and curvature for discrimination when using a sample of 10, 20, and 'All' components, respectively. Misclassification rates are 30\%, 15\% and 0\% for 10, 20 and 'All' components, respectively, when using only the ratio, 7.5\%, 2.5\% and 0\% when using only the curvature and 20\%, 12.5\% and 0\% when using both characteristics for a sample of 20 realisations that were osculated by a disc of radius $r=3$.}
    \label{fig:kmed_20_best_tissues_3}
\end{figure}

\begin{figure}[!ht]
    \centering
    \begin{minipage}{0.03\linewidth}\centering
        \rotatebox[origin=center]{90}{Ratio}
    \end{minipage}
    \begin{minipage}{0.93\linewidth}\centering
        \includegraphics[height=5.5cm, width=12.5cm]{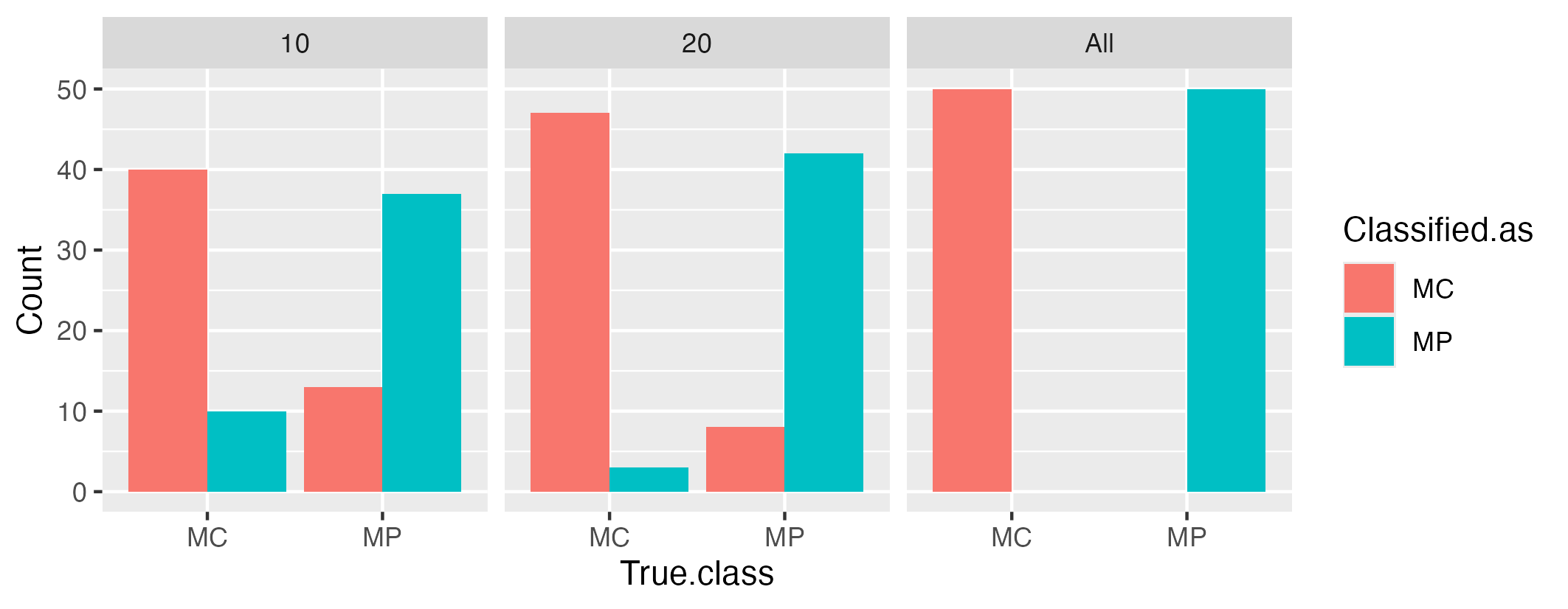}
    \end{minipage}
    \begin{minipage}{0.03\linewidth}\centering
        \rotatebox[origin=center]{90}{Curvature}
    \end{minipage}
    \begin{minipage}{0.93\linewidth}\centering
        \includegraphics[height=5.5cm, width=12.5cm]{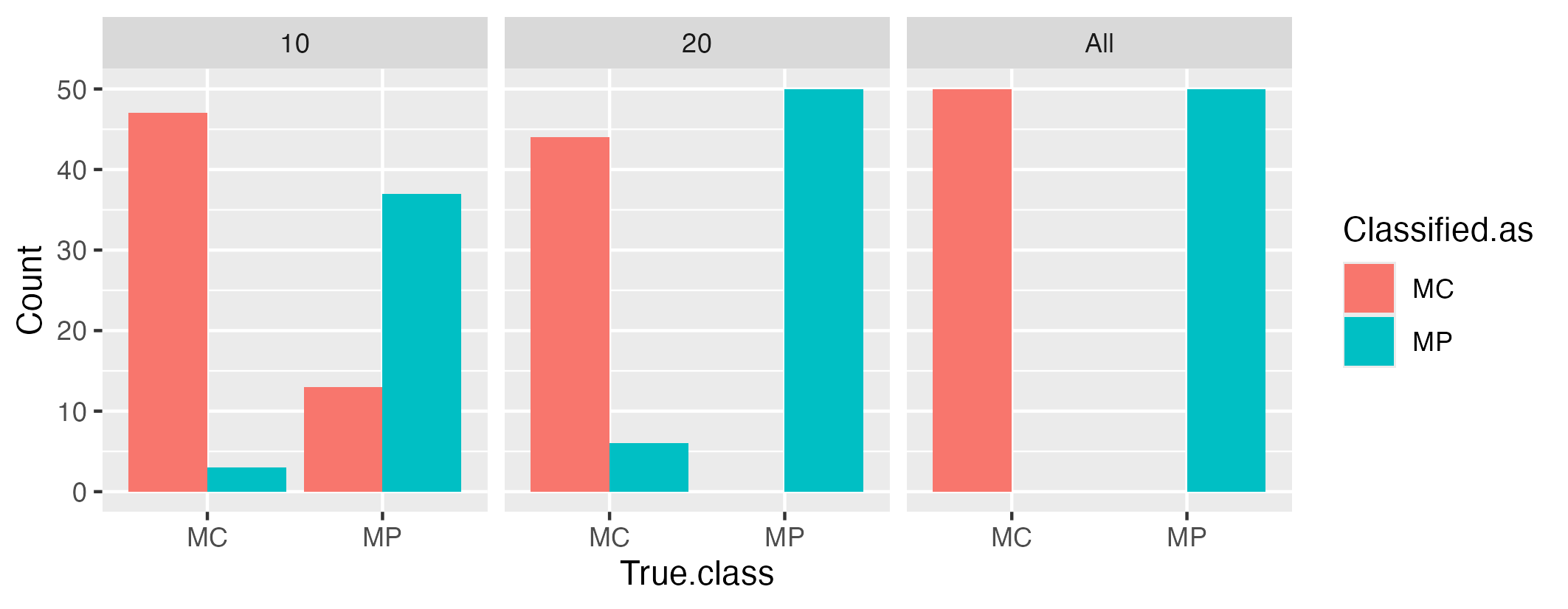}
    \end{minipage}
    \begin{minipage}{0.03\linewidth}\centering
        \rotatebox[origin=center]{90}{Both}
    \end{minipage}
    \begin{minipage}{0.93\linewidth}\centering
        \includegraphics[height=5.5cm, width=12.5cm]{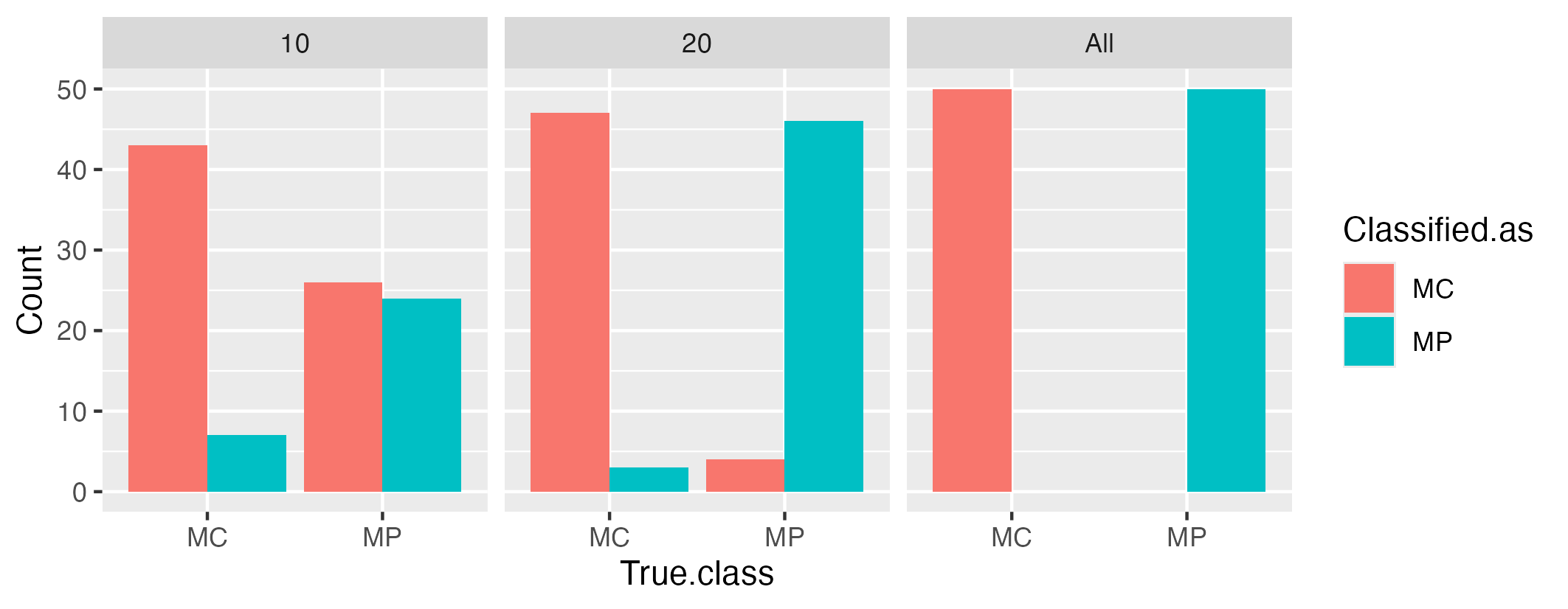}
    \end{minipage}
    \caption{Histograms of $k$-medoids classification accuracy using only the ratio, only the curvature and both ratio and curvature for discrimination when using a sample of 10, 20, and 'All' components, respectively. Misclassification rates are 23\%, 11\% and 0\% for 10, 20 and 'All' components, respectively, when using only the ratio, 16\%, 6\% and 0\% when using only the curvature and 33\%, 7\% and 0\% when using both characteristics for a sample of 50 realisations that were osculated by a disc of radius $r=3$.}
    \label{fig:kmed_50_best_tissues_3}
\end{figure}

\begin{figure}[!ht]
    \centering
    \begin{minipage}{0.03\linewidth}\centering
        \rotatebox[origin=center]{90}{Ratio}
    \end{minipage}
    \begin{minipage}{0.93\linewidth}\centering
        \includegraphics[height=5.5cm, width=12.5cm]{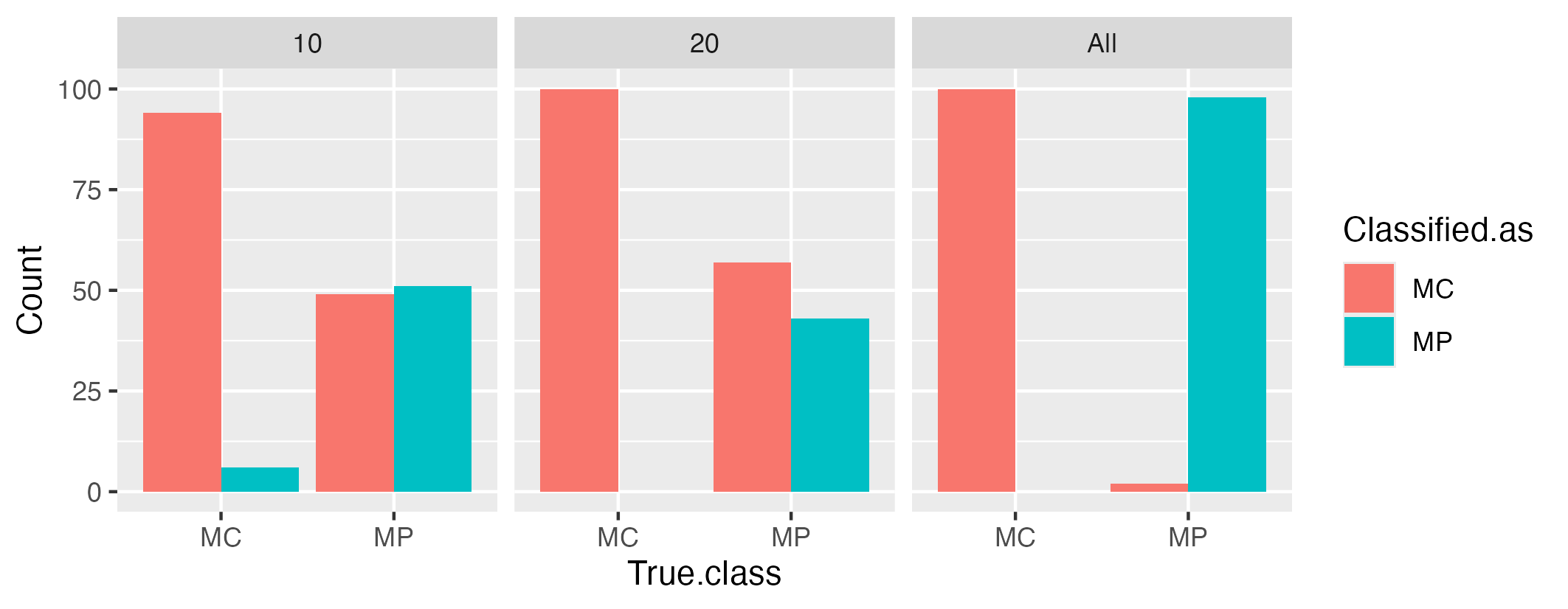}
    \end{minipage}
    \begin{minipage}{0.03\linewidth}\centering
        \rotatebox[origin=center]{90}{Curvature}
    \end{minipage}
    \begin{minipage}{0.93\linewidth}\centering
        \includegraphics[height=5.5cm, width=12.5cm]{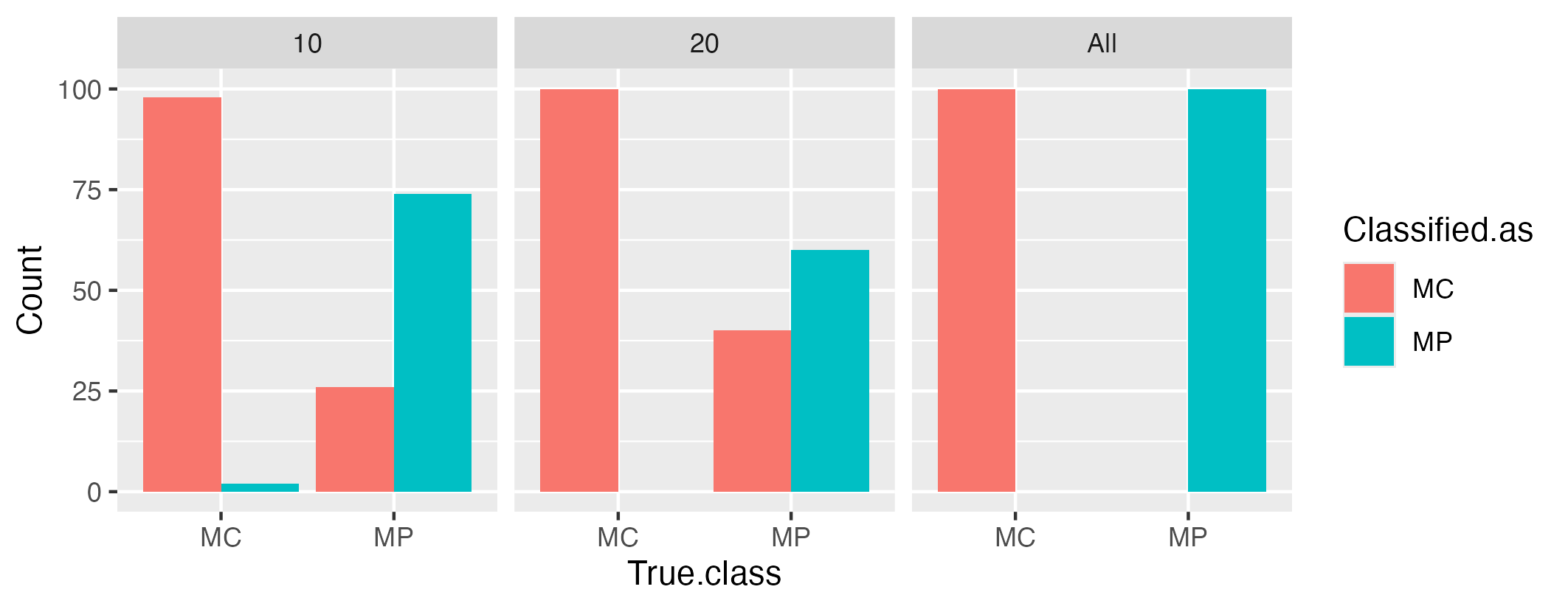}
    \end{minipage}
    \begin{minipage}{0.03\linewidth}\centering
        \rotatebox[origin=center]{90}{Both}
    \end{minipage}
    \begin{minipage}{0.93\linewidth}\centering
        \includegraphics[height=5.5cm, width=12.5cm]{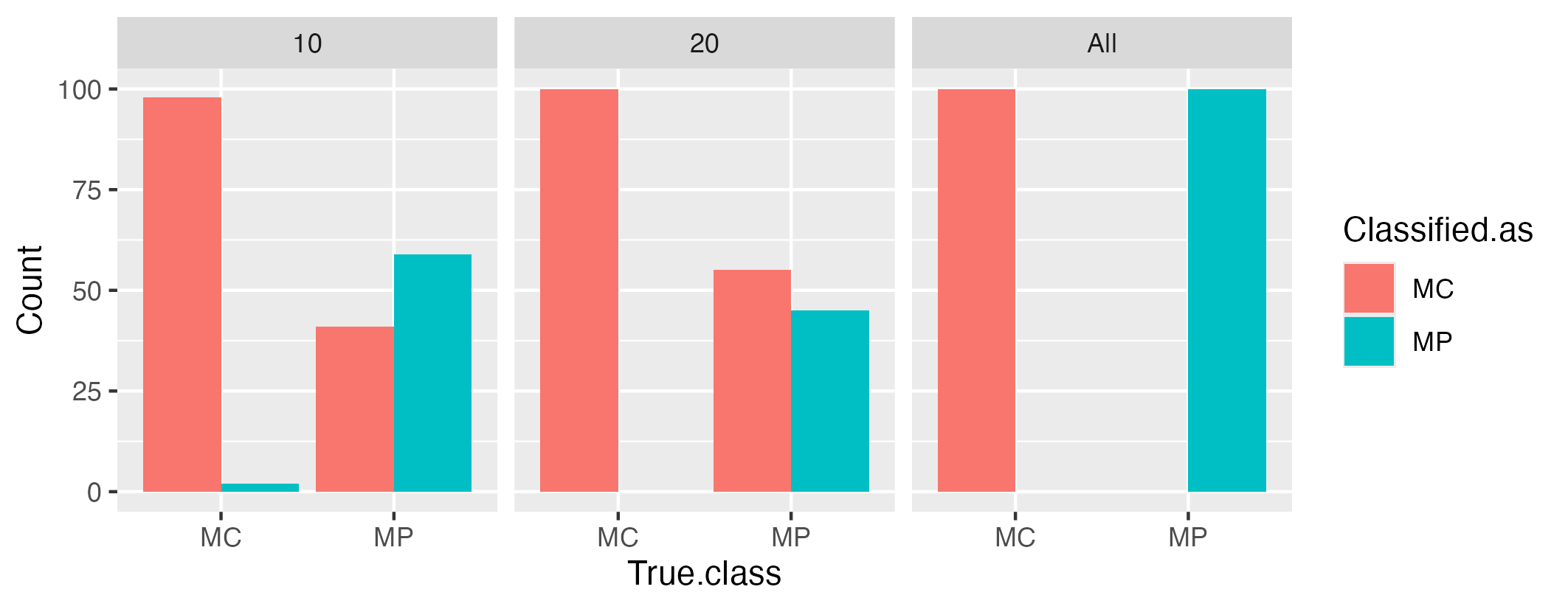}
    \end{minipage}
    \caption{Histograms of $k$-medoids classification accuracy using only the ratio, only the curvature and both ratio and curvature for discrimination when using a sample of 10, 20, and 'All' components, respectively. Misclassification rates are 27.5\%, 28.5\% and 1\% for 10, 20 and 'All' components, respectively, when using only the ratio, 14\%, 20\% and 0\% when using only the curvature and 21.5\%, 27.5\% and 0\% when using both characteristics for a sample of 100 realisations that were osculated by a disc of radius $r=3$.}
    \label{fig:kmed_100_best_tissues_3}
\end{figure}

\begin{figure}[!ht]
    \centering
    \begin{minipage}{0.03\linewidth}\centering
        \rotatebox[origin=center]{90}{20 Realisations}
    \end{minipage}
    \begin{minipage}{0.93\linewidth}\centering
        \includegraphics[height=5.5cm, width=12.3cm]{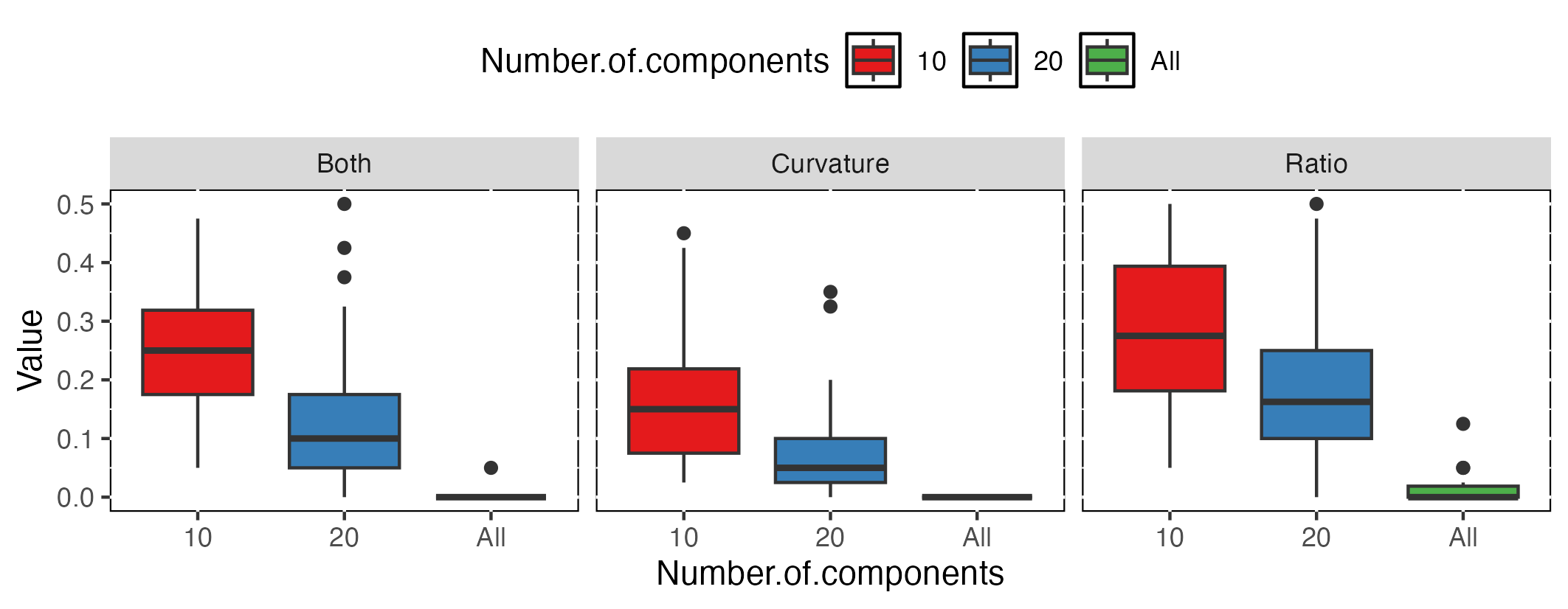}
    \end{minipage}
    \begin{minipage}{0.03\linewidth}\centering
        \rotatebox[origin=center]{90}{50 Realisations}
    \end{minipage}
    \begin{minipage}{0.93\linewidth}\centering
        \includegraphics[height=5.5cm, width=12.3cm]{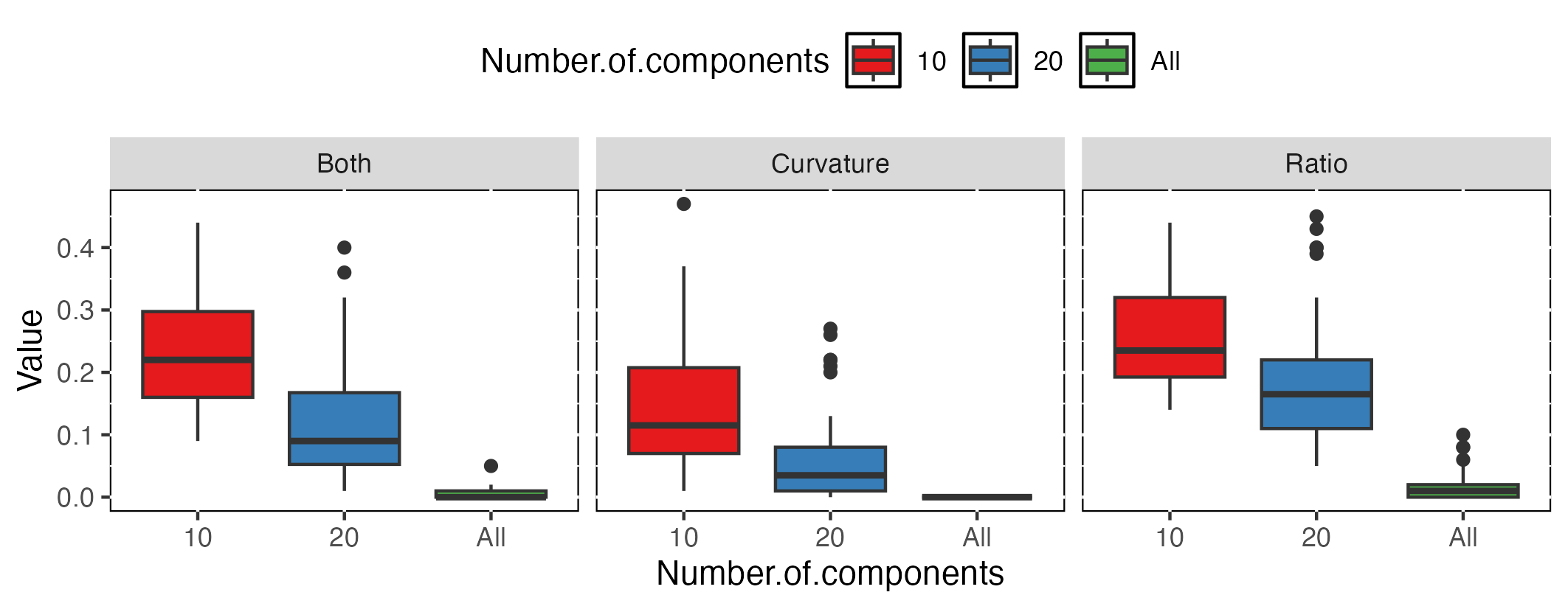}
    \end{minipage}
    \begin{minipage}{0.03\linewidth}\centering
        \rotatebox[origin=center]{90}{100 Realisations}
    \end{minipage}
    \begin{minipage}{0.93\linewidth}\centering
        \includegraphics[height=5.5cm, width=12.3cm]{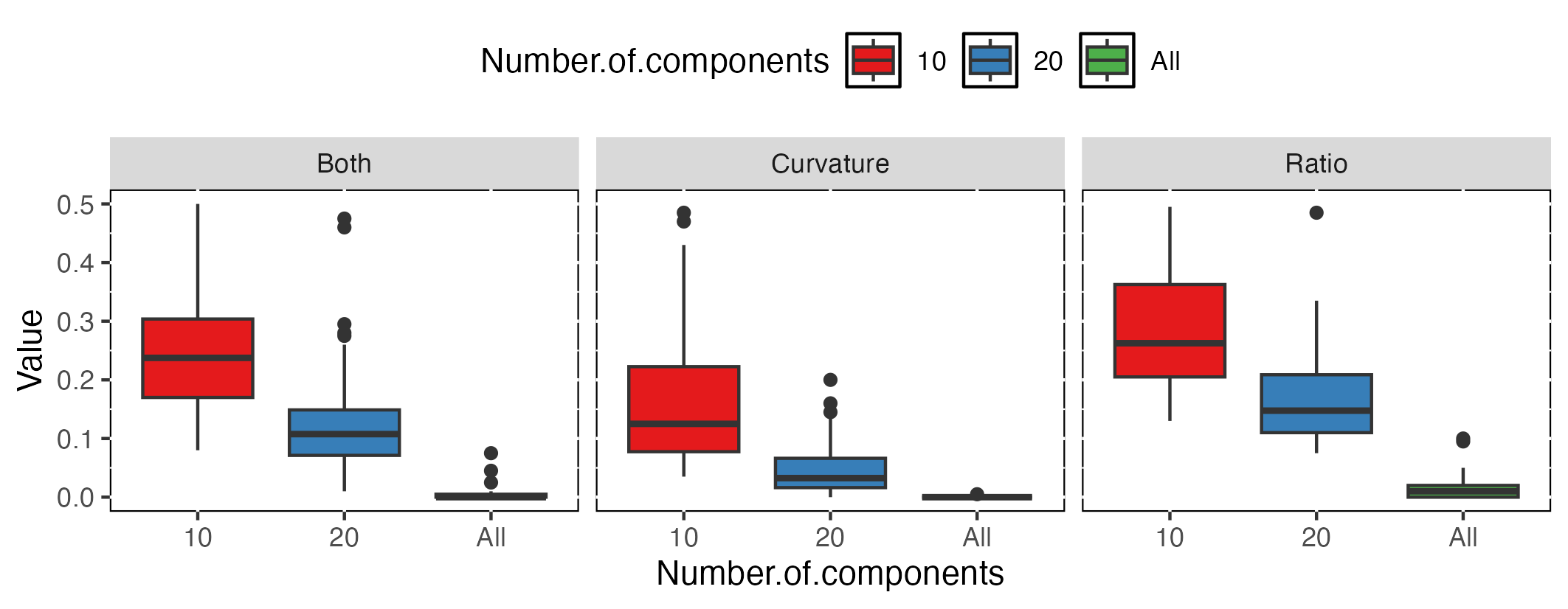}
    \end{minipage}
    \caption{Boxplots of misclassification rate for 50 runs of $k$-medoids algorithm when considering samples of 20 (top), 50 (middle) and 100 (bottom) realisations using both ratio and curvature, only the curvature and only the ratio for discrimination, respectively. For each setting, misclassification rates for different number of components considered (namely 10, 20, and 'All') are shown. Note that the characteristics were obtained using an osculating disc of radius $r=3$.}
    \label{fig:box_kmed_tissues_3}
\end{figure}

Figures \ref{fig:kmed_20_best_tissues_3}, \ref{fig:kmed_50_best_tissues_3} and \ref{fig:kmed_100_best_tissues_3} represent the histograms of classification accuracy based on $k$-medoids algorithm when 20, 50 and 100 realisations are considered, respectively. The corresponding boxplots of misclassification accuracy are shown in \ref{fig:box_kmed_tissues_3}.

\subsubsection{Hierarchical clustering}

\begin{figure}[!ht]
    \centering
    \begin{minipage}{0.03\linewidth}\centering
        \rotatebox[origin=center]{90}{Ratio}
    \end{minipage}
    \begin{minipage}{0.93\linewidth}\centering
        \includegraphics[height=5.5cm, width=12.5cm]{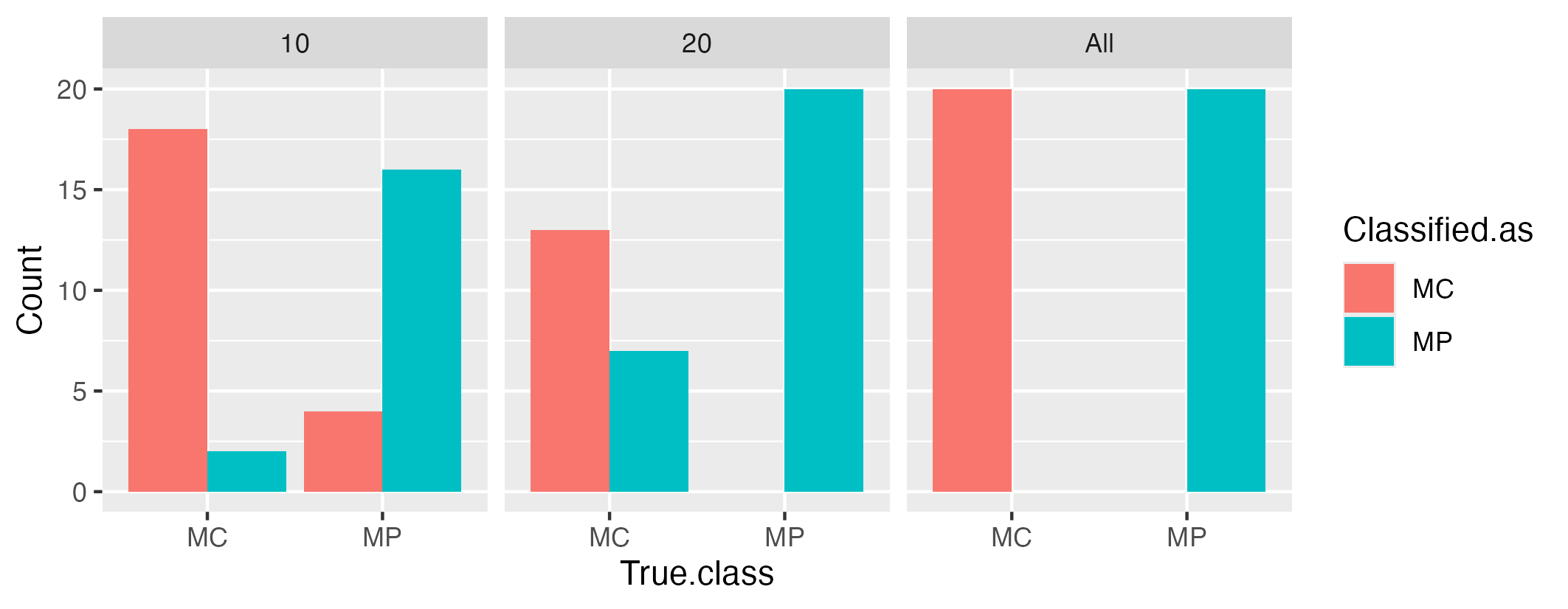}
    \end{minipage}
    \begin{minipage}{0.03\linewidth}\centering
        \rotatebox[origin=center]{90}{Curvature}
    \end{minipage}
    \begin{minipage}{0.93\linewidth}\centering
        \includegraphics[height=5.5cm, width=12.5cm]{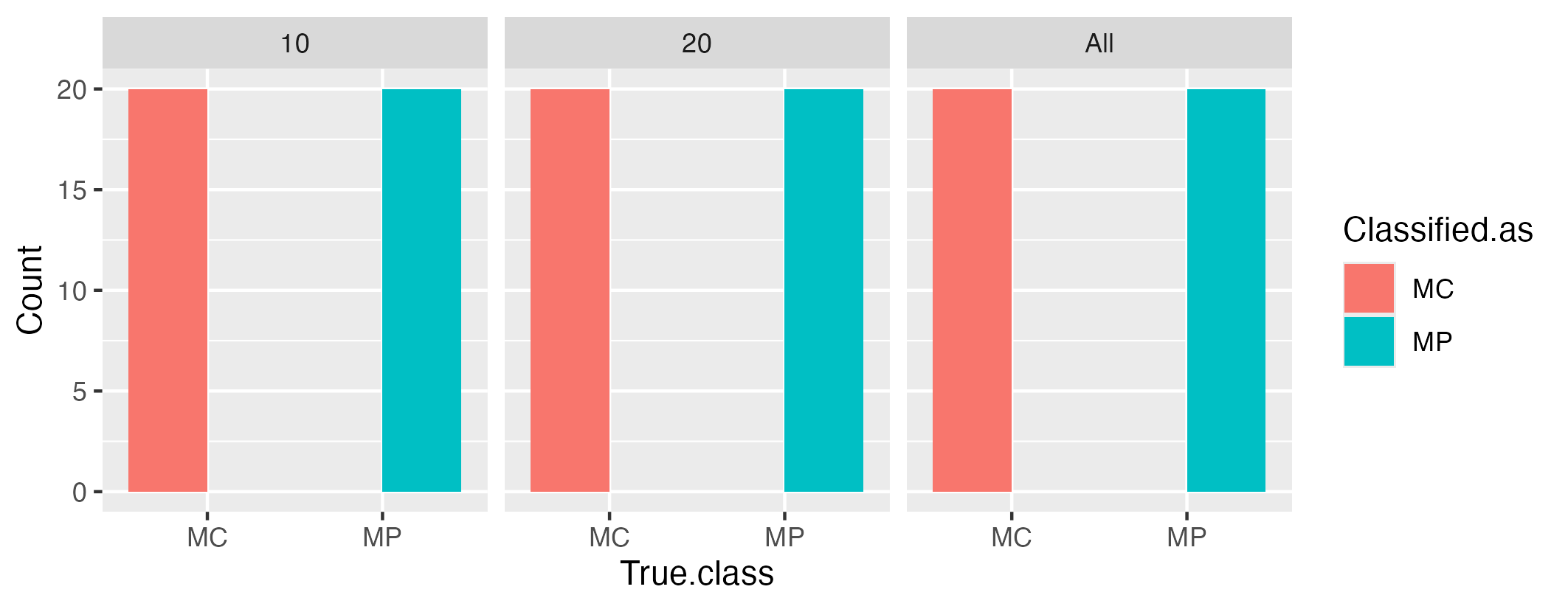}
    \end{minipage}
    \begin{minipage}{0.03\linewidth}\centering
        \rotatebox[origin=center]{90}{Both}
    \end{minipage}
    \begin{minipage}{0.93\linewidth}\centering
        \includegraphics[height=5.5cm, width=12.5cm]{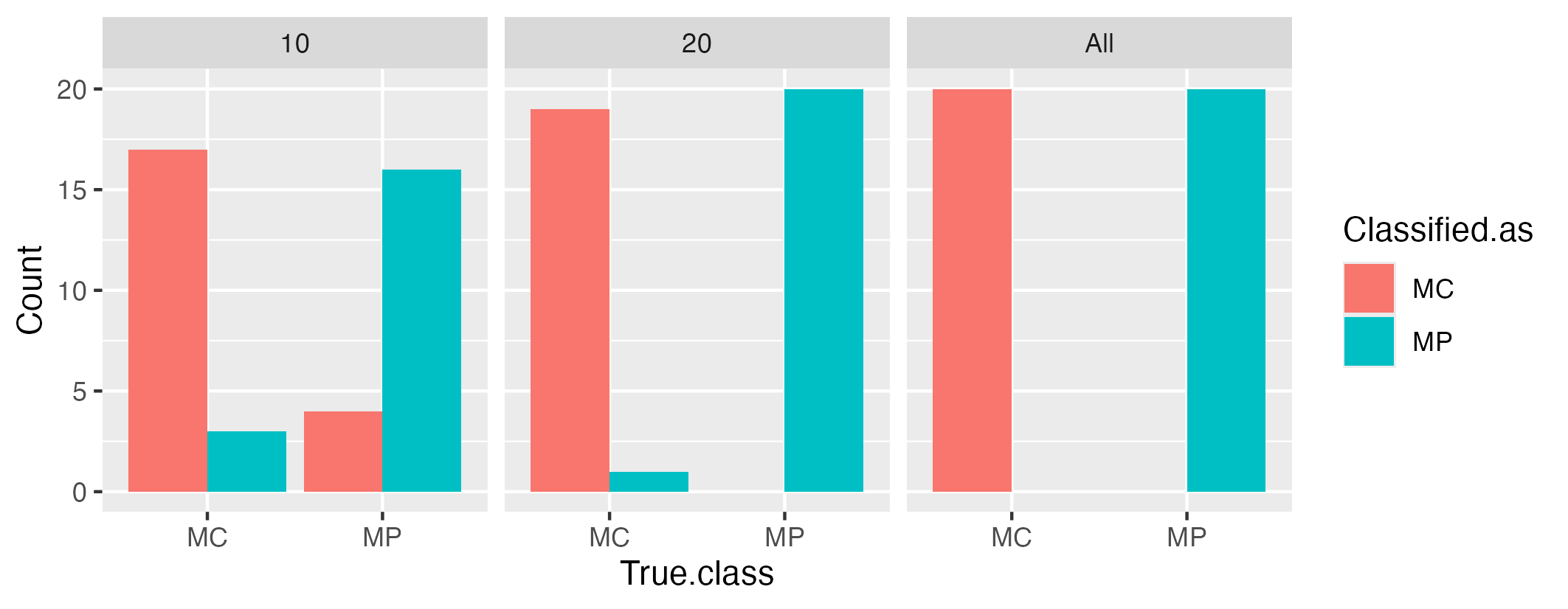}
    \end{minipage}
    \caption{Histograms of hierarchical clustering classification accuracy using only the ratio, only the curvature and both ratio and curvature for discrimination when using a sample of 10, 20, and 'All' components, respectively. Misclassification rates are 15\%, 17.5\% and 0\% for 10, 20 and 'All' components, respectively, when using only the ratio, 0\%, 0\% and 0\% when using only the curvature and 17.5\%, 2.5\% and 0\% when using both characteristics for a sample of 20 realisations that were osculated by a disc of radius $r=5$.}
    \label{fig:hc_20_best_tissues_5}
\end{figure}

\begin{figure}[!ht]
    \centering
    \begin{minipage}{0.03\linewidth}\centering
        \rotatebox[origin=center]{90}{Ratio}
    \end{minipage}
    \begin{minipage}{0.93\linewidth}\centering
        \includegraphics[height=5.5cm, width=12.5cm]{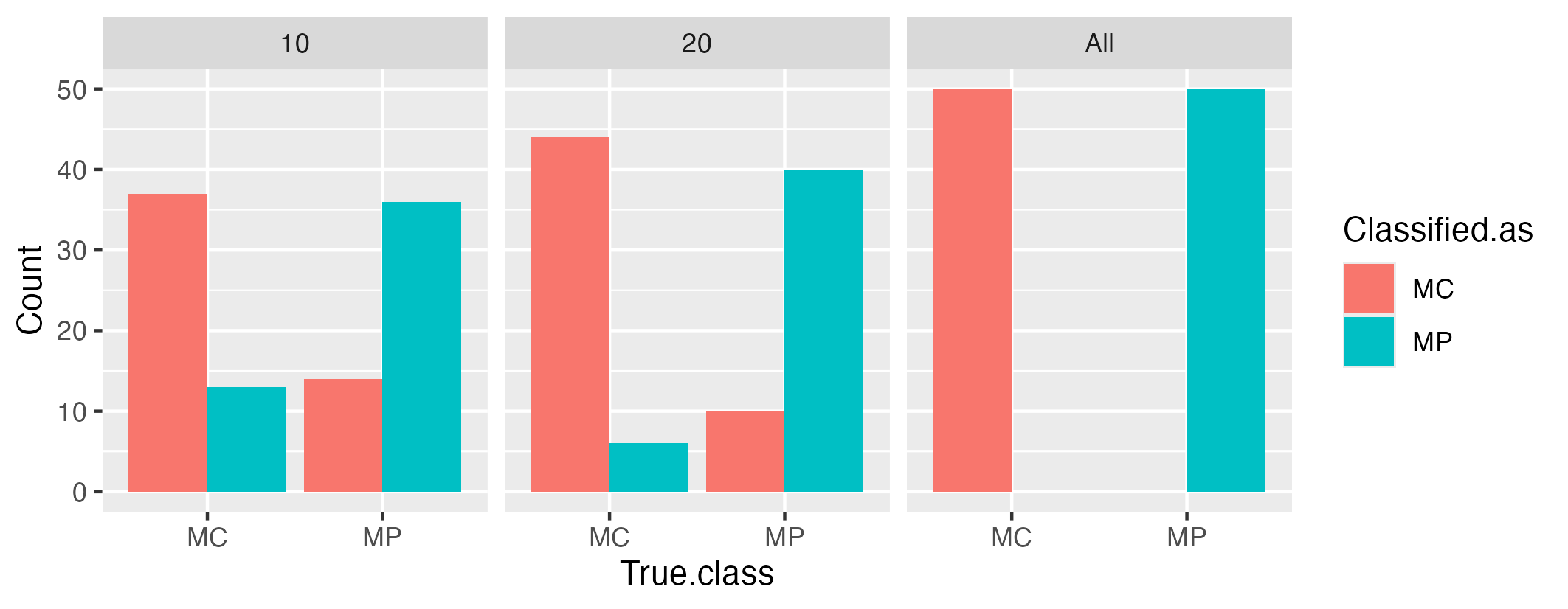}
    \end{minipage}
    \begin{minipage}{0.03\linewidth}\centering
        \rotatebox[origin=center]{90}{Curvature}
    \end{minipage}
    \begin{minipage}{0.93\linewidth}\centering
        \includegraphics[height=5.5cm, width=12.5cm]{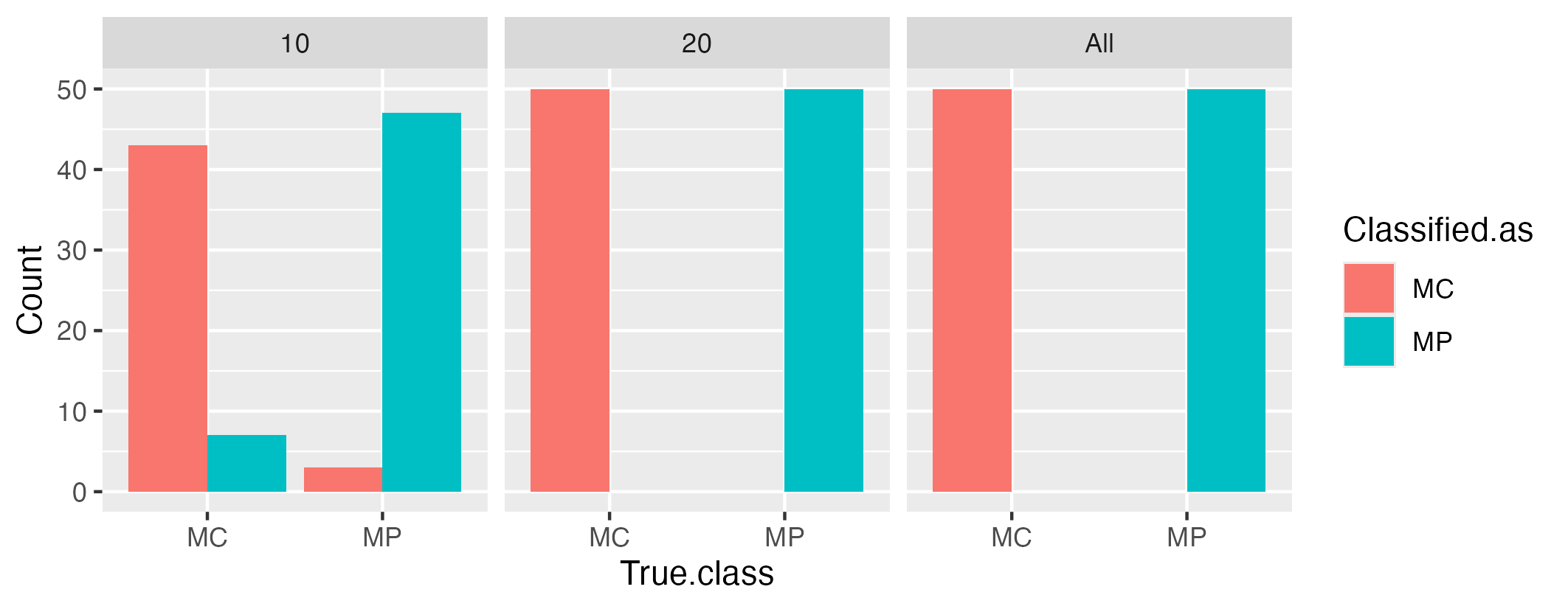}
    \end{minipage}
    \begin{minipage}{0.03\linewidth}\centering
        \rotatebox[origin=center]{90}{Both}
    \end{minipage}
    \begin{minipage}{0.93\linewidth}\centering
        \includegraphics[height=5.5cm, width=12.5cm]{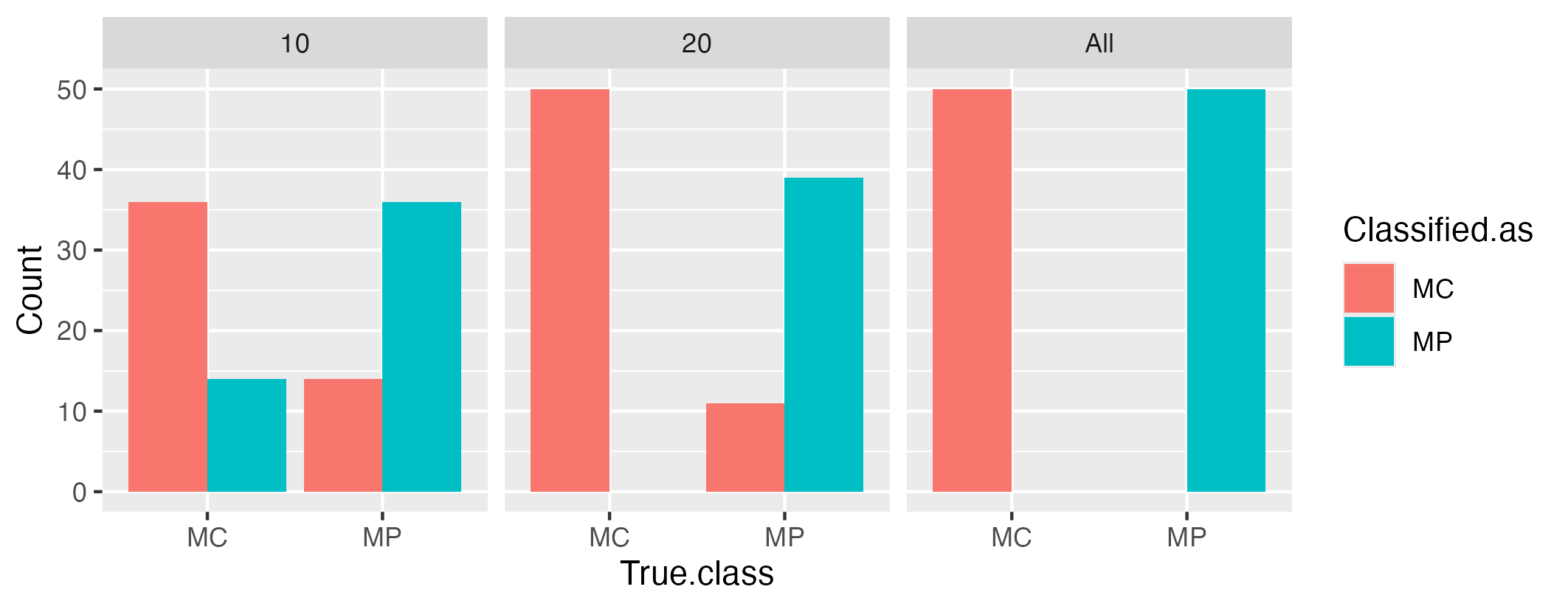}
    \end{minipage}
    \caption{Histograms of hierarchical clustering classification accuracy using only the ratio, only the curvature and both ratio and curvature for discrimination when using a sample of 10, 20, and 'All' components, respectively. Misclassification rates are 27\%, 16\% and 0\% for 10, 20 and 'All' components, respectively, when using only the ratio, 10\%, 0\% and 0\% when using only the curvature and 28\%, 11\% and 0\% when using both characteristics for a sample of 50 realisations that were osculated by a disc of radius $r=5$.}
    \label{fig:hc_50_best_tissues_5}
\end{figure}

Histograms of classification accuracy using Ward's agglomerative algorithm on data obtained with osculating circle with radius $r=5$, when 20 and 50 realisations are considered, are shown in Figures \ref{fig:hc_20_best_tissues_5} and \ref{fig:hc_50_best_tissues_5}, respectively.

\begin{figure}[!ht]
    \centering
    \begin{minipage}{0.03\linewidth}\centering
        \rotatebox[origin=center]{90}{Ratio}
    \end{minipage}
    \begin{minipage}{0.93\linewidth}\centering
        \includegraphics[height=5.5cm, width=12.5cm]{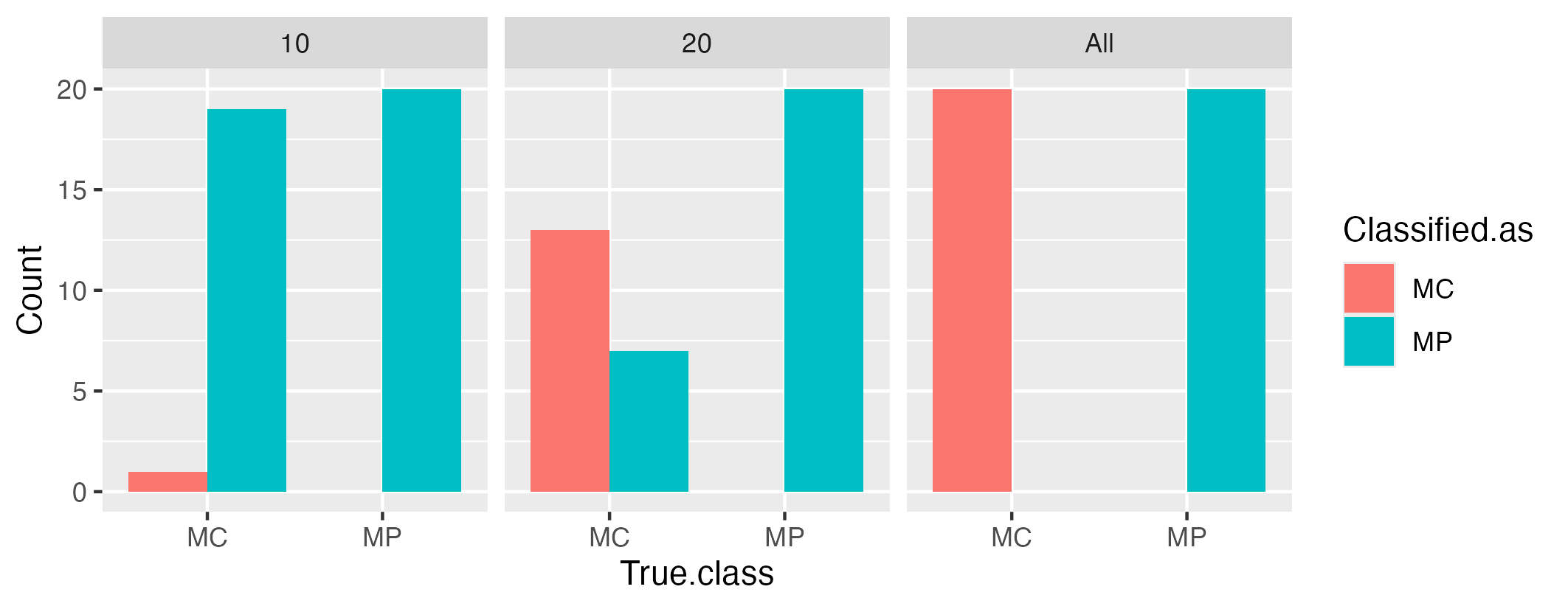}
    \end{minipage}
    \begin{minipage}{0.03\linewidth}\centering
        \rotatebox[origin=center]{90}{Curvature}
    \end{minipage}
    \begin{minipage}{0.93\linewidth}\centering
        \includegraphics[height=5.5cm, width=12.5cm]{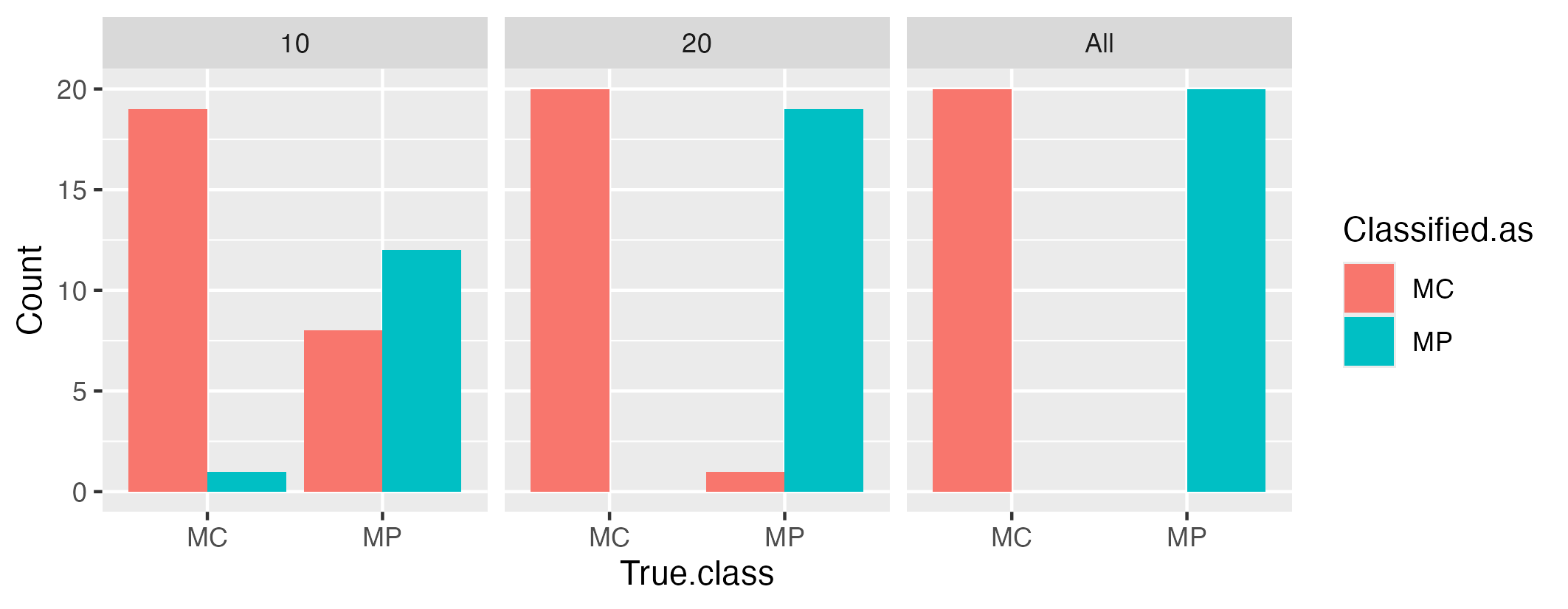}
    \end{minipage}
    \begin{minipage}{0.03\linewidth}\centering
        \rotatebox[origin=center]{90}{Both}
    \end{minipage}
    \begin{minipage}{0.93\linewidth}\centering
        \includegraphics[height=5.5cm, width=12.5cm]{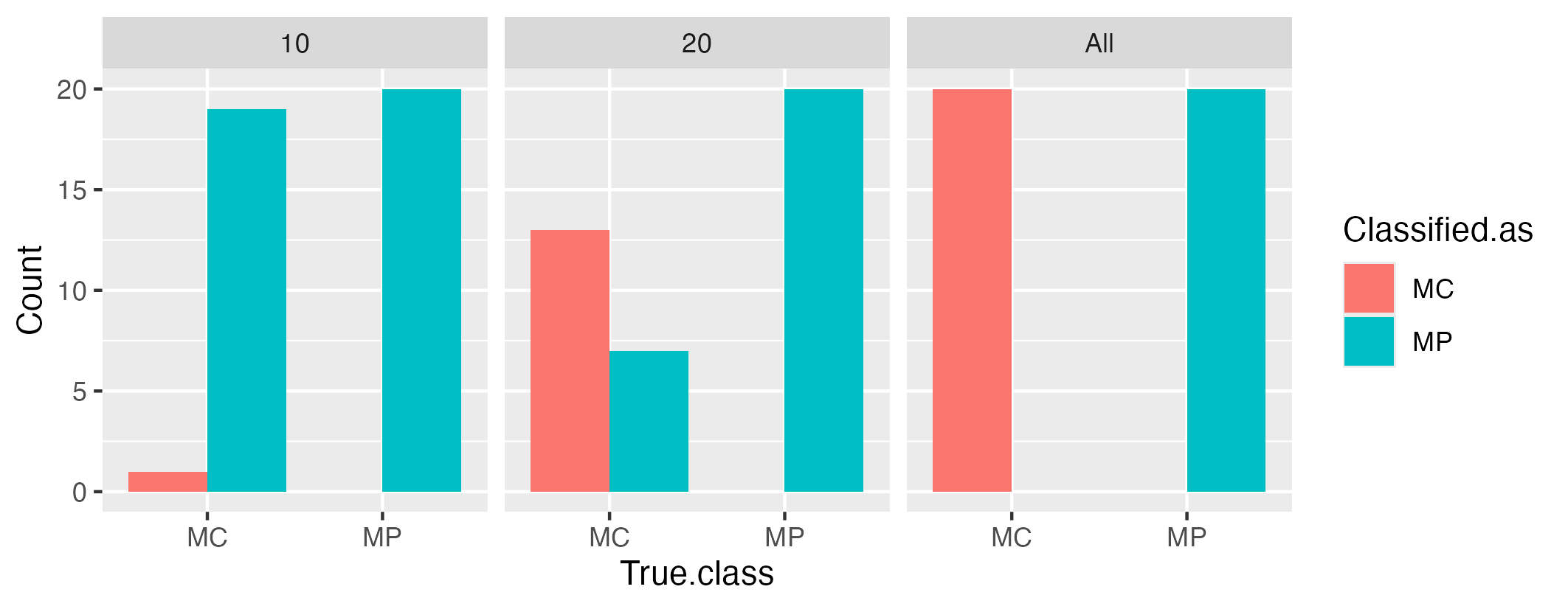}
    \end{minipage}
    \caption{Histograms of hierarchical clustering classification accuracy using only the ratio, only the curvature and both ratio and curvature for discrimination when using a sample of 10, 20, and 'All' components, respectively. Misclassification rates are 47.5\%, 17.5\% and 0\% for 10, 20 and 'All' components, respectively, when using only the ratio, 22.5\%, 2.5\% and 0\% when using only the curvature and 47.5\%, 17.5\% and 0\% when using both characteristics for a sample of 20 realisations that were osculated by a disc of radius $r=3$.}
    \label{fig:hc_20_best_tissues_3}
\end{figure}

\begin{figure}[!ht]
    \centering
    \begin{minipage}{0.03\linewidth}\centering
        \rotatebox[origin=center]{90}{Ratio}
    \end{minipage}
    \begin{minipage}{0.93\linewidth}\centering
        \includegraphics[height=5.5cm, width=12.5cm]{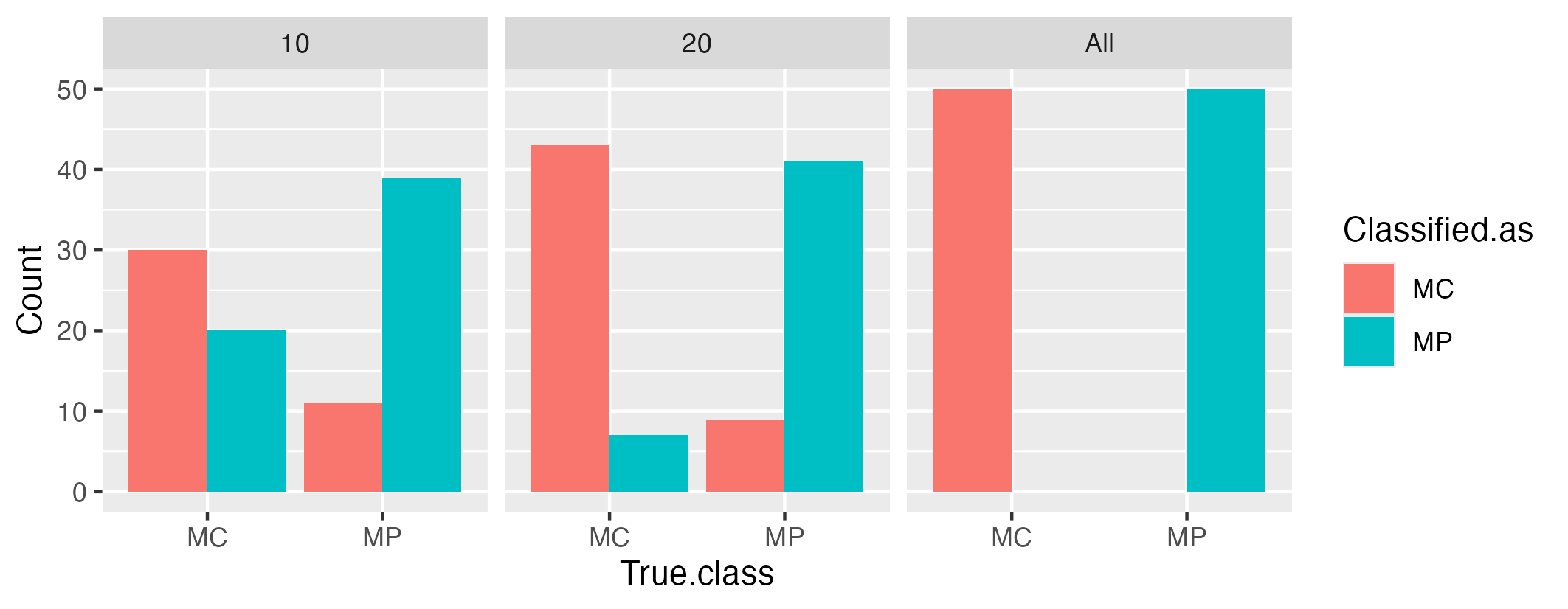}
    \end{minipage}
    \begin{minipage}{0.03\linewidth}\centering
        \rotatebox[origin=center]{90}{Curvature}
    \end{minipage}
    \begin{minipage}{0.93\linewidth}\centering
        \includegraphics[height=5.5cm, width=12.5cm]{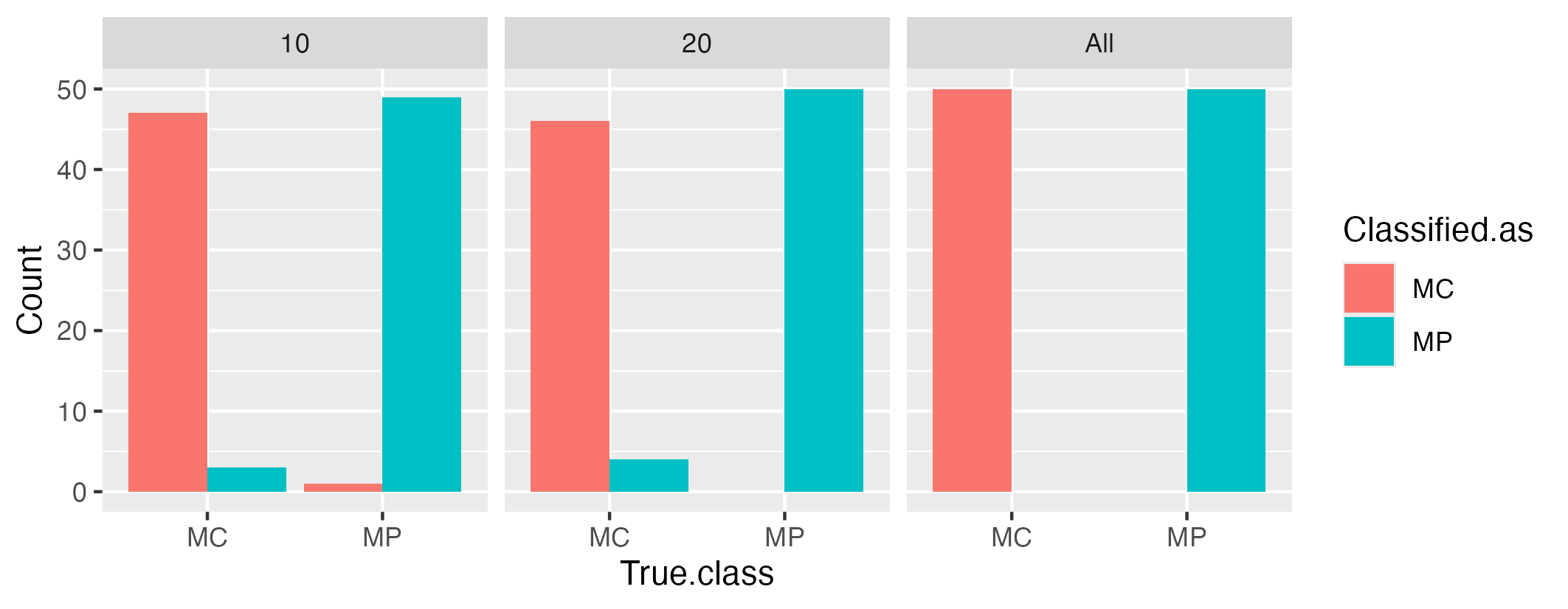}
    \end{minipage}
    \begin{minipage}{0.03\linewidth}\centering
        \rotatebox[origin=center]{90}{Both}
    \end{minipage}
    \begin{minipage}{0.93\linewidth}\centering
        \includegraphics[height=5.5cm, width=12.5cm]{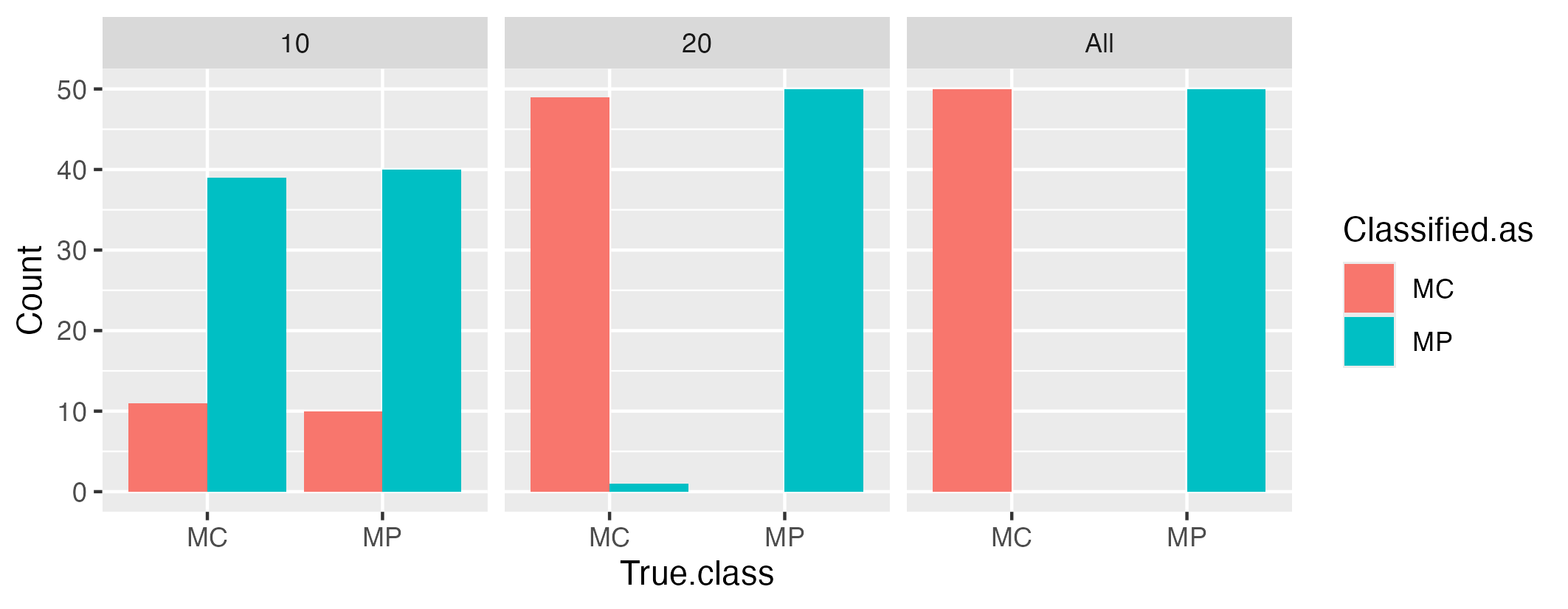}
    \end{minipage}
    \caption{Histograms of hierarchical clustering classification accuracy using only the ratio, only the curvature and both ratio and curvature for discrimination when using a sample of 10, 20, and 'All' components, respectively. Misclassification rates are 31\%, 16\% and 0\% for 10, 20 and 'All' components, respectively, when using only the ratio, 4\%, 4\% and 0\% when using only the curvature and 49\%, 1\% and 0\% when using both characteristics for a sample of 50 realisations that were osculated by a disc of radius $r=3$.}
    \label{fig:hc_50_best_tissues_3}
\end{figure}

\begin{figure}[!ht]
    \centering
    \begin{minipage}{0.03\linewidth}\centering
        \rotatebox[origin=center]{90}{Ratio}
    \end{minipage}
    \begin{minipage}{0.93\linewidth}\centering
        \includegraphics[height=5.5cm, width=12.5cm]{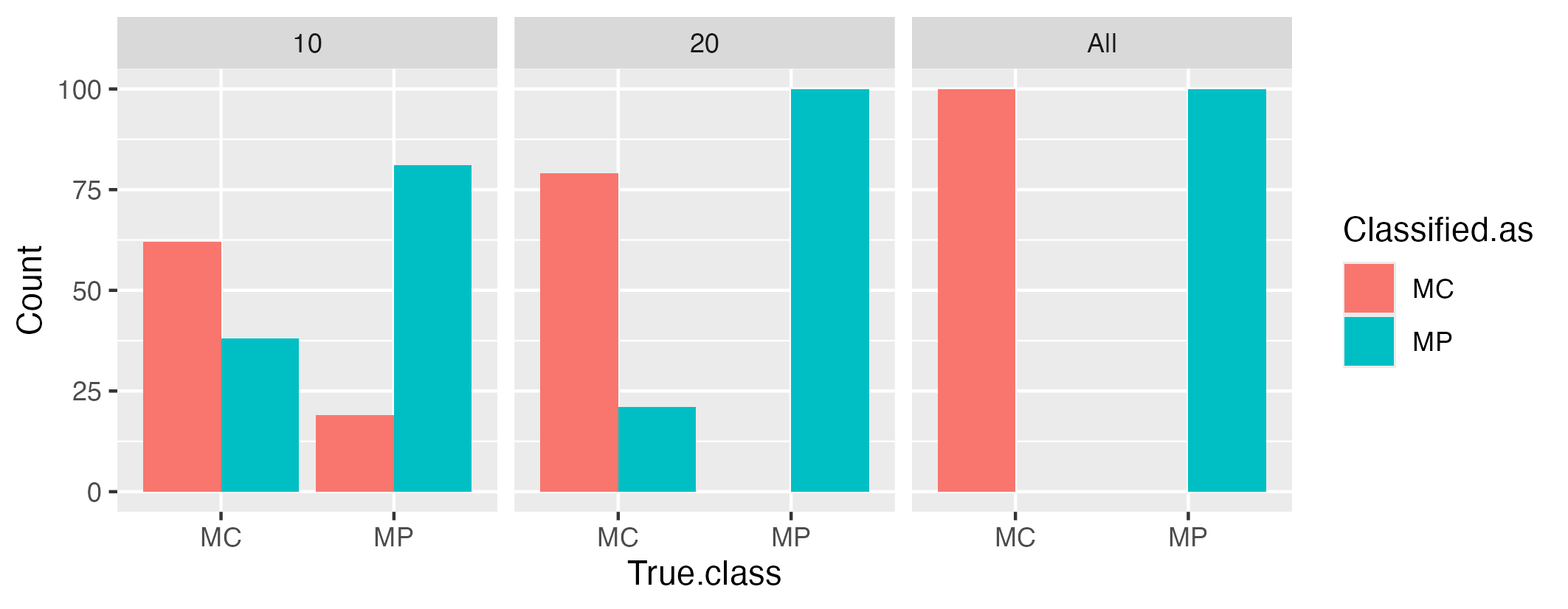}
    \end{minipage}
    \begin{minipage}{0.03\linewidth}\centering
        \rotatebox[origin=center]{90}{Curvature}
    \end{minipage}
    \begin{minipage}{0.93\linewidth}\centering
        \includegraphics[height=5.5cm, width=12.5cm]{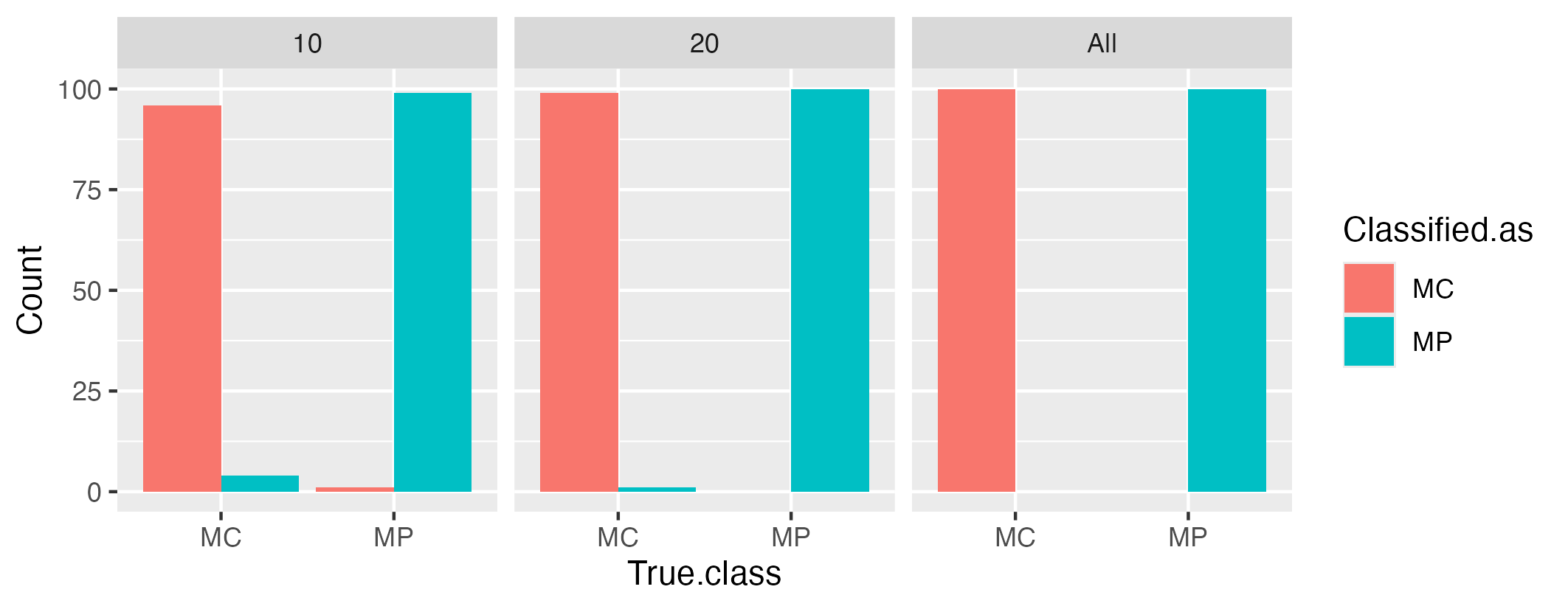}
    \end{minipage}
    \begin{minipage}{0.03\linewidth}\centering
        \rotatebox[origin=center]{90}{Both}
    \end{minipage}
    \begin{minipage}{0.93\linewidth}\centering
        \includegraphics[height=5.5cm, width=12.5cm]{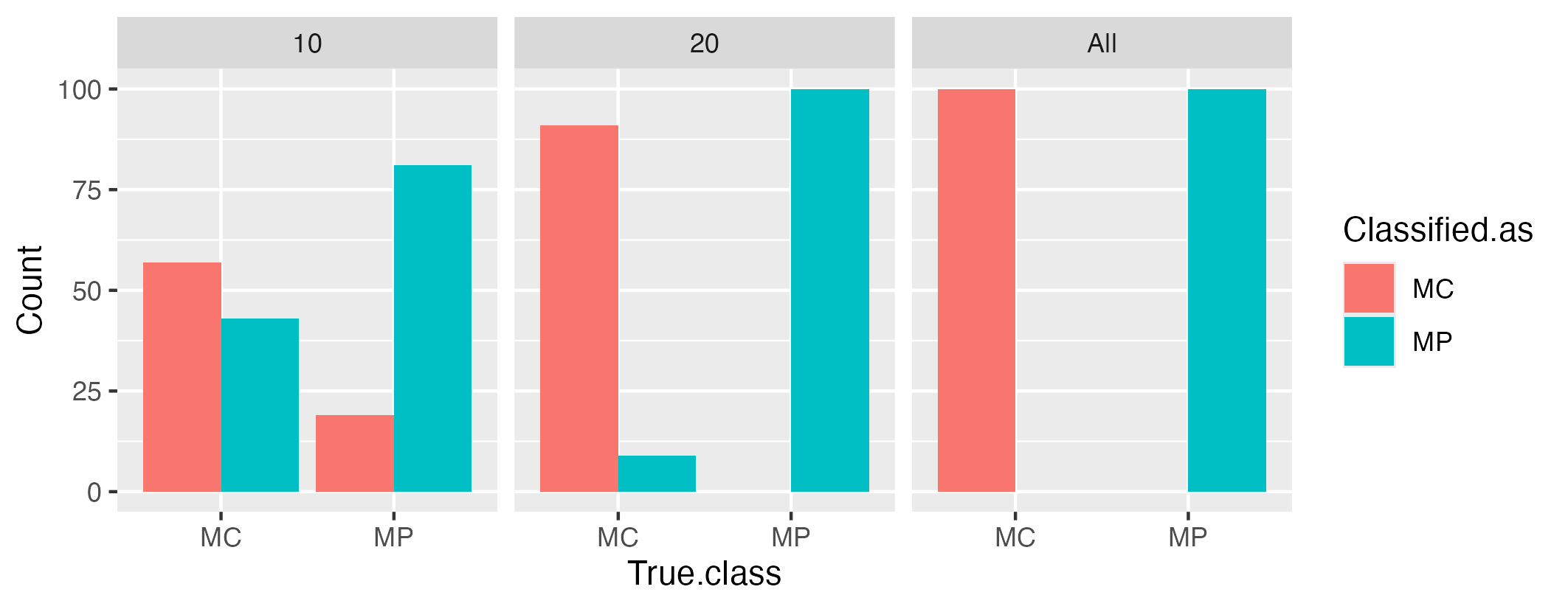}
    \end{minipage}
    \caption{Histograms of hierarchical clustering classification accuracy using only the ratio, only the curvature and both ratio and curvature for discrimination when using a sample of 10, 20, and 'All' components, respectively. Misclassification rates are 28.5\%, 10.5\% and 0\% for 10, 20 and 'All' components, respectively, when using only the ratio, 2.5\%, 0.5\% and 0\% when using only the curvature and 31\%, 4.5\% and 0\% when using both characteristics for a sample of 100 realisations that were osculated by a disc of radius $r=3$.}
    \label{fig:hc_100_best_tissues_3}
\end{figure}

\begin{figure}[!ht]
    \centering
    \begin{minipage}{0.03\linewidth}\centering
        \rotatebox[origin=center]{90}{20 Realisations}
    \end{minipage}
    \begin{minipage}{0.93\linewidth}\centering
        \includegraphics[height=5.5cm, width=12.3cm]{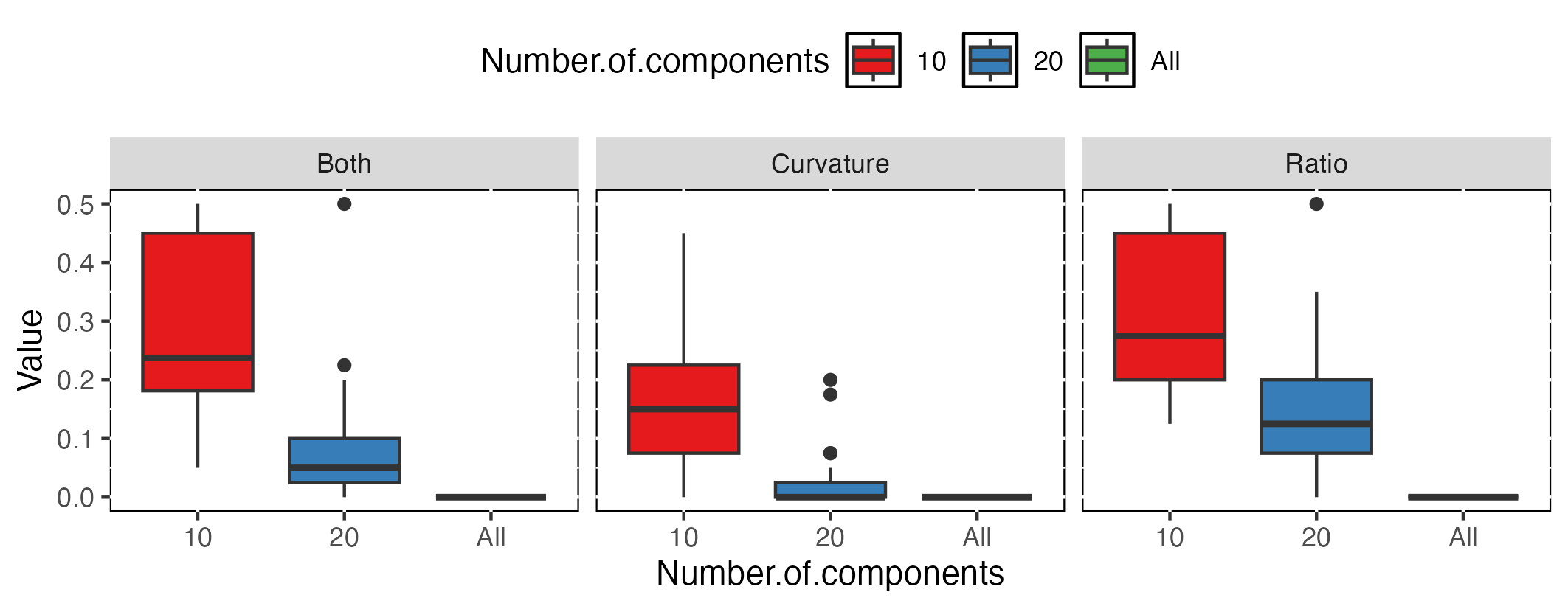}
    \end{minipage}
    \begin{minipage}{0.03\linewidth}\centering
        \rotatebox[origin=center]{90}{50 Realisations}
    \end{minipage}
    \begin{minipage}{0.93\linewidth}\centering
        \includegraphics[height=5.5cm, width=12.3cm]{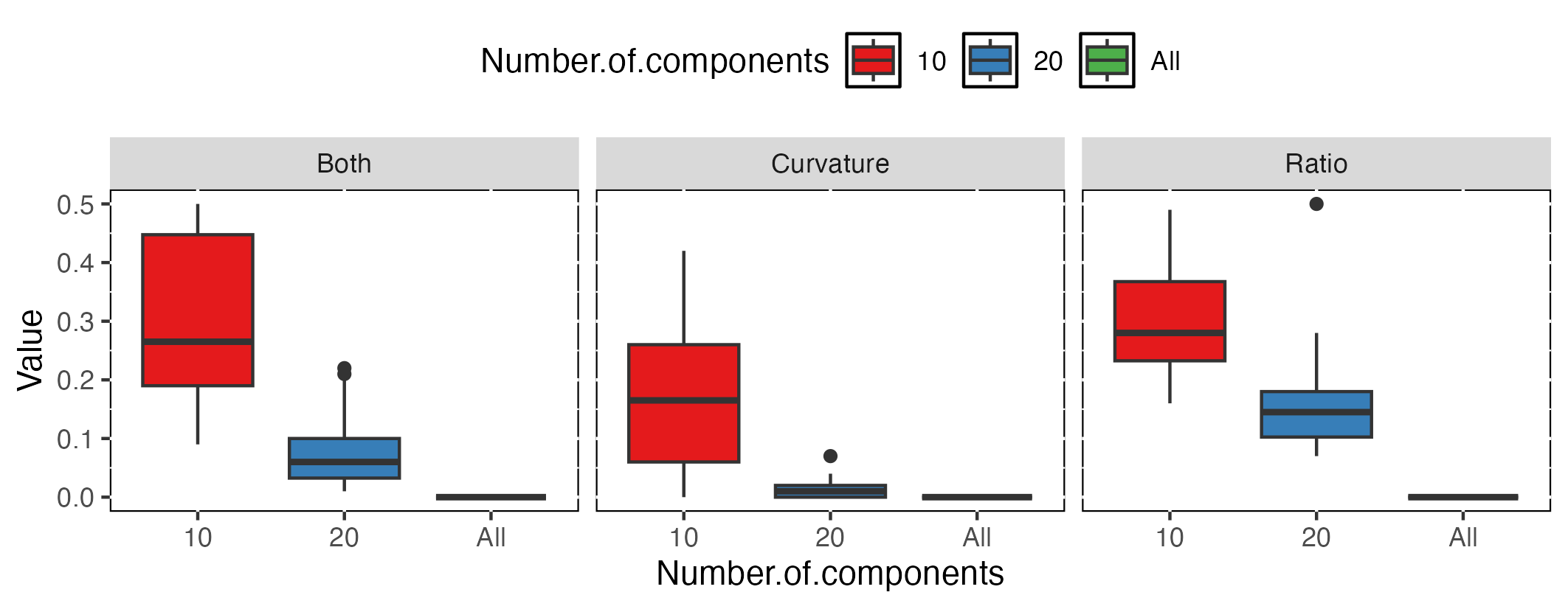}
    \end{minipage}
    \begin{minipage}{0.03\linewidth}\centering
        \rotatebox[origin=center]{90}{100 Realisations}
    \end{minipage}
    \begin{minipage}{0.93\linewidth}\centering
        \includegraphics[height=5.5cm, width=12.3cm]{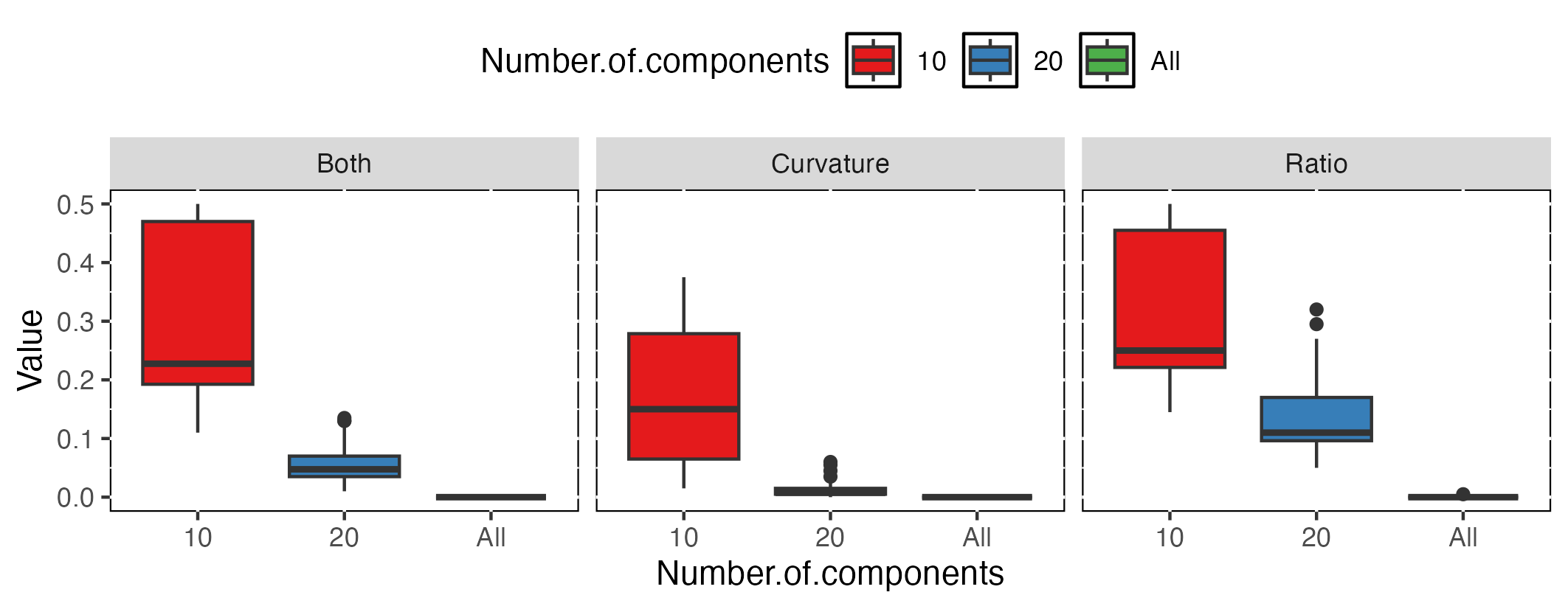}
    \end{minipage}
    \caption{Boxplots of misclassification rate for 50 runs of hierarchical clustering algorithm when considering samples of 20 (top), 50 (middle) and 100 (bottom) realisations using both ratio and curvature, only the curvature and only the ratio for discrimination, respectively. For each setting, misclassification rates for different number of components considered (namely 10, 20, and 'All') are shown. Note that the characteristics were obtained using an osculating disc of radius $r=3$.}
    \label{fig:box_hc_tissues_3}
\end{figure}

Figures \ref{fig:hc_20_best_tissues_3}, \ref{fig:hc_50_best_tissues_3} and \ref{fig:hc_100_best_tissues_3} represent the histograms of classification accuracy for hierarchical unsupervised clustering based on Ward's algorithm when 20, 50 and 100 realisations are considered, respectively. Corresponding boxplots of misclassification accuracy are shown in \ref{fig:box_hc_tissues_3}.

